\documentclass[review]{elsarticle}

\usepackage[colorlinks,citecolor=blue,linktoc=all,linkcolor=cyan]{hyperref}
\usepackage{graphicx}

\usepackage[T1]{fontenc}
\usepackage{dsfont}               
\usepackage{mathrsfs}             
\usepackage{slashed}              
\usepackage{amsmath}
\usepackage{amssymb}
\usepackage{amsbsy}
\usepackage{amsfonts}

\usepackage{natbib}

\numberwithin{equation}{section}
\numberwithin{table}{section}
\numberwithin{figure}{section}

\journal{Progress in Particle and Nuclear Physics}

\usepackage{titlesec}
\usepackage{sectsty}
\titleformat{\section}{\normalfont\Large\bfseries}{\thesection}{1em}{}
\titleformat{\subsection}{\normalfont\large\bfseries}{\thesubsection}{1em}{}
\titleformat{\subsubsection}{\normalfont\normalsize\bfseries}{\thesubsubsection}{1em}{}

\begin{document}

\begin{frontmatter}

\title{Binary Stars in the New Millennium}

\author[mymainaddress,mysecondaryaddress,mythirdaddress]{Xuefei CHEN\corref{Xuefei CHEN}}
\cortext[mycorrespondingauthor]{Xuefei CHEN}
\ead{cxf@ynao.ac.cn}

\author[mymainaddress,mysecondaryaddress,mythirdaddress]{Zhengwei LIU\corref{Zhengwei LIU}}
\cortext[mycorrespondingauthor]{Zhengwei LIU}
\ead{zwliu@ynao.ac.cn}

\author[mymainaddress,mysecondaryaddress,mythirdaddress]{Zhanwen HAN\corref{Zhanwen HAN}}
\cortext[mycorrespondingauthor]{Zhanwen HAN}
\ead{zhanwenhan@ynao.ac.cn}

\address[mymainaddress]{Yunnan Observatories, the Chinese Academy of Sciences, Kunming, 650296, P.R. China}
\address[mysecondaryaddress]{Key Laboratory for the Structure and Evolution of Celestial Objects, CAS, Kunming 650216, China}
\address[mythirdaddress]{International Centre of Supernovae, Yunnan Key Laboratory, Kunming 650216, China}

\begin{abstract}

Binary stars are as common as single stars. Binary stars are of immense importance to astrophysicists because that they allow us to determine the masses of the stars independent of their distances. They are the cornerstone of the understanding of stellar evolutionary theory and play an essential role in cosmic distance measurement, galactic evolution, nucleosynthesis and the formation of important objects such as cataclysmic variable stars, X-ray binaries, Type Ia supernovae, and gravitational wave-producing double compact objects. In this article, we review the significant theoretical and observational progresses in addressing binary stars in the new millennium. Increasing large survey projects have led to the discovery of enormous numbers of binary stars, which enables us to conduct statistical studies of binary populations, and therefore provide unprecedented insight into the stellar and binary evolution physics. Meanwhile, the rapid development of theoretical concepts and numerical approaches for binary evolution have made a substantial progress on the alleviation of some long-standing binary-related problems such as the stability of mass transfer and common envelope evolution. Nevertheless, it remains a challenge to have a full understanding of fundamental problems of stellar and binary astrophysics. The upcoming massive survey projects and increasingly sophisticated computational methods will lead to future progress.

\end{abstract}

\begin{keyword}
Binary stars\sep Population synthesis\sep Binary mass transfer \sep Common envelope \sep Nucleosynthesis

\end{keyword}

\end{frontmatter}

\newpage

\thispagestyle{empty}
\tableofcontents


\newpage
\section{Introduction}
\label{intro}

The term ``binary'' was first used by William Herschel in 1802 for double stars which are bound gravitationally and orbiting together. Since then, scientists started to pay much attention to discovering and studying binary stars. In the new millennium, we have realized that binary stars play important roles in various fields of astrophysics, including stellar formation, stellar physics, galaxies, and cosmology. For instance, the statistical properties of binary populations have been used to constrain stellar formation processes and their dependence on the environments. Also, binary stars (e.g. W~UMa stars) and/or products of binary evolution (e.g. type Ia supernovae, SNe Ia) have been used as accurate distance indicators to successfully measure cosmic distances.

Binary stars were almost exclusively studied case by case until the end of the 20th century. Depending on how binaries are discovered, they are generally classified into different categories, including visual binaries, spectroscopic binaries, eclipsing binaries, and astrometric binaries. 
Visual binaries usually have wide separations and the two components can be resolved by telescopes. In most cases, however, components of binaries are too close to be resolved and can only be discovered indirectly. For instance, eclipsing binaries (EBs) are identified based on their light curves from photometry observations, spectroscopic binaries (SBs) have been found by measuring the shifts of typical absorption/emission lines in their spectra, and astrometric binaries are recognized by the perturbation of their motions by an unseen companion that is too dim to be seen or is hidden in the shine of the visible star. There are some overlapping among different types of binaries, e.g. EBs could also be spectroscopic SBs. Binary stars provide a unique opportunity to measure stellar masses that are known as ``dynamical masses''. Moreover, the combined analysis of radial velocities of both components and light curves in a double-lined eclipsing binary (i.e. EB and SB binary) could provide a measurement for the absolute individual mass and radius of each star with an accuracy of $\lesssim 1\%$ (see \cite{Torres2010} for a review). The method for estimating the mass of the star using detached eclipsing binaries has been described in detail in Chapter~2 of a recent review of  \cite{Serenelli2021}.

Since the 1990s, large-scale multi-epoch photometric, spectroscopic, and astrometry surveys have been operating gradually (see section~\ref{sec:surveys}), leading to the discoveries of thousands of binaries and/or their higher-order counterparts and boosting the studies based on large databases. The largest binary sample to date contains about 530,000 candidates of EBs \cite{Mowlavi2022}. On the one hand, the big samples of various binaries make it possible for us to study stellar multiplicity and its dependence on fundamental stellar parameters and environments (e.g. \cite{Duchene2013ARA&A..51..269D,Moe2017ApJS..230...15M}). 
The properties of stellar multiplicity of underlying stellar populations and the statistic properties of binary populations (i.e. the multiplicity fraction, and the distributions of orbital period, mass ratio, and eccentricity) are highly valuable in different fields of astronomy and astrophysics. 
They play an essential role not only in the evolution of stars but also in the formation of important binary-related objects such as cataclysmic variable stars (CVs), X-ray binaries, SNe Ia, double compact objects etc. 
Growing studies have shown that the binary fraction and distributions of orbital parameters could significantly change with the primary mass, 
metallicity, and environments, although there are still many uncertainties in these studies and no consensus has been achieved yet (e.g. \cite{Raghavan2010ApJS..190....1R,Hettinger2015ApJ...806L...2H,Liu2019MNRAS.490..550L}). On the other hand, a large sample of a certain type of binaries can be used to test (or improve) some basic correlations widely used in astronomy such as the period-luminosity-color (PLC) relation for contact binaries. The PLC relation has been used for the distance measurement \cite{Eggen1967,Chen2016}, and this relation improves continually with the increase of the sample of contact binaries in the new Millennium \cite{Gettel2006,Jayasinghe2020,Petrosky2021,Green2022}. In addition, a large sample of wide binaries with dwarf stars is an ideal data set that can be used to test the chemical tagging as a method to identify co-eval stellar sub-populations in the Galaxy \cite{Andrews2017,Andrews2018}.

\begin{figure*}[t]
    \centering
    \includegraphics[width=0.98\textwidth, angle=0]{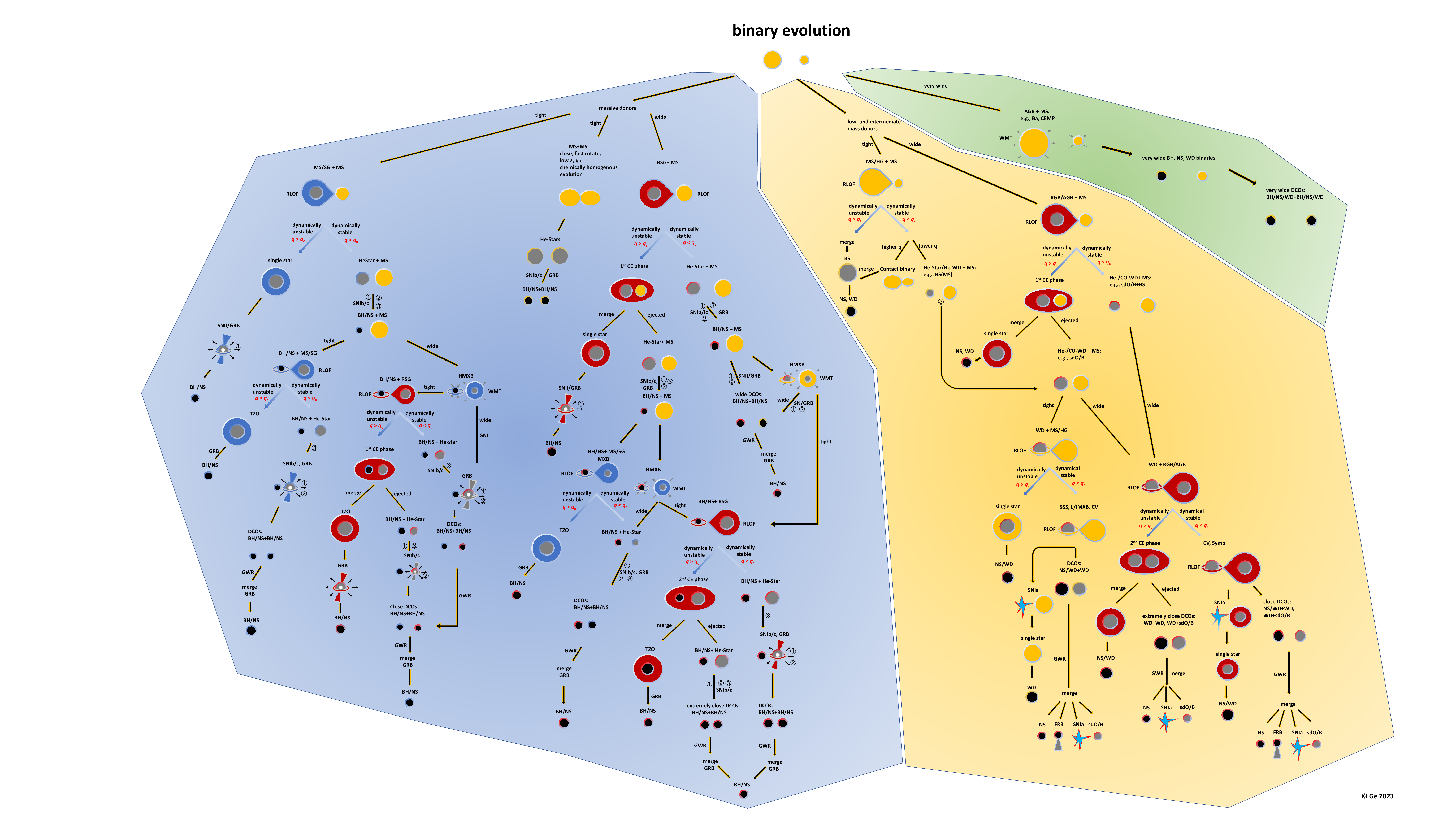}
    \caption{Binary evolution tree. The figure is reprinted from Figure~1 of \cite{Han2020RAA}.}
    \label{fig:binary_tree}
\end{figure*}

Binary interactions make stars evolve in a completely different way from that single stars. Based on the Roche model \cite{Orlov1961SvA.....4..845O} and the assumption of common envelope evolution (CEE) when a dynamical mass transfer occurs \cite{Paczynski1976}, the general picture of binary evolution has been established in the 1980s. This picture has been explored to account for many important findings in observations (e.g. chemical homogeneous evolution for GW~150914 and wind Roche lobe overflow for extremely metal-poor stars; see Section~\ref{sec:GW150914} and \ref{sec:CEMP}). Recently, Han et al. \cite{Han2020RAA} collected the extension of the general picture of binary evolution and summarized them (see also Fig.~1 of their paper), which is referred to as ``the binary evolution tree'' (see Fig.~\ref{fig:binary_tree}). Along the binary evolution tree, binary evolution is expected to lead to the formation of many intriguing objects such as Algols, cataclysmic variables (CVs), symbiotic stars, blue stragglers, barium (Ba) stars, hot subdwarfs (sdB/O), novae, supernovae (SNe), pulsars, X-ray binaries, gamma-ray bursts (GRB), and double compact binaries etc. These exotic stars can be used to probe stellar evolution processes, derive ages and metallicities of stellar populations, and measure cosmological distances. Therefore, they have been the key subjects in different studies in astronomy and astrophysics for many decades. In particular, double compact binaries have recently gained more and more positive attention since the successful detection of GWR from the merger of black hole-black hole (BH-BH) binaries in 2015 \cite{Abbott2016PhRvX...6d1015A}. 

In the new Millennium, binary population synthesis (BPS) models have been greatly developed and successfully used to explore the formation and evolution of large samples of exotic stars (for a review of BPS, see \cite{Han2020RAA}). In particular, since 2015 growing studies have used BPS method for the evolution of massive binary systems to address the formation of BH-BH, BH-NS, NS-NS binaries, making predictions on different properties of these objects such as their birthrates, space density, and distributions of mass ratios, etc. However, a large variation of the results among different BPS studies is seen because of long-standing unsolved problems in binary evolution such as the stability of mass transfer, CEE, and mass (and/or angular momentum) loss \cite{Han2020RAA}. In addition, there are also significantly uncertainties about the BPS method itself due to weak constraints on initial conditions adopted for BPS studies such as the initial mass function (IMF), star-formation rates, binary fraction, and the distributions of mass ratios and orbital periods (see Section~\ref{sec:bps}).

The stability of mass transfer in binary evolution determines the fate of a binary system (see Fig.~1 of \cite{Han2020RAA}). 
Given the importance of the stability of mass transfer in producing exotic objects, it has become attractive in the binary evolution community
for a long time, and significant progress has been made in the last ten years (e.g. \cite{Ge2010Ap&SS.329..243G,Ge2015ApJ...812...40G,Ge2020ApJS..249....9G}).
It is believed that the CEE follows soon after a dynamically unstable mass transfer, that is, the envelope of the mass donor expands dramatically, engulfing in the companion. The inner binary, composed of the core of the donor and the companion, spirals in the CE and eventually results in an ejection of the CE or the merger of two components (e.g. \cite{Ivanova2013A&ARv..21...59I,Roepke2022}). However, detailed physical processes of the CEE remain unclear, which leaves large uncertainties on both the number of merger events and the properties of short orbital-period binaries expected to be generated from the CE ejection. Despite substantial efforts on numerical simulations of the CEE in the last two decades, no strong conclusions have been made for this crucial process in binary evolution. As more and more samples of various types of binaries along the binary evolution tree have been obtained from observations, future studies on the evolution and formation of these binaries with BPS models and the statistical analysis of observed binary populations are expected to provide opportunities to place strong constraints on unclear physical processes at play throughout the lifetime of binary populations.

The CEE in binary evolution was originally proposed by Paczynski in 1976 \cite{Paczynski1976} to account for the formation of CVs and has been widely used to explain the formation of short-orbital period binaries which have experienced a dramatic orbital shrinkage in previous evolution since then. However, there has been a lack of observational evidence for the CEE until luminous red novae were suggested to be excellent candidates for this process about ten years ago \cite{Ivanova2013b,Blagorodnova2017,MacLeod2017}. More interestingly, the ejected CE has been observed around a hot subdwarf binary recently \cite{Li2022}, which was the first time to directly observe the ejected CE and thus provides strong evidence for the CE ejection in binary evolution.

Besides the close binaries introduced above, many binaries have very wide orbits (known as wide binaries) in which the two components do not interact strongly and evolve essentially as single stars. The two components of wide binaries have nearly the same distances, metallicities, and stellar ages. 
So wide binaries have been widely used for calibrating spectroscopic surveys to precisely constrain the ages of field stars and the initial-final mass relation of white dwarfs, and to dynamically probe the Galactic tidal field \cite{Moe2017,El-Badry2018,El-Badry2021,Barrientos2021}.

In this article, we will give a comprehensive review of the progress of the studies of binaries in the 21st. We will start with the large survey projects in section~\ref{sec:surveys}, including photometry and spectroscopy surveys and their contributions to the discoveries of binaries. The studies of the statistical properties of these binary samples as well as those from other techniques have been shown in the end part of this section (in section~\ref{sec:statistic}). Progress in binary evolution has been reviewed in Section~\ref{sec:progress} including the threshold of mass transfer, the common envelope evolution (e.g. the scenarios, numerical simulation, and observation constraints and evidence), and new evolutionary channels of binary stars. Eventually, we will focus on nucleosynthesis in binary stars in Section~\ref{sec:nucleosynthesis} in which studies on novae, kilonovae, type Ia supernovae, and neutron capture processes in metal-poor stars have been addressed.  A summary and outlook are given in Section~\ref{sec:sum}.

\section{Large Survey Projects and Contributions to Binaries}
\label{sec:surveys}

\subsection{Photometric surveys and eclipsing binaries}

Eclipsing binaries (EBs) are one of the most populated types of variable stars to be identified through photometry observations. Under certain conditions, we can accurately determine the fundamental properties of stars (e.g. mass and radius) through EBs, and obtain the orbit parameters of EBs as well. 
Well-studied EBs are extremely valuable in stellar astrophysics. 
For example, wide EBs have been used to supply stringent tests for stellar evolution theory, and close EBs are thought to be ideal laboratories for constraining physical processes in binary evolution. 
Unfortunately, however, a relatively small number of EBs have been well observed and studied. Only 45 EBs were collected in the catalog provided by \cite{Andersen1991}, and the total number recently increases to 305 from the website of DEBCat\footnote{DEBCat is the catalog of the physical properties of well-studied eclipsing binaries; see \url{http://www.astro.keele.ac.uk/jkt/debcat.}}.

The large-scale multi-epoch photometric surveys are expected to detect more EBs and thus expand the catalogue of well-observed EBs dramatically. The pioneered works for searches of EBs with large-scale multi-epoch photometric surveys started in the 1990s, including the ground-based ``Experience pour 1a recherche d'objects sombre'' (EROS, 1990-1995; \cite{Aubourg1993}), the ``Massive compact halo objec'' experiment (MACHO, 1992-1999; \cite{Alcock1997}), and the ``Optical gravitational lensing experiment'' (OGLE; \cite{Udalski1992}), and the space-based Hipparcos \cite{ESA1997}. These kinds of survey projects have sprung up in the new Millennium, leading to that more and more EBs have been reported with the automated classifications of variable stars, e.g.,  
773 sources from the Trans-Atlantic Exoplanet Survey (TRES; \cite{Devor2008}), 1,055 from the All Sky Automatic Survey for Supernovae (ASAS-SN; \cite{Pojmanski2002}), 
2,700 from the Lincoln Near-Earth Asteroid Research Survey (LINEAR; \cite{Palaversa2013}), $\sim$ 45,600 from 
the EROS2 survey \cite{Kim2014},
23,312 from the CATALINA survey \cite{Drake2017}, 
$\sim$ 110,000 from the Asteroid Terrestrial-impact Last Alert System survey \cite{Heinze2018}, and $\sim$350,000 \cite{Chen2020} from
the Zwicky Transient Facility (ZTF). 
In particular, the OGLE4 team provided the largest catalogue specifically dedicated to EBs by collecting 40,204 sources in the Large Magellanic Cloud (LMC), 8,401 in the Small Magellanic Cloud(SMC; \cite{Pawlak2016}), 
and 450,598 towards the Galactic Bulge \cite{Soszynski2016a,Soszynski2016b}.

Space missions dedicated to searches of exoplanets (e.g. \emph{Kepler} \cite{Borucki2010} and Transiting Exoplanet Survey Satellite [TESS] \cite{Ricker2014}) provide excellent data for the studies of EBs 
due to their high photometric precision and the continuous, high-cadence observations with long timescale. For instance, about 2878 and 4584 candidates of EBs were found from \emph{Kepler} \cite{Kirk2016} and TESS \cite{Prsa2022}, respectively. European Gaia space mission \cite{Gaia2016,Gaia2018} launched in 2013, and it was primarily aimed to determine the three-dimensional positions of over one billion stars in the Galaxy. Because the Gaia survey has great potential in terms of astrometry, photometry, and the number of sources surveyed, it has been expected to increase the number of EBs as well as the lever of completeness to an unprecedented degree. However, the detection, characterization, and classification of EBs in a big database are really challenging. 
Nevertheless, novel techniques for such data processing and analysis have been explored and developed by using both existing surveys and Gaia's simulated data \cite{Suveges2017,Kochoska2017,Mowlavi2017}. With these novel techniques, the first Gaia catalogue of EB candidates from Gaia-DR3 was provided by \cite{Mowlavi2022}, in which the two-Gaussian models were used. This catalogue contains about 530,000 EB candidates, which is the largest catalogue to date in the number of sources, sky coverage, and magnitude range. More importantly, the orbital periods, light curve model parameters, and global rankings of the sources in this catalogue have also been provided with their related uncertainties.

The physical parameters of EBs are fundamental for the sciences behind.  Several codes have been developed to derive the parameters of observed EBs, including the Wilson-Devinney program \cite{Kallrath1998ApJ...508..308K}, JKTEBOP \cite{Southworth2004a,Southworth2004b} and the PHOEBE \cite{Conroy2020ApJS..250...34C}. PHOEBE is developed based on the Wilson-Devinney program by combining it with the Markov chain Monte Carlo (MCMS) algorithm. Recently, PHOEBE has been widely used to obtain the posterior distributions of the parameters of EBs \cite{Foreman-Mackey2019}. For example,  the PHOEBE was recently employed to determine the sum of the fractional radii, the ratio of effective temperature, the inclinations, and the eccentricities for 35,464 detached EBs from ASAS-SN \cite{Rowan2022}. However, it is still a big challenge to derive parameters of binaries from the huge amount of data from the big photometric surveys by using traditional approaches. As a consequence, Artificial Intelligence (AI) has been developed and applied to derive the parameters of various binaries, in addition to automated classification of variable stars and phenomenological parameters as introduced above. The artificial neural network (ANN) approach has been described for fully automating the solution of light curves of EBs (i.e. so-called EBAI) and applied to the detached EBs \cite{Prsa2008ApJ...687..542P}, concluding that the success rate is approximately 90\% for the OGLE sample at that time (2580 light curves). By using EBAI, Prsa et al \cite{Prsa2011} also estimated the principal parameters of 1879 EBs from \emph{Kepler-Q0} and \emph{Kepler-Q1} data\footnote{The catalogue of EBs has been further updated due to the release of more \emph{Kepler} data \cite{Slawson2011,Matijevic2012,Conroy2014,Kirk2016,Abdul-Masih2016}.}, in which they presented statistics of EBs based on their determined periods
\footnote{The JKTEBOP code \cite{Southworth2004a,Southworth2004b} was used to examine the fidelity of parameters of EBs from the Kepler Eclipsing Binaries Catalog estimated by EBAI based on 78 detached EBs randomly chosen from over 1400 ones \cite{Holanda2018}, finding a good agreement between the two techniques for the sum of the fractional radii and a moderate agreement for $e {\rm cos}\omega$ and $e {\rm sin}\omega$. But it was found that orbital inclination is clearly underestimated by EBAI.}. Based on the data extracted from the Catalina Sky Survey, the physical properties of 2281 northern EBs with eclipsing Algol-type light curve morphology derived by EBAI were presented \cite{Papageorgiou2019}. More recently, the neural network (NN) models (with or without contamination of the third light) and MCMC algorithm was used to derive the parameters of contact binaries \cite{Ding2022}. It has been found that NN and MCMC algorithms can reduce the time consumption by about four orders of magnitude under the same running condition compared with PHOEBE. This may suggest that NN and MCMC algorithms have great potential for application in face of the flourishing of photometric surveys.

\subsection{Spectroscopic surveys}

In parallel, large spectroscopic surveys such as the Radial Velocity Experiment Wide-field spectroscopic surveys of the stellar content of the Galaxy (\emph{RAVE} \cite{Steinmetz2006,Steinmetz2020}), the Sloan Digital Sky Survey (\emph{SDSS} \cite{York2000}) and the Sloan Extension for Galactic Understanding and Exploration (\emph{SEGUE} \cite{Yanny2009}),
\emph{LAMOST} \cite{Cui2012,Deng2012,Zhao2012,Luo2015RAA,Liu2020}, \emph{Gaia} \cite{Katz2004,Gilmore2012,Gaia2016,Cropper2018} and \emph{Gaia-ESO} \cite{Merle2017,Merle2020}, \emph{GALAH} \cite{De-Silva2015}, and \emph{APOGEE} \cite{Majewski2017}, have yielded huge amount of data on stellar spectra and thus leading to numerous spectroscopic binaries being identified (known as spectroscopic binaries, SBs). The straightforward way to discover binaries (and variables) is to detect the varieties of RV from multi-epoch spectroscopic observations. However, this method is only valid for those with relatively short orbital periods of $\le5$ yr. If two components of a binary have comparable brightness, the signatures (e.g. different characteristic radial velocities) of the two components would be mixed together in the same spectrum. Such binaries (the so-called ``SB2'' systems) and their higher-order counterparts (SB3/SB4) can be identified in just a single epoch, either through the cross-correlation function (CCF) of a spectrum against a single model template or through a multi-model fit of the spectrum. 

With either spectral fitting or spectral disentangling, the stellar types and atmospheric parameters (effective temperature $T_{\rm{eff}}$, surface gravity $g$, and metallicity [M/H]) of each component of a binary can be determined. This would provide the most detailed insight into the binary theory by combining it with the information inferred from their light curves. Therefore some EBs (which are known as benchmark binaries), with well-determined metallicity, and the masses and radii of both components, have been used to test stellar evolution models. Very recently, Serenelli et al (\cite{Serenelli2021}, see their Table 2) collected all detached EBs with atmospheric parameters, in which the masses and radii of stars were determined with a precision of $2\%$ for high-mass, and gradually down to 1\% for low-mass stars. The metallicities of these detached EBs were also derived through spectroscopic analysis either from disentangled spectra or from double-lined composite spectra.

A large sample of various types of SBs with well-determined stellar parameters can provide a unique opportunity for large-scale statistical studies of stellar multiplicity and constraints on stellar/binary evolution theory, and for improving some basic relations used in astronomy as introduced below. The following subsections will mainly be based on the binaries from SDSS and APOGEE, and some results from other surveys will also be summarized.

\subsubsection{White dwarf + main sequence binaries from SDSS}

Binaries composed of a WD and a MS companion star (i.e. WDMS) are the majority of binaries that contain a compact object. Close WDMS binaries with massive WDs have been considered to be progenitors of SNe Ia. Very close WDMS binaries with orbital periods less than several days are believed to be produced from the CEE: the more massive star in a binary system evolves first and engulfs its companion when it is on the red giant branch (RGB) or AGB phase due to unstable mass transfer, forming a CE. After the CE is ejected, a WDMS binary system with a very tight separation is produced. These binaries produced from the CE ejection are widely known as post-CE binaries (PCEBs) in the literature. The PCEBs have been used to place stringent constraints on the CEE process in binary evolution such as the still-debated CE ejection efficiency (e.g. \cite{Zorotovic2010A&A...520A..86Z,Zorotovic2011A&A...536L...3Z,Zorotovic2014A&A...568A..68Z, Rebassa-Mansergas2012a,Rebassa-Mansergas2012b,Toonen2013A&A...557A..87T,Camacho2014A&A...566A..86C}; see more details in Section~3.3). Early studies on the PCEBs had been hindered by the limited number of WDMS binaries until SDSS unveiled thousands of such binaries. Since 2010, Rebassa-Mansergas et al had built up and updated the catalog of WDMS binaries from SDSS \cite{Rebassa-Mansergas2010,Rebassa-Mansergas2012a,Rebassa-Mansergas2013,Rebassa-Mansergas2016}, providing the most homogeneous and complete catalog of WDMS binaries of 3294 objects in the latest catalog released in 2016.  Among this sample, about 200 objects are identified as PCEBs according to their significant varieties of radial velocities \cite{Rebassa-Mansergas2007,Schreiber2008,Schreiber2010,Rebassa-Mansergas2011}, and 90 of which have measured orbital periods based on follow-up photometry observations \cite{Rebassa-Mansergas2008,Pyrzas2009MNRAS.394..978P,Nebot-Gomez-Moran2011A&A...536A..43N,Rebassa-Mansergas2012b}. Although this sample of PCEBs is still strongly biased against systems composed of cool WDs and/or early-type stars due to selection effects, numerous analyses have used this sample to give deep insights into stellar and binary evolution such as the formation of low-mass He WDs, the high mean mass of WDs in CVs, magnetic braking in low-mass binaries, and the CE ejection efficiency, etc. 

With successfully selecting 205 strong PCEB candidates from a sample of 670 WDMS binaries which have multiple spectroscopic observations spreading over at least two nights, the dependency of the relative number of PCEBs among WDMS binaries ($f=N_{\rm PCEB}/N_{\rm WDMS}$) on the secondary mass (the MS companion) was further explored \cite{Schreiber2010}. It leads to the finding of that the value of $f$ peaks around a secondary mass of $M_{\rm sec} \sim 0.25M_\odot$ and drops steeply towards higher mass secondaries in the range of $0.25-0.4M_\odot$ \cite{Schreiber2010}. This therefore suggests that the evolutionary timescales of PCEBs with fully convective secondaries are significantly longer than those for secondaries with a radiative core, which is consistent with the prediction by the disrupted magnetic braking scenario of that the magnetic wind braking can be significantly reduced for fully convective stars. In addition,  the distributions of WD masses in both PCEBs and wide WDMS binaries was investigated \cite{Rebassa-Mansergas2011}. It has been shown that WDs in PCEBs have a clear concentration towards the low-mass end, while the mass distribution in wide WDMS binaries is similar to that of single WDs \cite{Rebassa-Mansergas2011}. This indicates that the majority of low-mass WDs of $<0.5\,M_\odot$ may be formed through binary evolution rather than single-star evolution.

By selecting pre-CVs that would evolve into semi-detached binaries undergoing stable mass transfer within the Galaxy age among a sample of PCEBs, the previous study has found that the masses of WDs in selected pre-CVs peak around $0.67\pm 0.21\,M_\odot$, which is much smaller than the mean WD mass among CVs of $0.83\pm 0.23\,M_\odot$ \cite{Zorotovic2011A&A...536A..42Z}. Also, two possible explanations were further suggested for higher WD masses among CVs. First, the WDs in CVs grow their masses through accretion during the mass transfer process. Alternatively, most CVs are formed above the orbital-period gap which requires a high WD mass to stabilize the mass transfer process or initiate a previous phase of thermal timescale mass transfer. Both explanations seem to indicate that CVs are possible progenitors of SNe Ia.

\subsubsection{Main-sequence binaries from APOGEE and beyond}
\label{sec:APOGEE}

The Apache Point Observatory Galactic Evolution Experiment (APOGEE) can provide high-resolution spectra with a resolution $R\sim$ 22,500 on average in the wavelength range of 1.51-1.70\,$\mu$m \cite{Wilson2010SPIE.7735E..1CW,Wilson2019MNRAS.485.4492W,Majewski2017}. Therefore, the APOGEE has great potential in stellar astrophysics in addition to its importance in studies of the Galaxy. 
In particular, according to the design of the APOGEE survey, most of the survey stars have multiple visits, with at least one visit separated by at least one month, which is efficient to identify binary stars through the detection of radial velocity variations. The integrated spectra over multiple visits can provide very precise RV measurements with a level of $\le 1 {\rm km/s}$, which thus gives a unique opportunity for discovering a large homogeneous sample of spectroscopic binaries through the detection of RV variations.

If components of binaries have comparable brightness,
the signatures of both components would be captured in the same spectrum at the same time, giving different characteristic radial velocities in a system (SB2). Such binary systems can be identified in just a single epoch, either through the cross-correlation function (CCF) of a spectrum against a single model template or through a multimodel fit of the spectrum. Both stellar parameters of each component in a binary system and abundances of 15 individual elements can be derived from the spectra at the same time. 
 Thus, the APOGEE provides an excellent opportunity for the investigations of statistical properties of samples of binary stars (especially for MS stars) and their dependencies on stellar parameters or/and environment, then help us to deepen our understanding of stellar multiplicity and stellar formation theories. 
 
 Multiple pipelines have been developed for deriving stellar parameters from the APOGEE spectra such as ASPCAP \cite{Garcia-Perez2016AJ....151..144G} and APOGEE~Net \cite{Olney2020AJ....159..182O}. The ASPCAP uses the $\chi^2$ minimization in multidimensional parameter space to identify the best-matching theoretical template spectrum for the stars with an effective temperature (which is denoted by the symbol $T_{\rm eff}$. It is defined as the temperature of a Black Body of the same size as the star and that would radiate the same total amount of electromagnetic power as emitted by the star) of $T_{\rm eff} < 8,000$\,K. The effective temperature is one of the most important parameters to determine the characteristics of a spectrum. Therefore, the primary spectrum in a binary system is generally expected to be observationally different from that of a single star due to the contribution from a cooler secondary. This fact has been exploited to identify binaries that could not be detected through radial velocity variations \cite{El-Badry2018MNRAS.473.5043E,El-Badry2018MNRAS.480.4884E,El-Badry2018MNRAS.476..528E, Kovalev2022MNRAS.517..356K}. More importantly, this method does not rely on radial velocity offset between two components of a binary and is therefore more efficient to detect relatively wider binary systems with periods of hundreds or thousands of years. By fitting a binary model to the spectra of APOGEE 13, El-Badry et al. \cite{El-Badry2018MNRAS.476..528E} identified $\sim 2,500$ MS binaries spectroscopically, in which the atmospheric parameters and abundances of both the primary and secondary stars are also obtained simultaneously. In addition, they identified $\sim 200$ triple systems, as well as about $700$ systems with variable radial velocities in which the secondary does not contribute detectably to the spectrum.

By deconvolving custom cross-correlation functions (CCFs), \cite{Kounkel2019AJ....157..196K,Kounke2021AJ....162..184K} developed and improved a pipeline that autonomously identified SB2 and higher-order multiples (SB3/SB4) in the APOGEE spectra. They applied this new pipeline to the analysis of the APOGEE DR16 and DR17 data and discovered 7273 candidates SB2s, 813 SB3s, and 19 SB4s. They further estimated the mass ratios for binaries and performed orbital fit for a subset of binary systems with sufficient numbers of measurements, confirming that most systems with periods of less than 10 days have circularized. At the same time, numerous SB2s have been discovered by other spectroscopic surveys, e.g., 123 from the RAVE \cite{Matijevic2010AJ....140..184M}, 342 from the Gaia-ESO survey \cite{Merle2017A&A...608A..95M}, 12,760 from GALAH (LALactic Archaeology with HERMES \cite{Traven2020A&A...638A.145T}), and 3133 SB2 candidates from the LAMOST-MRS (LAMOST DR6 and DR7 \cite{Li2021ApJS..256...31L}).

In summary, most identified SBs do not have orbital parameters being measured, except for those with sufficient follow-up photometry observations. Nevertheless, a statistical analysis of RVs of these stars have yielded valuable insights on the stellar multiplicity and its dependence on the stellar parameters and environments, which will be reviewed in detail in Section~\ref{sec:statistic}.

\begin{figure*}[t]
    \centering
    \includegraphics[width=0.98\textwidth, angle=0]{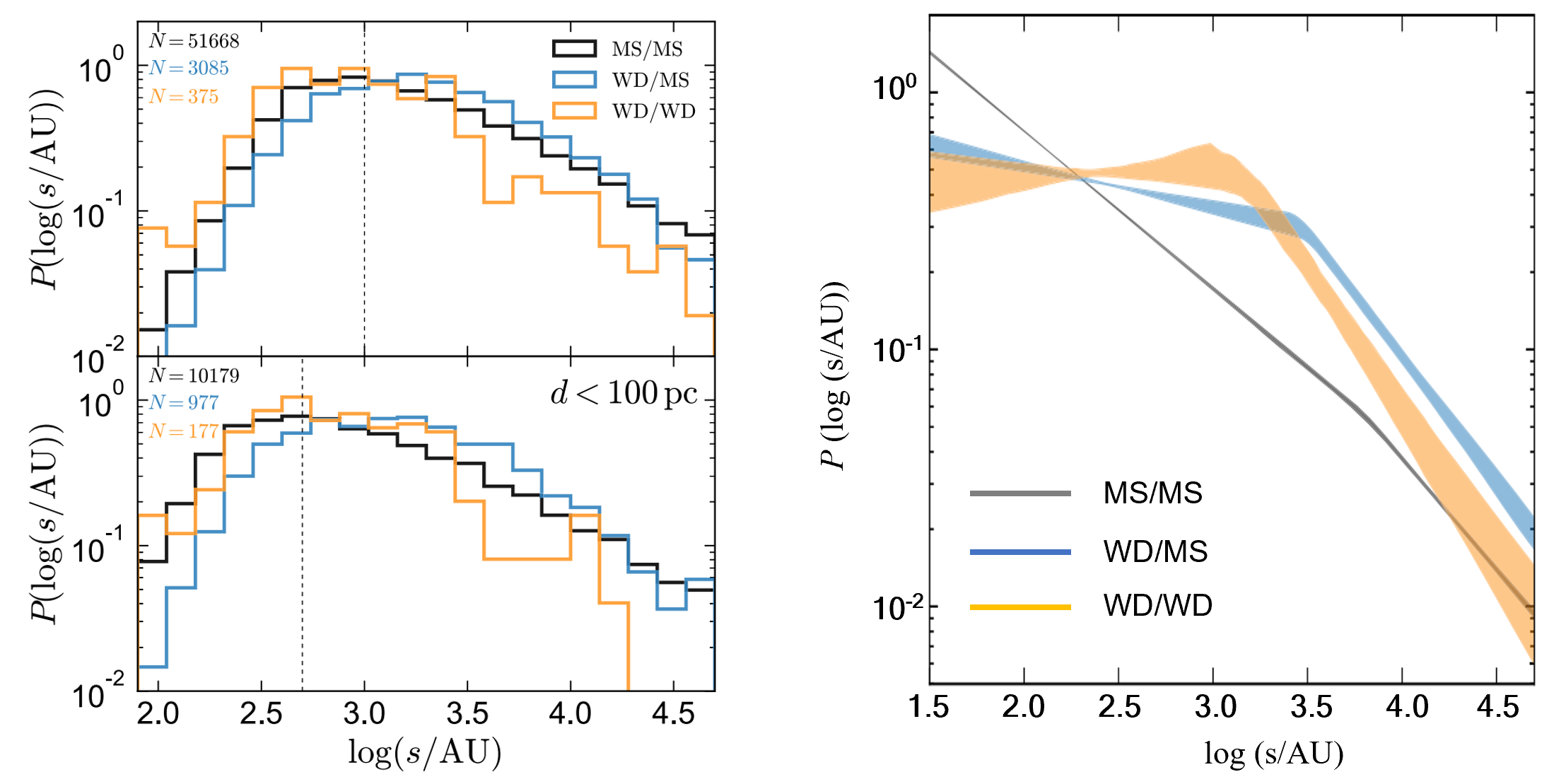}
    \caption{\textit{Left panel:} Normalized distributions of projected physical separation for MS/MS binaries (\textit{black}), WD/MS binaries (\textit{blue}), and WD/WD binaries (\textit{yellow}) within $200\,\rm{pc}$ (\textit{top panel}) and $100\,\rm{pc}$ (\textit{bottom panel}), respectively. The dearth of systems with small separations reflects the fact that binaries with small angular separations are spatially unresolved, leading to incompleteness in the sample at $\rm{log}\,(s/\rm{au})\lesssim3$ (i.e., $1000\,\rm{au}$; \textit{top panel})  and $\rm{log}\,(s/\rm{au})\lesssim2.7$ (i.e., $500\,\rm{au}$; \textit{bottom panel}). The separation distribution of MS/MS binaries falls off as $dN/dS\sim s^{-1.6}$. The separation distributions of WD/WD and WD/MS binaries fall off more sharply than that of MS/MS binaries at large separations. \textit{Right panel:} Broken power-law fits to the intrinsic separation distributions of MS/MS, WD/MS, and WD/WD binaries, after accounting for incompleteness and selection effects. The figure is reprinted from Figures~7  and 8 of \cite{El-Badry2018MNRAS.480.4884E}.}
    \label{fig:power-law}
\end{figure*}

\subsection {Wide binaries from the Gaia}
\label{sec:wide}

Wide binaries are gravitationally bound pairs of stars that have a separation ($s$) up to about $1\,\mathrm{pc}$. It is believed that two components of wide binaries were born at roughly the same time and generated from very similar pre-stellar material, and have nearly the same distances. However, the exact origin of wide binaries is still unclear. Several scenarios have been proposed for the formation of wide binaries. For example, they have been suggested to be possibly produced from either the disruption of triple systems \cite{Reipurth2012Natur.492..221R}, the association of stellar pairs during the dissolution phase of young star clusters \cite{Kouwenhoven2010MNRAS.404.1835K,Moeckel2011MNRAS.410.2799M,Moeckel2011MNRAS.415.1179M}, the turbulent fragmentation \cite{Lee2017NatAs...1E.172L}, or the gravitational attraction of nearby pre-stellar cores \cite{Tokovinin2017MNRAS.468.3461T}.

The co-eval and co-chemical characteristics of both components in wide binaries make them to be widely used in different fields of modern astrophysics. In particular, they have been used for the calibration of M-dwarf metallicities \cite{Lepine2007ApJ...669.1235L,Rojas-Ayala2010ASPC..430..528R,Montes2018MNRAS.479.1332M}, age-magnetic activity (or metallicity) relation \cite{Garces2011A&A...531A...7G,Chaname2012ApJ...746..102C,Rebassa-Mansergas2016}, and the initial-final mass relation \cite{Zhao2012,Andrews2015ApJ...815...63A}. Meanwhile, wide binaries can provide a test for the presence of dark matter in the Galaxy \cite{Bahcall1985ApJ...290...15B,Yoo2004ApJ...601..311Y}, the chemical tagging \cite{Andrews2018MNRAS.473.5393A} and the modified-gravity theories \cite{Pittordis2018MNRAS.480.1778P,Pittordis2019MNRAS.488.4740P,Pittordis2022arXiv220502846P}.

Based on Gaia DR2, an extensive catalogue of wide binaries within $200\,\rm{pc}$ of the Sun was constructed \cite{El-Badry2018}, including more than 50,000 MS+MS, more than 3000 WD+MS, and nearly 400 WD+WD binaries with projected separations in a range of $50-50,000\,\rm{au}$ \cite{El-Badry2018}. To ensure the purity of the catalogue and let the catalogue have an estimated contamination rate of $\sim0.1$ per percent, relatively strict cuts on astrometric and photometric quality and signal-to-noise (SNR) have been applied in this study. As shown in Fig.~\ref{fig:power-law}, it has been found that the separation distributions after corrections of completeness and selection effects show significant differences between the MS+MS binaries and WD+MS systems. The separation distribution of MS+MS binaries is nearly consistent with a single power law of slope $-1.6$ over at least $500\le s \le 50,000\,\mathrm{au}$, with marginal steepening at $s\ge10\,000\rm{au}$,  while those of MS+WD and WD+WD binaries show distinct breaks at $\sim3000\,\rm{au}$ and $\sim1500\,\rm{au}$, respectively, and are flatter at small separation. The difference in small separation could be explained by a certain degree of asymmetric mass loss (a kick of about $0.75\,\mathrm{km\,s^{-1}}$) during the formation of WDs, and that in large separations implies that most wide binaries with separations exceeding a few thousand au become unbound during post-MS evolution.

With less stringent cuts on astrometric SNR, the catalogue of wide binaries has been extended to 4\,kpc for systems with $s\le 1$ pc using data from Gaia DR2 by \cite{Tian2020ApJS..246....4T}. This catalogue contains 807,611 candidates, and the pairs with $s \ge$ 20,000 au have high contamination rates. The authors then used the tangential velocity as an additional selection criterion to define three kinematic sub-samples, i.e., disk-like, intermediate, and halo-like binaries, each of which has thousands of wide binaries. Using the SUPERBLINK high proper catalog \cite{Lepine2005AJ....129.1483L,Lepine2011AJ....142..138L} and the data from Gaia DR2, \cite{Hartman2020ApJS..247...66H} presented a catalog of 99,203 wide binaries. In this work, they did not use a strict distance cut but limited the study to high-proper-motion pairs ($\ge$ $40\,\mathrm{mas\,yr^{-1}}$, equivalent to a distance limit of order 200 pc for the Galactic disc stars typically).

The catalogue of wide binaries was further extended to 1\,kpc by adopting wide binaries selected from Gaia EDR3 and with projected separations ranging from a few au to 1\,pc \cite{El-Badry2021}, in which 1.1 million binaries with a 99\% probability of being bound, including 16,000 WD+MS binaries and 1400 WD+WD binaries. This catalogue has been used to calibrate the published Gaia DR3 parallax uncertainties by assuming that binary components have near-identical parallaxes. Furthermore, this binary catalogue was cross-matched with LAMOST survey, yielding 91,477 binaries in which at least one component has a LAMOST spectrum. This sample will be useful for studying the dependence of binary fractions on metallicity \cite{El-Badry2019MNRAS.482L.139E,Hwang2021MNRAS.501.4329H}. With a sample of binaries with both components have a LAMOST spectrum and the angular separation is at least 3 arcseconds\footnote{The latter cut is to avoid cases where both stars fall inside a single fiber, leading to potentially biased stellar parameters and abundances \cite{El-Badry2018}.}, one can verify whether most binary candidates are bound because that genuine-wide binaries are expected to have similar RVs (or metallicities) of the two components (see Fig.~6 of \cite{El-Badry2021}). This can also be used to calibrate the abundances and uncertainties derived by LAMOST (see Fig.~14 of \cite{El-Badry2021}).

\subsection{Statistical properties of binary population}
\label{sec:statistic}

The statistical properties of binary populations (including binary fraction, distributions of orbital-period, mass-ratio, and orbital eccentricity) and their dependence on stellar types, environments, and/or metallicities are essential for understanding stellar formation and evolution. For example, if the mass-ratio distribution is consistent with random pairings drawn from the initial mass function (IMF,  \cite{Abt1990ApJS...74..551A,Tout1991MNRAS.250..701T,McDonald1995MNRAS.275..671M}), it would suggest that two components of a binary system may form independently. In addition, if binary components show correlated masses as observed in close binaries by different studies \cite{Tokovinin2000A&A...360..997T,Raghavan2010ApJS..190....1R,Sana2012Sci...337..444S}, which would indicate that two components are likely to experience co-evolution during pre-MS phase through physical processes such as fragmentation, fission, competitive accretion or mass transfer \cite{Bonnell1992ApJ...401..654B,Kroupa1995MNRAS.277.1507K,Kroupa1995MNRAS.277.1491K,Clarke1996MNRAS.278L..23C,Bate1997MNRAS.285...33B,Bate2012MNRAS.419.3115B,Kratter2006MNRAS.373.1563K,Kouwenhoven2009A&A...493..979K,Marks2011MNRAS.417.1702M}. Meanwhile, as the basic inputs of BPS calculations, the statistical properties of binary populations would inevitably affect the BPS results. Therefore, by comparing the BPS results with the observations will provide some constraints on the stellar and binary evolution theory \cite{de-Mink2015ApJ...814...58D}.  

\begin{figure*}[t]
    \centering
    \includegraphics[width=0.95\textwidth, angle=0]{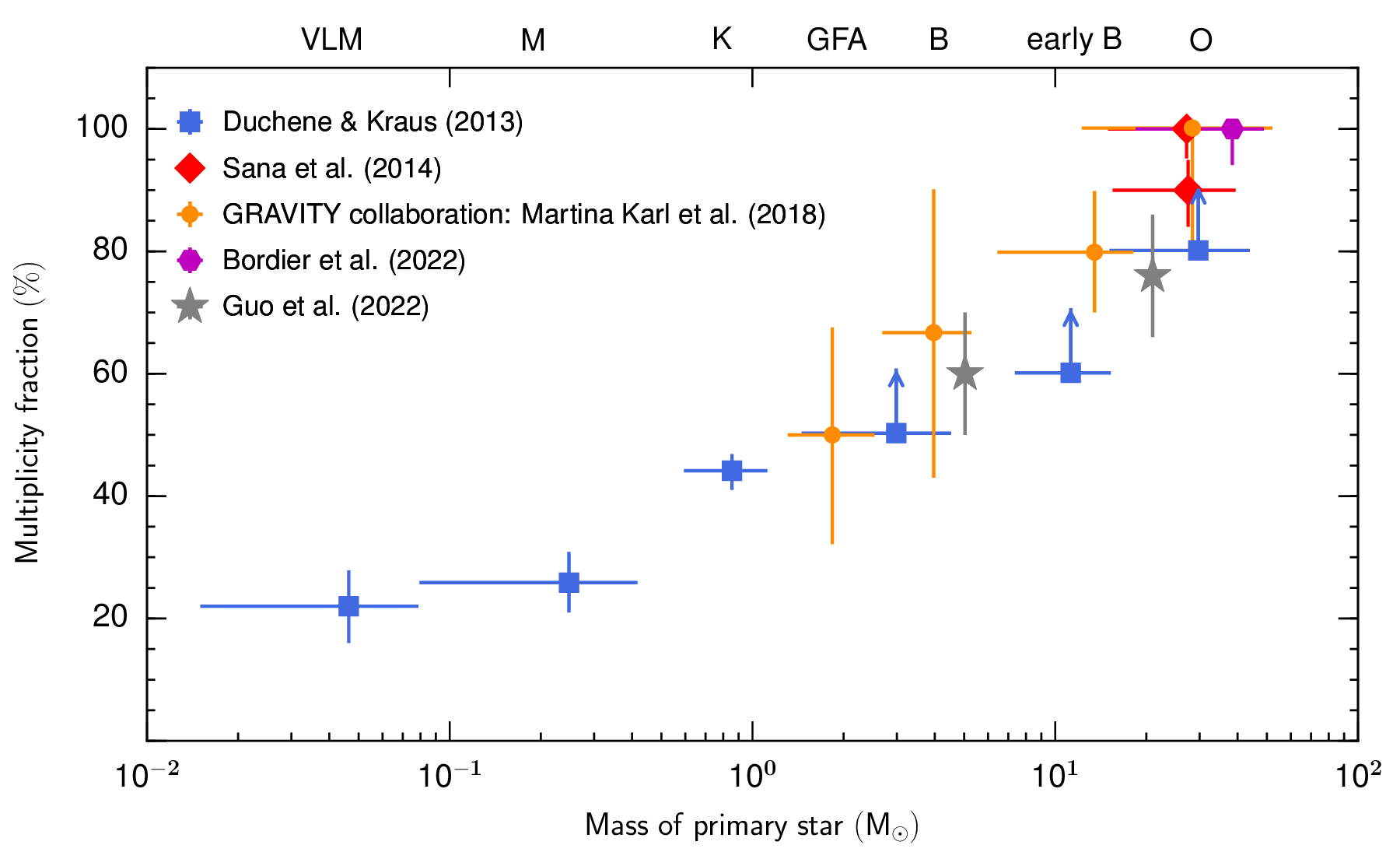}
    \caption{Multiplicity fraction in various binary systems as a function of the primary mass. The blue squares, red diamonds, orange circles,  pink hexagon, and gray stars represent the results from  \cite{Duchene2013ARA&A..51..269D}, \cite{Sana2014}, \cite{Gravity2018}, \cite{Bordier2022}, and \cite{Guo2022A&A...667A..44G}, respectively. Note that because that the results (gray stars) of \cite{Guo2022A&A...667A..44G} do not have information of the mass of primary star, only the multiplicity faction as function of stellar types is shown here.}
    \label{fig:bf}
\end{figure*}

The binary fraction, $f_{\rm b}$, is one of the most crucial parameter among the statistical properties of binary populations and has been explored extensively in the past few decades (for a review, see \cite{Duchene2013ARA&A..51..269D} and references therein), in which most studies focused on massive stars because the detected massive stars increase significantly over the past decades and they are the major contributor to formations of the black hole-black hole (BH-BH), black hole-neutron star (BH-NS), and NS-NS binaries, which are the dominant gravitational wave sources for the Laser Interferometer Gravitational-wave Observatory(LIGO), Virgo and KAGRA detectors \cite{Abbott2016PhRvX...6d1015A,Abbott2016PhRvL.116x1103A,Abbott2016PhRvL.116f1102A,Han2020RAA,Langer2020A&A...638A..39L}. 

Substantial progress on the binary fraction of massive stars has been made in the last few years thanks to the huge and homogenous samples provided by medium and large surveys. For example, our understanding of stellar multiplicity for massive stars has been brought to a new level due to two kinds of Galactic surveys. First, surveys focusing on either specific young clusters \cite{Sana2012Sci...337..444S} or OB associations \cite{Kiminki2012ApJ...751....4K,Kobulnicky2014ApJS..213...34K}. These surveys have acquired a large number of observational epochs for each target, which allows us to obtain orbital solutions for most of the detected spectroscopic binaries in the sample. Second, Galaxy-wide surveys with various spectral resolutions and epochs, including (i) The Galactic O-Star Spectroscopic Survey (GOSSS; \cite{Sota2011ApJS..193...24S,Sota2014ApJS..211...10S,Maiz-Apellaniz2016ApJS..224....4M}, with a resolution $R=$ 2500); (ii) The BESO Survey of Galactic O and B stars (\cite{Chini2012MNRAS.424.1925C}, $R=$ 50\,000); (iii) The IACOB survey which is a multi-epoch high-resolution spectroscopic survey of Northern Galactic O- and B-type stars (\cite{Simon-Diaz2015hsa8.conf..576S}, and references therein); (iv) The OWN survey which is high-resolution spectroscopic monitoring of Southern O and WN stars with enough multi-epoch observations to measure the orbital properties of most detected binaries \cite{Barba2010RMxAC..38...30B}. 

A benchmark work on the statistical properties of massive stars is given by \cite{Sana2012Sci...337..444S}, in which they homogeneously analyzed 71 O-type stars from six nearby Galactic open clusters and investigated the relevant intrinsic multiplicity properties based on the spectroscopy observations of Ultraviolet and Visible Echelle Spectrograph at Very Large Telescope (VLT/UVES). They identified 40 spectroscopic binaries, 85\% of which have constrained orbital periods and 78\% have constrained mass ratios as well. They further performed a set of Monte Carlo simulations to correct any biases in observations by adopting power laws for the probability density functions of orbital period $P$ (in log P), mass ratio $q$, and eccentricities $e$, with exponents $\pi$, $\gamma$, and $\eta$, respectively, that is,

\begin{equation}
f({\rm log}_{10}\,P)\propto ({\rm log}_{10}\,P)^{\pi},
f(q)\propto q^{\gamma},
f(e)\propto e^{\eta}
\end{equation}

The power-law exponents ($\pi$, $\gamma$, and $\eta$) and the intrinsic binary fraction $f_{\rm b}^{\rm in}$ were simultaneously determined by a comparison of simulated populations of stars with the sample allowing for the observational biases. Finally, they obtained an intrinsic binary fraction of $f_{\rm b}^{\rm in} = 0.69 \pm 0.09$, a strong preference for close pairs ($\pi = -0.55 \pm 0.2$), and uniform distribution of mass ratio ($\gamma = -0.1 \pm 0.6$) for binaries with periods up to about 9 years. In comparison to previous studies, the work of \cite{Sana2012Sci...337..444S} gives a steeper period distribution, and thus shows a larger intrinsic binary fraction for short-period systems than previously thought. The stellar multiplicity from very low-mass to high-mass stars among main-sequence and pre-main sequence populations has been reviewed by \cite{Duchene2013ARA&A..51..269D}.

The time-domain survey operating on the LAMOST-MRS provides an opportunity to study the statistical properties of massive stars using a large homogeneous sample. The study based on the LAMOST low-resolution survey (LAMOST~DR5) has obtained a binary fraction of $40\%$ for OB stars by using spectra with more than three observations. More recently, 886 early-type stars are collected, each of them having more than six observations from the LAMOST~DR8 \cite{Guo2022A&A...667A..44G} to perform a detailed investigation of the statistical properties of their binary populations. The sample was further  divided into subgroups based on the derived effective temperature $T_{\rm eff}$, metallicity ([M/H]), and projected rotational velocity (i.e., $v\,{\rm sin}\,i$, which gives the rotational velocity of the star on the line of sight, where $i$ is the inclination). 

A method similar to that of \cite{Sana2012Sci...337..444S} was then used to identify high-confidence binaries and to correct the observational biases by performing a set of Monte Carlo simulations, 
in which power-law possibility distributions of orbital period and mass ratio have been adopted i.e. $f (P) \propto P^{\rm \pi}$ and $f (q) \propto q^{\rm \gamma}$. The study shows that the intrinsic binary fractions $f_{\rm b}^{\rm in}$ are $48\% \pm 10\%$, $60\% \pm 10\%$, and $76\% \pm 10\%$ for stars in groups of B8-A, B4-B7, and O-B3, respectively, suggesting that  $f_{\rm b}^{\rm in}$ increases as $T_{\rm eff}$ goes up as in \cite{Sana2012Sci...337..444S}. The authors also found that $f_{\rm b}^{\rm in}$ is positively correlated with metallicity, with derived values of $44\% \pm 10\%$, $60\% \pm 10\%$, and $72\% \pm 10\%$ for the metallicity ranges of $\mathrm{[M/H] < -0.55}$,  $-0.55 \le {\rm [M/H]} < -0.1$, and ${\rm [M/H]} \ge -0.1$, respectively.  
They presented that $\pi = -0.9 \pm 0.35$, $-0.9 \pm 0.35$, and $-0.9 \pm 0.35$, $\gamma = -1.9 \pm 0.9$, $-1.1 \pm 0.9$, and $-2.0 \pm 0.9$ for stars of types O-B3, B4-B7, and B8-A, respectively. However, no correlations have been found between $\pi$ (or $\gamma$) and $T_{\rm eff}$, nor for $\pi$ (or $\gamma$) and [M/H]. In Fig.~\ref{fig:bf}, we present the observational results of the stellar multiplicity fractions in various binary systems from different studies in the literature as function of the primary mass and stellar types \cite{Duchene2013ARA&A..51..269D,Sana2014,Gravity2018,Bordier2022,Guo2022A&A...667A..44G}. It has clearly shown that the stellar multiplicity fraction increases as the primary mass increases (see Fig.~\ref{fig:bf}).   

Discovering massive binaries in nearby galaxies and studying the binary fraction and their orbital properties have also been explored.
For example, an ESO Large Programme, the VLT-Flames Tarantula Survey (VFTS; \cite{Evans2011A&A...530A.108E}) obtained multi-epoch spectroscopy of over 800 massive OB and WR stars in the 30 Dor region in the Large Magellanic Cloud (LMC) and showed a corrected close binary fraction of $\sim$ 50\% (e.g. \cite{Sana2012Sci...337..444S,Vink2017IAUS..329..279V}), providing direct observational constraints on the multiplicity of massive stars at low metallicity environment. As the successor of VFTS, the Tarantula Massive Binary Monitoring (TMBM; \cite{Almeida2017A&A...598A..84A}) was designed to obtain multi-epoch spectroscopy of massive binaries (OB stars) in the 30 Dor region with sufficient time sampling in order to measure the orbital parameters for mass binary systems with orbital periods up to about one year. 
The first paper of the series of TMBM \cite{Almeida2017A&A...598A..84A,Shenar2017A&A...598A..85S,Shenar2021A&A...650A.147S,Shenar2022A&A...665A.148S,Mahy2020A&A...634A.118M,Mahy2020A&A...634A.119M} was presented by \cite{Almeida2017A&A...598A..84A}, in which they analyzed 32 FLAMES/GIRAFFE observations of 93 O-type and 7 B-type binaries and obtained orbital solutions for 82 systems including 51 SB1 and 31 SB2. The atmospheric parameters and photometry observations of the SB2 in their sample have been published in a later study done by \cite{Mahy2020A&A...634A.119M,Mahy2020A&A...634A.118M}. They found that the overall binary fraction and orbital properties of their sample in the 30 Doradus are similar to that in the Galactic sample, although the orbital-period distribution in their sample seems to be slightly flatter (in log P) than that in the Galaxy. In addition, the stars in 30 Doradus can have an orbital period as short as 1.1 d, which is shorter than that observed in the Galactic samples. This study indicates a relative universality of the binary fraction and orbital properties for massive stars in a metallicity range from solar ($Z_\odot$) to about half solar, and thus provides the first direct constraints on massive binary properties in massive star-forming galaxies at redshifts of $z \sim$ 1 to 2 (the star formation peak in our Universe where have an estimated metallicity of $\sim 0.5 Z_\odot$). Recently, \cite{Shenar2022A&A...665A.148S} have explored the characteristics of hidden companions 51 SB1 O-type and evolved B-type binaries identified in LMC, establishing a virtually flat mass-ratio distribution ($\kappa =0$ for O-type stars at LMC metallicity, with $0.05\ge q\ge 1$ and $0\le {\rm log} P {\rm (d)}\le 3$.

In addition to massive (or early-type) stars, the statistical properties of different types of stars have also been systematically investigated with the boosting of survey database. Since the majority of SBs do not have orbital solutions, such studies can only be performed based on the variations of radial velocities or disentangling/fitting the spectra. 
For example, using noisy radial velocity data with as few as two randomly spaced epochs per object, a previous study has presented a method to characterize statistically the parameters of a binary sample \cite{Maoz2012ApJ...751..143M}.  They analyzed the distribution of the maximum radial velocity difference ($\Delta RV_{\rm max}$) between any two epochs for the same object, finding that the distribution is a Gaussian function with $<\Delta RV>=0$ and the center of this distribution is dominated by measurement errors at low values of $\Delta RV_{\rm max}$. For large enough samples, however, there is a high-velocity tail that can effectively constrain the parameters of the binary population. They employed Monte Carlo techniques to produce grids of simulated $\Delta RV_{\rm max}$ distributions with specific binary population parameters, and the same sampling cadences and radial velocity errors as the observations. They then compared the synthetic distributions with the real $\Delta RV_{\rm max}$ distribution to constrain the properties of the binary population investigated. This method has been applied to $\sim 4000$ WDs in the SDSS \cite{Badenes2012ApJ...749L..11B}, finding 15 WDs with high $\Delta RV_{\rm max}$. This gives an estimated WD merger rate of $1.4^{\rm +3.4}_{\rm -1.0} \times 10^{\rm -13} {\rm yr}^{\rm -1} M^{\rm -1}_\odot $ ($1 \sigma$ limits). This rate is similar to the measured rate of SNe Ia in Milky-Way-like Sbc galaxies, while the super-Chandrasekhar merger rate is only $1.0^{\rm +1.6}_{\rm -0.6} \times 10^{\rm -14} {\rm yr}^{\rm -1} {\rm M}^{\rm -1}_\odot$. This seems to indicate that sub-Chandrasekhar mergers in double-degenerate systems might be a major contribution to the overall SN Ia rate if the sub-Chandrasekhar mergers could indeed lead to SNe Ia as what has been recently suggested by some studies (e.g.  \cite{Guillochon2010ApJ...709L..64G,Pakmor2011A&A...528A.117P}).

Based on the multi-epoch radial velocities acquired by the $APOGEE$,  a large-scale statistical study of stellar multiplicity was performed for $\sim 90,000$ field stars in the Galaxy that span from MS to the red clump \cite{Badenes2018ApJ...854..147B}. Given the distribution of APOGEE temporal baselines, the method is sensitive to binaries with ${\rm log} P {\rm (d)} \le 3.3$. As a consequence, it has been found that the $\Delta RV_{\rm max}$ distribution is a strong function of surface gravity ${\rm log} g$. The $\Delta RV_{\rm max}$ can be as high as $\sim 300\,\rm{km\,s^{\rm-1}}$ for MS stars and steadily drops down to $\sim 30\,\rm{km\,s^{\rm-1}}$ for ${\rm log}\,g \sim 0$ (close to the tip of RGB), but the red clump stars show $\Delta RV_{\rm max}$ values comparable to that of stars at the tip of the RGB \cite{Badenes2018ApJ...854..147B}. This effect is accompanied by a decrease of binaries at all values of $\Delta RV_{\rm max}$ as stars climb up the RGB. The lack of high-$\Delta RV_{\rm max}$ systems in the RGB is consistent with a log-normal period distribution in the MS and a multiplicity fraction of 0.35, which is truncated at the critical period for RLOF for each ${\rm log} g$. In addition, the $\Delta RV_{\rm max}$ distributions were found to strongly depend on metallicity. The metal-poor (${\rm [Fe/H]} \le -0.5$) stars have a higher multiplicity fraction than that of metal-rich (${\rm [Fe/H]}\ge 0.0$) stars by a factor of 2–3 higher in the sample, which was suggested to have profound implications for the formation rates of interacting binaries observed by astronomical transient surveys and gravitational wave detectors, as well as the habitability of circumbinary planets.

By assuming a simple model distribution for a binary population with circle orbits (i.e. the Salpeter IMF, a uniform mass-ratio distribution between 0.05 and 1, a log-normal distribution of the orbital period, and a uniform distribution of the orbital orientations in 3D space), a Bayesian method was employed to infer the maximum likelihood of the binary fraction for binary stars with orbital periods of 1000 days or shorter for FGK stars based on multi-epoch stellar spectra from the SDSS and LAMOST \cite{Gao2014ApJ...788L..37G}. It was found that the overall inferred close binary fraction for the samples from SDSS SEGUE and LAMOST is $43.0\% \pm 2.0\%$ and $30\% \pm 8.0\%$, respectively, and the apparent close binary fraction decreases with stellar effective temperatures. By dividing the SEGUE and LEGUE data into three subsamples with different metallicities of ${\rm [Fe/H]} \le -1.1$; $-1.1 \le {\rm [Fe/H]} \le -0.6$ and $-0.6 \le {\rm [Fe/H]}$), it was found that the inferred close binary fractions are $56 \pm 5.0\%$, $56.0 \pm 3\%$, and $30 \pm 5.7\%$, respectively in these three subsamples. This result is opposite to that of \cite{Guo2022A&A...667A..44G} for early-type stars.
Therefore, it could be concluded that metal-rich stars in their sample are substantially less likely to possess a close companion than otherwise similar stars drawn from metal-poor populations. This further suggests that population ages, stellar formation environments, metallicity, and binary evolution could all contribute to accounting for such observational results. Soon after, \cite{Hettinger2015ApJ...806L...2H} performed a similar statistical analysis for $\sim 14,000$ F-type dwarfs in the Galaxy through time-resolved spectroscopy with sub-exposures archived in the SDSS. They obtained absolute radial velocity (RV) measurements through template cross-correlation of individual sub-exposures with temporal baselines varying from minutes to years and analyzed the sparsely sampled RV curves by using the Markov chain Monte Carlo technique. From their work, they showed that metal-rich disk stars are 30\% more likely to have companions with periods shorter than 12 days than metal-poor halo stars. 

\begin{figure*}[t]
    \centering
    \includegraphics[width=0.95\textwidth, angle=0]{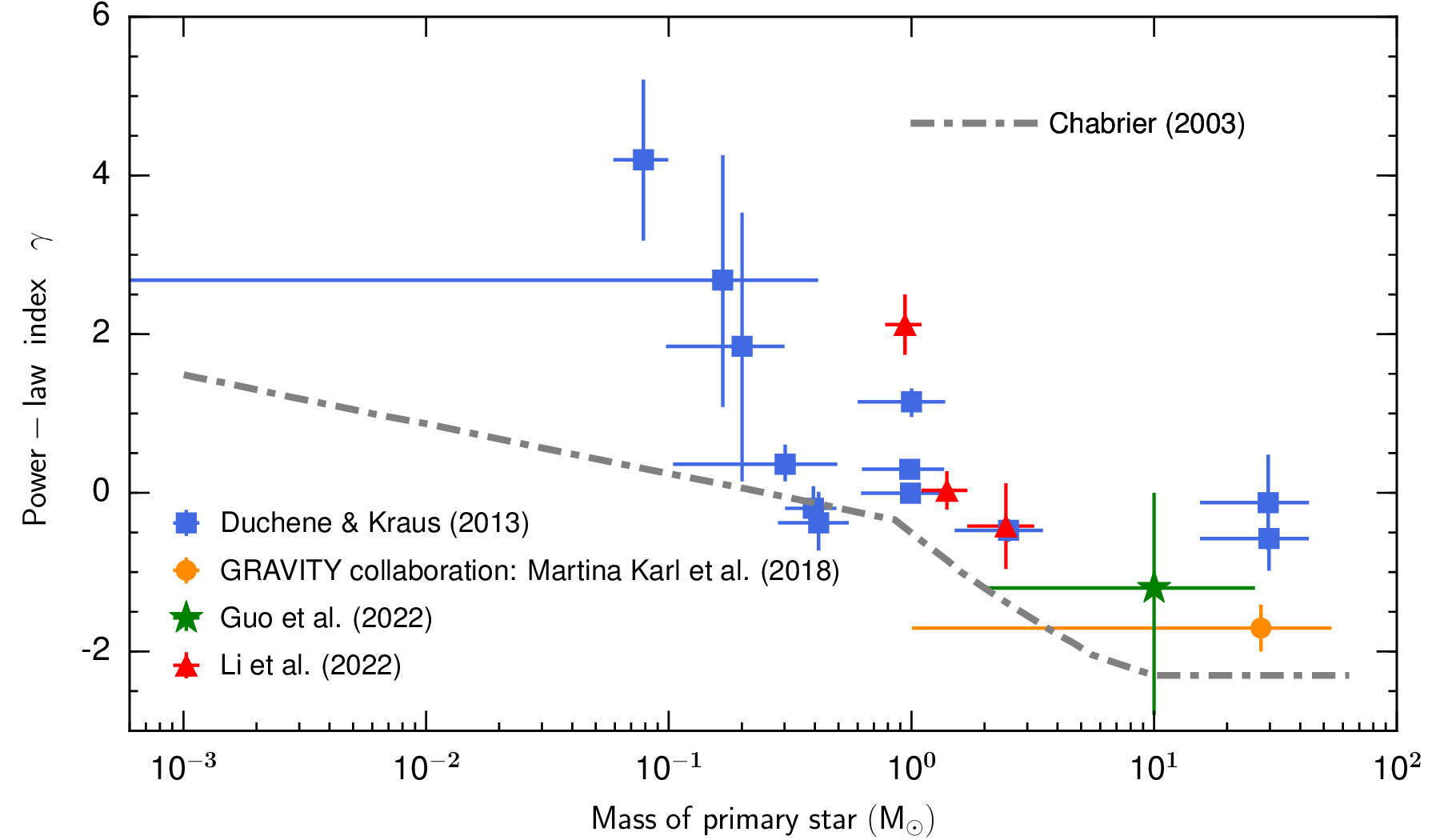}
    \caption{Power-law index ($\gamma$) of the mass ratio distributions with $dN/dq\, \propto\, q^{\gamma}$ as a function of the primary mass. The blue squares, orange circles and red stars represent the results from  \cite{Duchene2013ARA&A..51..269D},  \cite{Gravity2018} and \cite{Guo2022A&A...667A..44G}, respectively. The results given by the initial mass function (IMF) of \cite{Chabrier2003} is also shown in a gray dash-dotted line for a comparison.}
    \label{fig:gamma}
\end{figure*}

Recently, using a custom Monte Carlo sample (the Joker) to discover and characterize binary systems through radial velocities in sparse, noisy, and poorly sampled multi-epoch data, 19,635 high-confidence close binaries (${\rm P} \le $ few yrs, $s\le$ few au)  were identified from 232,495 sources in APOGEE~DR16 \cite{Price-Whelan2020ApJ...895....2P}. On the one hand, it was found that the observed close-binary fraction $f_{\rm b}^{\rm o}$ depends on the stellar parameters of the primary star, as shown in Fig.~5 of \cite{Price-Whelan2020ApJ...895....2P}. For example, it was found that the value of $f_{\rm b}^{\rm o}$ increases with effective temperature along the main sequence (i.e., stellar mass), which is consistent with previous studies. In addition, $f_{\rm b}^{\rm o}$ was found to decrease gradually when the primaries ascend upward the giant branch, and there is a notable dearth of close companions around the red clump, which is similar to that presented in \cite{Badenes2018ApJ...854..147B}. This indicates that companions likely have been engulfed gradually by the primaries when they evolve along the giant branch and become larger. On the other hand, they also found that the observed binary fraction is anti-correlated with metallicity, i.e., it decreases linearly with metallicity with a slope of $-0.1 {\rm dex}^{\rm -1}$ in the range of $-1\le[M/H]\le +0.4$. They also analyzed bout $\sim 8600$ FGK stars to infer the eccentricity distribution of long-period ($P>100$ days) for intermediate-mass MS star binary systems, showing that long-period binaries prefer moderate eccentricities.

In 2022, a peak amplitude ratio (PAR) approach was devised  to derive the mass ratio of a binary system based on one-epoch spectrum from LAMOST-MRS survey \cite{Lijiangdan2022}. 
By computing a cross-correlation function (CCF) of a spectrum against a single model template, a relation between the mass ratios and the PARs of the CCFs of the binary spectra was established. Based on the spectra of the SB2 sample obtained from LAMSOT~DR6 and DR7, the authors found that the power index $\gamma$ values of the mass-ratio distributions (i.e. $f\,(q) \propto q^{\rm \gamma}$) are $-0.42 \pm 0.27$, $0.03 \pm 0.12$, and $2.12 \pm 0.19$ for the A, F, and G type stars identified in the sample, respectively. The derived $\gamma$ values display an increasing trend toward lower primary star masses, and G type binaries tend to be twins more frequently (see Fig.~\ref{fig:gamma}). It was also found that the close binary fractions in the sample (for $P \le 150 $ days and $q \ge 0.6$) are $7.6\pm0.5\%$, $4.9\pm0.2\%$, and $3.7\pm0.1\%$, for A-, F- and G-type binaries, respectively \cite{Lijiangdan2022}. Note that the PAD approach is valid only for binaries with mass ratio in the range of $0.6-1.0$. In Fig.~\ref{fig:gamma}, we summarize the power-law index values of the mass-ratio distributions in various binary systems derived from different studies in the literature \cite{Duchene2013ARA&A..51..269D,Gravity2018,Guo2022A&A...667A..44G,Lijiangdan2022}. The observed distributions are also compared with distribution functions of \cite{Chabrier2003}. 

The methods described above based on radial velocity variations are biased to close binaries and sensitive to the assumed orbital period distribution. In reality, radial velocity variations of high mass ratio binaries may not be as simple as assumed in the analysis. A stellar locus outlier (SLOT) method (i.e. using the color-color diagram) was also proposed to determine binary fractions of MS stars statistically \cite{Yuan2015ApJ...799..135Y} proposed. The SLOT method is not sensitive to the period or mass-ratio distributions of binaries, therefore providing model-free estimates of binary fractions for large numbers of stars of different populations in large survey volumes. The SLOT method has been applied to FGK field stars from the SDSS Stripe 82, in which the samples are constructed by combining the re-calibrated SDSS photometric data with the spectroscopic information from the SDSS and LAMOST surveys. For the SDSS spectroscopic sample, an average binary fraction of $41\% \pm 2\%$ was found, which decreases toward late spectral types. In addition, it was found that the binary fractions for stars with $g-i$ color in the range $0.3-0.6$ mag, $0.6-0.9$ mag, $0.9-1.2$ mag, and $1.2-1.6$ mag are $44\% \pm 5\%$, $43\% \pm 3\%$, $35\% \pm 5\%$, and $28\% \pm 6\%$, respectively. Moreover, the binary fraction was found to decrease with the increasing of metallicity, which gives the binary fraction of $f_{\rm b}= 37\% \pm 3\%$, $39\% \pm 3\%$, $50\% \pm 9\%$, and $53\% \pm 20\%$ for stars with $\rm [Fe/H]$ between -0.5 and 0.0 dex, -1.0 and -0.5 dex, -1.5 and -1.0 dex, and -2.0 and -1.5 dex, respectively. This leads to a conclusion of that stars of Galactic thin and thick disks have comparable binary fractions, whereas that in the Galactic halo has a significantly larger binary fraction. Note that the LAMOST spectroscopic sample\cite{Guo2022A&A...667A..44G} gives consistent results.

In summary, the studies of statistic properties of star populations have generally shown that the binary fraction $f_{\rm b}$ increases with the stellar mass (see Fig.~\ref{fig:bf}). The binary fraction can be up to 70\% for massive O-type stars, but the exact value of $f_{\rm b}$ has significant uncertainties. In addition, there is no consensus on how $f_{\rm b}$ depends on metallicity. Different works have shown an anti-correlation between $f_{\rm b}$ and metallicity for FGK and M stars based on the SDSS and LAMOST data. However, the study indeed found that $f_{\rm b}^{\rm in}$ is positively correlated with metallicity for early-type stars from LAMOST \cite{Guo2022A&A...667A..44G}. In addition, the studies based on TMBM samples indicate a relative universality of the binary fraction and orbital properties for massive stars in the metallicity range from solar metallicity ($Z_\odot$) to about half of that ($0.5\,Z_\odot$). It seems that the formation environment would affect the statistical properties of binary stars. Stars in Galactic thin and thick disks seem to have comparable binary fractions, whereas that in the Galactic halo has a significantly larger binary fraction.

\section{Progress in Binary Evolution}
\label{sec:progress}

\begin{figure*}[t]
    \centering
    \includegraphics[width=0.48\textwidth, angle=0]{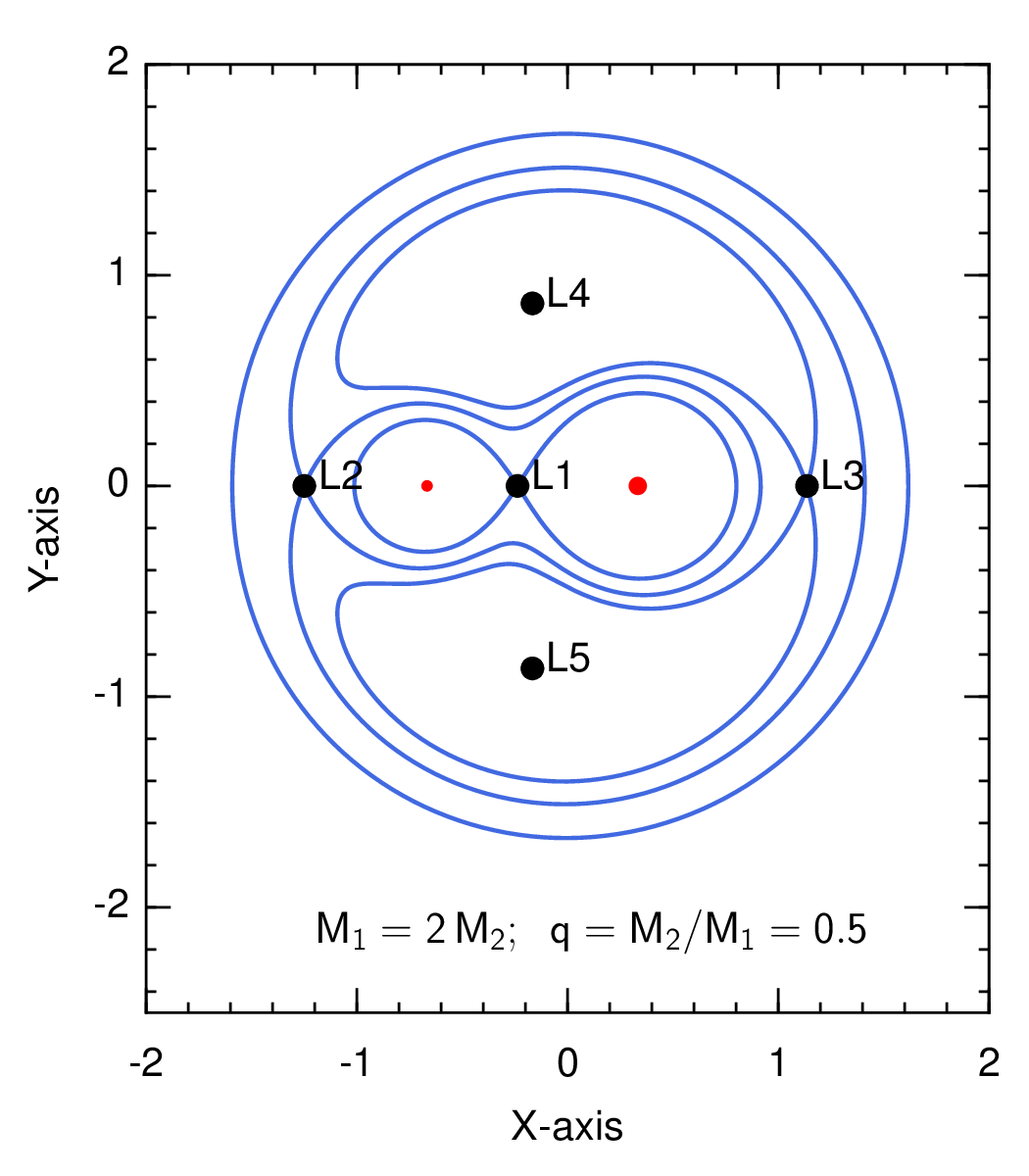}
    \hspace{2pt}
        \includegraphics[width=0.48\textwidth, angle=0]{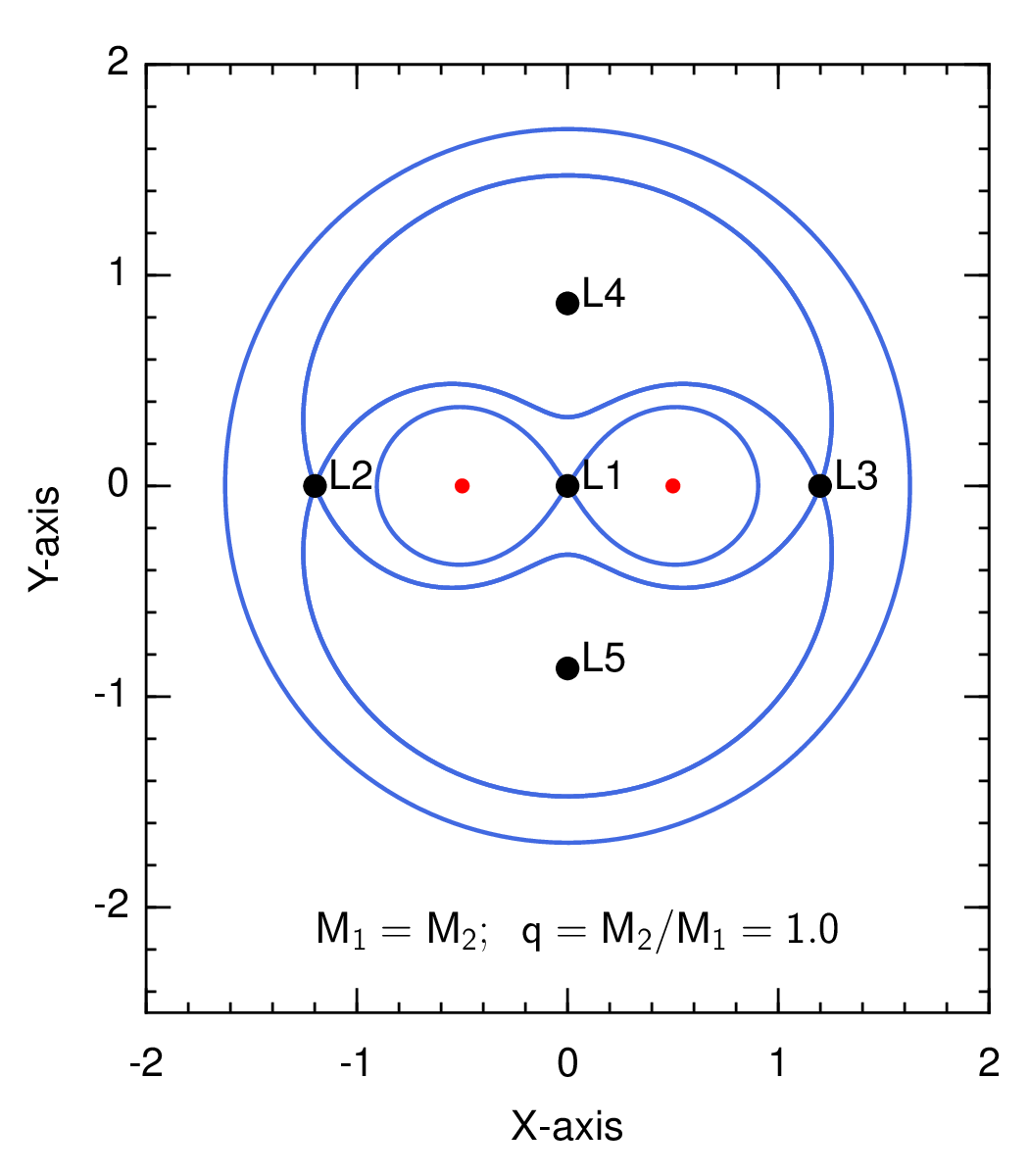}
    \caption{Contour map of equipotentials in the orbital plane of binary systems with mass ratio $q=M_{2}/M_{1}=0.5$ (left panel) and $q=M_{2}/M_{1}=1.0$ (right panel), respectively. The origin of rotating reference is the centre of mass of the binary system, and the positions of the two stars are indicated by red circles. The Lagrangian equilibrium points, $L_{1}$ to $L_{5}$ are indicated in black dots.}
    \label{fig:roche}
\end{figure*}

The studies of binary evolution are based on the ``Roche model'' \cite{Kuiper1941ApJ....93..133K,Kopal1954MNRAS.114..101K,Orlov1961SvA.....4..845O}. In the classical Roche model, one can define an effective potential in a co-rotating frame that includes the gravitational potential of the two stars and the centrifugal force acting on a zero-mass test particle. As shown in Fig.~\ref{fig:roche}, the potential has 5 Lagrangian points where the gradient of the effective potential is zero (i.e. there is no force in the co-rotating frame), among which the most important one is the inner Lagrangian point, L1. The equipotential surface passing through this point is known as the critical Roche-lobe potential which connects the gravitational spheres of influence of the two stars. If one star in a binary fills its Roche lobe, its material can flow through L1 into the Roche lobe of the other star. This process is referred as to Roche-lobe overflow (RLOF), which is the most important mode for mass transfer between two components of a binary.

\begin{figure*}[ht]
    \centering
    \includegraphics[width=0.95\textwidth, angle=0]{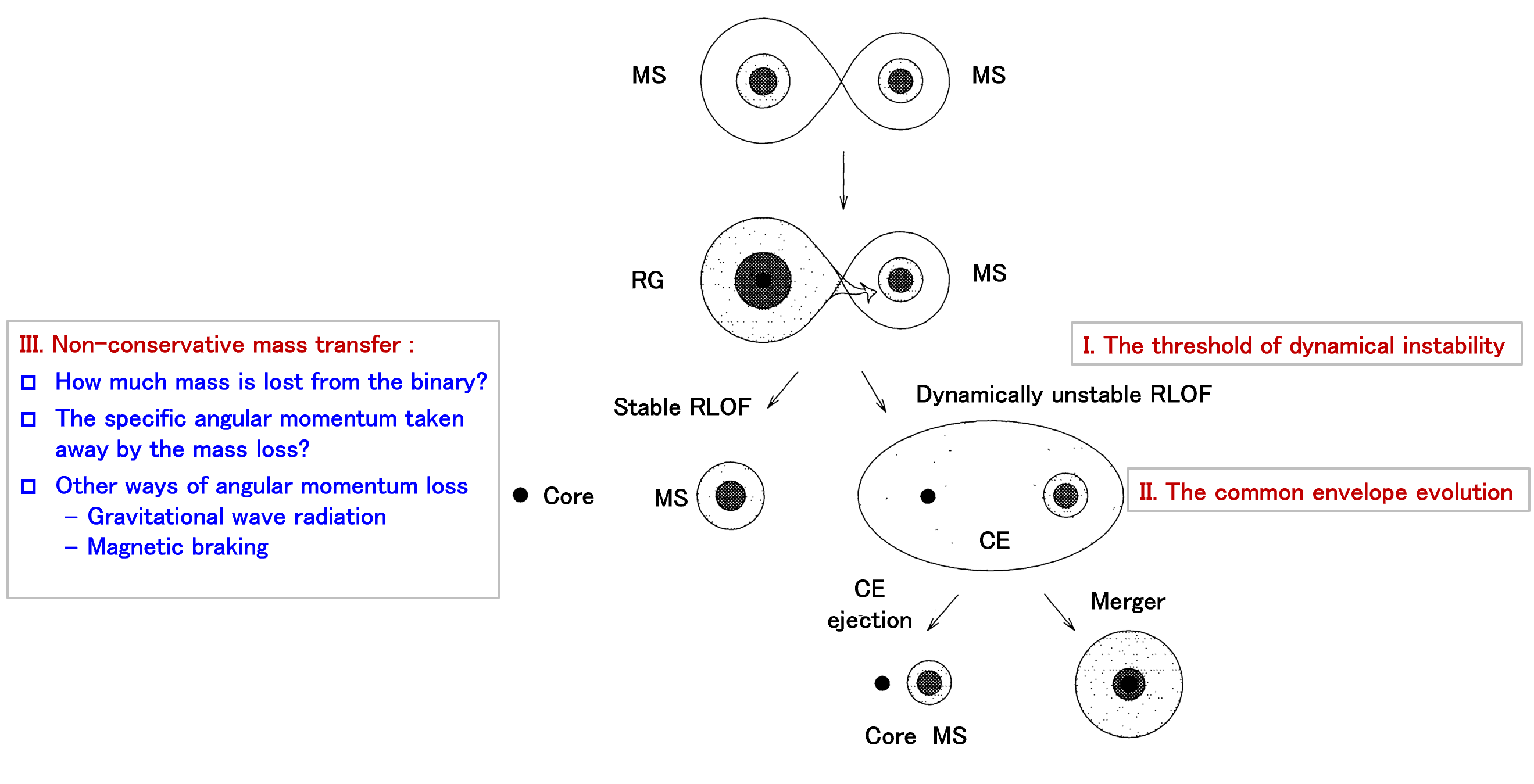}
    \caption{The schematic plot for basic processes in binary evolution and major uncertainties in modelling binary evolution indicated with roman characters I, II and III. The abbreviations are explained as follows: MS-main sequences, RG-red giant, RLOF-Roche lobe overflow, CE-common envelope.}
    \label{fig:Binary-evolution}
\end{figure*}

Figure~\ref{fig:Binary-evolution} illustrates the schematic picture of binary evolution based on the Roche model and the main uncertainties in modeling binary evolution.  In an initial binary system composed of two zero-age MS stars, the more massive star (the primary) evolves first and expands to fill its Roche lobe, and then transferring material to the less massive companion (the secondary) through L1. Depending on whether the mass transfer is dynamically stable or not, the binary system has different evolutionary paths and fates. On the one hand, if the mass transfer is dynamically stable, the primary star would transfer most of its envelope to the secondary star through RLOF, and a new binary with a relatively long orbital period is formed after the RLOF. 
The secondary is likely to become more massive due to mass accretion. 
On the other hand, if the mass transfer is dynamically unstable, the primary would expand dramatically to result in a sharp increase in the mass transfer rate. This leads to that the secondary cannot accrete all the transferred material, which subsequently piles up on the surface of the secondary and starts to expand, ultimately overfilling the secondary’s Roche lobe. As a consequence, a CE is formed, and the core of the primary and the secondary spiral towards one another inside the CE. As time goes on, the orbital energy of the inner binary is reduced and deposited in the CE due to friction with the CE. If parts of the reduced orbital energy ($\Delta E_{\rm orb}$) could be used to eject the CE, a binary with a very short orbital period would be produced after the CE ejection. 
In contrast, if the CE cannot be ejected successfully, the inner binary would merge into a single star that is likely with a high spin velocity.

Although binary evolution plays a very important role in different fields of astrophysics, there are still many uncertainties in modeling binary evolution. The main uncertainties are assigned to three aspects as indicated in Fig.~\ref{fig:Binary-evolution}, i.e., the criteria of dynamically unstable mass transfer, the CEE, and the non-conservative mass transfer. These three main uncertainties are referred to as long-standing unresolved problems in binary evolution. In the following parts, we will review substantial efforts and progresses that have been made to place constraints on these uncertainties in the new Menilium.

\subsection{The threshold of dynamical instability}
\label{sec:threshold}

In general, we can examine the stability of mass transfer by comparing the responses of the radius of the mass donor with its Roche lobe radius during the mass transfer. The variation of stellar radius can be written as follows:

\begin{equation}
d {\rm ln} R_{\rm 1}=(\frac{\partial {\rm ln} R_{\rm 1}}{\partial t})_{M_{\rm 1}} dt+(\frac{\partial {\rm ln} R_{\rm 1}}{\partial {\rm ln} M_{\rm 1}})_t d{\rm ln} M_{\rm 1}
\label{eq:dr1}
\end{equation}

where $M_{\rm 1}$ and $R_{\rm 1}$ are the mass and radius of the mass donor. The first term on the right-hand side of the equation represents the expansion (or contraction) of the mass donor due to both its evolution without mass loss and the thermal relaxation if it is out of thermal equilibrium. The second term is the hydrostatic response to the mass loss under the condition of that thermal, chemical diffusion and nuclear evolution are suppressed, which is also referred to as the adiabatic response to the mass loss. This logarithmic derivative can therefore be labeled by the subscript `ad' instead of `t'. One can further define a mass-radius index as:
\begin{equation}
\zeta_{\rm ad} \equiv (\frac{\partial {\rm ln} R_{\rm 1}}{\partial {\rm ln} M_{\rm 1}})_t
\end{equation}

If a binary with a circular orbit, the Roche lobe radius of the primary is a function of its mass $M_{\rm 1}$, the total mass of the binary M (or mass ratio $q$), and the orbital angular momentum of the binary $J$, that is,
\begin{equation}
    R_{\rm L}=r_{\rm L}(M_{\rm 1}/M)\frac{J^2M}{GM_{\rm 1}^2(M-M_{\rm 1})^2}
\end{equation}

and we have 

\begin{equation}
    d {\rm ln} R_{\rm L}=(\frac{\partial {\rm ln} R_{\rm L}}{\partial {\rm ln} M_{\rm 1}})_{M,J} d {\rm ln} M_{\rm 1}+
    (\frac{\partial {\rm ln} R_{\rm L}}{\partial {\rm ln} M})_{M_{\rm 1},J} d {\rm ln} M+
   (\frac{\partial {\rm ln} R_{\rm L}}{\partial {\rm ln} J})_{M_{\rm 1},M} d {\rm ln} J.
\label{eq:dRL0}
\end{equation}

In the case of conservative mass transfer, the latter two terms on the right-hand side of Equation \ref{eq:dRL0} vanish and $M$ and $J$ are constant. However, a binary generally may experience mass and/or angular momentum losses through, for example, stellar winds or gravitational wave radiation. The systemic mass and angular momentum losses also contain contributions arising from a loss of the partly mass transfer stream between binary components i.e. the consequential losses induced by the variation of $M_{\rm 1}$. We may therefore decompose the terms in $d {\rm ln} M$ and $d {\rm ln} J$ into terms linearly independent of $d {\rm ln} M_{\rm 1}$ (with subscript $M_{\rm 1}$) and the consequential losses into terms linearly proportional to $d {\rm ln} M_{\rm 1}$ (with subscript $t$), by analogy with Equation \ref{eq:dr1}, that is,
\begin{equation}
d {\rm ln} R_{\rm L}=(\frac{\partial {\rm ln} R_{\rm L}}{\partial t})_{M_{\rm 1}} dt+(\frac{\partial {\rm ln} R_{\rm L}}{\partial {\rm ln} M_{\rm 1}})_t d{\rm ln} M_{\rm 1}.  
\label{eq:dRL}
\end{equation}

We further define 
\begin{equation}
\zeta_{\rm L} \equiv (\frac{\partial {\rm ln} R_{\rm L}}{\partial {\rm ln} M_{\rm 1}})_t,
\end{equation}
then, by combining equation \ref{eq:dr1} and \ref{eq:dRL}, we have,
\begin{equation}
d {\rm ln} R_{\rm 1}/R_{\rm L}=[(\frac{\partial R_{\rm 1}}{\partial t})_{M_{\rm 1}}-(\frac{\partial R_{\rm L}}{\partial t})_{M_{\rm 1}})]dt+
(\zeta_{\rm ad}-\zeta_{\rm L})d {\rm ln }M_{\rm 1}.
\label{eq:dr1vdrl}
\end{equation}

We assume that the mass donor exactly fills its Roche lobe during mass transfer. To ensure that mass transfer continues, $R_{\rm 1}$ needs to increase faster (or decrease slower) than $R_{\rm L}$, which requires the coefficient of $dt$ in Equation \ref{eq:dr1vdrl} to be positive. If the mass transfer rate is self-limiting, the second term on the right-hand side of Equation \ref{eq:dr1vdrl} needs to be negative to ensure that the stellar radius is parallel to the Roche lobe radius as mass transfer proceeds. 
Because $d {\rm ln} M_{\rm 1} < 0$, the self-limiting (stable) mass transfer only occurs when $\zeta_{\rm ad}>\zeta_{\rm L}$. In this case, mass transfer is driven typically on a timescale $R_{\rm 1}/{\dot M}_{\rm 1}$. 
For example, if the mass donor remains in thermal equilibrium, the mass transfer would proceed on a nuclear timescale. The donor would usually expand rapidly on its thermal timescale to accordingly proceed with the mass transfer on the thermal timescale if the mass donor is out of thermal equilibrium due to mass loss. In some circumstances, such as in short-period cataclysmic variables (CVs) or double-degenerate (DD) binaries, the driving timescale is characterized by the contraction of the Roche lobe induced by angular momentum losses.

Dynamical timescale mass transfer occurs when $\zeta_{\rm ad}<\zeta_{\rm L}$, in which both terms on the right-hand side of Equation \ref{eq:dr1vdrl} are positive. The radius of the mass donor is therefore driven by simple hydrostatic relaxation far beyond its Roche lobe radius. Three-dimensional (3D) simulations of this process (e.g. \cite{Rasio1996ApJ...471..366R,Taam2006astro.ph.11043T,Ricker2008ApJ...672L..41R}) have shown that the envelope of the donor is tidally disrupted, engulfing the companion star while tidal torques from the lagging, distorted envelope plunge the companion and the core of the donor into orbits deeply embedded in that envelope. The binary then enters CEE as proposed by \cite{Paczynski1976IAUS...73...75P}. 

The response of stars to adiabatic mass loss based on the polytropic models with power-law equations of state (i.e., $p\propto \rho ^{(1+1/n)}$, where $p$ is pressure, $\rho$ is density and $n$ is the polytropic index) has also been investigated \cite{Hjellming1987ApJ...318..794H}, in which three models such as complete polytropes ($n=3$), composite polytropes ($n=3$ cores with $n=3/2$ envelopes), and condensed polytropes ($n=3/2$ envelopes with point mass cores) were considered. These three models are relevant for some MS stars, giant branch stars, and WDs, respectively.  
Generally, we model a fully convective star or a WD with a polytropic index of $n=1.5$. Such stars expand in response to mass loss, while its Roche lobe radii decrease if the donors are more massive than the accreting star. As a consequence, the mass donor in this case will overfill their Roche lobe by an ever-increasing amount, leading to dynamical mass transfer and the formation of a CE. The conservative mass transfer requires the critical mass ratio to be $q_{\rm c}\sim 2/3$. This means that the mass transfer would be dynamically unstable if the mass donor has a mass larger than 2/3 of the mass of its companion. Giant stars have degenerate cores and convective envelopes, and thus they are generally modeled with the composite polytrope ($n=3$ for the core and 1.5 for the envelope). The degenerate cores in giant stars may lead to an increase in the critical mass ratio substantially  (e.g. \cite{Hjellming1987ApJ...318..794H}). They conform to the formula as follows (see also \cite{Webbink1988covp.conf..403W}):
\begin{equation}
    q_{\rm c}=0.362+\frac{1}{3(1-M_{\rm c}/M_{\rm 1})}
\end{equation}
where $M_{\rm c}$ and $M_{\rm 1}$ are the core mass and the total mass, respectively, of the donor as RLOF begins.

The criteria of dynamical instability based on the polytropic models have been widely used in BPS studies of the formation of peculiar stars, but there are widely held theoretical misconceptions for this criteria as discussed in \cite{Podsiadlowski2001ASPC..229..239P}. On the theoretical side, several full binary evolution calculations have shown that this criterion of dynamical instability based on the polytropic models in most BPS studies is not really appropriate. The major argument comes from the mass transfer in binary systems containing a giant star. For example, there are evidence that mass transfer is dynamically stable for all giants up to a mass of $\sim 2 M_\odot$ in the case of (sub-)giants transferring mass to a neutron star of 1.3/1.4 $M_\odot$ (e.g. \cite{Tauris1999A&A...350..928T,Podsiadlowski2002ApJ...565.1107P,Podsiadlowski1994AIPC..308..403P}) for an earlier example involving massive stars). Meanwhile, it has long been clear that quite a few systems that should experience dynamical mass transfer and a CE phase appear to be able to avoid it. Also, many observed intermediate-period binaries composed of compact components have been found to locate in the period gap predicted by the criteria from the polytropic models \cite{Podsiadlowski1992ApJ...391..246P}.

\begin{figure*}[t]
    \centering
    \includegraphics[width=0.9\textwidth, angle=0]{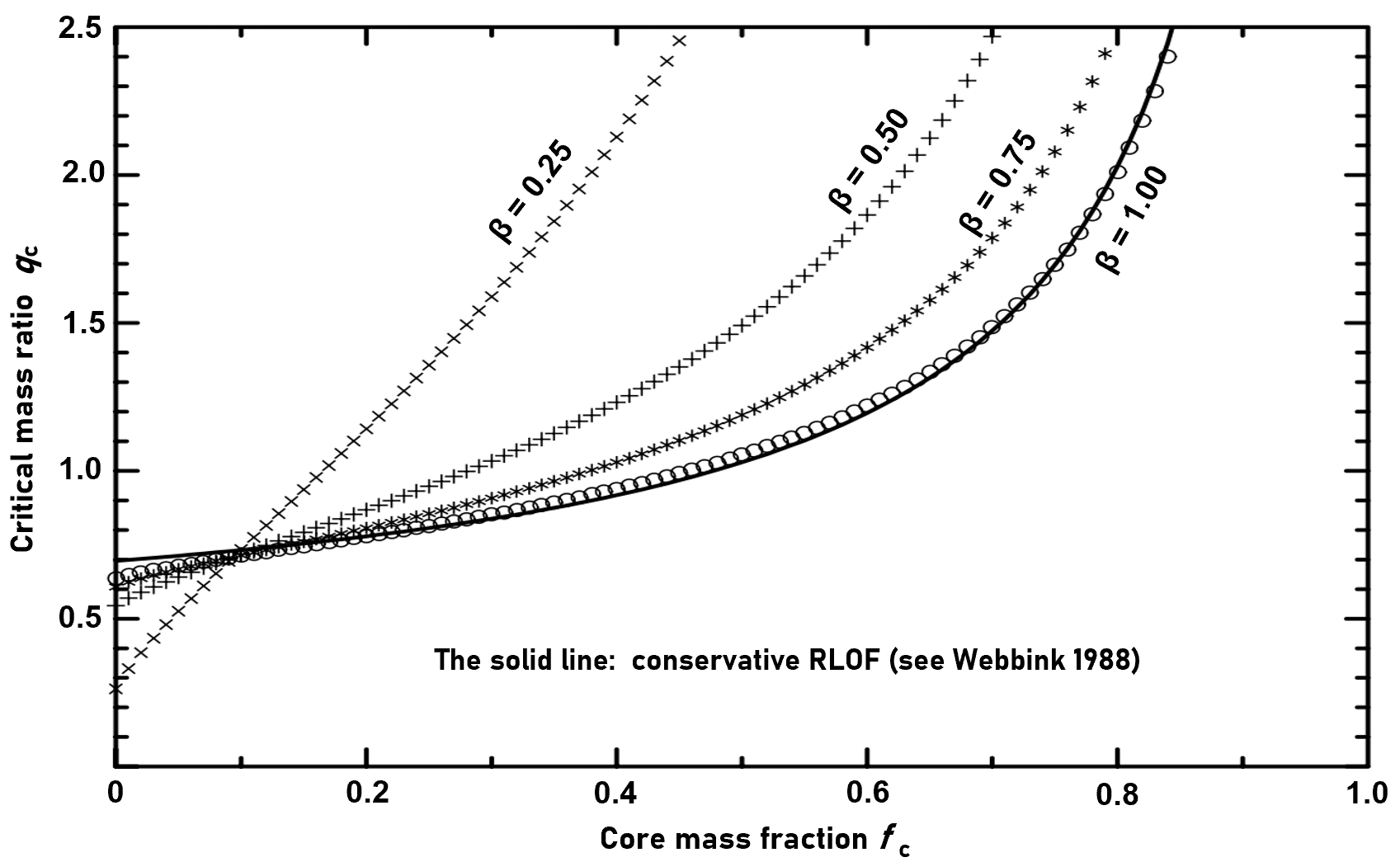}
    \caption{Dependence of critical mass ratio $q_{\rm c}$ on the core mass fraction $f_{\rm c}$ for dynamically unstable RLOF, based on the polytropic model. The cross, plus, asterisk and circle lines are for mass-transfer efficiency $\beta = 0.25$, 0.5, 0.75 and 1.0, respectively. The lost mass is assumed to carry away the same specific angular momentum as pertaining to the mass donor. The solid line is from \cite{Webbink1988covp.conf..403W} for conservative RLOF, i.e. Equation~(4.8) in the text. The figure is reproduced based on Figure~1 of \cite{Chen2008MNRAS.387.1416C}.}
    \label{fig:qc-polytrope}
\end{figure*}

There are some severe simplications for the criteria from the polytropic models such as that the non-conservative mass transfer and mass loss due to a stellar wind prior to the onset of mass transfer may significantly reduce the mass of the giant and increase the fractional mass of the degenerate core. This mass loss could be significantly enhanced due to the tidal interaction with the companion \cite{Eggleton1989ApJ...347..998E}. 
A fitted formula for Roche lobe mass–radius exponent was given by  \cite{Soberman1997A&A...327..620S} based on the assumption that the lost mass carrying away the same specific angular momentum as pertaining to the mass donor, i.e. 
$\zeta_{\rm L}=-1.7\beta +(2.4\beta-0.25)q$, 
where $\beta$ is the mass fraction of the lost mass from the primary accreted by the secondary. We may then obtain the critical mass ratio $q_{\rm c}$ by setting the adiabatic mass–radius exponent $\zeta_{\rm ad} =\zeta_{\rm L}$, where $\zeta_{\rm ad}$ can be fitted from the data of numerical calculations of (e.g. \cite{Hjellming1987ApJ...318..794H, Han2001ASPC..229..205H}). In Fig.~\ref{fig:qc-polytrope}, it clearly shows that there is a strong dependence of $q_{\rm c}$ on the value of $\beta$.
In the study of the formation channels of three double heilum WDs with known component masses and orbital periods\cite{Nelemans2000A&A...360.1011N}, the authors found that the orbital periods of the progenitors of helium WDs after the first mass transfer phase cannot be explained by either the standard CE ejection nor the stable conservative mass transfer. By employing non-conservative mass transfer in binary evolution, however, the other work ten years later \cite{Woods2012ApJ...744...12W} successfully reproduced evolutionary histories of these double helium WDs by employing non-conservative mass transfer.

The more fundamental problem of the criteria from the polytropic models may be that it does not consider the detailed dynamics of the mass transfer process. In particular, a substantial amount of mass may have already been lost during the turn-on phase before the dynamical instability occurs. On the other hand, the polytropic models leave many important evolutionary stages of a donor unexplored. A more sophisticated approach, using realistic stellar models, is both desirable and possible, as shown by \cite{Hjellming1989SSRv...50..155H,Hjellming1989PhDT.........7H}.

Considering the limitations of the criteria from the polytropic models, several detailed binary evolution simulations have been performed to derive the critical mass ratio $q_{\rm c}$ for special proposals such as the formation of sdB stars and blue stragglers with long orbital periods \cite{Han2002MNRAS.336..449H,Chen2008MNRAS.387.1416C}. The results of these studies demonstrated that mass transfer may be dynamically stable even if the mass donor is substantially more massive than the secondary. For each zero-age MS star, the critical mass ratio tends to decrease for the systems that evolved more initially. This is because the evolutionary timescale in such systems is shorter, hence giving higher mass transfer rates than those of the less-evolved systems. Since the core is not completely degenerate and the envelope is not fully convective when the star is at (or near) the base of the FGB, $q_{\rm c}$ at this evolutionary phase is much larger than that in the following evolutionary phases for each given mass-transfer efficiency $\beta$. Typically , both sdB stars and blue stragglers are formed from RLOF during which giants stably transferred material to MS or WD companions by following the unique mass-period relation (which is common in wide pulsars with WD companions; \cite{Rappaport1995MNRAS.273..731R}) without any non-standard assumptions (\cite{Chen2009MNRAS.395.1822C,Chen2013MNRAS.434..186C}. This has been confirmed with the discoveries of numerous long-period sdBs with M-type dwarfs and blue stragglers with hot WD companions, and their well-determined periods as well \cite{Vos2019MNRAS.482.4592V}.

Depending on whether the donor experiences overflow through its outer Lagrangian point, a new criterion for mass transfer instability was proposed by \cite{Pavlovskii2015MNRAS.449.4415P}, that is, 
the dynamical mass transfer is unavoidable when the radius of the mass donor overfills its outer Lagrangian point $L_2$, since the specific angular momentum taken away by the lost material is likely as that pertains to $L_2$ and the orbit will shrink dramatically due to this.  
They described an optically thick nozzle method to calculate the mass-transfer rate via the Lagrangian point $L_{\rm 1}$, and with which one can evolve binary systems for a substantial Roche lobe overflow. They performed a series of calculations with donor mass ranging from $1\, M_\odot$ to $50\,M_\odot$, finding that the critical initial mass ratio for stable conservative mass transfer varies from 1.5 to 2.2 under the new criterion, which is about twice as large as previously believed if the donor star has a well-developed outer convective envelope. The critical mass ratio might be even larger if the stars are underdeveloped giants and have shallow convective envelopes. As the growth of the convective envelope, and particularly for most cases of massive donors, the critical mass ratio gradually decreases from that of radiative donors (normally larger than 3) to a value of $1.5-2.2$. 
By using the new criteria, the group further investigated mass transfer between massive donors and compact companions\cite{Pavlovskii2017MNRAS.465.2092P}, which was previously postulated to have dynamically unstable mass transfer to subsequently lead to the CEE. However, they found that the mass transfer is stable for a large range of binary orbital separations as expected. This significantly reduces the merger rate of double BHs predicted by BPS calculations, and then reconciled the theoretical rate with the empirical rate obtained by LIGO, 9-240 ${\rm Gpc}^{\rm -3} {\rm yr}^{\rm -1}$. Moreover, they found that the new stability may lead to the formation of ultraluminous X-ray sources, and the predicted rates of bright ultraluminous X-ray sources which are expected to be powered by stellar mass BHs, are high enough to explain the observations.

Following the works of \cite{Hjellming1987ApJ...318..794H}, further significant progress on the stability of mass-transfer has been made through the establishment of the adiabatic mass-loss models (e.g., \cite{Ge2010ApJ...717..724G,Ge2015ApJ...812...40G,Ge2020ApJ...899..132G,Ge2020ApJS..249....9G}). Compared with the studies of \cite{Hjellming1987ApJ...318..794H}, the adiabatic mass-loss models are based on realistic stellar models rather than polytropic models. The response of the donors to very rapid mass loss was examined by constructing model sequences, which begin with a donor filling its Roche lobe at an arbitrary evolutionary point. The specific entropy and composition profiles are fixed while the mass is removed from the donor surface. Meanwhile, the stellar interior remains in hydrostatic equilibrium and the luminosity profiles are reconstructed from the specific entropy profiles and their gradients. 
These adiabatic mass-loss sequences have been used to quantify threshold conditions for dynamical timescale mass transfer in a binary system. They found that dynamical instability occurs early in mass transfer for donor stars with surface convection zones of any significant depth. For MS stars with radiative envelopes, however, dynamical instability commonly occurs after a prolonged phase of thermal timescale mass transfer ( which is known as ``delayed dynamical instability''). Finally, the critical mass ratios for the onset of dynamical timescale mass transfer were obtained when the adiabatic response of the stellar radius of the donor to mass loss matches that of its Roche lobe at some point during the mass transfer. An example of such calculations has been given in Fig.~4 of \cite{Ge2015ApJ...812...40G}).

Based on the new established adiabatic models, investigated the adiabatic mass loss sequences of Population I stars ($Z= 0.02$) with masses ranging from $0.10M_\odot$ to 100 $M_\odot$  have been investigated in detail by covering their evolutionary stages from the ZAMS to the tip giant branches/AGB (e.g., \cite{Ge2015ApJ...812...40G,Ge2020ApJS..249....9G}). In these works, the logarithmic derivatives of the radius with respect to mass along adiabatic mass-loss sequences have been translated into critical mass ratios for dynamical timescale mass transfer under the assumption of conservative mass transfer. The results were found to agree well with the behavior of time-dependent models presented by \cite{Chen2008MNRAS.387.1416C} for intermediate-mass stars that initiate mass transfer in the Hertzsprung gap. The main results of these works can be summarized as follows: (i) For intermediate- and high-mass donors with radiation envelopes, dynamical mass transfer is preceded by an extended phase of thermal timescale mass transfer as most of the envelope of the donor has been stripped. Most intermediate- or high-mass binaries with non-degenerate accretors probably evolve into contact before manifesting the dynamic instability. The critical mass ratio $q_{\rm c}$ increases with the advancing evolutionary age of the donor star. As these stars approach the base of the giant branch and begin developing a convective envelope, $q_{\rm c}$ drops dramatically to values of order unity, and a prompt dynamical instability occurs. (ii) For low-mass stars, the prompt instability prevails throughout MS evolution, with $q_{\rm c}$ declining with decreasing mass, and asymptotically approaching 2/3, close to that of a classical isentropic n = 3/2 polytrope. (iii) For RGB and AGB stars, the mass transfer tends to be more stable than previously believed,  which is similar to that given by \cite{Pavlovskii2015MNRAS.449.4415P}. This may be helpful to explain the abundance of observed post-AGB stars with orbital periods of around 1000 days.

The results presented above (\cite{Ge2015ApJ...812...40G,Ge2020ApJS..249....9G}) were given under the assumption that the donor response to rapid mass loss is an adiabatic expansion throughout the interior of the donor to its surface. This approximation is valid throughout the bulk of the interior once the mass-loss rate significantly surpasses the thermal timescale rate, but it may break down near the stellar photosphere, where radiative relaxation becomes extremely rapid. More significantly, it is crucial to notice that the growth to supra-thermal mass transfer rates generally extends over many thermal timescales, while the donor response even in these circumstances is still approximated to being purely adiabatic. For RGB and AGB stars with moderately deep surface convection zones, thermal relaxation during this acceleration phase is probably of little consequence because convection zones tend to respond as coherent entities (specific entropy rises or falls more or less uniformly throughout the convection zone), but more significantly, because such stars are subject to prompt dynamical instabilities that cut short the acceleration phase. Binaries containing these kinds of donors may therefore evolve into a CE phase even when donor stars are dynamically stable. This can be explained if the thermal timescale of these donor stars is short enough, and the rapid mass transfer process caused the donor star to overfill its Roche lobe by a significant margin (10\%–20\% of its radius in some cases), even overfilling its outer Lagrangian point. If the donor star is transferring material to a degenerate companion, the latter may not accrete all transferred mass quickly, leading to the formation of an extended structure. Consequently, both of the companions will overfill their Roche lobes, and a CE forms even on thermal timescales. The condition for this to occur requires thermal-equilibrium mass-loss models, which can be used to study the response of donor stars suffering a thermal timescale mass transfer. Furthermore, by establishing the thermal-equilibrium mass-loss models to derive the critical mass ratios for both thermal-timescale mass transfer and unstable mass transfer, finding that unstable mass transfer occurs when the outer Lagrangian point, $L_{\rm 2}$, is overfilled \cite{Ge2020ApJS..249....9G}.

\begin{figure*}
    \centering
    \includegraphics[width=0.95\textwidth, angle=0]{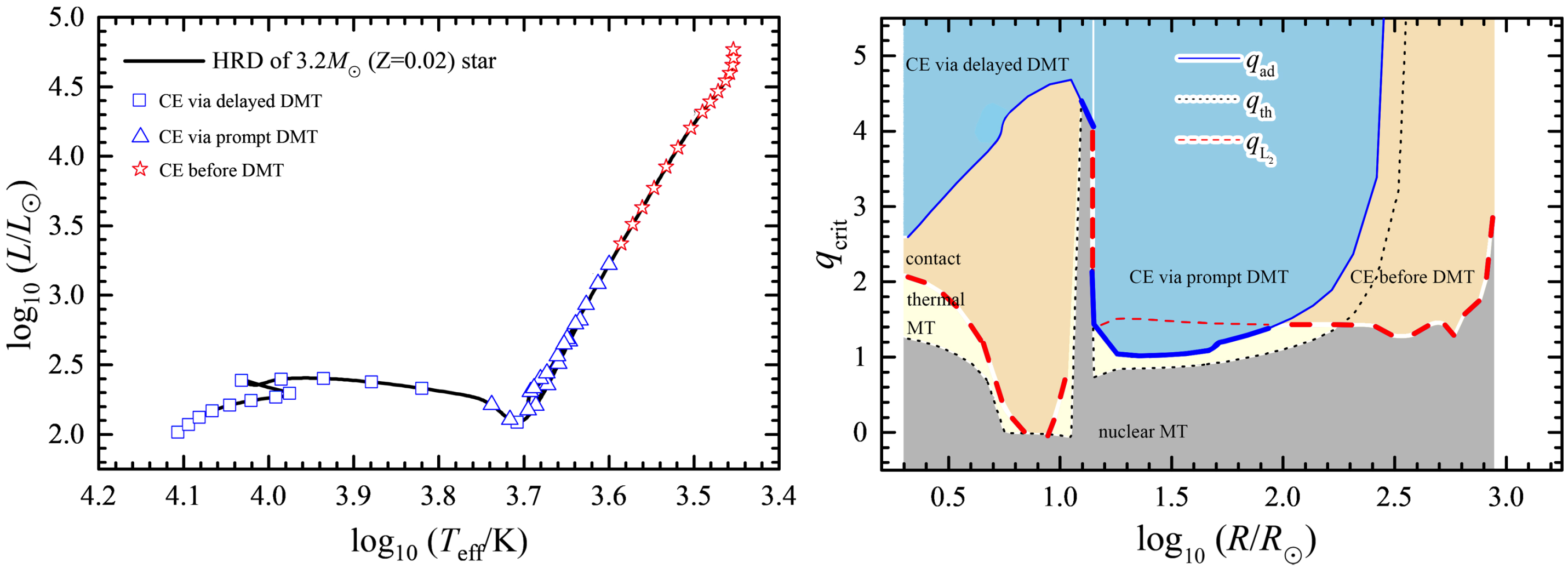}
    \caption{Hertzsprung Russell diagram (upper panel) for a $3.2M_\odot$ star and possible evolutionary channels (marked with colors) on the mass ratio vs. stellar radius plane (lower panel) for binary systems containing a $3.2M_\odot$ donor star. Three critical initial mass ratios derived from various models are presented in the bottom panel with different line styles, that is,  $q_{\rm ad}$, $q_{\rm th}$, and $q_{\rm L2}$, for dynamical-timescale mass transfer, thermal-timescale mass transfer, and overfilling of the outer Lagrangian surface $L_{\rm 2}$, respectively. The thick red dashed line and thick solid blue line show the more effective constraint on the mass ratio which may lead to a common-envelope phase, and the corresponding kinds of CE are indicated in the upper panel with various symbols. The red squares correspond to the thick red dashed line on the left of the lower panel and indicate the CE possibly happened in MS and HG donors. The blue triangles correspond to the thick solid blue line in the middle of the lower panel and show the possible CE that happened in RGB and early AGB donors. And the red stars reveal the possible CE of the late AGB donors, and they correspond to the thick red dashed line on the right of the lower panel. For clarity, the stellar models at evolutionary phases in which the donor star’s radius is smaller than in a preceding phase, such as during core helium burning, or just beyond the terminal main sequence. The figure is reproduced by H. Ge based on Figure~7 of \cite{Ge2020ApJS..249....9G} \copyright AAS. Reproduced with permission.}
    \label{fig:q-th-dy}
\end{figure*}

The possible mass-transfer channels of binary systems containing a $3.2\,M_\odot$ donor star in calculations with  the thermal-equilibrium mass-loss models  have been shown in Fig.~\ref{fig:q-th-dy} (see also Fig.~7 of \cite{Ge2020ApJS..249....9G}), where thermal-timescale mass transfer, unstable mass transfer through $L_{\rm 2}$, and dynamical-timescale mass transfer are all taken into account. A comprehensive study for such simulations for a grid of donor stars with different masses (from 0.1 to 100 $M_\odot$ with $Z=0.02$) at different evolutionary stages has been done by \cite{Ge2020ApJS..249....9G}), showing that unstable mass transfer due to the overfilling of the outer Lagrangian point may also play an essential role in the formation of CEs for late RGB and AGB donors. 

\subsection{The common envelope evolution}
\label{sec:CEE}

As mentioned before, the common envelope evolution (CEE) has been first proposed in 1976 to account for the origin of cataclysmic variables (CVs).  CVs are binaries consisting of a WD with a typical mass of around $0.8\,M_\odot$ and an MS companion star with a typical mass of $0.6\,M_\odot$, in which the MS star fills its Roche lobe and undergoes the mass transfer. They typically have an orbital period of 1 to 10 hours. Regarding the origin of CV systems, the question is how to form such massive WDs in a such close binary system with an orbital period of a few hours. According to the stellar evolution theory, massive WDs could be formed from cores of red giants (or supergiants) before their envelopes are removed. In this case, the orbital period of a binary system needs to be long enough to keep the red giant (or supergiant) within its Roche lobes until a massive core is formed. Then, a mechanism is needed to drastically reduce the orbital period to a few hours consequently to match the observed periods of CVs. This therefore leads to the proposal of CEE (e.g. \cite{Ostriker1973ApJ...180..171O,Webbink1975MNRAS.171..555W,Paczynski1976IAUS...73...75P}). By adopting the CE scenario, the origin of V~471~Tau was successfully explained \cite{Paczynski1976IAUS...73...75P}, i.e., a detached binary in the Hyades cluster which has a $0.8\,M_\odot$ WD and a $0.8M_\odot$  K-type dwarf, with an orbital period of $12\,\rm{h}$. This binary system is expected to evolve into a CV system within a few $10^{10}$ yr.

In the following years, a phenomenological description based on energy conservation (known as $\alpha$-mechanism) has been proposed to deal with the crucial CE process in binary evolution (e.g. \cite{van-den-Heuvel1976IAUS...73...35V,Webbink1984ApJ...277..355W}). A brief introduction to the $\alpha$ scenario has been given at the beginning of this section (the second paragraph in Sec.\ref{sec:progress}). In this scenario, an efficiency of CE ejection, $\alpha_{\rm CE}$, is defined as the fraction of reduced orbital energy $\Delta E_{\rm orb}$ that is used in ejecting the CE. The CE will be successfully ejected when $\alpha_{\rm CE}$ fulfils the below condition:

\begin{equation}
\alpha_{\rm CE}(\frac{GM_{\rm 2}M_{\rm c}}{2a_{\rm f}}-\frac{GM_{\rm 1}M_{\rm 2}}{2a_{\rm i}}) \geq \frac{GM_{\rm 1}(M_{\rm 1}-M_{\rm c})}{\lambda R_{\rm 1}}
\label{eq:alpha}
\end{equation}
where $M_{\rm 1}$ and $M_{\rm c}$ are the mass of the donor and its core mass at the onset of the CE; $M_{\rm 2}$ is the mass of the secondary, and G is the gravitation constant. The right-hand term of the equation is the binding energy, and $\lambda$ is the structure parameter of the envelope.

The formula of $\alpha$-mechanism and its modifications that will be introduced below have been used throughout all BPS studies for the CEE of binary systems. However, there are still uncertainties about this formula. For example, (i) The exact values of $\alpha_{\rm CE}$ and $\lambda$ are still unknown. (ii) The exact boundary between the envelope and the core of the star (especially for the slightly evolved star) is poorly constrained. (iii) Whether the internal energy (or enthalpy) needs to be included in the binding energy is unclear. (iv) Whether there are additional energy sources in addition to the orbital energy contributing to the CE ejection. However, some constraints on these uncertainties can still be placed by comparing theoretical predictions of BPS calculations to the observations. Typically, the value of $\alpha_{\rm CE}$ needs to be larger than unity (sometimes as large as 3, but which is unphysical) to reproduce the observational rates of some particular stars such as SNe Ia (e.g. \cite{Toonen2013A&A...557A..87T,Claeys2014A&A...563A..83C}).

Besides the $\alpha$-description (i.e. the standard energy mechanism), there are also other mechanisms that have been proposed for the CEE in binary evolution. To explain the origin of three double He~WDs with well-determined orbital parameters and component masses, different formation scenarios were investigated (i.e. CE+CE and RLOF+CE channels) in the framework of the standard energy mechanism \cite{Nelemans2000A&A...360.1011N}. However, none of the reasonable choices for $\alpha$ and $\lambda$ were found to successfully reconstruct the formation of these three objects. They therefore proposed a new mechanism for the CEE description based on angular momentum conservation, which is referred to as ``$\gamma$-mechanism''. The $\gamma$-mechanism assumes that the entire envelope of the donor is lost completely from the binary system, taking away the angular momentum of the system in a linear way. The $\gamma$-mechanism can be written as: 
\begin{equation}
\frac{\Delta J}{J}=\gamma \frac{\Delta M}{M_{\rm tot}}
\end{equation}
where $\Delta J$ is the decrease of the angular momentum due to the CE ejection and $\Delta M$ is the envelope mass lost from the binary; $J$ and $M_{\rm tot}$ represent the total angular momentum  mass of the binary before the formation of a CE. Furthermore, by using the $\gamma$-formula for the first unstable mass transfer and the $\alpha$-mechanism for the second CEE process, the evolutionary histories of three double He WDs successfully reconstructed with $\gamma$ in the range of 1.4 to 1.7 \cite{Nelemans2000A&A...360.1011N}. As mentioned before, the completely non-conservative mass transfer can also explain the formation of the three double He~WD systems. This is not surprising because the $\gamma$ formula is actually initiated from the general case of non-conservative mass transfer proposed by \cite{Paczynski1967AcA....17....7P}.

Based on the fact that massive red supergiants (RSGs) have a sizable radiative layer between the dense He core and the convective envelope, a new simple formalism for the CEE was recently proposed to predict the orbital separations after the CE phase for massive binary systems \cite{hirai2022}. In this new formalism, the CE phase is separated into two stages, which are dynamical inspiral through the outer convective envelope and thermal timescale mass transfer from the radiative intershell. As expected, this new formalism typically predicts much wider separations in comparison to the standard energy mechanism with fiducial choices of parameters. Meanwhile, post-CE binary orbital separations predicted by this new formalism strongly depend on the actual evolutionary stage of the donor star and its companion mass. This new formalism also provides a physically motivated alternative option for the treatment of CEE in BPS studies, especially for massive binary systems.

All the above-mentioned descriptions for CEE are proposed especially for binary evolution, which cannot be directly applied to the study of triple systems and higher-order multiples. Recently, a new CE formalism was presented to deal with the CEE in triple systems and higher-order multiples,  which is referred to as ``\emph{Single Components’ Angular momenTum TransfER (SCATTER)}'' \cite{stefano2022}. In this formalism, each mass in the system transfers angular momentum to material in the CE. Similar to that of the $\gamma$ formalism, the authors have not considered the exact mechanism that ejects the envelope. They focused entirely on the angular momentum transferred from or to the individual stellar components of the multiple-star system. In addition, they assumed that it is the transfer of angular momentum to and from the individual stars that change their orbits, which allows us to establish a mapping between an orbit at the start of the CE and the orbit at the end of the CE phase.

\subsubsection{The binding energy $E_{\rm bind}$}
\label{sec:Eb}

The internal energy ($E_{\rm th}$) was first included in the calculations for the binding energy ($E_{\rm bind}$) of the envelope of an RGB/AGB star (see \cite{Han1994MNRAS.270..121H} and Equation~2 in that paper).  Since then, it has been widely adopted by different BPS studies for the formation of binary-related objects that have undergone the CEE. If we define $\alpha_{\rm th}$ as the thermal contribution to the ejection of the envelope, the condition for the CE ejection could be written as follows:
\begin{equation}
\alpha_{\rm CE} |E_{\rm orb}| \geq |E_{\rm gr}+\alpha_{\rm th} E_{\rm th}|
\label{eq:interU}
\end{equation}

 By using this CE-ejection condition, a binary model was developed for the formation of sdBs and successfully reproduced the observational properties of sdBs with short orbital periods (e.g. \cite{Han2002MNRAS.336..449H,Han2003MNRAS.341..669H}). In addition, it was found that the best fitting CE-ejection parameters for these sdBs are $\alpha_{\rm CE}=0.75$ and $\alpha_{\rm th}=0.75$. It should be noticed that the internal energy here involves contributions of different terms, including the ionization of H, the dissociation of ${\rm H_2}$, the basic $\frac{3\Re T}{2\mu}$ for a simple perfect gas, and the Fermi energy of degenerate electron gas, although they may work in different regions of the CE and/or at evolutionary phases during the CEE (e.g. \cite{Ivanova2013A&ARv..21...59I}).  The inclusion of internal energy during the CEE can make a significant difference in the energetic ease of the CE ejection for some evolutionary phases of binaries, especially when the donors are on the upper part of giant branches where the internal energy becomes comparable to that of the gravitational potential energy.

Because that the internal energy can be combined within the parameter of $\lambda$, Equation~\ref{eq:alpha} can be modified as 
\begin{equation}
    \alpha_{\rm CE}\lambda \Delta E_{\rm orb} \geq \frac{GM_{\rm 1}(M_{\rm 1}-M_{\rm c})}{R_{\rm 1}}
\label{eq:alpha-mod}
\end{equation}
where the combined parameter, $\alpha_{\rm CE}\lambda$, is used in BPS studies in general. By systematically calculating the binding energy for both Population~I and Population~II stars with masses of $1-20\,M_\odot$, the study has obtained the values of $\lambda$ in cases with or without the contribution from the internal energy accordingly \cite{Xu2010ApJ...716..114X}. It is found that $\lambda$ is generally around 1.0 for low-mass MS stars in calculations without including the internal energy, and it drops significantly when these stars evolve into giant branches. In addition, $\lambda$ is far less than 1.0 for intermediate and massive stars when they are on the MS, and $\lambda$ becomes  $\lesssim 0.1$ when the stars leave off the MS. However, the inclusion of internal energy makes the $\lambda$ value increase dramatically, particularly when the stars evolve to the upper parts of giant branches.

Different BPS studies in the literature have commonly used Equation \ref{eq:alpha-mod} for the CEE in binary evolution calculations, and $\alpha_{\rm CE}\lambda$ is generally set to be 1. However, if we use the accurate $\lambda$ values determined from detailed stellar structure calculations for massive stars as introduced above, the choice of $\alpha_{\rm CE}\lambda \approx 1$ would give a very large and unphysical value of $\alpha_{\rm CE}$. This leads to difficulties in the formation of some particular stars such as BH low-mass X-ray binaries composed of a BH and a low-mass companion of $<2.0\,M_{\odot}$. This is because the low-mass companion stars are not likely to provide the sufficient gravitational potential to unbind the envelope of the massive progenitor of the BH during a prior CE phase, eventually leading to a merger during the CEE rather than a BH low-mass X-ray binary. It is suggested that this difficulties may be relieved if magnetic braking of Ap/Bp stars is applied to BH LMXBs \cite{Justham2006MNRAS.366.1415J}. Interestingly, it has been noticed that, in some cases, stellar envelopes undergo stationary mass outflows likely during the slow spiral-in stage of the CE event \cite{Ivanova2011ApJ...731L..36I}. As a result, the condition for such outflows was proposed \cite{Ivanova2011ApJ...731L..36I}. This condition is in a manner similar to Equation \ref{eq:interU} but with an addition of $P/\rho$ term in the energy balance equation, accounting therefore for the enthalpy of the envelope rather than the internal energy. This may further reduce the $\alpha_{\rm CE}$ values required for the study of the formation of BH low-mass X-ray binaries \cite{Ivanova2011ApJ...731L..36I}.

The exact boundary between the core and the envelope of the donor star in a binary system could also significantly affect the final results of the CEE. In the literature, the mass fraction of H of $ X=0.1$, 0.01, or 0.0001 has been used to define the core and the envelope. However, there is no consensus on the exact mass fraction of H for the boundary between the core and the envelope of the star. On the one hand, there is a sharp decrease in both density and H abundance profiles at the interface of the core and the envelope of an evolved star. In this case, adopting different H mass fractions could result in a change of $E_{\rm bind}$ up to several orders of magnitude. On the other hand, unevolved (or slightly evolved) stars have relatively flatter and smoother profiles of density and H abundance. As a consequence, different X values would lead to large differences in the position where the companion can be involved, thus giving large differences of $\Delta E_{\rm orb}$. To give consistent results, previous studies defined the boundary a bit  outside the interface of the core and the envelope \cite{Han2003MNRAS.341..669H,Han2004MNRAS.350.1301H}. In this way, the binding energy would not be affected significantly due to the offset of the definition among different stellar models. 
By studying the evolution of post-CE remnant, a methodology for unambiguously defining the post-CE remnant mass after it has been thermally readjusted was proposed \cite{Ivanova2011ApJ...730...76I}, namely, the core boundary, which is the radius in the hydrogen shell corresponding to the local maximum of the sonic velocity. Very recently,  the final stages of the CEE for massive stars as progenitors of neutron star (NS) binaries have been explored \cite{Vigna-Gomez2022MNRAS.511.2326V}. By considering an instantaneously stripped donor star as a proxy for the CEE, 1D single-stellar evolution was used to investigate the subsequent radial evolution of the star. Finally, a range of stripping boundaries that allow the star to avoid significant rapid re-expansion were determined, which have been used to represent plausible boundaries for the termination of the CEE. These boundaries have been found to lie above the maximum compression point, a commonly used location of the core/envelope boundary. This gives a conclusion of that massive donors may retain fractions of a solar mass of H-rich material even after the CEE. This further suggests that all of the models would overfill their Roche lobes after ejecting the envelope if the orbital energy is the only energy source available.

It has been pointed out that, due to the fact that the stellar core and envelope contribute mutually to each other’s gravitational potential energy, proper evaluation of the total energy of a star requires integration over the entire stellar interior, and not just over the ejected envelope alone as commonly assumed \cite{Ge2010ApJ...717..724G}. Therefore, the change in total energy of the donor star as a function of its remaining mass along an adiabatic mass-loss sequence can be calculated either by integration over initial and final models or by a path integral along the mass-loss sequence. Moreover, it has been suggested that the combination of the change in total energy of the donor star with the requirement that both the remnant of the donor and its companion star fit within their respective Roche lobes can circumscribe energetically possible survivors of the CEE (e.g., \cite{Ge2010ApJ...717..724G,Deloye2010ApJ...719L..28D}).

More recently, a new method based on the adiabatic mass loss model has been further developed to calculate the binding energy of the donor star $E_{\rm bind}$ by tracing the change of total energy \cite{Ge2022ApJ...933..137G}. In this new method, the binding energy is calculated by integrating the binding energy from the core to the surface, and it assumes that the outcome of CEE is determined by energy conservation and when both components shrink within their Roche lobes. The results with this new method have shown that the remnant core radius can vary by orders of magnitude although the remnant core mass is almost the same. Also, the remnant core mass is slightly larger than the He core, as shown in Fig.~4 of \cite{Ge2022ApJ...933..137G}. These two methods have been further applied to 142 sdB binaries to estimate the CE efficiency parameters. For shorter orbital-period sdBs, it has been found that the binding energies ($E_{\rm bind}$ ) given by the two methods are almost the same, and the difference in CE efficiency parameter between the two methods is very small. For longer orbital-period sdBs, the binding energy can differ by up to a factor of 2, and the CE efficiency parameter derived from the new method becomes smaller compared with that given by the previous method when the final orbital period ${\rm log} P {\rm (d)} > -0.5$ \cite{Ge2022ApJ...933..137G}.

\subsubsection{The parameter of common envelope efficiency $\alpha_{\rm CE}$}
\label{sec:efficiency}

 Whether we can obtain an accurate value of $\alpha_{\rm CE}$ is one of the major challenges to the standard $\alpha-$mechanism for the CEE. The value of $\alpha_{\rm CE}$ should be less than 1.0 in physical. However, BPS studies need to set it to be larger than 1 to reproduce observational properties of some peculiar objects such as the observationally inferred rates of BH low mass X-ray binaries. According to the physical meaning of the $\alpha_{\rm CE}$ itself, it is believed that its accuracy depends on the companion type and detailed structures of the envelope, because both of which would significantly affect the friction efficiency between the inner binary and the envelope. Despite substantial efforts that have been done, constraints on the $\alpha_{\rm CE}$ value are still quite poor from the theoretical side. As growing surveys have been carried out in the new Menilium, increasing numbers of various types of binaries that are expected to be produced from the CE ejection (i.e. Post-CE binaries) have been discovered, including tight WD+MS/BD binaries, binary central stars in planetary nebulae, CVs, double WDs, and sdB binaries with short periods. This leads to that constraining the value of $\alpha_{\rm CE}$ from these post-CE binaries becomes popular in the community. In particular, the WD+MS/BD binaries are the most extensively investigated because the different properties of a large number of such objects have been well-characterized (e.g. \cite{Raymond2003AJ....125.2621R,Silvestri2006AJ....131.1674S,Heller2009A&A...496..191H,Zorotovic2010A&A...520A..86Z,Zorotovic2011A&A...536L...3Z,Zorotovic2011A&A...536A..42Z,Zorotovic2014A&A...568A..68Z,Zorotovic2022MNRAS.513.3587Z,Rebassa-Mansergas2012a,Rebassa-Mansergas2012b,Toonen2013A&A...557A..87T,Camacho2014A&A...566A..86C})

By investigating the different prescriptions for the CEE with analyzing a sample of PCEBs with known periods among WD+MS binaries from SDSS, it has been found that all the PCEBs in their sample could be reconstructed with the standard parametrized energy equation if the realistic internal energy of the envelope is included \cite{Zorotovic2010A&A...520A..86Z}. In addition, simultaneous solutions for all the PCEBs could be achieved with setting $\alpha_{\rm CE}=0.2-0.3$. This may suggest that the energy mechanism seems to place stronger constraints on the outcome of the CEE compared with that given by the alternative angular momentum balance. By further analyzing the orbital distribution of the PCEBs containing either He-core WDs or CO-core WDs, it was found that the orbital period distribution of PCEBs with He-core WDs has a median value of $0.28\,\mathrm{d}$, while those with CO WDs have a median value of $P=0.57$\,d \cite{Zorotovic2011A&A...536A..42Z}. This seems to indicate that the envelope of AGB stars could be ejected easier compared with that of RGB stars. Also, it has been found that the systems with more massive companions have longer orbital periods, which is consistent with the expectation from the energy equation. Meanwhile, the analysis of two WDMS binaries with relatively long orbital periods (i.e., SDSS J121130.94-024954.4 with $P=7.818\pm0.002$\,d and SDSS J222108.45+002927.7 with $P=9.588\pm0.002$\,d) had found that the long orbital periods in these two systems and the evolution of both systems can be well understood by using the energy mechanism without including the recombination energy \cite{Rebassa-Mansergas2012a}.

By performing BPS studies with the Seba code, the populations of PCEBs of the Galaxy, i.e., detached WDMS-binaries with periods less than 100\,d that underwent a CE phase were simulated \cite{Toonen2013A&A...557A..87T}. By incorporating the selection effects, these simulations could reproduce the statistical properties of the observed post-CE WDMS binaries such as space density, the fraction of visible PCEBs amongst WDMS binaries, and the WD mass versus MS mass distribution \cite{Toonen2013A&A...557A..87T}. However, it was found that the fraction of PCEBs with He~WDs was overestimated. This study showed that selection effects have a limited impact on the period distribution, and a low $\alpha_{\rm CE}$ value is required to explain the observed dearth of long-period systems of PCEBs from the SDSS. Similar results have also been found by other studies \cite{Camacho2014A&A...566A..86C}, in which they obtained that the models with a low value of CE efficiency (less than 0.3) and a moderate fraction (less than 10\%) of the internal energy seem to match the observations better. However, it seems that the distributions of orbital periods and WD masses of PCE WDMS binaries could also be reproduced with the combination of other free parameters and mass ratio distribution $f(q)$ when the observation biases are carefully considered. In addition, it was found that more binary systems containing a He WD are observed than what we expected. To particularly examine the influence of the contribution of the recombination energy to the CE ejection, it was found that a large number of the long orbital period ($>10$ d) systems mostly with high-mass WDs are expected to be observed if a fraction of recombination energy is included \cite{Zorotovic2014A&A...568A..68Z}. Also, the fraction of systems with He WDs increases with $\alpha_{\rm CE}$, and the very low-mass He WD can be produced only for a high CE efficiency of $\alpha_{\rm CE} >0.5$. They presented that all models on average predict longer orbital periods for binaries with CO-core WDs than those containing He WDs, and binary systems with more massive secondary stars usually have longer orbital periods after the CE ejection.

Very recently,  the evolutionary history of eight observed PCEBs containing a WD and a brown dwarf (BD) companions was investigated \cite{Zorotovic2022MNRAS.513.3587Z} by using the similar method previously adopted for WDWS binaries. This finds that a low $\alpha_{\rm CE}$  (0.24-0.41) is required to explain the formation of these WD+BD binaries, suggesting that the contribution from the recombination energy may be not necessary. It seems to conclude that the vast majority of PCEBs formed from the CEE can be parameterized with a small CE efficiency and without considering any additional energy sources \cite{Zorotovic2022MNRAS.513.3587Z}.

The majority of the WDMS sample has a low-mass M dwarf companion because these kinds of systems are relatively easier to be identified from optical colors and spectra compared with the WD binaries with an FGK-type companion. The two components in WD binary systems with a FGK-type companion have different optical brightness, leading to that the MS companions may completely outshine WDs in optical observations. This means that the combination of ultraviolet (UV) and optical observations is required for detecting such WD binaries. The WD binaries with FGK-type companions are more promising for the studies of some important objects such as SNe Ia progenitor \cite{Maoz2012ApJ...751..143M} compared with those with an M-dwarf companion. Despite efforts have been made to discover WD binaries with FGK-type companions (e.g. \cite{maxted09}), only a small sample of these systems has been obtained. In SDSS sample, FGK stars generally are observed at a very far distance at which almost all WDs (except for the very hottest one) could not be detected in UV surveys. Nevertheless, some previous studies of far-UV and soft X-ray spectra of MS stars (e.g. \cite{Landsman93,Landsman96,Barstow94,Vennes1995,Vennes1997,Chris1996,Burleigh97,Burleigh98a,Buleigh98b,Buleigh99}) have revealed the presence of WD binaries with FGK-type companions.  However, the vast majority of these systems are visually resolved binaries with long orbital periods, and such systems are very unlikely to experience the CEE during their formation.

To establish a large sample of PCEBs with early-type components,  a project of the White Dwarf Binary Pathways Survey (WDBPS) started \cite{parsons15}. This project aims at using the data from surveys targeted to the detection of bright FGK stars in combination with UV data from the Galaxy Evolution Explorer (GALEX) to identify MS stars with substantial UV-excess colors that are attributed to the existence of nearby WD companions. Follow-up spectroscopic observations are then used to distinguish the samples between the close PCEBs and wide systems that are unlikely to be produced from the CE ejection. Based on this survey, TYC~6760-497-1 is the first identified WD binary with an F-type star and an orbital period of 12\,h. This system is also the first known super-soft X-ray source, in which the WD is expected to grow up in mass through accretion from the companion star in its following evolution \cite{parsons15}. To date, there are a series of works about the WDBPS have been published (e.g. \cite{WDFGK-i,WDFGK-ii,WDFGK-iii,WDFGK-iv,WDFGK-v,WDFGK-vi,WDFGK-vii,WDFGK-viii,WDFGK-ix}), in which the discoveries of some WD binaries with a FGK-type companion have been reported, including five systems with G-type companions and one system with a K-type companion. According to their new discoveries, different studies have concluded that the CE efficiency $\alpha_{\rm CE}$ is likely to be comparable between close WD binaries with AFGK companions and PCEBs with M-dwarf companions, and their evolutionary histories can be well reproduced with $\alpha_{\rm CE}=0.2-0.3$ \cite{WDFGK-iv,WDFGK-vi}. However, six potential PCEBs containing a WD and an early-type companion star found by earlier works (i.e. IK Peg, KOI-3278, SLB1, SLB2, SLB3, and KIC 8145411) have much longer orbital periods  (which are $P\sim21$, 88, 683, 728, 419, and 455\,d, respectively). It was found that their formations can not be explained with such low $\alpha_{\rm CE}$ values. The origin of these six objects still needs to be further investigated \cite{WDFGK-iv}.

Binary central stars in planetary nebulae (PNe) have most likely descended from the CEE. Obviously, the orbital period distribution of binary central stars of PNe is a strong function of $\alpha_{\rm CE}$. Adopting a value of $\alpha_{\rm CE}=0.1$ given by the study of the post-CE eclipsing binary V471 Tauri and O'Brien \cite{OBrien2001ApJ...563..971O}, it has been shown that the orbital periods of binaries inside PNe are likely to be evenly distributed over a range of $0.3-30$\,d \cite{Yungelson1993ApJ...418..794Y}. However, other studies also  indicated a very high value of $\alpha_{\rm CE}$ in order to match the observations of PNe such as the fraction of PNe with close binary nuclei, and the birth rates of CVs and DDs \cite{Han1995MNRAS.277.1443H}, and suggested that the best-fitting model has CE parameters of $\alpha_{\rm CE}=1.0$ and $\alpha_{\rm th}=1$ with either a constant or a rising mass ratio distribution. Observationally, the photometric monitoring has revealed that about 10\% of PN nuclei (PNNi) are very close binaries with orbital periods from a few hours up to a few days (e.g. \cite{Bond1990ApJ...355..568B,Bond2000ASPC..199..115B}). Therefore, the orbital period distribution of central binaries is expected to give a deep insight into the value of $\alpha_{\rm CE}$.

\begin{figure*}[t]
    \centering
    \includegraphics[width=0.90\textwidth, angle=0]{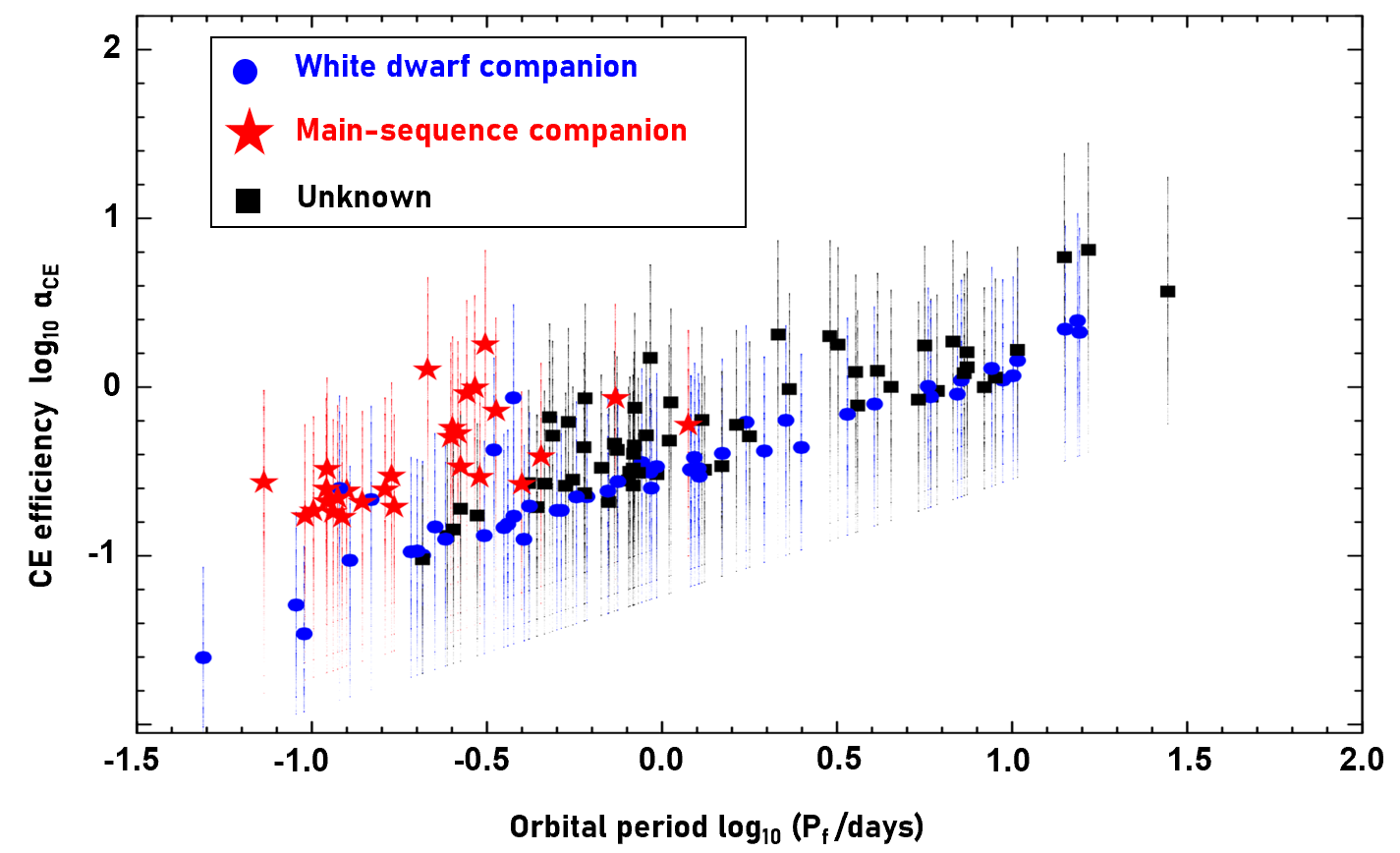}
    \caption{The derived $\alpha_{\rm CE}$ versus final orbital periods for a sample of sdB stars from \cite{Kupfer2015A&A...576A..44K}. Different colors are for different types of companions as indicated in the plot. The small dots are the results from various donor masses and core masses at the onset of the CEE while the big filled symbols are the mean values weighted by the Salpter's initial mass function and flat distribution of the separation in logarithmic. Inter energy has been included with $\alpha_{\rm th}=1.0$ in the study. The tendency that $\alpha_{\rm CE}$ increases with the final orbital period is a consequence of the method itself as described in the text. The figure shows a wide range of  $\alpha_{\rm CE}$ values which should be different for each individual binaries. 
In particular, the type of companion has a non-negligible effect on the $\alpha_{\rm CE}$ i.e. the main sequence companions obviously have larger $\alpha_{\rm CE}$ in comparison to the WD companions for similar final orbital periods.}
    \label{fig:p-alpha}
\end{figure*}

 An RV survey of PNNi with the WIYN 3.5~m telescope and Hydra spectrograph was initiated \cite{De-Marco2004ApJ...602L..93D}, which is aimed at discovering intermediate-period binaries. In this project, the short-period binary fraction is found to be 10–15\% from a survey of $\sim 100$ central stars for photometric variability indicative of irradiation effects, ellipsoidal variability, or eclipses. The orbital periods of all binaries discovered by this survey are less than three days. This is because that this survey technique is biased against the binaries with long orbital periods. This project further assessed the status of knowledge of binary central stars discovered because of irradiation effects (e.g. \cite{De-Marco2008AJ....136..323D}), and determined that, for average parameters, the method should be biased against periods longer than 1–2 weeks. Therefore, this project concluded that it is surprising that no binaries were found with periods longer than 3 days. Furthermore, it has been found that 9 out of 12 irradiated binaries have periods smaller than 1 day, which is at odds with the BPS predictions. This suggests that either all CE models tend to overestimate post-CE periods or this binary survey might have suffered from additional, unquantified biases. By detecting the central stars in the I band to search for cool companions, a new survey for binary-related variability was then introduced. This survey technique would give better-quantified biases and is independent of orbital periods. This leads to the determination of a debiased binary fraction of $67-78$ per cent for all companions at all separations, with uncertainty between 10 and 30 per cent. This fraction is consistent with that predicted by \cite{Han1995MNRAS.272..800H}. However, the distribution of orbital periods is lacking and thus cannot be used to further constrain the value of $\alpha_{\rm CE}$.

Using the existing photometry observations to cross with the sample of PNe would lead to more central stars being discovered.  For example, Miszalski et al. \cite{Miszalski2009A&A...496..813M} combined the photometry from the OGLE microlensing survey with the largest sample of PNe at that time towards the Galactic bulge to systematically search for new central binaries. A total of 21 periodic binaries were found, doubling the known sample of central binaries of PNe at that time. By comparing the observed period distribution of these newly discovered binaries with the BPS predictions,  they found that the model~3 from \cite{de-Kool1992A&A...261..188D} provides the best fit, in which $\alpha_{\rm CE}$ was set to be 1.0 and a random initial mass ratio distribution was adopted. The derived close binary fraction is 10-20\% in the sample, which agrees well with that given by previous photometric observations (10-15\%), and is consistent with the BPS predictions from the CE scenario (17\%; \cite{Han1995MNRAS.272..800H}).

Close double WDs and CVs may also provide some clues to the CEE. Nelemans et al.  BPS studies for double WDs were performed \cite{Nelemans2001A&A...365..491N}, in which the treatment of the first unstable mass transfer was based on $\gamma$-formula and the secondary unstable mass transfer using the standard energy mechanism. Given an initial binary fraction of $\sim 50\%$ and setting $\gamma \approx 1.75$ and $\alpha\lambda \approx 2$, a satisfactory agreement between theoretical predictions and observations of the local sample of double WDs were obtained \cite{Nelemans2001A&A...365..491N}. Two planets are discovered to orbit around the CV the CV-system HU~Aquarii (i.e., HU~Aqr; \cite{Portegies-Zwart2013MNRAS.429L..45P}). The measured eccentricities and orbital separations of the planets in HU~Aqr enable us to place constraints on the CEE. By reconstructing the evolution of the CV system in HU~Aqr, which gives the CE parameters of $\alpha \lambda =0.45\pm0.17$ or $\gamma=1.77\pm0.02$ \cite{Portegies-Zwart2013MNRAS.429L..45P}.

Most current studies reviewed above give large uncertainties on the values of $\alpha_{\rm CE}$ and/or $\alpha_{\rm CE}\lambda$ since the details of  the donors before the CEE are poorly known. For example, the progenitors of WDs may spread on the whole giant branches, along which the detailed structures of the envelope of the star could be different significantly, leading to the determined values of $\alpha_{\rm CE}$ and/or $\alpha_{\rm CE}\lambda$ have a wide range. One promising way to give strong constraints on CE parameters is to have a binary sample in which the properties of progenitors are known clearly. The sdBs with short orbital periods are potentially excellent objects for limiting CE parameters strictly (in particular those in clusters) due to three advantages: (i) Detailed binary evolution calculations have shown that the short-orbital-period sdBs are mainly produced from the CEE when their progenitors are at the tip of RGB. This dramatically reduces uncertainties of parameter spaces for progenitors of the donors. (ii) The internal energy becomes comparable to that of the gravitational energy only when the stars are at the top of giant branches. (iii) If sdBs are in clusters, their age and progenitor mass could be well determined.    

For example, Fig.~\ref{fig:p-alpha} presents strong constraints on the $\alpha_{\rm CE}$ values by using the sdB samples from \cite{Kupfer2015A&A...576A..44K}. There are three subtypes of sdBs in the catalog of \cite{Kupfer2015A&A...576A..44K} according to the type of companions (i.e. WDs, dM stars, and the unknown companion type). In Fig.~\ref{fig:p-alpha}, it clearly shows that $\alpha_{\rm CE}$ increases as the observed orbital period gets longer. This is because that a wider final separation corresponds to less orbital energy that can be used to eject the CE, which gives a larger $\alpha_{\rm CE} $ based on the $\alpha_{\rm CE}$ formula. In addition, it is found that the value of $\alpha_{\rm CE}$ seems to be dependent on the type of companion. This is consistent with previous studies for individual objects introduced above. The systems with dM companions give a larger $\alpha_{\rm CE}$ compared with that derived from systems with WD companions. This might be partly explained by the fact that the dM stars have a much larger radius than WDs, which corresponds to a relatively larger cross section with the CE.

Recently, some studies suggested that the size of the surface convection region could play a role in determining the value of $\alpha_{\rm CE}$ (e.g., \cite{Wilson2019MNRAS.485.4492W,Wilson2022MNRAS.516.2189W}). By defining two timescales, i.e., the convection timescale $t_{\rm conv}$ and the spiral-in timescale $t_{\rm insp}$, one could compare these two timescales in typically low-mass and high-mass giant envelopes to determine the value of $\alpha_{\rm CE}$. If $t_{\rm insp} < t_{\rm conv}$, the energy could be deposited in the envelope and contribute to the ejection.  Otherwise, the deposited energy would be brought up to the surface by the convection and radiates away, i.e., $\alpha_{\rm CE}=0$ in the region with $t_{\rm insp} > t_{\rm conv}$. It has been found that the region with low $\alpha_{\rm CE}$ in massive stars is quite small whereas most parts of the envelope of the low-mass giants have low $\alpha_{\rm CE}$ a due to the difference in the role of convection in high-mass and low-mass envelope \cite{Wilson2019MNRAS.485.4492W,Wilson2022MNRAS.516.2189W}. According to such calculations, more than 70 percent of massive stars have $\alpha_{\rm CE}$ of 0.9-1. In addition, about 30\% of low-mass donors have $\alpha_{\rm CE}$ of 0-0.1, and this fraction gradually decreases to 1 percent when $\alpha_{\rm CE}$ is in the range of 0.8-0.9 and becomes $\sim 20$ percent for $\alpha_{\rm CE}$ in the range of 0.9-1. Note that this study is the first attempt to understand CE efficiency physically \cite{Wilson2019MNRAS.485.4492W,Wilson2022MNRAS.516.2189W}.

\subsubsection{Numerical simulations of common envelope evolution}

\begin{figure*}[t]
    \centering
    \includegraphics[width=0.95\textwidth, angle=0]{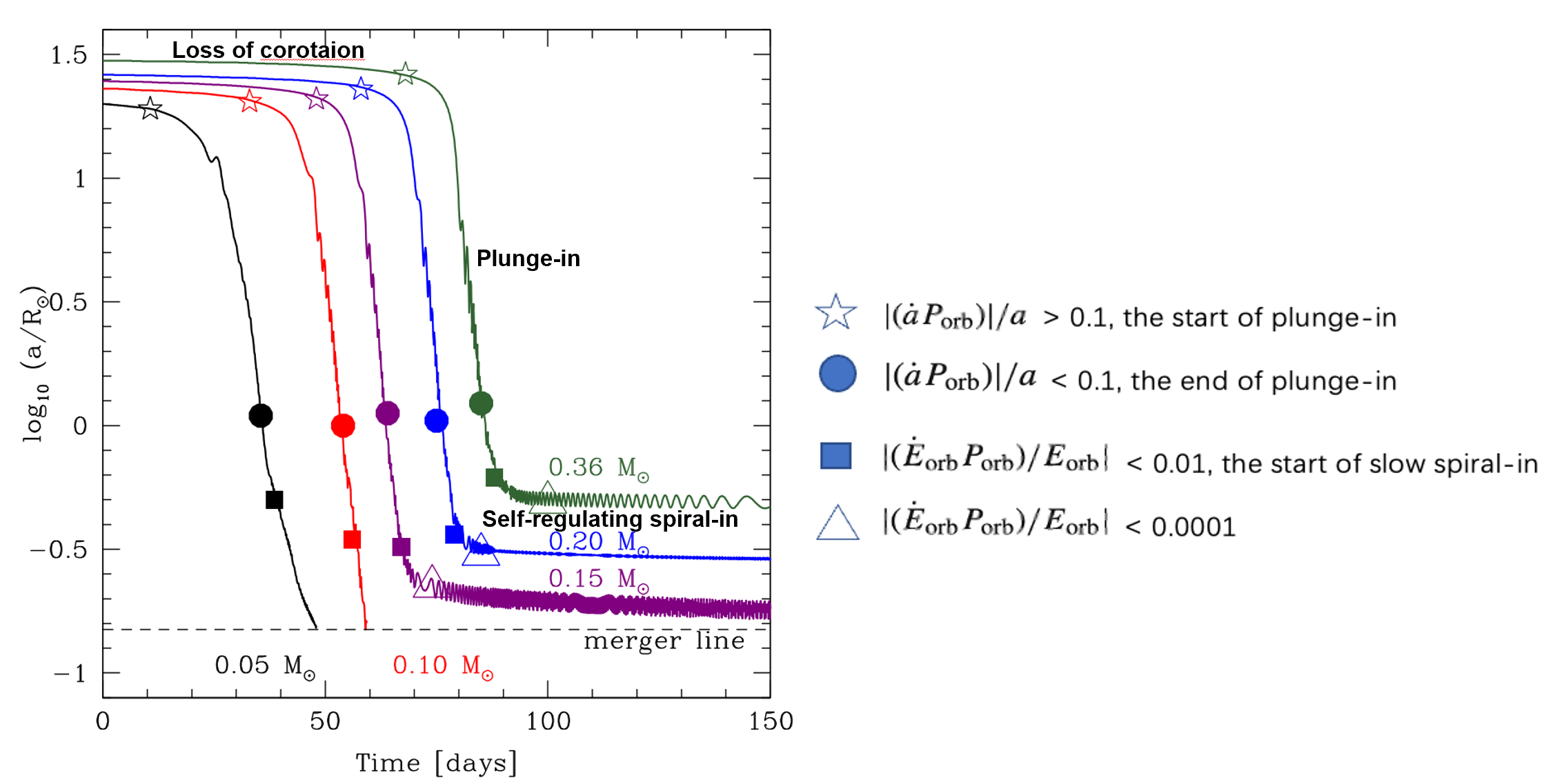}
    \caption{Evolution of orbital separations during the CEE for binaries with a $1.8M_\odot$ red giant donor, which has a core mass of $0.318M_\odot$ and radius of $16.3R_\odot$. The companion masses are 0.05, 0.10, 0.15, 0.20, and $0.36M_\odot$, of which the time axis are shifted by 520, 550, 558, 510, and 270 d, respectively, to show the relative orbital evolution in more detail. The dashed line indicates the position the companion and the core will definitely merge. Symbols of stars, circles, squares, and triangles are for some transition points that occurred in the CEE, as indicated in the bottom part of the plot. The first three phases i.e. loss of corotation, plunge-in, and self-regulating spiral-in, are clearly shown in the figure. During the self-regulating spiral-in, each binary has a constant, albeit slowly decreasing, orbital period, as expected, and the orbital separation shows oscillations due to the nonzero eccentricity. The figure is reproduced based on Figure~1 of \cite{Ivanova16}. }
    \label{fig:CEE-process}
\end{figure*}

Due to the importance of the common envelope evolution (CEE) in binary evolution, a number of hydrodynamical studies have been undertaken by different groups to provide a theoretical understanding of the CEE. Such study starts as early as the CE scenario was proposed, including 1D  \cite{Taam1978ApJ...222..269T,Meyer1979A&A....78..167M}, two-dimensional (2D) axisymmetric \cite{Taam1989ApJ...337..849T,Taam1991ApJ...373..246T} and 3D studies based on either the smoothed particle hydrodynamics (SPH) method (e.g. \cite{de-Kool1987PhDT.......112D,Terman1994ApJ...422..729T,Terman1995ApJ...445..367T,Terman1996ApJ...458..692T,Rasio1996ApJ...471..366R}) or a finite differencing numerical technique with Eulerian nested grid (e.g. \cite{Sandquist1998ApJ...500..909S,Sandquist2000ApJ...533..984S}). However, modelling the CEE is numerically challenging because the detailed description of the CE process requires modeling different diverse physical processes in detail over wide length scales ranging from less than $1\,R_{\odot}$ (e.g. about $10\,\mathrm{km}$ for neutron stars) to $>1\,\rm{AU}$ and a wide orbital period from an hour to several years (for a recent review, see \cite{Roepke2022}). Therefore, most earlier hydrodynamical simulations of CEE were only done only for a certain stage of the CE process. Nevertheless, these simulations are still very helpful to reveal important insights into the CEE. For example, some early simulations have shown that the initial evolution of CE could be very rapid, which is on a timescale comparable to the orbital period. In this case, the deposition of the reduced orbital energy in the CE is unlikely isotropic, and the mass ejection would be asymmetric eventually. This leads to that the majority of the CE would be lost in the orbital plane of the binary. This has important implications for the shaping of PNe. There is evidence that the companion has been significantly spinning up during the CEE if its mass is comparable to that of the CE, which may aid the decay of the orbit and lead to the CE ejection eventually. The degree of orbital shrinkage depends on both the evolutionary stages of the donors and the masses of components, ranging from modest factors (of the order of 20) to dramatic factors (more than 100). A weaker orbital shrinkage would be possible if the companion mass is comparable to that of the CE, which is possibly due to the fact that internal energy plays a role in this case. The estimated value of $\alpha_{\rm CE}$ is about 0.4 in such a case.

3D hydrodynamical simulations for the CEE have been extensively developed in the new Menilium. A wide variety of numerical tools have been used in these simulations, including Lagrangian codes, Eulerian codes, and a combination of the two.  In addition, much higher resolution modelling has become possible thanks to the significant development of modern computational sources. A detailed introduction to these codes is beyond the scope of this paper. Recent comprehensive reviews of the challenges and developments of numerical simulations of CEE, and the strengths and weaknesses of different numerical techniques have been given by \cite{Ivanova2013A&ARv..21...59I} and \cite{Roepke2022}. Here we only give a brief summary of important physics included in the simulations and the main results obtained from them.

{\bf I. Main phases of the common envelope evolution}

The progression of an idealized common envelope evolution (CEE) can be divided into the following phases (see Fig.~\ref{fig:CEE-process}): 
\vspace{-\topsep}

\begin{itemize}\itemsep5pt 

\item[(i)] \textbf{Loss of corotation}, in which a stable binary is transformed into a spiraling-in binary on a dynamical timescale. The spiral-in can be caused by the dynamical instability of the donors (see section~4.1), the Darwin or secular tidal instability (e.g.  \cite{Hut1980A&A....92..167H,Lai1993ApJ...406L..63L,Eggleton2001ApJ...562.1012E}), and that the reaction of the accretors leads to the matter filling the binary orbit (i.e. $L_{\rm 2}/L_{\rm 3}$ overflow in section~4.1). 

\item[(ii)] \textbf{Plunge-in}, in which the companion is plunged deep inside the CE. It is a purely dynamic process and has been most extensively studied by current hydrodynamical simulations. During this stage, the reduced orbital energy is deposited in the envelope and drives its expansion, possibly leading to either a dynamical ejection of the CE or a rapid merger of the binary straightway.

\item[(iii)] \textbf{Self-regulating spiral-in}, in which the spiral-in might slow down and become a self-regulating state after the envelope expands enough if non-local energy dissipation (e.g. \cite{Ricker2008ApJ...672L..41R,Passy2012ApJ...744...52P}) does not dominate the entire spiral-in phase. The frictional luminosity accumulated in the spiral-in phase is transported to the surface and radiated away in this stage, as first suggested by \cite{Meyer1979A&A....78..167M}. 

\item[(iv)] \textbf{Termination of the self-regulating phase}. The self-regulating stage ends with the ejection of the envelope, or when either the secondary or the primary’s core overfills its Roche lobe (resulting in a merger eventually). This phase will take several dynamical timescales.  

\item[(v)] \textbf{Post-CE evolution}, in which the CE is ejected and a post-CE binary system is produced. This is the final phase of CEE. The properties of post-CE systems are also very important because they could be affected by the specific circumstances of the CEE. For example, any remaining circumstellar matter, likely in the form of a circumbinary disk, may change the orbit and eccentricity of a surviving binary. In addition, the remnant cores might further overfill its Roche lobe during the thermal evolution, driving mass transfer again.

\end{itemize}

It is unclear whether the self-regulating spiral-in phase occurs or not if non-local energy dissipation is common in the entire spiral-in phase. Either the CE ejection or the merger of the binary may occur rapidly following the plunge-in phase. On the other hand, as proposed by \cite{Ivanova2013a}, a self-regulated spiral-in could also be followed by another dynamical plunge if the mechanism maintaining self-regulation somehow ends, and this second plunge could in turn be followed by another self-regulated phase in some cases.

{\bf II. Role of the recombination energy and loss of orbital energy}

The energy budget during the CEE is the most important aspect to determine the outcomes of the CEE. In Section~\ref{sec:Eb}, we have discussed the terms of energy included in the standard energy mechanism. Numerical simulations have shown evidence that the recombination energy (one term of internal energy) could play a crucial role in the final ejection of the CE because it gives the final push to make a loose envelope unbound \cite{Ivanova16,Nandez}. Four types of ejection processes from the simulations were identified in the simulation of \cite{Ivanova16}, that is, the initial ejection (pre-plunge-in), the plunge-in outflow, the recombination runaway outflow, and the shell-triggered ejection. The first two processes take place in all the simulations, including those ending up with a merger. The amount of the pre-plunge-in ejection can be estimated roughly by using the energy formalism on the material outside the orbit. The prompt plunge-in ejection carries away substantially more mass, but it is hard to be estimated properly. The shell-triggered ejection is caused by the hollow shell fallback when the envelope is decoupling with the core-companion binary, providing another prompt ejection process as a part of the CE. The recombination runaway outflow starts during the slow spiral-in when the expanded envelope is cooled down to start hydrogen recombination. With detailed energy analysis, the authors further gave a description for the final ejection phase as follows \cite{Nandez}: At the end of the plunge-in, the outer envelope 
hosts most of the unbound mass and has some kinetic energy obtained from the previous plunge-in, leading to an expansion of the envelope on the dynamical time-scale. In this process, material in the outer envelope has a parabolic (bound) trajectory with respect to the binary and will gain enough energy to become unbound once the material expands and cools down enough to start recombination. The recombination radius at which the released recombination energy can remove material out of the potential well has been derived in their simulations. They found that the recombination takes place at large optical depths which are at least 10 in 3D models and could be above 100 in 1D models \cite{Ivanova2015MNRAS.447.2181I}.

Hydrodynamical simulations of CEE showed that a significant fraction of the released orbital energy is taken away by the ejected material. The appearance of large-scale waves with tidal arms could trail the binary orbit. Spiral shocks transfer angular momentum to the matter in the envelope and some of the energy in the shocks will be dissipated as heat \cite{Ricker2012ApJ...746...74R}. Obviously, this kind of energy dissipation is far away from the secondary, which is different from that normally expected, i.e., the orbital energy is thermalized locally by viscous dissipation in the region of the in-spiraling companion. Even if the energy thermalization is local, a significant fraction of energy would be transferred to the surface and radiated away during the following self-regulating phase. Furthermore, it is found that about 25 percent of the CE could be flung out by the spiral waves, taking away a significant part of orbital energy \cite{Ricker2012ApJ...746...74R}.  The simulations of 25 3D hydrodynamical interactions between a low-mass RG and a WD companion by \cite{Nandez2016MNRAS.460.3992N} have obtained the total energy taken away by the ejecta, $E_{\rm tot,unb}^{\infty}$. The fraction of energy loss related to the total energy can calculated via a fitting formula as: 
\begin{equation}
    \alpha_{\rm unb}^{\infty}=-\frac{E_{\rm tot,unb}^{\infty}}{\Delta E_{\rm orb}}=-0.16-0.30M_{\rm d}+0.49M_{\rm 2}+2.27M_{\rm c}
\end{equation}
where $M_{\rm d}$, $M_{\rm c}$ and $M_{\rm 2}$ are the donor mass, the core mass of the donor and the companion mass in solar mass, respectively. The value of $\alpha_{\rm unb}^{\infty}$ ranges from about 50 percent to 20 percent when the donor increases in mass from 1.2 to 1.8\,$M_\odot$. The energy formalism is therefore preferred to be revised as  
\begin{equation}
    \Delta E_{\rm orb} (1-\alpha_{\rm unb}^{\infty})+E_{\rm bind}+\eta (M_{\rm d}-M_{\rm c})=0,
\end{equation}
where the parameter $\eta$ is used to assess the recombination energy stored in the envelope, which is proportional to the mass of the envelope. The exact value of $\eta$ varies with chemical composition of the envelope, and which is $\eta=1.5\times 10^{\rm 13}$ erg ${\rm g}^{\rm -1}$ for fully ionized gas that consists of 30\% He and 70\% H.

In order to make the results being used conveniently by BPS studies, a fitted formula for $\alpha_{\rm CE}\lambda$ value, as a function of the $M_{\rm d}$, $M_{\rm c}$ and $M_{\rm 2}$ is given \cite{Nandez} as below:
\begin{equation}
    \alpha_{\rm CE}\lambda=0.92+0.55M_{\rm d}-0.79M_{\rm 2}-1.19M_{\rm c}.
\end{equation}
They obtained that the maximum value of $\alpha_{\rm CE}\lambda$ is less than 1.03 in their studies. We should always keep in mind that this formula is based on limited models and thus may be only valid in narrow parameter space. Note that low-mass RG stars in these studies have an initial mass of 1.2, 1.4, 1.6, or $1.8\,M_{\odot}$ and a He core mass of $0.32$ or $0.36\,M_\odot$, and the WD companions have a mass of 0.32, 0.36 or $0.40M_\odot$.

Despite the challenges of spatial scales, CE simulations for massive donors have emerged gradually in recent two years \cite{lau2022,moreno2022}. These simulations also demonstrated the importance of internal energy, particularly the recombination energy in the final CE ejection. The study of \cite{moreno2022} presented the values of CE parameters for the standard energy formulae. They performed 3D magnetohydrodynamic (MHD) simulations of the CE evolution of a binary system composed of an initially $10M_\odot$ red supergiant primary star and either a BH ($5M_\odot$) or NS ($1.4M_\odot$) companion. 
The simulations showed that about half of the CE in these systems is dynamically ejected, and the ejected mass is still increasing at the end of the simulations, which leaves some uncertainties on the true final ejecta mass. 
The authors suggested that if all the remaining ionization energy is released by the expanding envelope and can be used for unbinding the material, it is possible that almost the full envelope will be eventually ejected. Assuming that the full envelope is ejected, they found a value of $\alpha_{\rm CE}\lambda =1.17$ and 1.32 in the cases with an NS and a BH companion, respectively. The inclusion of the thermal and ionization energies (i.e. $\alpha_{\rm th}= 1.0$) gives $\lambda = 1.91$ from the relaxed dynamical model, resulting in an ejection efficiency of $\alpha_{\rm CE} =0.61$ and 0.69 in the NS and BH companion case, respectively\footnote{However, the energy formalism with these values can only reproduce the outcomes of their simulations if the core–envelope boundary is adjusted to their setup and one does not apply the usual hydrogen mass fraction of 0.1 or maximum compression point.}.

{\bf III. Additional energy sources}

Most detailed hydrodynamical models of CEE encounter difficulty in ejecting most of the envelope following the inspiral, showing that most material of the CE remains bound to the binary though the CE expands to larger separations. In addition to the internal energy, more possibilities for extra energy sources have been considered in numerical simulations to reduce the difficulty in ejecting the CE, including nuclear energy, accretion and jets, long pulsations, dust-driven winds, and radiation energy for massive donors \cite{Roepke2022}.

It is possible that nuclear energy come from either the burning at the base of the CE or the ignition on the surface of the accretor. The former could be explosive if H-rich material penetrates deep into the core of the giant star and reaches the He-burning shell during the final self-regulating spiral-in phase (e.g. \cite{Ivanova2002MNRAS.334..819I}). The energy released in the explosion could exceed the binding energy of the He shell and lead to a successful CE ejection immediately. This CE ejection mechanism may help a less massive companion to survive the CE and thus has an advantage for the formation of low-mass black-hole X-ray binaries.

Mass transfer during the CEE will change gravitational potential energy into thermal energy. Also, the accretion luminosity may be another energy source for the CE ejection. In most cases of CEE, the Bondi-Hoyle-Lyttleton (BHL)\footnote{The Bondi-Hoyle-Lyttleton (BHL) model was first developed by \cite{Bondi1944MNRAS.104..273B} for spherical accretion onto a point mass traveling through a gas cloud (or the interstellar medium) that is assumed to be free of self-gravity, and to be uniform at infinity.} prescription predicts a super-Eddington accretion rate. However, hydrodynamical simulation of \cite{Ricker2012ApJ...746...74R} showed that the true accretion rate given by BHL prescription has been significantly overestimated and that the contribution of the accretion luminosity could be negligible. Furthermore, some authors argued that the accretion during the CEE is commonly not significant at least for non-degenerate companions. This is because the CE has typically a much higher specific entropy than that of the surface of the accretor, which would cause a temperature inversion or roughly isothermal layer between the accretor and the CEE and thus prevent further accretion. In the cases with degenerate companions, the ignition of nuclear burning on the surface of the accretor might be inhibited by the high-entropy accreted material which is buoyant and hard to be compressed to the ignition conditions. The jets driven by the accretion of compact companions (WDs or NSs) during spiraling-in could be an effective way for assisting the CE ejection. This has been observed by the simulations of \cite{Shiber19} for binaries with a $0.88\,M_\odot$ RG star with a radius of $83\,R_\odot$ and a $0.3\,M_\odot$ compact companion. The simulations showed that the inclusion of jets could unbind higher envelope mass approximately three times that of identical simulations without jets. Meanwhile, the authors suggested that the jets would play a crucial role in shaping the final outcomes of CEE by generating high-velocity outflows in the polar directions, increasing the final core-companion orbital separation, 
and causing a kick velocity of the core-companion binary. However, small-scale simulations for the same binary by \cite{jet2022} showed that the broken-out of the RG envelope in large-scale simulations is choked and that the accreted angular momentum onto the companion is not high enough to form a disk. The mass accretion rate onto the MS star is about 1-10\% of the BHL rate.

Using the stellar evolution code MESA, 1D hydrodynamic simulations 
of a low-mass red giant undergoing a synthetic CE event in the final slow spiral-in phase has been carried out \cite{clay17}. In these calculations, they investigated the response of the giant’s envelope by including the heating of the envelope due to frictional dissipation in multiple configurations, finding that the envelopes of their models become dynamically unstable and develop large-amplitude pulsations with periods in the range of 3–20 years. In some cases, the shocks and associated rebounds that emerged with the pulsations are strong enough to eject the matter up to 0.1\,$M_\odot$ (which is about 10\% of the envelope mass) at a speed exceeding the escape velocity of the surface. Such ejections repeat within a few decades, leading to an averaged mass-loss rate of the order of $10^{\rm -3}\,M_\odot {\rm yr}^{\rm -1}$. They therefore suggested that long-period, large-amplitude pulsations could be a candidate mechanism for removing the entire envelope over the duration of the slow spiral-in phase.

It has been proposed that the dust-driven winds can be produced following the CEE \cite{glanz18}. In this case, such winds might evaporate the CE through similar processes operating in the envelope ejection of AGB stars \cite{glanz18}. In AGB stars, the pulsations drive the envelope expansion, leading to the material in the envelope being cooled down to low temperatures to enable dust condensation. The radiation pressure then accelerates the dust and drives winds when the dust is coupled to the gas. As a consequence, the dust-driven winds eventually erode the whole envelope of AGB stars. It has been shown that the inspiral phase of CEE can effectively replace the role of stellar pulsations that appeared in AGB stars and drive the CE expansion to the scales comparable with those of AGB envelopes, giving rise to efficient mass-loss through dust-driven winds \cite{glanz18}. By modelling the CEE on a moving mesh, it was found that the model with setting the red giant into corotation with the orbital motion ejects more mass from the system than the non-rotating model, resulting in a larger final orbital separation \cite{prust19}. Massive stars are expected to be significantly different from low-mass stars because their envelopes have significantly been supported by radiation pressure. Very recently, 3D hydrodynamical simulations of a CE event involving a $12\,M_\odot$ massive red supergiant donor have found that the inclusion of radiation energy leads to that 60\% more mass being ejected (which is up to two-thirds of the total envelope mass becoming unbound, or more) and thus a final separation is 10\% larger than that in models with an ideal gas \cite{lau2022}. In addition, the inclusion of recombination energy could eject almost the entire envelope (at least three-quarters of the envelope), and lead to the post-plunge-in separation increase of a further 20\%. This seems to suggest that the He recombination energy is the major source that contributes to the additional mass ejection. More importantly, these simulations further demonstrated the roles played by the internal energy, radiation transport, and convection in the ejection of the CE for binaries with massive donors.

{\bf IV. Magnetic fields in the common envelope evolution}

It is generally believed that magnetic WDs and CVs are related to the CE evolution. The magnetic fields could be generated or amplified in the differentially rotating envelope during the dynamical spiral-in (e.g. \cite{Regos1995MNRAS.273..146R}) or in an accretion disc around the giant core formed from the tidally disrupted companion in merging phases (e.g. \cite{Nordhaus2011PNAS..108.3135N}). 
The MHD simulations for the CEE were preformed for the first time by
\cite{ohl2016b}. In these simulations, a $1\,M_\odot$ companion has been spiraled into the envelope of a $2\,M_\odot$ RG star with a $0.4\,M_\odot$ He core. Various initial surface magnetic field strengths of $10^{\rm -10}$, $10^{\rm -6}$, and $10^{\rm -2}$\,G were adopted in their simulations to investigate the influence of magnetic fields. 
The simulations showed that magnetic fields are strongly amplified in the accretion stream that appeared around the $1\,M_\odot$ companion on a dynamical timescale, leading to 10-100\,kG field strengths throughout the entire envelope after 120\,d. The temporal and spatial scales are compatible with the magnetorotational instability in the accretion flow around the companion, thus providing a new way of generating magnetic fields during the dynamical spiral-in of the CE phase. This short-timescale dynamical amplification of magnetic fields, however, is difficult to anchor at the WD surface as pointed out by \cite{Potter2010MNRAS.402.1072P}  because the strengths of generated magnetic fields during the CE phase are several orders of magnitude weaker than those in WDs (up to $10^{\rm 9}$\,G).

Similarly, other studies further followed the 3D MHD simulations for an AGB star engulfing a companion by \cite{ond2022}. In their simualtions, the AGB star has a mass of $0.970\,M_\odot$, with a core of $0.545\,M_\odot$ and a radius of $173\,R_\odot$ at the onset of the CE evolution. 
The companion is unresolved and represented by a point particle, with a mass of 0.243, 0.486, and $0.729\,M_\odot$, respectively.
The simulation starts from the plunge-in of the companion into the envelope and covers hundreds of orbits of the binary system until the envelope has been ejected completely. In this study, magnetic fields are strongly amplified in two consecutive episodes, i.e., when the companion spirals in the envelope, and when the companion forms a contact binary with the core of the former giant star. A magnetically driven, high-velocity gas outflow is launched self-consistently in the second episode. The outflow is bipolar and is collimated by the ejected CE. They also fund that initial parameters of the simulated system have significant impacts on the morphology of the ejected CE material and therefore provide an explanation for the variety of shapes observed in PNe. Their simulations could well reproduce typical morphologies and velocities of young PNe, which leads to an establishment of a direct link between observed bipolar PNe and the enigmatic CE phase of binary evolution. As proposed by the authors, the magnetic driving mechanism could be a universal consequence of CE evolution responsible for a substantial fraction of observed PNe, and it is likely to also exist in CEE of other binary stars that lead to the formation of SNe Ia, X-ray binaries, and GW merger events.

\subsubsection{Evidence for the common envelope evolution}
\label{sec:evidence}

Although the CEE plays an important role in binary evolution, direct observational evidence for this process has been lacking for several decades. The existence of extremely short-orbital period binaries only means that they are very likely to be produced from the CEE. There are two major reasons for this situation. First, the timescale for the CEE is very short ($\sim 1000-10,000$\,yr), leading to such events having a low possibility of being discovered. Second, the binaries and the merge products (e.g. TZO) during the CEE stage have similar observational properties as some normal stars.


At the termination of the CEE, substantial materials are expected to be ejected, even when the CEE leads to a merger of the binary eventually\footnote{Partial ejection is likely to happen for merger events because the orbital energy deposited into the CE at an early stage during the merger may exceed the binding energy of the outer layers and the upper layers may also contain a substantial amount of the orbital angular momentum of the companion (i.e. the merger timescales is too short to transport the angular momentum across the whole envelope).}. Motivated by the models of type IIP SNe \footnote {In the model of type IIP SNe, the ejected stellar plasma expands and cools, and the recombination occurs and changes the opacity, leading to the propagation of a photosphere that moves inwards with respect to the mass variable.}, observational signature of the CEE was proposed by \cite{Ivanova2013Sci...339..433I} in which the radiation from the ejected matter is assumed to be dominated by a recombination front as the ejected matter cools. According to this description, the recombination defines a photosphere as the radiation surface which does not grow with the ejected matter expanding, therefore resulting in a plateau shape in the light curves. The plateau luminosity $L_{\rm p}$ and the timescale $t_{\rm p}$ were estimated by applying an analytic model for the luminosity and duration of Type II SNe light curves from \cite{Popov1993ApJ...414..712P}. The results are presented in Figure~1 of \cite{Ivanova2013Sci...339..433I} and compared with a sample of luminous red novae (LRNe; i.e. V1309 Sco, M85 OT, M31-RV, and V838 Mon), a subset of intermediate luminosity red transients (ILRTs). The effective photospheric temperature is expected to be $\sim 5,000$K for thick ejecta, leading to that the ejection events (outburst) being red in color naturally. When the ejected envelope has fully recombined, the material may become transparent suddenly. Such expected characteristics, as well as the ejection velocities and event rate, are consistent with those of LRNe.

After the pioneered work of \cite{Ivanova2013Sci...339..433I}, lots of attention has been paid to the LRNe \cite{MacLeod2017,Blagorodnova2017,Pastorello2019,Cai2019,Matsumoto2022}. In particular, the properties of two LRNe in nearby galaxies (M101~OT2015-1 and M31LRN~2015) which have pre-outburst sources being observed were presented \cite{Blagorodnova2017,MacLeod2017}. From archival data spanning 15-8 years before the outburst, it has found that an F-type yellow supergiant with a luminosity of $8.7\times 10^4\,L_\odot$ and an effective temperature of $\sim 7000$K is consistent with the observations of M101~OT2015-1 \cite{Blagorodnova2017}. M101~OT2015-1 has an estimated mass of $18\pm1\,M_\odot$ and is crossing the Hertzsprung gap after H has been exhausted in the core. Based on the observed properties, this object has been argued to be a binary system, in which the more evolved system is overfilling the Roche lobe \cite{Blagorodnova2017}. By analyzing the optical photometry and spectroscopy along with the color and magnitude of the pre-outburst source detected by HST for stellar-merger transient M31LRN~2015 in the Andromeda galaxy, it was found that its primary star is a $3-5.5\,M_\odot$ subgiant-branch star with a growing radius of $30-40\,R_\odot$, which is consistent with a picture where a subgiant primary star engulfs its companion \cite{MacLeod2017}. Also, it has been suggested that transient M31LRN~2015 has lasted less than 10 orbits of the original binary, which had a pre-merger period of $\sim 10$ days \cite{MacLeod2017}.

Very recently,  a sample of 14 LNRe was presented, among which 6 have pre-transient observations  \cite{Matsumoto2022ApJ...938....5M}.
It has been found that most LRN progenitor stars in this sample are moderately evolved, with radii between $\sim 1-10$ times that of MS stars of their mass. It is a bit strange since giant donors are more likely to experience unstable mass transfer and step into the CEE in binary evolution. It may mean that the progenitors of the CEE are more likely to be giant stars or supergiant stars rather than MS stars, which is, however, inconsistent with the observations from these 14 LNRe. Two reasons might be responsible for this discrepancy. First, the sample is still small and giant-like donors are expected to be found in the future. Second, the mass ejection events of the CEE from giant-like progenitors have properties different from that of LRNe. However, a final answer for the reasons behind the discrepancy can only be given by a better understanding of the observational properties of ejection events of the CEE and LRNe. Considering the fact that the ejecta of the CEE has lower velocities, higher density, and lower specific thermal energy in comparison to that of type II SNe and that recombination could be an important source of luminosity, a new model for synthetic light curves of LRNe was developed  \cite{Matsumoto2022ApJ...938....5M}. In this new model, theoretical light curves of LRNe typically present two peaks, which are consistent with observations. The early phase light curves are powered by the initial thermal energy of the hot, fastest ejecta layers, and the second peak (the plateau) is powered by H recombination from the bulk of the ejecta. This new model was applied to the sample of LRNe to infer the ejecta properties such as the mass, velocity, and launching radius \cite{Matsumoto2022ApJ...938....5M}. These properties are expected to be helpful for comparison with the progenitor properties from pre-transient imaging in future observations, which will give us a better understanding of mass ejection.

Using 9 LRNe with known masses of the quiescent progenitors, a linear fit between the progenitor mass (in logarithmic, ${\rm log} M_{\rm pro}$) and V-band second peak (or the plateau) magnitude ($M_{\rm V}$) has been given \cite{Cai2022A&A...667A...4C}, which can be written as 
\begin{equation}
    {\rm log}(M_{\rm pro}/M_{\odot})=(-0.162\pm0.020)M_{\rm v}-(0.701\pm0.048),
\end{equation}
or 
\begin{equation}
    M_{\rm v}=(-5.56\pm0.68){\rm log}(M/M_{\odot})-(4.91\pm0.26).
\end{equation}

This linear fit could be used to infer the progenitor mass or predict the plateau luminosity conveniently.

Some peculiar single stars (such as magnetic stars, blue stragglers, rapid rotators, T CrB stars, etc) have been considered to be likely formed by the merging of binaries through the CEE. An unusual, ring-shaped ultraviolet nebula with the star at its center, TYC 2597-731-1, has been discovered \cite{Hoadley2020Natur.587..387H}. It is likely an unobstructed stellar merger in an evolutionary stage between its dynamic onset of the CE ejection and the theorized final equilibrium state. Both the spectrum and the position suggest TYC 2597-731-1 as an old star. But TYC 2597-731-1 has an abnormally low surface gravity and a detectable long-term luminosity decay which is uncharacteristic for its evolutionary stage. Meanwhile, TYC 2597-735-1 exhibits many features that can be accounted for by stellar mergers, including ${\rm H_\alpha}$ emission, radial velocity variations, enhanced ultraviolet radiation, and excess infrared emission as well as signs of dusty circumstellar disks, stellar activity, and accretion. The comparison between observations and stellar evolution models suggests that TYC 2597-735-1 merged with a lower-mass companion about several thousand years ago. Therefore, this object could be used to infer how two stars merge into a single star directly.

On the other hand, a recent study claimed that they likely observed the ejected CE around a short-period sdO+WD binary J1920-2001 \cite{Hoadley2020Natur.587..387H}. The binary system J1920-2001 has an orbital period of 3.5\,hr and a mass ratio (the WD mass over the sdO mass) of 0.738. In this binary, the sdO star has a mass of $0.55M_\odot$ and it is overfilling its Roche lobe to likely transfer mass to the WD via an accretion disk. Based on the binary model for sdB stars \cite{Han2002MNRAS.336..449H,Han2003MNRAS.341..669H}, this binary was likely produced from a CE ejection channel, in which the progenitor of the sdO star is either an RGB star or an early AGB star. In addition, strong Ca H\&K lines have been observed to blue-shifted by $\sim 200\,\rm km\,s^{-1}$ in J1920-2001. This outstanding feature likely originated from the CE ejected about 10,000 years ago \cite{Hoadley2020Natur.587..387H}. It seems not surprising that the ejected CE has been detected around short-period sdB stars because such binaries are mainly produced through the CE ejection. Applying a similar method to the LAMOST survey is expected to discover much more candidates of sdB binaries like J1920-2001, which will be very helpful for us to better understand the CE process and thus examine the current binary evolution theory.

The discovery of Thorne-Zytkow objects (TZOs) that contain an NS inside an RG would be direct evidence of stellar mergers from the CEE. However, none of such objects have been discovered yet even though growing survey projects are being carrying out. Detecting the inner structures of RGs directly would provide some clues on this.

\subsection{Non-conservative mass transfer}

The treatment of non-conservative mass transfer is one of the key uncertainties in binary evolution. There are some important questions associated with the non-conservative mass transfer that remain unclear: How much mass can be lost from the binary system during the stable mass-transfer phase? How much specific angular momentum can be taken away by the mass loss? What is the dominant mechanism to drive the angular momentum loss? The non-conservative mass transfer was first described in 1970s (e.g. \cite{Refsdal1974A&A....36..113R,Warner1978AcA....28..303W,Han2001ASPC..229..205H}) and it was referred to as ``liberal'' \cite{Eggleton2000NewAR..44..111E} in very early studies to distinguish binary evolution with conservative mass transfer. The non-conservative mass transfer has been widely studied in detail by recent studies from both theoretical and observational sides. On the one hand, theoretical studies have shown that mass transfer in binary systems could be non-conservative, causing the mass and angular momentum loss from the system (e.g. \cite{Warner1978AcA....28..303W,Sytov2007ARep...51..836S,Nanouris2015A&A...575A..64N,De-Marco2017PASA...34....1D,Lu2023MNRAS.519.1409L}). For instance, it has been suggested that the accretion disk around the donor star could become geometrically thick at near the disk outer radius if the mass transfer rate in a binary system is sufficiently high, which leads to that a large fraction of the transferred mass can be lost from the system through the outer Lagrangian (L2) point \cite{Lu2023MNRAS.519.1409L}. On the other hand, various studies have concluded that non-conservative mass transfer seems to be required to explain observational properties of classes of particular binaries, including classical Algols \cite{van-Rensbergen2006A&A...446.1071V,van-Rensbergen2008A&A...487.1129V,van-Rensbergen2011A&A...528A..16V,Erdem2014MNRAS.441.1166E}, sdB stars \cite{Han2002MNRAS.336..449H,Han2003MNRAS.341..669H},  accreting millisecond X-ray pulsars \cite{Marino2019A&A...627A.125M}, and ultra-compact
X-ray source \cite{Iaria2021A&A...646A.120I}, etc. For instance, 
some studies have concluded that about 60 per cent of transferred masses from $\beta\,\mathrm{Per}$ loses from the system and takes away about 30 per cent of the angular momentum during mass transfer through RLOF \cite{Sarna1993MNRAS.262..534S}. 
In addition, in order to reproduce the short-period sdB binaries with WD companions, the first phase of mass transfer seems to has to be stable and non-conservative \cite{Han2002MNRAS.336..449H,Han2003MNRAS.341..669H}. Also, the total number of Algol binaries predicted by the non-conservative mass transfer model has been found to match the observation better \cite{van-Rensbergen2011A&A...528A..16V}. Recently, it has been suggested that the large orbital period derivative in the ultra-compact
X-ray source XB~1916-053 is very likely due to a high non-conservative mass transfer \cite{Iaria2021A&A...646A.120I}.

Different treatments of the mass and angular momentum loss in non-conservative mass transfer play an important role in determining the distributions of binary properties and their fates. Three mechanisms are generally adopted to deal with orbital angular momentum loss in a binary system: (i) \emph{gravitational radiation}, (ii) \emph{magnetic braking} (MB), and (iii) \emph{mass loss}. Gravitational radiation has been well studied theoretically, which is thought to be dominant in angular momentum loss of tight binary systems. In particular, the gravitational wave signals have been confirmed by the observations \cite{Abbott2016PhRvL.116f1102A,Abbott2017PhRvL.119p1101A}. The standard prescription from \cite{Landau1975ctf..book.....L} is commonly used for the calculation of gravitational radiation in binary evolution. The prescription for MB is still uncertain. Earlier theoretical
models generally employed the standard ``Skumanich MB'' law that derived from observations of the time-dependence of rotational braking of Sun-type stars \cite{Skumanich1972ApJ...171..565S,Verbunt1981A&A...100L...7V,Rappaport1983ApJ...275..713R}:

\begin{equation}
	\frac{{\rm d}J_{\rm MB,Sk}}{{\rm d}t} = -3.8 \times 10^{-30}\;M_2\,R_\odot^4\,\left(\frac{R_2}{R_\odot}\right)^{\gamma_{\rm mb}}\Omega_{2}^{3}\quad{\rm dyne\,cm}\;,
\end{equation}
where $\gamma_{\rm mb}$ ( which is from 0 to 4) is a dimensionless parameter; $M_{2}$, $R_{2}$ and $\Omega_{2}$ are the mass, radius and angular velocity of the donor star, respectively. Recently, it is found that the standard ``Skumanich MB'' law gives mass transfer rates too weak to explain
most observed persistent low-mass X-ray binaries (LMXBs) \cite{Van2019ApJ...886L..31V}. The MB law was therefore modified by considering a scaling of the magnetic field strength with the convective turnover time, and scaling of MB with the wind mass-loss rate, i.e.

\begin{equation}
    \frac{{\rm d}J_{\rm MB}}{{\rm d}t} = \frac{{\rm d}J_{\rm MB,Sk}}{{\rm d}t}\;\left(\frac{\Omega_2}{\Omega_\odot}\right)^{\beta}\left(\frac{\tau_{\rm conv}}{\tau_{\rm \odot,conv}}\right)^{\xi}\left(\frac{\dot{M}_{\rm 2,wind}}{\dot{M}_{\rm \odot,wind}}\right)^{\alpha}\;,
\label{eq:mb_vih}
\end{equation}
where $\Omega_{\odot} = 3.0 \times 10^{-6}\;{\rm s}^{-1}$, $\tau_{\rm conv}$ is the turnover time of convective eddies and $\tau_{\rm \odot,conv} = 2.8 \times 10^{6}\;{\rm s}$, $\dot{M}_{\rm 2,wind}$ is the wind mass-loss rate of the donor star and $\dot{M}_{\rm \odot,wind} = 2.54 \times 10^{-14}\;M_{\odot}\,{\rm yr}^{-1}$ \citep{Carroll2006ima..book.....C}. Four cases based on different combinations of values of $\beta$, $\xi$, and $\alpha$ were defined (see also Table~1 of \cite{Van2019MNRAS.483.5595V}): 

\begin{equation}
  (\beta,\,\xi,\,\alpha) = \, \left\{ 
  \begin{array}{ll}
  (0,\,0,\,0) & \mbox{\hspace{0.1cm} standard Skumanich MB}\\
  (0,\,2,\,0) & \mbox{\hspace{0.1cm} convection-boosted MB}\\
  (0,\,2,\,1) & \mbox{\hspace{0.1cm} intermediate MB}\\
  (2,\,4,\,1) & \mbox{\hspace{0.1cm} wind-boosted MB}\\
  \end{array}
  \right.
  \label{eq:MB-laws}
\end{equation}

The models with ``intermediate MB'' law (i.e. $\beta=0$, $\xi=2$ and $\alpha=1$) has been found to reproduce the observed orbital characteristics of LMXBs. 

By considering the dependence of the magnetic field strength and the Alfv\`{e}n radius on the convective turnover time and the donor's rotation,  another new MB prescription, the so-called ``Convection And Rotation Boosted (CARB)'', was further presented in the same year \cite{Van2019ApJ...886L..31V}. 
It is found that the CARB MB prescription can well reproduce the observed mass transfer rates of all observed Galactic persistent NS~LMXBs. There are also more MB prescriptions have also been presented by other studies \cite{Matt2008ApJ...678.1109M,Matt2012ApJ...754L..26M,Reiners2012ApJ...746...43R}. The strength and weaknesses of different MB prescriptions presented by various studies in explaining the observational characteristics of LMXBs have been discussed in detail by \cite{Deng2021ApJ...909..174D}. They showed that convection-boosted MB prescription by \cite{Van2019MNRAS.483.5595V} seems to provide the best match with the observations. With using the newly suggested MB prescriptions, different studies \cite{Van2019MNRAS.483.5595V, Chen2021MNRAS.503.3540C} have investigated their effects on the formation and evolution of LMXBs, radio binary millisecond pulsars (BMSPs), extremely low-mass helium WDs and ultra-compact X-ray binaries (UCXBs). Interestingly, it has been found that the initial orbital period range of LMXBs which can evolve via detached NS+WD binaries into UCXBs becomes significantly wider if the ``intermediate MB prescription'' of \cite{Van2019MNRAS.483.5595V} is adopted. This therefore helps to relieve the long-standing fine-tuning problem in producing UCXBs presented by earlier studies \cite{van-der-Sluys2005A&A...431..647V,van-der-Sluys2005A&A...440..973V,Istrate2014A&A...571A..45I}. But it was found that the MB prescription of \cite{Van2019MNRAS.483.5595V} fails to explain the wide-BMSPs \cite{Chen2021MNRAS.503.3540C}. Further study showed that the CARB MB prescription can reproduce the LMXB phase as well as being able to form UCXBs and wide BMSPs systems \cite{Soethe2021MNRAS.506.3266S}.

There are also other ways that could affect mass transfer in a binary system such as the spin of the accretor. For example, by calculating the maximum accreted mass of an NS in the recycling process in a binary system with considering the effect of NS spin evolution, it has been found that massive NSs can accrete more material than low-mass NSs with other initial parameters fixed. In addition, the spin evolution of an NS can affect the maximum NS accreted masses due to the propeller effects \cite{Li2021ApJ...922..158L}. In summary, despite substantial efforts on the theoretical and observational sides, the constraints on the mass and angular momentum loss in the non-conservative mass transfer of a binary system are still quite weak. Future observations and multi-dimensional hydrodynamical simulations of mass transfer in a binary system are still needed to give a better understanding of non-conservative mass transfer.

\subsection{Extended binary evolution channels}

\subsubsection{The chemically homogeneous evolution of massive binaries}
\label{sec:GW150914}

\begin{figure*}[t]
    \centering
    \includegraphics[width=0.95\textwidth, angle=0]{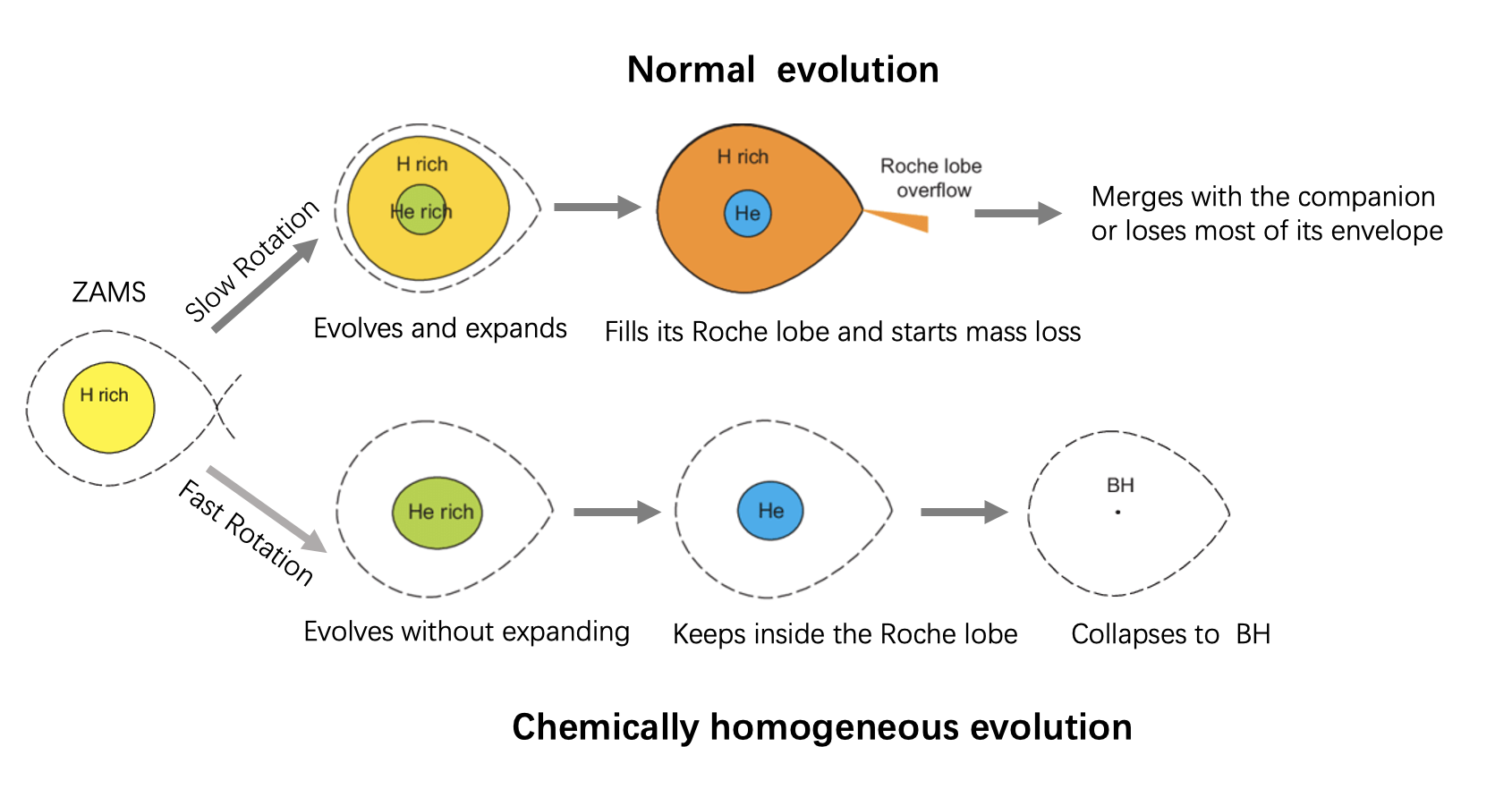}
    \caption{Schematic overview of the chemically homogeneous evolution (bottom tracks) in a close binary system. For a comparison, the `normal' evolution (top tracks) of a binary system is also presented. The figure is reprinted from \cite{Mandel2016} (see also \cite{deMink2008}).}
    \label{fig:chemical}
\end{figure*}

The first direct detection of gravitational waves GW150914 was announced by the LIGO and Virgo scientific collaboration on 2015
September 14, which has been confirmed to come from the merger of binary black holes (BBHs) with masses of about $30\,\mathrm{M_{\odot}}$ \cite{Abbott2016PhRvL.116f1102A,Abbott2016ApJ...832L..21A,Abbott2016PhRvL.116x1103A}. This reveals the existence of BHs with such heavy masses in the form of binaries. To form such heavy BHs, it requires the star to form in sub-solar metallicity environments because a higher metallicity can cause the strong stellar winds to lead to that the remnant mass of the star being below $30\,\mathrm{M_{\odot}}$ \cite{Belczynski2010ApJ...715L.138B,Spera2015MNRAS.451.4086S}. 
In the literature, several possible formation scenarios have been proposed to produce such heavy BBHs (see \cite{Abbott2016PhRvL.116f1102A}, and references therein), including (i) the dynamical formation in dense stellar clusters \cite{Rodriguez2016ApJ...824L...8R,OLeary2016ApJ...824L..12O,Askar2017MNRAS.464L..36A}, (ii) the classical isolated binary evolution channel involves non-conservative mass transfer or common envelope ejection \cite{Belczynski2016Natur.534..512B,Eldridge2016MNRAS.462.3302E}, (iii) the chemically homogeneous evolution in tidally locked close binaries \cite{Marchant2016A&A...588A..50M,Mandel2016MNRAS.458.2634M,de-Mink2016MNRAS.460.3545D}, and (iv) the coalescence of primordial black holes \cite{Sasaki2016PhRvL.117f1101S}. In this section, we will only focus on the discussion of the chemically homogeneous evolution scenario.

The so-called ``chemically homogeneous evolution'' was originally described for \emph{rotating single stars} by Maeder in 1987 \cite{Maeder1987A&A...178..159M}. If the stars rotate rapidly, they may experience efficient mixing throughout core H burning, which allows 
nuclear-burning products produced in the center of the stars to transport to their envelopes significantly \cite{Maeder1987A&A...178..159M,Maeder2000ARA&A..38..143M,Langer1992A&A...265L..17L,Heger2000ApJ...544.1016H}. This leads to that the build-up of a chemical gradient between core and envelope is prevented and the star evolves chemically homogeneously to avoid the strong post-MS expansion. As a consequence, the  star stays compact and their envelopes become He-rich gradually during their MS evolution. Meanwhile, the stars become hotter and more luminous and slowly evolve towards the He MS. In Fig.~\ref{fig:chemical}, we show a comparison between chemically homogeneous evolution in a close binary system and its nomral evolution. Different studies on the chemically homogeneous evolution of the stars have shown that this channel is favoured at low metallicity environments, in which a strong angular-momentum loss is avoided and the stars remain fully mixed during their MS evolution (e.g. \cite{Yoon2005A&A...443..643Y,Woosley2011ApJ...734...38W,Brott2011A&A...530A.115B,Koehler2015A&A...573A..71K,Szecsi2015A&A...581A..15S}). The chemically homogeneous evolution of rapidly rotating single stars have been suggested as progenitors of long-duration gamma-ray bursts \cite{Woosley2006ApJ...637..914W,Yoon2005A&A...443..643Y} because that this channel can produce rapidly spinning iron cores. 

Interestingly, it has been suggested that the tidal interaction of both stars in massive tight binary systems can force the stars to spin rapidly, which is sufficient to reach the conditions for chemically homogeneous evolution (\cite{de-Mink2009A&A...497..243D}; see also \cite{Song2016A&A...585A.120S,Marchant2016A&A...588A..50M}). By considering the mixing induced by rotation and angular momentum transport by magnetic fields, detailed studies have provided the initial binary parameter space that allows for chemically homogeneous evolution in massive close binaries (e.g. \cite{de-Mink2009A&A...497..243D,Song2016A&A...585A.120S,Marchant2016A&A...588A..50M}). The chemically homogeneous evolution could force two stars in a massive close binary to shrink and remain inside their Roche lobes to avoid the CE phase. This leads to the two stars gradually converting nearly all their H into He. As a consequence, two massive He stars are expected to be formed in this evolutionary scenario. These two massive He stars may eventually collapse to form two massive BHs that can explain BH masses inferred from GW150914 \cite{de-Mink2009A&A...497..243D,Marchant2016A&A...588A..50M}. Merging binary black holes formed through chemically homogeneous
evolution in massive tight binary systems have been investigated with both detailed binary stellar evolution calculations (e.g. \cite{Marchant2016A&A...588A..50M,Song2016A&A...585A.120S}) and Monte Carlo simulations (e.g. \cite{Mandel2016MNRAS.458.2634M,de-Mink2009A&A...497..243D}). It has been suggested that about 500 merging BHs events per year could be detected with advanced ground-based detectors operating at full sensitivity in the chemically homogeneous evolution channel, and the typical total masses of two BHs in this channel are $50$-$110\,\mathrm{M_{\odot}}$. Also, this channel seems to favour producing binary BHs with comparable masses with mass ratios $>0.6$ \cite{Mandel2016MNRAS.458.2634M,de-Mink2009A&A...497..243D}. Furthermore, it has been found that the chemically homogeneous evolution channel can fully explain the inferred parameters and merger rates of the GW150914 \cite{Mandel2016MNRAS.458.2634M,de-Mink2009A&A...497..243D}. 

However, whether all of the detected GW events were produced through a single formation scenario is still unclear \cite{Stevenson2017NatCo...814906S,Belczynski2020A&A...636A.104B}. In addition, as for other proposed formation channels, the chemically homogeneous evolution channel also has some uncertainties regarding the predictions of the formation of BBHs such as the simplified treatment of the mixing processes. Also, solid observational evidence for the existence of chemically homogeneously evolving stars is still missing \cite{de-Mink2009A&A...497..243D,de-Mink2016MNRAS.460.3545D,Mandel2016MNRAS.458.2634M}, although a few observed binary systems have been suggested to undergo chemically homogeneous evolution such as HD~5980 \cite{Koenigsberger2014AJ....148...62K} and VFTS~352 \cite{Almeida2015ApJ...812..102A}. Finally, it seems that the chemically homogeneous evolution channel has difficulties in explaining relatively low-mass BBHs such as GW151226 with a total mass of about $21.8\,\mathrm{M_{\odot}}$.

\subsubsection {Wind Roche-lobe overflow mass transfer}
\label{sec:CEMP}

\begin{figure*}[t]
    \centering
    \includegraphics[width=0.95\textwidth, angle=0]{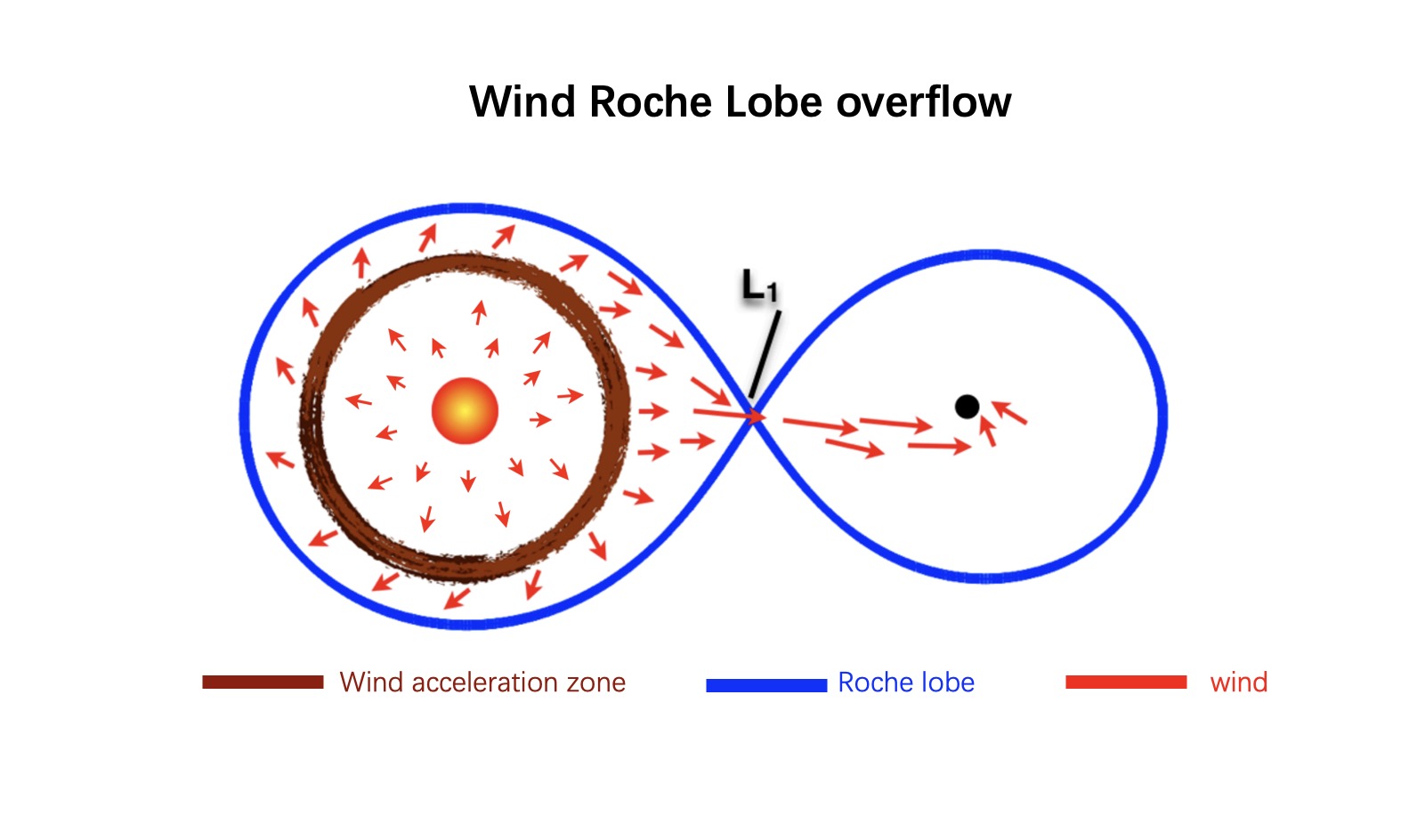}
    \caption{Schematic illustration of the wind Roche-lobe overflow (WRLOF) in a binary system (reproduced from the talk of Ph. Podsiadlowski, see also \cite{Mohamed2007ASPC..372..397M}). The WRLOF occurs when the radius of the wind acceleration zone (red circle) of the donor star is larger than (or comparable to) the Roche-lobe radius (blue curve) of the mass-losing star. The accreting star is given in a black dot.} 
    \label{fig:WRLOF}
\end{figure*}

The wind mass transfer plays an important role in the formation of peculiar stars formed in the binary scenario, in particular for the formation of Carbon-enhanced metal-poor (CEMP) stars . Early studies on the formation of CEMP stars generally adopted the canonical Bondi-Hoyle-Lyttleton (BHL) model \cite{Bondi1944MNRAS.104..273B} for the treatment of wind mass transfer from an AGB companion. The BHL model assumes that the wind speed is much faster than the orbital speed of the accreting star. However, AGB stars in wide binary systems are expected to have a slow wind speed of $5-35\,\mathrm{km\,s^{-1}}$ that is comparable with the orbital velocity their orbital velocities (e.g. \cite{Vassiliadis1993ApJ...413..641V,Knapp1998ApJS..117..209K,Hoefner2018A&ARv..26....1H}). A number of recent hydrodynamical simulations of mass transfer in a binary system with an AGB donor have shown that the accretion efficiency of AGB winds and the amount of angular-momentum loss can be significantly different from those described by the BHL model (e.g. \cite{Theuns1993MNRAS.265..946T,Theuns1996MNRAS.280.1264T, Mohamed2007ASPC..372..397M,de-Val-Borro2009ApJ...700.1148D,Shagatova2016A&A...588A..83S,Liu2017ApJ...846..117L,Chen2017MNRAS.468.4465C,Saladino2018A&A...618A..50S,Siess2022A&A...667A..75S}, and references therein).

In particular, there are some studies have presented a new mass transfer mode in binary systems with an evolved giant donor - the so-called ``wind Roche-lobe overflow (i.e. wind-RLOF, or WRLOF)'' \cite{Mohamed2007ASPC..372..397M,Mohamed2012BaltA..21...88M} - in which wind material fills the donor's Roche-lobe and transfers to the accretor through the inner Lagrangian point, as for the standard RLOF. As shown in Fig.~\ref{fig:WRLOF}, the WRLOF occurs when the radius of the wind acceleration region of the donor star is larger than the Roche-lobe radius of the donor star. The accretion rate in WRLOF model could be 100 times higher than that given by the canonical BHL model \cite{Mohamed2007ASPC..372..397M,Mohamed2012BaltA..21...88M,Abate2013A&A...552A..26A}. This new mass transfer mode (i.e. WRLOF) between RLOF and wind mass-transfer is expected to have important implications for stellar and galactic chemical evolution such as the formation of SNe Ia \cite{Abate2017MmSAI..88..308A,Ikiewicz2019MNRAS.485.5468I}, CEMP stars \cite{Abate2013A&A...552A..26A} and X-ray binaries \cite{El-Mellah2019A&A...622L...3E,Zuo2021A&A...649L...2Z}, etc. Very interestingly, by investigating the effect of WRLOF model on the formation of CEMP stars, it has been found that the number of CEMP stars predicted by the WRLOF model can increase by a factor of $1.2$-$1.8$ compared to that given by the classical BHL model \cite{Abate2013A&A...552A..26A}. 

To comprehensively investigate how the WRLOF affects the evolution and fates of binary systems, one need to provide a realistic prescription of the WRLOF that can be implemented into 1D stellar evolution and population-synthesis codes. However, how mass and angular momentum accretion (or loss) efficiencies depend on binary properties (e.g. binary orbital period, mass ratio, mass-loss rate and wind velocity, and so on) is still quite uncertain, which requires future improvements of two- or three-dimensional hydrodynamial simulations of WRLOF to cover a wider range of binary parameters, and to better consider the detailed wind acceleration mechanism due to star pulsation and radiation pressure on dust \cite{Liu2017ApJ...846..117L,Mohamed2007ASPC..372..397M}.

\subsection{Binary population synthesis}
\label{sec:bps}

The traditional stellar evolution theory generally focuses on the evolution of a single object (a star and/or binary) at a time. This leads to that it is not able to present the statistical properties of stellar and/or binary populations. However, a galaxy consists of thousands of millions of stars,  and more than half of stars are in binaries. Understanding the properties of galaxies and different types of stars requires a statistical investigation of stellar populations. In addition, growing large surveys such as the Sloan Digital Sky Survey (SDSS) (\cite{York2000}), Large Area Multi-Object Fiber Spectroscopic Telescope (LAMOST) survey (\cite{Zhao2012}), Gaia (\cite{Gaia2016}), \emph{Kepler} (\cite{Borucki2010}) and TESS (\cite{Ricker2014}), have revealed the statistical properties of stars and galaxies from the observational side, which makes significant progress in astrophysics. Therefore, to provide insight into the formation and evolution properties of different stars and binaries rely on not only individual studies of each object but also systematical investigations of stellar and/or binary populations from the theoretical side, and then comparing theoretical predictions with the observations. This leads to the proposal of ``binary population synthesis (BPS)'' approach.

BPS approach was developed by astronomers in  the 1990s to rapidly evolve a large sample of binary stars over the course of their lifetimes to study their evolution and properties of various types of close binary populations with a simplified treatment for some physical processes such as the CEE, the stability and rate of mass transfer (see \cite{Han2020RAA}, for a recent review). BPS has been very successfully used for studies of many aspects of astrophysics. For instance, it has been widely used to investigate the formation and evolution of exotic objects, including double black holes (BBHs; \cite{Kruckow2018MNRAS.481.1908K,Stevenson2017NatCo...814906S,Breivik2020ApJ...898...71B}), double neutron stars (e.g. \cite{Belczynski2002ApJ...572..407B,Oslowski2011MNRAS.413..461O,Kruckow2020A&A...639A.123K}), double white dwarfs (e.g. \cite{Han1995MNRAS.272..800H,Han1998MNRAS.296.1019H,Li2023A&A...669A..82Z}), Type Ia supernovae (e.g. \cite{Yungelson1994ApJ...420..336Y,Nelemans2001A&A...365..491N,Han2004MNRAS.350.1301H,Ruiter2009ApJ...699.2026R,Wang2009MNRAS.395..847W,Meng2009MNRAS.395.2103M,Toonen2012A&A...546A..70T,Claeys2014A&A...563A..83C,Liu2018MNRAS.475.5257L}), gamma ray bursts (e.g. \cite{Chrimes2020MNRAS.491.3479C,Belczynski2006ApJ...648.1110B}), X-ray binaries (e.g. \cite{Chen2021MNRAS.503.3540C}), pulsars (e.g. \cite{Willems2002MNRAS.337.1004W,Zhu2015MNRAS.454.1725Z}), novae (e.g. \cite{Chen2014MNRAS.445.1912C,Chen2015MNRAS.453.3024C,Chen2016MNRAS.458.2916C}), cataclysmic variables (e.g. \cite{Nelemans2001A&A...368..939N,Goliasch2015ApJ...809...80G}), hot subdwarfs stars (e.g. \cite{Han2002MNRAS.336..449H,Han2003MNRAS.341..669H,Chen2013MNRAS.434..186C,Wu2018A&A...618A..14W}), barium stars (e.g. \cite{Han1995MNRAS.277.1443H,Izzard2009A&A...508.1359I}), blue stragglers (e.g. \cite{Chen2008MNRAS.387.1416C,Chen2009MNRAS.395.1822C}), etc. In addition, it is a robust approach to studying the spectral energy distribution and the chemical evolution of galaxies (e.g. \cite{Zhang2004A&A...415..117Z,Han2007MNRAS.380.1098H}). 
Many BPS codes have been developed, e.g. Scenario Machine (\cite{Lipunov1996A&A...310..489L,Lipunov2009ARep...53..915L}), SeBa (\cite{Portegies-Zwart1996A&A...309..179P,Nelemans2001A&A...365..491N,Nelemans2001A&A...368..939N,Toonen2012A&A...546A..70T}),  
Yunnan Model (\cite{Han1998MNRAS.296.1019H,Han2002MNRAS.336..449H,Han2003MNRAS.341..669H,Zhang2002MNRAS.334..883Z,Zhang2004A&A...415..117Z,Zhang2005MNRAS.357.1088Z}), BSE (\cite{Hurley2002MNRAS.329..897H}), 
StarTrack (\cite{Belczynski2002ApJ...572..407B}), BiSEPS (\cite{Willems2002MNRAS.337.1004W}), BPASS (\cite{Eldridge2008MNRAS.384.1109E}), 
SYCLIST (\cite{Georgy2014A&A...566A..21G}), COMPAS (\cite{Stevenson2017NatCo...814906S}), MOBSE (\cite{Giacobbo2018MNRAS.474.2959G}), Combine (\cite{Kruckow2018MNRAS.481.1908K}), dart-board (\cite{Andrews2018ApJS..237....1A}), COSMIC (\cite{Breivik2020ApJ...898...71B}),
etc.

\begin{figure*}[t]
    \centering
    \includegraphics[width=0.7\textwidth, angle=0]{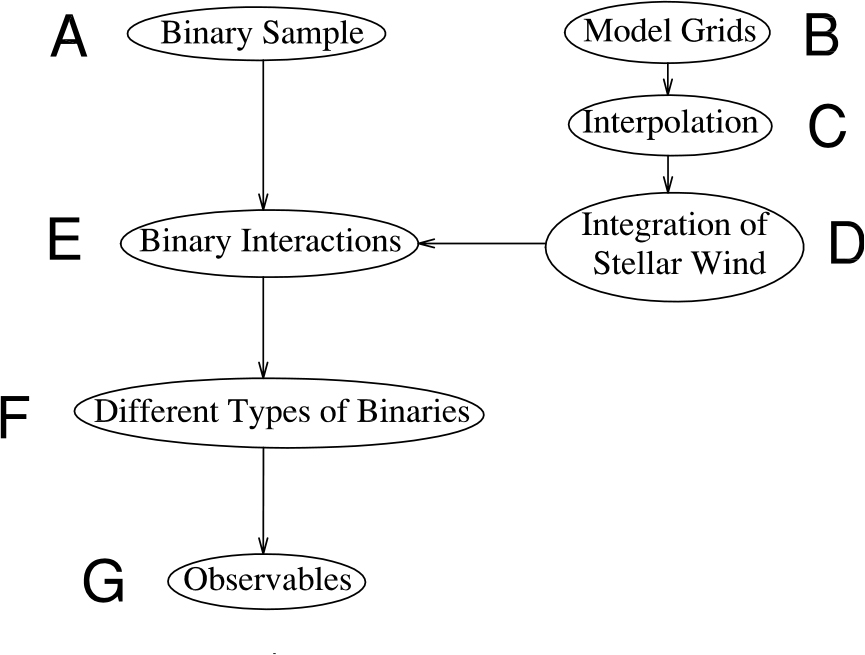}
    \caption{Major steps in BPS studies. The figure is adopted from \cite{Han2003MNRAS.341..669H}, see also Figure~2 of \cite{Han2020RAA}.}
    \label{fig:bps-step}
\end{figure*}

As shown in Fig.~\ref{fig:bps-step}, BPS studies generally involve several major steps as follows (see also Fig.~2 of \cite{Han2020RAA}): (i) Generating a large sample of binaries by setting initial distributions of some parameters (e.g. masses, periods, mass ratios); (ii) Following the evolution of both components of every binary system and their interactions over the course of the binary lifetime; (iii) Deriving the diverse properties and observables of binary populations such as the frequency of an important astrophysical event; (iv) Comparing the results of BPS calculations with the corresponding observations.

The details of the BPS approach have been reviewed by Han et al \cite{Han2020RAA}, including the role of BPS, its general picture and main ingredients, and the various components that comprise it. Here, only skim the surface of it. First, to generate sample binaries, we need to set some basic input parameters and assumptions, which are described below.

\begin{enumerate}
\item[(1)] \textbf{\emph{The star formation rate (SFR)}} --- One can simply assume a constant SFR over the last $15\,\mathrm{Gyr}$ in most cases. 

\item[(2)] \textbf{\emph{The initial mass function (IMF)}} --- The IMFs of \cite{Salpeter1955ApJ...121..161S} and  \cite{Miller1979ApJS...41..513M} are usually adopted for analytic studies and BPS studies, respectively (see also \cite{Kroupa1993MNRAS.262..545K}). A canonical discussion on the IMF has been given by \cite{Kroupa2001MNRAS.322..231K}. Generally, the primary mass is generated with the formula of \cite{Eggleton1989ApJ...347..998E},
\begin{equation}
M_1={0.19X\over (1-X)^{0.75}+0.032(1-X)^{0.25}}, 
\end{equation}
where $X$ is a random number uniformly distributed between 0 and 1, and the adopted ranges of primary masses are 0.8 to $126.0\,M_\odot$.

\item[(3)] \textbf{\emph{The mass-ratio distribution}} --- BPS studies mainly take a constant mass-ratio distribution (e.g. \cite{Mazeh1992ApJ...401..265M,Duchene2013ARA&A..51..269D}),
\begin{equation}
n(1/q)=1,\qquad  0\leq 1/q \leq 1, 
\end{equation}
where $q=M_1/M_2$. An alternative distribution of mass-ratio is the case where the masses of both binary components are chosen randomly and
independently from the same IMF.

\item[(4)] \textbf{\emph{The orbital period distribution}} --- We simply assume that all stars are members of binary systems and that the distribution of separations is constant in $\log a$, where $a$ is the separation, for wide
binary separations, and falls off smoothly at close separations (for a discussion, see \cite{Duquennoy1991A&A...248..485D,Duchene2013ARA&A..51..269D}),
\begin{equation}
an(a)=
\begin{cases} 
           \alpha_{\rm sep}({a \over a_0})^m, \& a\leq a_0;\cr
           \alpha_{\rm sep}, \&                  a_0 < a < a_1\cr
\end{cases}
\end{equation}
where $\alpha_{\rm sep} \approx 0.070$, $a_0=10\,R_\odot$,
$a_1=5.75\times 10^6\,R_\odot=0.13\,{\rm pc}$, and $m\approx 1.2$.
This distribution means that there is an equal number of wide binary
systems per logarithmic interval and that approximately 50\%
of stellar systems are binaries with orbital periods less than
100\,yr.

\item[(5)] \textbf{\emph{The orbital eccentricity distribution}} --- Under an assumption of that all binaries are circularized, the eccentricity can set to be $e=0$.

\end{enumerate}

\begin{figure*}[t]
   \centering
   \includegraphics[width=0.95\textwidth, angle=0]{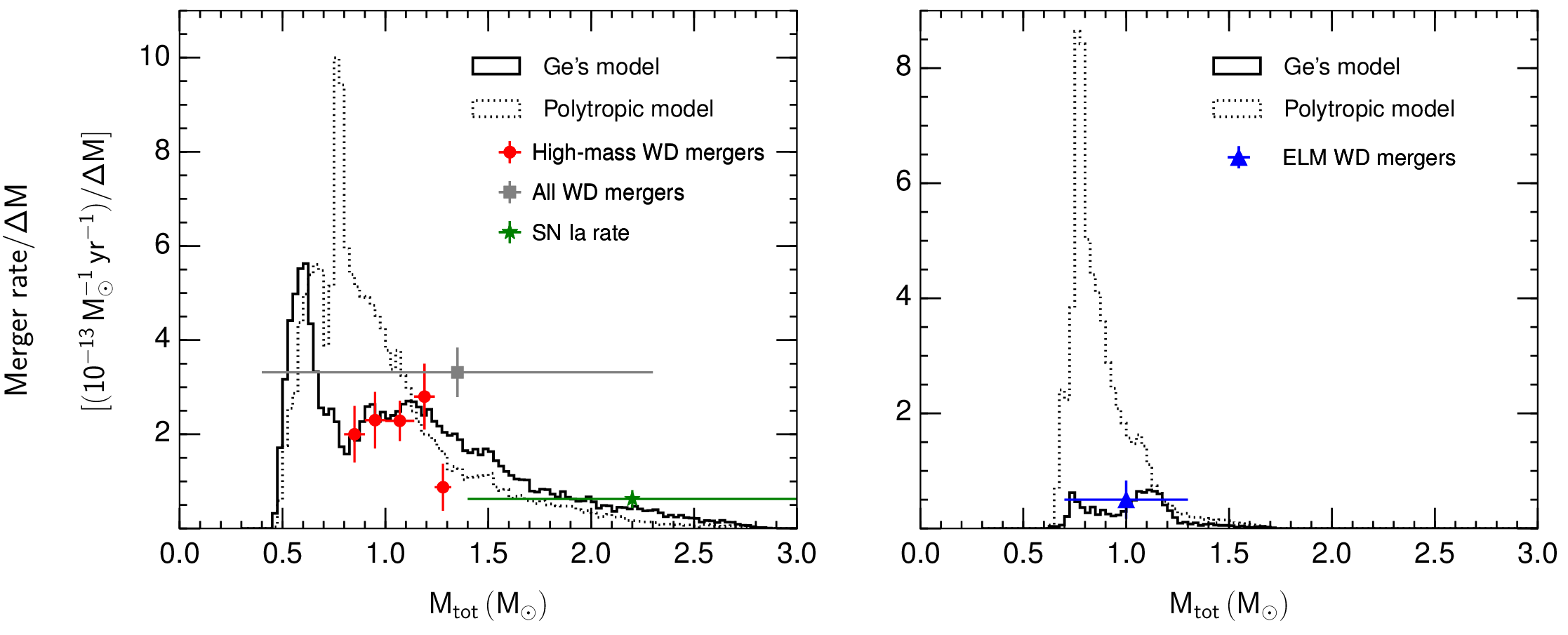}
   \caption{Comparison between merger rate distributions inferred from the observations and theoretical predictions from the polytropic (\emph{solid curves}) and Ge's (\emph{dotted curves}) models. Different coloured points with error bars present the observational rates of DWD mergers that produce $0.8-1.32\,M_{\odot}$ WDs (\emph{red points}; \cite{Cheng2020ApJ...891..160C}), SNe Ia (\emph{green points}; \cite{Li2011MNRAS.412.1473L}), ELM WD mergers (\emph{blue point}; \cite{Brown2020ApJ...889...49B}), and all DWD mergers (\emph{gray point}; \cite{Maoz2018MNRAS.476.2584M}), respectively. The figure is reproduced from \cite{Li2023A&A...669A..82Z} with permission \copyright\ ESO.}
   \label{Fig:DWDs}
   \end{figure*}

Once a large sample of initial binary systems is generated, we need to follow the evolution of both components of every binary system until the studied important objects are formed. The studies of binary populations in principle need to follow the evolution of every binary system in detail. However, it is not feasible to evolve a population of binary stars with a detailed stellar evolution code. Therefore, the evolution of each component in a binary system in BPS studies is commonly done by using either the rapid fit evolutionary calculations or interpolations in available stellar evolution model grids. This simplifying treatment for the evolution of the single star might introduce some uncertainties in the results although these are typically small compared to other uncertainties of BPS studies.

Moreover, there are different physical processes that need to deal with for the interactions between two components of a binary system, including mass-transfer mechanisms such as stellar wind and RLOF, mass transfer rate, prescriptions for the stability of mass transfer, tidal evolution, CEE, and mass and angular momentum loss from the binary system, etc. These physical processes are generally treated with some simplifying assumptions in BPS studies. 
For instance, prescriptions for the stability of mass transfer in BPS studies are commonly treated as the parametrization (e.g. \cite{Hurley2002MNRAS.329..897H,Han2020RAA}). This also introduces more uncertainties in BPS results. However, most of these physical processes remain quite uncertain and cannot be modelled in detail with a stellar evolution code.   

In summary, BPS is a robust approach to deriving the observables of binary populations in the studies of many aspects of astrophysics. However, one should keep in mind that theoretical predictions from BPS studies still have uncertainties due to weak constraints on the initial inputs and assumptions, and some physical processes in binary evolution. For instance, a series of BPS calculations were performed by adopting two different descriptions for mass transfer stability in a binary system (i.e. the adiabatic mass loss model \cite{Ge2015ApJ...812...40G} --- Ge's model and the composite polytropic model) to investigate the influence of a mass transfer stability criterion on double white dwarf (DWD) populations\cite{Li2023A&A...669A..82Z}. The study suggested that mass transfer stability plays an important role in the formation and properties of DWD populations as well as in the progenitors of SNe Ia and detectable GW sources. In addition, they found that Ge's model seems to provide a better match with observational properties of DWD populations such as the merger rate distribution per Galaxy and the space density in the Galaxy (Fig.~\ref{Fig:DWDs}; \cite{Li2023A&A...669A..82Z}).

\section{Stellar Nucleosynthesis}
\label{sec:nucleosynthesis}

\begin{figure*}[t]
    \centering
    \includegraphics[width=0.95\textwidth, angle=0]{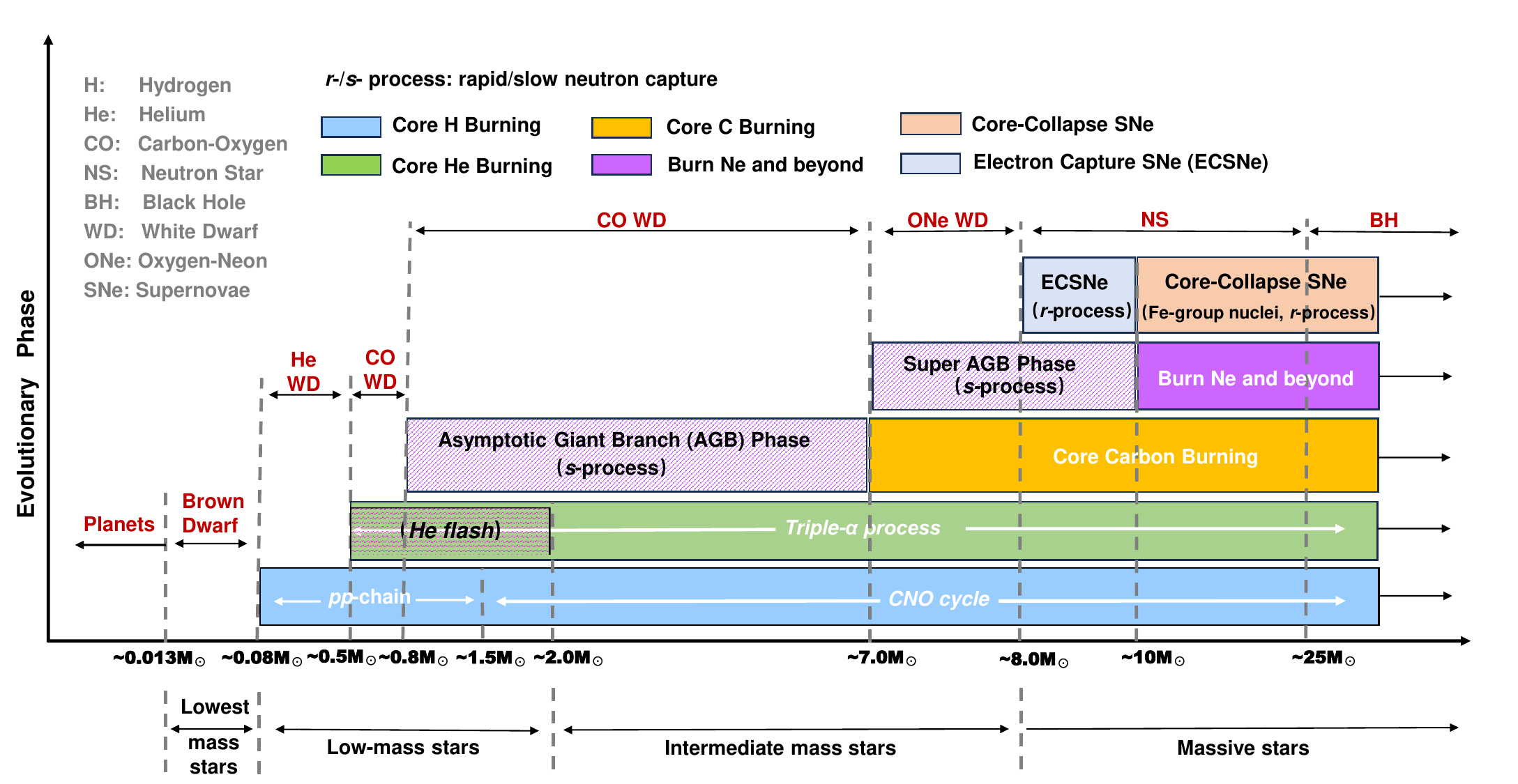}
    \caption{Schematic overview of how the main nuclear processes in the star depend on its masses at solar metallicity. The figure is reproduced base on Figure~1 of \cite{Karakas2014}.}
    \label{fig:stellar_nuclear}
\end{figure*}

Nucleosynthesis first occurred within a few minutes of the Big Bang, which produced light elements (hydrogen, helium, and lithium). 
Elements heavier than helium (which are known in astronomy as metals) are thought to be produced in the interiors of stars during their evolution and in the final explosion. The evolution of nucleosynthesis of a single star primarily depends on its initial mass and metallicity (which is denoted by the symbol $Z$). The first-generation stars formed after the Big Bang had abundances of about 76\% hydrogen, 24\% helium, and a tiny fraction of heavier elements (i.e. an extremely low Z). 

\begin{figure*}[t]
    \centering
    \includegraphics[width=0.95\textwidth, angle=0]{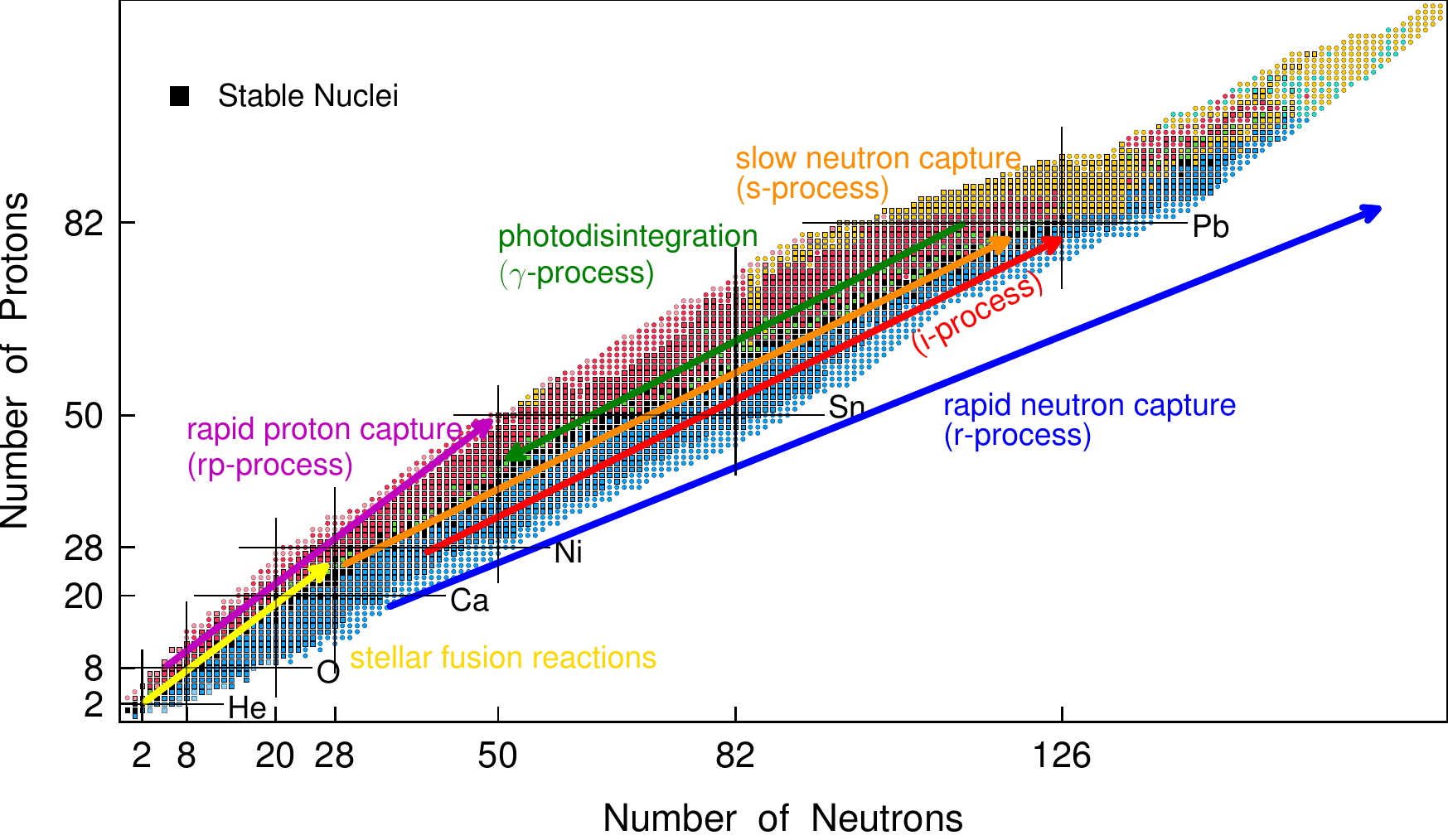}
    \caption{Schematic overview of the nuclear processes in the Universe on the chart of nuclides. Here, r-, s-, i-, rp-, and $\rm \gamma-$process respectively represent the rapid neutron capture, slow neutron capture, intermediate neutron capture, rapid proton capture, and photodisintegration process.}
    \label{fig:nuclear_chart}
\end{figure*}

Figure~\ref{fig:stellar_nuclear} presents the evolution of nucleosynthesis in stars. 
It starts from the hydrogen core burning with the $pp$ chains or the CNO cycle, then to helium-core burning with triple-$\alpha$ reactions, 
and to carbon and oxygen core burning, and so on. 
Finally, an onion-like structure has been formed with the iron core in the center. For low and intermediate-mass stars, the carbon- and oxygen-core burning and the following series of reactions may not happen at all. The envelope would be ejected to space via superwind, leaving white dwarfs as remnants. An important process for those stars is the slow neutron capture process (s-process) when the stars are in the AGB stage, leading to the production of several elements heavier than Fe. An intermediate neutron process (known as i-process) has been proposed to occur in AGB stars to account for the abundant pattern observed in metal-poor carbon-enhanced stars(see Section \ref{sec:CEMP-neutron}). Those elements may be dredged up to the surface to be observed by us. In fact, the productions in the core or burning shell have several chances to be taken onto the surface when the stars climb upwards the giant branches, which are known as the first/second/third dredge-up due to the development of the surface convection zone. 
Massive stars end their lives via supernovae or direct collapse to black holes. Abundant nucleosynthesis will take on in the supernovae. The rapid neutron capture process (r-process) is believed to occur in core-collapse supernovae, producing some other heavier elements. Fig.~\ref{fig:nuclear_chart} summarizes the element origin from stellar evolution.

Two components in a binary system could interact with each other. As a result, the surface material of one star (i.e., the donor star) can be transferred to the other star (i.e., the accretor) through either RLOF, wind, or WRLOF. This leads to the formation of different important objects such as novae, kilonovae, type Ia supernovae, carbon-enhanced metal poor stars, barium stars, and so on. Therefore, one can expect that nucleosynthesis in binary stars are significantly different from that in single star as discussed in the above section. Here, we briefly discuss the nucleosynthesis of some peculiar objects that are widely thought to be formed in binary systems.

\subsection{Novae}

It is commonly understood that a nova outburst results from a thermonuclear runaway (TNR) on the surface of a WD (usually CO- or ONe-WD) accreting hydrogen- or helium-rich matter from a low-mass companion in a close binary system \cite{Gehrz1998,Yaron2005,Denissenkov2013}. The TNR is triggered by the accumulation of a hydrogen-rich envelope on the WD continues until a critical pressure is achieved at the base of the accreted envelope, which requires a mass-arrection rate of $\lesssim10^{-9}\,\rm M_{\odot}\,yr^{-1}$ onto the WD. Once the TNR starts, the temperature rises rapidly to 10-100 million K and the entire envelope is burnt within a few minutes, ejecting mass and elements. Nova outbursts play an important role in the chemical evolution of Galaxy. Although it is generally thought that novae can provide about 0.3 per cent of the interstellar medium in the Galaxy, growing evidence both in observational and theoretical sides have suggested that novae may be the most important contributors of $\rm ^{7}Li$, $\rm ^{13}C$, $\rm ^{15}N$, $\rm ^{17}O$, as well as the radioactive isotopes like  $\rm ^{7}Be$, $\rm ^{22}Na$ and $\rm ^{26}Al$ in the Galaxy \cite{Jose1998,Denissenkov2013}. However, more observations are still needed to confirm such processes in novae. 

The nucleosynthesis in novae is thought to operate dominantly by both the \textit{pp} chains as well as the CNO cycle. Meanwhile, the NeNa and MgAl cycles are also thought to operate in nova nucleosynthesis to alter the isotopic compositions of Ne and Mg. However, the details of nucleosynthesis in novae are very hard to model because of short timescale of the event, determination of density and temperature of the burning region. In addition, convection could play a critical role in the nova explosion. However, it is unclear that how the accreted shell is mixed with the outer layers of the underlying WD, which also poses challenges to modelling nova nucleosynthesis \cite{Denissenkov2013, Guo2022aa}. The observations have detected the enrichments of intermediate-mass elements (e.g., C, N, O, Ne, Na, Mg, and Al) in nova ejecta \cite{Gehrz1998,Jose1998}, which is thought to be strong evidence of mixing between the material of WD and the accreted matter. Nonetheless, the exact mixing fraction in novae remain unknown. In addition, the large uncertainties of some challenging reactions such as $\rm ^{18}F(p,\alpha)\rm ^{15}O$, $\rm ^{25}Al(p,\gamma)\rm ^{26}Si$ and $\rm ^{30}P(p,\gamma)\rm ^{31}S$ also have a strong impact on nova nucleosynthesis \cite{Jose2006, Jose2017}, which still requires improvements in these related reaction rates by future experimental investigations.

\subsection{Merge of double neutron stars and kilonovae}

Theoretical studies on the inspiral and merger of double neutron stars have been carried out since the 1970s (e.g. \cite{Lattimer1974ApJ...192L.145L}), showing that neutron-rich matter that undergoes rapid neutron capture (i.e. r-process) nucleosynthesis is ejected in the violent merger of two neutron stars (e.g. \cite{Eichler1989Natur.340..126E,Freiburghaus1999ApJ...525L.121F,Rosswog1999A&A...341..499R,Goriely2011ApJ...738L..32G,Perego2014MNRAS.443.3134P,Just2015MNRAS.448..541J,Sekiguchi2016PhRvD..93l4046S,Kasen2017Natur.551...80K,Pian2017Natur.551...67P,Cowan2021RvMP...93a5002C}). Therefore, the mergers of double neutron stars have long been proposed to be one of the main astrophysical sites for the occurrence of the r-process which can assemble into rare heavy elements such as gold and platinum. The radioactive decay of unstable nuclei synthesized during the merger of double neutron stars can power a rapidly evolving electromagnetic transient called a ``kilonova'' (e.g. \cite{Li1998ApJ...507L..59L,Metzger2010MNRAS.406.2650M,Roberts2011ApJ...736L..21R,Kasen2013ApJ...774...25K,Kasen2019ApJ...876..128K}). Detecting the electromagnetic spectrum of a kilonova is therefore expected to provide the direct evidence for the r-process nucleosynthesis in the merger of double neutron stars. However, there has been a lack of direct observational evidence for this hypothesis.

The first gravitational wave (GW) signal of the merger of double neutron stars was detected by the Laser Interferometer Gravitational-Wave Observatory (LIGO) in August of 2017, i.e. GW170817 \cite{Abbott2017PhRvL.119p1101A}, which opens the new era of multimessenger astronomy (e.g. \cite{Abbott2017ApJ...848L..12A,Abbott2017ApJ...848L..13A,Pian2017Natur.551...67P,Margalit2017ApJ...850L..19M,Radice2018ApJ...852L..29R}). More importantly, the discovery of an electromagnetic counterpart (i.e. a kilonova AT2017gfo, \cite{Coulter2017Sci...358.1556C,Soares-Santos2017ApJ...848L..16S,Arcavi2017Natur.551...64A}) to GW170817 for the fist time provides the important and direct opportunity to probe the nucleosynthesis in merger of double neutron stars \cite{Kasen2017Natur.551...80K,Pian2017Natur.551...67P}. Theoretical studies have shown that the ejected matter of neutron star mergers contains two distinct 
components: (i) The neutron-rich ejecta composed of heavy r-process elements ($\mathrm{A\ge140}$), which radiate red/infrared light, producing a ``red'' kilonova; (ii) The less neutron-rich ejecta composed of light r-process elements ($\mathrm{A\leq140}$), which radiate blue/optical light, producing a ``blue'' kilonova (e.g. \cite{Kasen2017Natur.551...80K,Cowperthwaite2017ApJ...848L..17C,Drout2017Sci...358.1570D}). The behaviour of the AT2017gfo has been found to be in good agreement with theoretical predictions for kilonovae. Furthermore, the ejecta mass for the red and blue components of the kilonova light curve of AT2017gfo has been roughly estimated to be about $5\times10^{-2}\,\mathrm{M_{\odot}}$ and  $2\times10^{-2}\,\mathrm{M_{\odot}}$, respectively \cite{Kasen2017Natur.551...80K,Curtis2023MNRAS.518.5313C}. Given that a substantial ejecta mass has been inferred from AT2017gfo, the question is whether or not the binary neutron-star mergers is the dominant producer of heavy r-process elements in the Universe. Early studies have suggested that a mass production rate of r-process nuclei of $\sim10^{-6}\,\mathrm{M_{\odot}\,yr^{-1}}$ is required to explain the r-process abundances in the Milky Way \cite{Qian2000ApJ...534L..67Q,Kasen2017Natur.551...80K}. Under an  assumption of that all neutron star mergers have the same productions of r-process elements as for AT2017gfo and by taking a Galactic binary neutron-star merger rate of $8\times10^{-5}\,\mathrm{yr^{-1}}$ (which is however still uncertain; \cite{Kalogera2004ApJ...601L.179K,Abadie2010CQGra..27q3001A,Abbott2016ApJ...832L..21A,Kim2015MNRAS.448..928K}), the mass production rates of heavy and light r-process elements from neutron star mergers are expected to be $\sim4\times10^{-6}\,\mathrm{M_{\odot}\,yr^{-1}}$ and $\sim2\times10^{-6}\,\mathrm{M_{\odot}\,yr^{-1}}$, respectively. This is approximately consistent with the Galactic r-process production rates. As a consequence, Kasen et al (2017) \cite{Kasen2017Natur.551...80K} suggested that the binary neutron-star mergers may be the dominant contributors to the production of heavy r-process elements in the Galaxy. However, there are still some uncertainties on the estimates of binary neutron-star rate and ejecta mass derived from the observation of a kilonova such as the radioactive heating rate and viewing angle
effects (for a detailed discussion on these uncertainties, see \cite{Kasen2017Natur.551...80K}). To give a stronger conclusion on the question of whether the binary neutron-star mergers is the dominant producer of heavy r-process elements in the Universe, future GW detections of more neutron star mergers and follow-up multi-messenger observations of their electromagnetic counterparts are still needed.

\subsection{Type Ia supernovae}

SNe Ia are widely accepted to be thermonuclear explosions of WDs. SNe Ia have been used as the standard candles to measure the accelerating expansion of the Universe and constrain the properties of Dark Energy \cite{Riess1998AJ....116.1009R,Schmidt1998ApJ...507...46S,Perlmutter1999ApJ...517..565P}. SNe Ia are the important elements factories in the universe \cite{Arnett1969Ap&SS...5..180A,Nomoto1984ApJ...286..644N,Thielemann1986A&A...158...17T,Seitenzahl2017hsn..book.1955S}. They are responsible for the production of iron-group elements (IGEs; e.g. Ti, Cr, Mn, Fe, Co, Ni, Cu and Zn) and intermediate-mass elements (IMEs; e.g. Mg, Si, S, Ar and Ca). The nucleosynthetic outcomes of SNe Ia strongly depend on the local fuel density of the exploding WD. This means that different explosion mechanisms (which are still mysterious) of SNe Ia may lead to various nucleosynthetic outcomes \cite{Hillebrandt2013FrPhy...8..116H,Seitenzahl2017hsn..book.1955S,Lach2020A&A...644A.118L}. For instance, stable IGEs can only be synthesized when the densities are higher a few $10^{8}\,\mathrm{g\,cm^{-3}}$, which requires a CO WD above $1.2\,\mathrm{M_{\odot}}$. Depending on whether SNe Ia are triggered when the WDs approach the near Chandrasekhar mass or the sub-Chandrasekhar mass, a number of explosion models which involve deflagrations and/or detonations have been proposed in the literature, such as near Chandrasekhar-mass deflagrations, near Chandrasekhar-mass detonations, near Chandrasekhar-mass delayed detonations, gravitationally-confined detonations, sub-Chandrasekhar-mass double detonations, and violent mergers, etc (for a review of SN Ia explosion models, see \cite{Hillebrandt2013FrPhy...8..116H, Liu2023RAA}). 

Because a detonation with supersonic speed does not allow the WD to expand to lower densities, the near Chandrasekhar-mass detonation model produces too much $^{56}\mathrm{Ni}$ and IGEs but too little IMEs (e.g. Si, S and Ti) to match the observations of normal SNe Ia \cite{Arnett1969Ap&SS...5..180A}. As a consequence, deflagrations of near Chandrasekhar-mass WDs have been proposed for SNe Ia. A subsonic deflagration allows the WD to expand to lower densities, which reduces the production of $^{56}\mathrm{Ni}$ and gives an efficient IME yields. One widely discussed near Chandrasekhar-mass deflagration mode is the 1D `W7 model'  which was originally proposed by Nomoto et al \cite{Nomoto1984ApJ...286..644N}. The nucleosynthetic yields 
 of the W7 model such as $^{56}\mathrm{Ni}$ mass and IMEs are  consistent with those of normal SNe Ia. However, the W7 model uses a parametrized description for the thermonuclear burning process. 3D near Chandrasekhar-mass deflagration model has difficulties in producing enough $^{56}\mathrm{Ni}$ mass for normal SNe Ia ($\sim0.5\,\mathrm{M_{\odot}}$), and it leaves too much-unburned material such as C, O and Mg. However, weak deflagrations of near Chandrasekhar-mass WDs have recently suggested to well reproducing the observational features of a sub-luminous subclass of SNe Ia, the so-called SNe Iax (e.g. \cite{Kromer2013MNRAS.429.2287K,Kromer2015MNRAS.450.3045K}). The near Chandrasekhar-mass delayed detonation model has been developed to overcome the shortcoming of nucleosynthetic yields of pure deflagration or detonation model \cite{Khokhlov1991A&A...245L..25K,Khokhlov1991A&A...245..114K}. In this model, an initial subsonic deflagration allows the WD to expand at an early phase, leading to the later-on detonation triggered at lower densities. On the one hand, the delayed detonation model enhances the yields of IMEs but reduces the production of $^{56}\mathrm{Ni}$ mass compared with those in the pure detonation model \cite{Hillebrandt2000ARA&A..38..191H,Hillebrandt2013FrPhy...8..116H}. On the other hand, it produces more $^{56}\mathrm{Ni}$ mass and IGEs than those in the pure deflagration model.

\begin{figure*}[t]
    \centering
    \includegraphics[width=0.95\textwidth, angle=0]{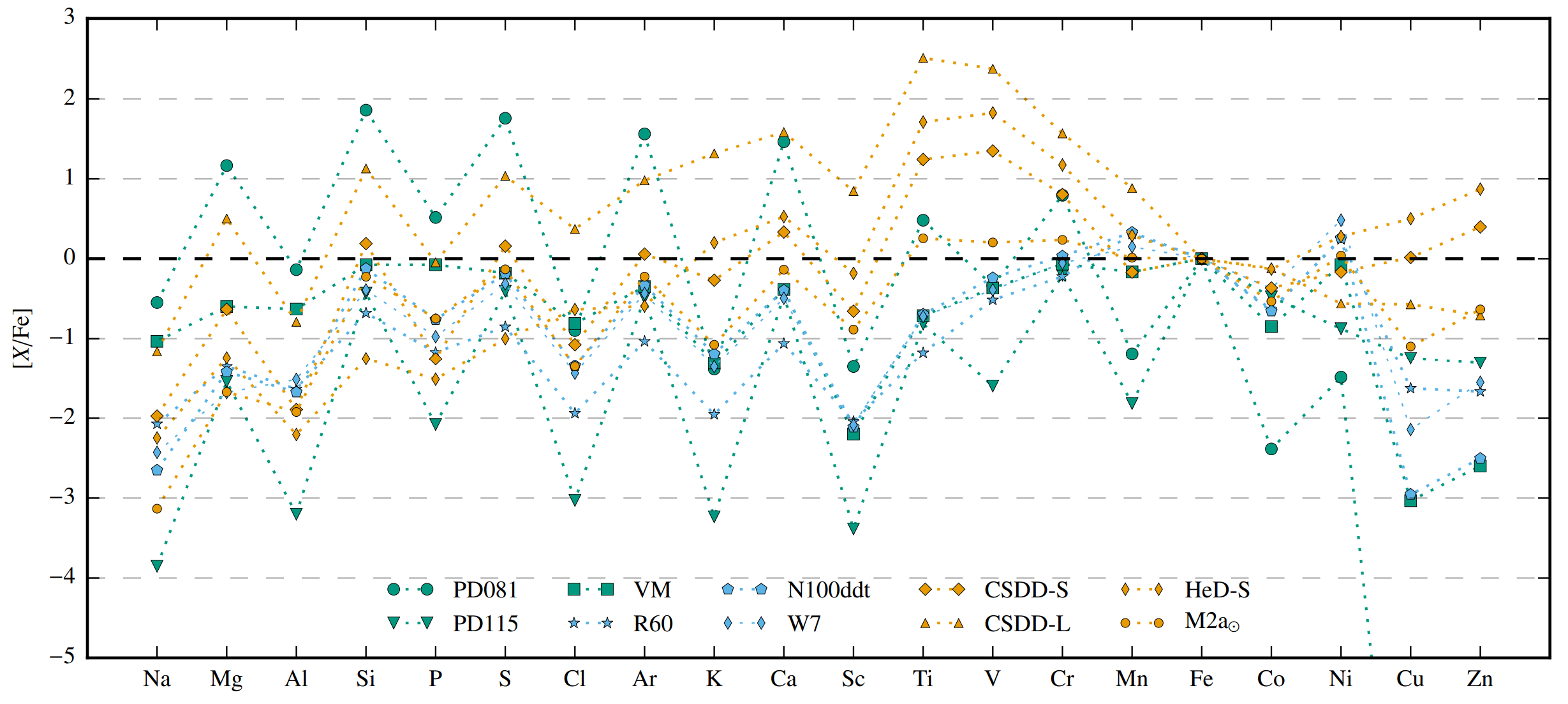}
    \caption{Nucleosynthesis yields of a variety of SN Ia explosion models, including one C-ignited violent merger (VM), two sub-Chandrasekhar pure detonations (PD081, PD115), three double detonations ($\rm M2a_{\odot}$, CSDD-S, CSDD-L), one helium detonation (HeD-S), two Chandrasekhar-mass pure deflagrations (R60, W7), and a Chandrasekhar-mass delayed detonation (N100ddt). Here, it presents elemental ratios to Fe (with radioacitve isotopes decayed to $\rm 2\times10^{9}\,yr$) compared to their solar ratios. The figure is reproduced from Figure~1 of \cite{Lach2020A&A...644A.118L} with permission \copyright\ ESO.}
    \label{fig:sne1a}
\end{figure*}

The density structure of a sub-Chandrasekhar mass WD is different from that of a near Chandrasekhar mass WD (see Fig.~1 of \cite{Seitenzahl2017hsn..book.1955S}). This is expected to give different nucleosynthetic yields. The sub-Chandrasekhar mass WDs can also trigger SN Ia explosion through a double-detonation mechanism (i.e. an initial detonation of the He shell and a subsequent core carbon detonation), in which a He shell is accumulated and ignited on the surface of the WD (e.g. \cite{Taam1980,Livne1990,Woosley1994,Bildsten2007,Fink2010,Gronow2021A&A...656A..94G,Shen2021ApJ...922...68S}). Comparing to explosive C-O burning, the initial detonation in a thick He shell ($\sim0.1$-$0.2\,\mathrm{M_{\odot}}$) typically leads to the over-production of $\alpha$-isotopes such as $^{36}\mathrm{Ar}$, $^{40}\mathrm{Ca}$, $^{44}\mathrm{Ti}$, $^{48}\mathrm{Cr}$ and/or $^{52}\mathrm{Fe}$, which seems to be inconsistent with the observations of normal SNe Ia (e.g. \cite{Kromer2010ApJ...719.1067K,Woosley2011ApJ...734...38W,Sim2012MNRAS.420.3003S}). However, the nucleosynthetic products of the outer layer on the top of the WD strongly depend on the mass, and chemical condition of the He shell. Recent studies have suggested that a thin He shell ($<0.1\,\mathrm{M_{\odot}}$) with pollution of CO seems to hold promise for explaining SNe Ia (e.g. \cite{Pakmor2013ApJ...770L...8P,Townsley2019,Gronow2021A&A...656A..94G,Shen2021ApJ...922...68S}). The double detonation mechanism could also occur during the violent merger of two WDs, which is known as ``\emph{He-ignited violent merger model}'' or the ``\emph{dynamically
driven double-degenerate double-detonation (D6)}'' model (e.g. \cite{Guillochon2010ApJ...709L..64G,Pakmor2013ApJ...770L...8P,Roy2022ApJ...932L..24R}). In D6 model, because the He shell on the top of primary WD is generally quite thin and the He detonation happens in the dynamic stage rather than in the hydrostatic equilibrium state, the influence of He detonation on the final nucleosynthesis products of the D6 model is not significant. This brings the D6 model to have the potential to explain the bulk of normal SNe Ia \cite{Pakmor2013ApJ...770L...8P,Shen2021ApJ...922...68S}. The violent merger of two WDs can also directly trigger a C-detonation in primary WD to cause an SN Ia explosion, which is called as ``\emph{C-ignited violent merger model}'' (e.g. \cite{Pakmor2010Natur.463...61P,Pakmor2011A&A...528A.117P,Pakmor2012ApJ...747L..10P}). Generally, the violent merger model enhances the production of neutron-rich electron capture nuclei such as $^{54}\mathrm{Fe}$ and $^{58}\mathrm{Ni}$ \cite{Seitenzahl2017hsn..book.1955S}. There are other explosion models have been suggested for SNe Ia such as the ``WD-WD collisions model''. The nucleosynthesis yields of a variety of SN Ia explosion models\footnote{A detailed review of nucleosynthetic yields of different explosion models of SNe Ia have been given by \cite{Seitenzahl2017hsn..book.1955S}.} have been presented in Fig.~\ref{fig:sne1a}. Very recently, by analyzing the nucleosynthesis yields of various SN Ia explosion models, it has concluded that both the Chandrasekhar-mass explosion models and the models including a He detonation can significantly contribute to the production of Mn, leading to super-solar values of [Mn/Fe] \cite{Lach2020A&A...644A.118L}.  In addition, they found that double-detonation models are able to produce Zn and Cu in super-solar ratios with respect to Fe \cite{Lach2020A&A...644A.118L}.

\subsection{Neutron-capture process in carbon-enhanced metal-poor stars}
\label{sec:CEMP-neutron}

Carbon-enhanced metal-poor stars (which are usually referred to as CEMP stars) are a subclass of chemically peculiar stars which show enhancement of carbon ($\mathrm{[C/Fe]}\gtrsim+1.0$; \cite{Aoki2007ApJ...655..492A}). CEMP stars contribute to a substantial fraction  (about $20\%$; \cite{Lucatello2006ApJ...652L..37L,Allen2012A&A...548A..34A,Lee2013AJ....146..132L,Placco2014ApJ...797...21P}) of the observed very metal-poor stars with $\mathrm{[Fe/H]}\lesssim-2.0$ \cite{Beers2005ARA&A..43..531B}, and the fraction of CEMP stars increases as the decrease of metallicity (e.g. \cite{Lee2013AJ....146..132L,Placco2014ApJ...797...21P,Arentsen2022MNRAS.515.4082A}; and references therein).  However, their exact origins are still poorly understood. CEMP stars are generally divided into four subclasses according to the abundance of barium (Ba) and europium (Eu) which are produced by slow ($s-$process) and rapid ($r-$process) neutron-capture processes (but the exact definitions for various subclasses of CEMP stars vary for different studies; \cite{Beers2005ARA&A..43..531B,Lugaro2012ApJ...747....2L,Abate2013A&A...552A..26A}): (i) \textbf{\emph{CEMP-s stars}}  with $\mathrm{[C/Fe]}>1.0$, $\mathrm{[Ba/Fe]}>1.0$ and $\mathrm{[Ba/Eu]}>0.5$. This subtype accounts for at least $80\%$ of all CEMP stars \cite{Aoki2007ApJ...655..492A}. (ii) \textbf{\emph{CEMP-r stars}} with $\mathrm{[C/Fe]}>1.0$ and $\mathrm{[Eu/Fe]}>1.0$. (iii) \textbf{\emph{CEMP-r/s stars}} with $\mathrm{[C/Fe]}>1.0$, $\mathrm{[Ba/Fe]}>1.0$, $\mathrm{[Eu/Fe]}>1.0$ and $\mathrm{[Ba/Eu]}>0$. (iv) \textbf{ \emph{CEMP-no stars}} with $\mathrm{[C/Fe]}>1.0$, $\mathrm{[Ba/Fe]}\le1.0$ and $\mathrm{[Eu/Fe]}\le1.0$. One of the key questions of CEMP stars is that how different subclasses of CEMP stars are formed, and whether they share the same formation scenario. A variety of scenarios are invoked to explain the observed abundance patterns of CEMP stars in the literature (e.g. \cite{Jonsell2006A&A...451..651J,Aoki2007ApJ...655..492A,Lugaro2009PASA...26..322L}). However, no single published scenario can consistently explain the observational signatures of CEMP stars. It has been suggested that CEMP-$r$ and CEMP-$no$ stars may result from pre-enhancement of the
interstellar medium when these stars formed (e.g. \citep{Cooke2014ApJ...791..116C,Frebel2015ARA&A..53..631F,Hansen2015A&A...583A..49H}), whereas CEMP-$r$ stars were also claimed to originate from the binary evolution (e.g. \cite{Komiya2007ApJ...658..367K}). CEMP-$s$ stars are generally thought to result from mass transfer from an AGB companion in a binary system because this formation scenario can explain the simultaneous enrichments of C, Ba, and other $s$-elements in CEMP-$s$ stars. In addition, this is supported by the fact of that most CEMP-$s$ stars show radial-velocity variations (e.g. \cite{Lucatello2006ApJ...652L..37L,Aoki2007ApJ...655..492A,Masseron2010A&A...509A..93M,Starkenburg2014MNRAS.441.1217S,Hansen2016A&A...588A...3H}; and references therein). However, the formation scenario of CEMP-$r/s$ stars is puzzling. Especially, the origin of enrichment of $r$-elements in CEMP-$r/s$ is strongly debated by different studies in the literature (e.g. \cite{Jonsell2006A&A...451..651J,Lugaro2009PASA...26..322L,Abate2015A&A...581A..62A,Abate2016A&A...587A..50A,Hampel2016ApJ...831..171H}, and references therein). The current AGB models do not reach sufficiently high neutron density to produce enough $r$-elements that match CEMP-$r/s$ stars. The majority of CEMP-$r/s$ stars are found in binary systems \cite{Lucatello2005ApJ...625..825L,Hansen2016A&A...588A...3H}. It is therefore thought that CEMP-$r/s$ stars may be formed due to both pre-enrichment with $r$-process material and pollution of $s$-process matter from an AGB companion \cite{Jonsell2006A&A...451..651J,Lugaro2009PASA...26..322L,Bisterzo2011MNRAS.418..284B,Abate2016A&A...587A..50A}. However, this model seems to have difficulties to explain the observed number of CEMP-$r/s$ stars \cite{Abate2016A&A...587A..50A}. Interestingly, the nucleosynthesis of the \emph{intermediate neutron-capture process} (i.e. $i$-process; see Section~\ref{sec:CEMP-neutron}) in AGB models can fit the observed abundances
of CEMP-$s/r$ stars very well \cite{Dardelet2014nic..confE.145D,Abate2016A&A...587A..50A,Hampel2016ApJ...831..171H,Karinkuzhi2021A&A...645A..61K}. However, whether or not the $i$-process can be active in AGB stars remains uncertain. By performing population synthesis calculations, different studies have investigated the formation of CEMP stars and attempted to reproduce the observed fraction ($9\%$-$25\%$; \cite{Frebel2006ApJ...652.1585F,Marsteller2005NuPhA.758..312M,Lucatello2006ApJ...652L..37L}) and distributions of different elements (e.g. C, Ba and other heavy elements) of CEMP stars (e.g. \cite{Lucatello2005ApJ...625..825L,Komiya2007ApJ...658..367K,Izzard2009A&A...508.1359I,Abate2013A&A...552A..26A,Abate2015A&A...581A..62A,Abate2015A&A...581A..22A}). However, only weak constraints have been placed on the origins of CEMP stars due to the uncertainties on the nucleosynthesis in AGB stars, initial mass function (IMF) at low metallicity, binary fraction, wind mass transfer process etc.

CEMP stars often show clear enhancements in C and neutron-capture elements such as Ba and Eu \cite{Aoki2007ApJ...655..492A}. Three neutron-capture processes are involved in the formation of different subclasses of CEMP stars: (i) slow ($s-$process) neutron-capture process; (ii) rapid ($r-$process) neutron-capture process; (iii) intermediate ($i-$process) neutron-capture process. The $r-$, $s-$ and $i-$process require completely different physical conditions (e.g. \cite{Karakas2014PASA...31...30K,Arcones2023A&ARv..31....1A}). The $s$ process requires low neutron densities of $10^{6}-10^{10}\,\mathrm{cm^{-3}}$, and the AGB stars and He-burning core of massive stars are commonly thought to be the predominant producer of $s$-process elements (e.g. \cite{Truran1981A&A....97..391T,Gallino1998ApJ...497..388G,Busso1999ARA&A..37..239B,Frischknecht2012A&A...538L...2F,Karakas2014PASA...31...30K,Arcones2023A&ARv..31....1A}). Most CEMP-$s$ stars are found in the binary systems, the pollution of $s$-process elements from an AGB donor is therefore suggested to be the most likely origin of CEMP-$s$ stars (see Section~\ref{sec:CEMP}). High neutron densities of $\gtrsim10^{20}\,\mathrm{cm^{-3}}$ is required to trigger $r$-process. The $r$-process is generally thought to occur during supernova explosions, accretion-induced collapses and merger of double neutron stars (e.g. \cite{Karakas2014PASA...31...30K,Thielemann2017ARNPS..67..253T,Arcones2023A&ARv..31....1A}, and references therein). Because such extreme conditions are unlikely to reach in AGB stars, the enhancement of $r$-process elements in CEMP-$r$ and/or CEMP-$r/s$ stars are suggested to result from the pre-enhancement of $r$-elements material when these stars formed \cite{Cooke2014ApJ...791..116C,Frebel2015ARA&A..53..631F}. Despite substantial efforts in the study of the effect of $s$- and $r$-processes on the formation of CEMP-$r/s$ stars, it is still difficult to explain the observed $s/r$ abundance pattern and other properties (e.g. the number, the observed correlation between the enrichment in the $s$- and $r$-process elements, the ratios of heavy $s$-process elements to light $s$-process elements) of CEMP-$r/s$ stars with different formation scenarios that involve  $s$- and $r$-processes (e.g. \cite{Jonsell2006A&A...451..651J,Lugaro2012ApJ...747....2L,Abate2015A&A...581A..22A,Abate2016A&A...587A..50A}). The $i$-process is expected to happen when neutron densities are in the order of $\sim10^{15}\,\mathrm{cm^{-3}}$. There are possible physical sites in stellar evolution that may allow $i$-process to occur such as when hydrogen is mixed into a convective helium-burning zoo during the evolution of low-mass AGB stars (e.g. \cite{Cowan1977ApJ...212..149C, Choplin2021A&A...648A.119C,Choplin2022A&A...667A.155C}). The $i$-process is able to produce sufficient $r$-process elements and a high ratio of heavy $s$-process elements to light $s$-process elements, it therefore has been used to successfully explain the observed abundance patterns of CEMP-$r/s$ stars in the context of wind mass transfer from an AGB or TPAGB donor in binary systems \cite{Dardelet2014nic..confE.145D,Abate2016A&A...587A..50A,Hampel2016ApJ...831..171H,Hampel2019ApJ...887...11H,Karinkuzhi2021A&A...645A..61K}. This is supported by most CEMP-$r/s$ belong to binary systems \cite{Hansen2016A&A...586A.160H,Karinkuzhi2021A&A...645A..61K}. However, how physical sites for the $i$-process can be identified in stellar evolution is still quite uncertain. More future studies are still needed to carry out the detailed calculations of the nucleosynthesis of the $i$-process and the evolution of low-mass AGB stars, which will provide insight into the importance of $i$-process in the formation of CEMP stars and the related nucleosynthesis.

\section{Summary and outlook}
\label{sec:sum}

In this review article, we have tried to highlight recent advances in our theoretical and observational understanding of binary stars.  Substantial progress has been made in binary star research due to both the increasing number of binary stars from different large surveys like SDSS, SEGUE, Gaia, GALAH, LAMOST, APOGGE and \emph{Kepler}, and because of the rapid development of numerical approaches and theoretical models. Nevertheless, there remain many uncertainties. Future progress in binary stars, both observational and theoretical, will enable us to have a better understanding of the fundamental problems of stellar and binary astrophysics. Here, we point to some directions that may lead to future progress.

\begin{enumerate}

\item[(1)]\textbf{Further simulations of binary processes}. A complete understanding of the evolution of binary stars firmly relies on whether the related fundamental physical processes are clearly determined. Binary mass transfer and common envelope evolution (CEE) are the two most important physical processes in binary evolution. Despite our attempt to model these two processes with a number of multi-dimensional simulations over the years, it is still difficult to reach a consensus due to the inherent multi-physics and multi-scale challenges. As a consequence, there are still many open questions. For example, what is the exact outcome of CEE? How much mass and angular momentum are lost from the binary system in non-conservative mass transfer? What are the exact mass-accretion efficiency and angular-momentum loss through wind mass transfer (or wind Roche-lobe overflow) in binary systems? Given the rapid development of numerical techniques and present-day computational power, we expect that simulations in near future will have sufficient accuracy to firmly constrain and/or solve these remaining fundamental problems of binary evolution. This will enable us to give answers to the above questions.

\item[(2)]\textbf{Further observations of exotic stars}. In this article, we have reviewed the recent discoveries of different types of binary stars and exotic objects related to various evolutionary processes in binary evolution. Clearly, the discoveries of these stars have significantly advanced our understanding of binary star evolution. For example, the observed luminous red novae (LRNe) are thought to be very likely candidates of CEE events. The discovery of samples of post-CE binaries have placed constraints on the CE ejection efficiency, although its exact value is still unknown. In addition, reproducing the observational properties of short-period sdB binaries with WD companions have required non-conservative mass transfer. In general, the strict constraints on the fundamental evolutionary processes of binary stars is firmly based on the discoveries of different new types of binary systems and exotic objects that predicted from binary evolution. Relying on a number of present-day and future large surveys, in the nearest future we expect the discovery of a much bigger sample of such exotic events, including post-CE binaries, LRNe, sdB binaries, chemical peculiar stars, extremely low-mass WDs, runaway and hypervelocity stars, double compact objects, and so on. In particular, as pointed out above, the discovery of Thorne-Zytkow objects (TZOs) that contain an NS inside a RG is thought to be the direct evidence of stellar mergers from the CEE. However, none of such objects have been discovered yet.


\item[(3)]\textbf{Statistical properties of binary populations}. The statistical properties of binary populations, including binary fraction, orbital-period distribution, mass-ratio distribution and orbital eccentricity distribution, as well as their dependence on stellar types, environment and/or metallicities, are essential for our understanding of different aspects of star formation to galactic evolution. Moreover, the statistics of binaries are basic inputs of binary population synthesis studies, which will thus have a major effect on the formation and evolution of binary-related objects such as CVs, X-ray binaries, SNe Ia, millisecond pulsars, compact binary systems that emit detectable gravitational waves. Over the last few years, increasingly large survey projects have led to the discovery of a huge number of binary stars, which allows us to provide statistical properties of binary populations. Indeed, significant progress in statistics of binaries such as binary fraction, mass ratio distributions have been made. In particular, growing observational evidence has suggested the non-universality of the initial mass function (IMF) and the existence of correlations between different binary properties. Nevertheless, the statistical properties of binaries are still controversial, and their dependence on stellar types, environments or metallicities is only poorly constrained. Different ongoing surveys introduced in this review and upcoming massive survey projects like the \emph{James Webb Space Telescope (JWST)} \cite{Gardner2006SSRv..123..485G}, the Vera C. Rubin Observatory \emph{Legacy Survey of Space and Time (LSST) \cite{Ivezic2019ApJ...873..111I}}, the \emph{Chinese Survey Space Telescope (CSST)} \cite{Zhan2011SSPMA..41.1441Z}, and the \emph{Euclid space telescope} \cite{Laureijs2011arXiv1110.3193L} are expected to discover orders of magnitude more binary stars in the future, from wide non-interacting pairs to very close compact binaries. This will advance the studies of observational properties of binary populations a significant step further.

\item[(4)]\textbf{Development of binary population synthesis}. Binary population synthesis (BPS) provides the bridge between the theoretical properties of binary evolution models and observational properties of different binary-related events such as their rates, mass and orbital period distributions. However, the results given by BPS studies still have uncertainties. The BPS method is developed to rapidly evolve large samples of zero-age MS binaries initially generated with given IMF, binary fraction, and distributions of orbital period, mass ratio and eccentricity. On the one hand, uncertainties of these initial parameter distributions have a significant influence on the overall population properties of present-day binary populations. On the other hand, the evolution of binary systems needs to deal with different fundamental physical processes such as the CEE and mass transfer. As mentioned before, however, these binary processes remain unsolved, and they are treated in a simplified way in BPS studies, generally parameterized. This brings certain uncertainties on the formation and evolution of binary-related events, and therefore leads to that the comparison between BPS predictions and observations cannot give an undisputed conclusion. For instance, studies have shown that different choices of initial parameter distributions and binary evolution models could lead to the predicted rates of double compact objects changed by more than an order of magnitude. Future numerical simulations, as well as the discoveries of more exotic objects predicted by different evolutionary phases of binary systems, will give much stronger constraints on the fundamental processes of binary evolution. Meanwhile, statistical properties of binary populations are expected to be well-determined by much larger samples of binary stars from ongoing and upcoming massive survey projects. Therefore, there is hope that an accurate and universal BPS model may be developed in near future, which will be very helpful for examining and testing the stellar and binary evolution theory.

\item[(5)] \textbf{Nucleosynthesis in binary stars}. Stars are the main cosmic factories that produce elements heavier than He. Most stars spend their lives in binary systems or multiple systems, the nucleosynthesis related to binary stars therefore plays an important role in the production of different elements. We have reviewed the importance of the nucleosynthesis of binary-related objects in the past decade, both in observation and theory. For example, GW170817 has provided a great impact on the studies of r-process nucleosynthesis. Recent observations have revealed that novae should be the promising sites of $\rm{^{7}Li}$ production. In addition, the accretion of $i-$process elements from an AGB star onto its companion in a binary system has been found to provide a good explanation for the observed abundance patterns of CEMP-$r/s$ stars. On the observational side, we expect that future observations will discover a big sample of chemically peculiar stars such as CEMP stars, Ba stars and extremely metal-poor stars (Population~III stars) enriched with r/s-process elements. The statistical analysis of the chemical abundance properties of these peculiar stars gives essential insights into r-process, s-process, and big-bang nucleosynthesis, as well as binary evolution models. Moreover,  with more GW detection from mergers of double compact objects and follow-up observations of their electromagnetic counterparts such as the spectroscopic observations of r-process elements in kilonovae, we expect to greatly deepen our understanding of nucleosynthesis of heavy nuclei and EoS of neutron stars. On the theoretical side, future powerful computing resources will allow us to remarkably advance the theoretical modelling of binary processes to make binary models more realistic, which will provide more reliable calculations of nucleosynthesis of different binary-related objects. In addition, future computational power will enable us to consistently model SN explosion, dynamics of mergers of double compact objects, explosive nucleosynthesis, and synthetic outcome, and therefore eventually reveal the origin of elements.

\item[(6)] \textbf{Impacts on the cosmology}. SNe Ia have been used as the standard candles to successfully measure the cosmic distance and to place constraints on the properties of Dark Energy. It is widely accepted that progenitors of SNe Ia are in binary systems. However, the nature of SN Ia progenitors and their explosion mechanism remains a mystery, which significantly affects the precision of SN Ia cosmology. Future theoretical and observational progress on the binary stars that have been mentioned above will allow us to narrow down the constraints on SN Ia progenitor systems and explosion mechanisms. This will help to reduce the uncertainty of SN Ia cosmology and therefore deepen our understanding of Dark Energy.

\end{enumerate}

\section*{Acknowledgements}

We would like to thank the anonymous referee for useful comments that helped to improve this article. The authors would like to thank Hongwei Ge for fruitful discussions and for his contributions to some figures of this article. This work is supported by the National Natural Science Foundation of China (NSFC, No.\ 12125303, 12288102, 12090040/3), the National Key R\&D Program of China (Nos.\ 2021YFA1600403/1 and 2021YFA1600400), the Chinese Academy of Sciences (CAS) and the Natural Science Foundation of Yunnan Province (Nos.\ 202201BC070003, 202001AW070007), the International Centre of Supernovae, Yunnan Key Laboratory (No. 202302AN360001) and the ``Yunnan Revitalization Talent Support Program''—Science \& Technology Champion Project (NO.~202305AB350003).

\bibliography{ref}

\begin{thebibliography}{557}
\expandafter\ifx\csname natexlab\endcsname\relax\def\natexlab#1{#1}\fi
\providecommand{\url}[1]{\texttt{#1}}
\providecommand{\href}[2]{#2}
\providecommand{\path}[1]{#1}
\providecommand{\DOIprefix}{doi:}
\providecommand{\ArXivprefix}{arXiv:}
\providecommand{\URLprefix}{URL: }
\providecommand{\Pubmedprefix}{pmid:}
\providecommand{\doi}[1]{\href{http://dx.doi.org/#1}{\path{#1}}}
\providecommand{\Pubmed}[1]{\href{pmid:#1}{\path{#1}}}
\providecommand{\bibinfo}[2]{#2}
\ifx\xfnm\relax \def\xfnm[#1]{\unskip,\space#1}\fi
\bibitem[{{Torres} et~al.(2010){Torres}, {Andersen}, and
  {Gim{\'e}nez}}]{Torres2010}
\bibinfo{author}{G.~{Torres}}, \bibinfo{author}{J.~{Andersen}},
  \bibinfo{author}{A.~{Gim{\'e}nez}},
\newblock \bibinfo{title}{{Accurate masses and radii of normal stars: modern
  results and applications}},
\newblock \bibinfo{journal}{\aapr} \bibinfo{volume}{18} (\bibinfo{year}{2010})
  \bibinfo{pages}{67--126}. \DOIprefix\doi{10.1007/s00159-009-0025-1}.
\bibitem[{{Serenelli} et~al.(2021){Serenelli}, {Weiss}, {Aerts}, {Angelou},
  {Baroch}, and et~al.}]{Serenelli2021}
\bibinfo{author}{A.~{Serenelli}}, \bibinfo{author}{A.~{Weiss}},
  \bibinfo{author}{C.~{Aerts}}, et~al.,
\newblock \bibinfo{title}{{Weighing stars from birth to death: mass
  determination methods across the HRD}},
\newblock \bibinfo{journal}{\aapr} \bibinfo{volume}{29} (\bibinfo{year}{2021})
  \bibinfo{pages}{4}. \DOIprefix\doi{10.1007/s00159-021-00132-9}.
\bibitem[{{Mowlavi} et~al.(2022){Mowlavi}, {Holl}, {Lec{\oe}ur-Ta{\"\i}bi},
  {Barblan}, {Kochoska}, {Pr{\v{s}}a}, {Mazeh}, {Rimoldini}, {Gavras},
  {Audard}, {Jevardat de Fombelle}, {Nienartowicz}, {Garcia-Lario}, and
  {Eyer}}]{Mowlavi2022}
\bibinfo{author}{N.~{Mowlavi}}, \bibinfo{author}{B.~{Holl}},
  \bibinfo{author}{I.~{Lec{\oe}ur-Ta{\"\i}bi}}, et~al.,
\newblock \bibinfo{title}{{Gaia Data Release 3. The first Gaia catalogue of
  eclipsing binary candidates}},
\newblock \bibinfo{journal}{arXiv e-prints}   (\bibinfo{year}{2022})
  \bibinfo{pages}{arXiv:2211.00929}.
\bibitem[{{Duch{\^e}ne} and {Kraus}(2013)}]{Duchene2013ARA&A..51..269D}
\bibinfo{author}{G.~{Duch{\^e}ne}}, \bibinfo{author}{A.~{Kraus}},
\newblock \bibinfo{title}{{Stellar Multiplicity}},
\newblock \bibinfo{journal}{\araa} \bibinfo{volume}{51} (\bibinfo{year}{2013})
  \bibinfo{pages}{269--310}.
  \DOIprefix\doi{10.1146/annurev-astro-081710-102602}.
\bibitem[{{Moe} and {Di Stefano}(2017)}]{Moe2017ApJS..230...15M}
\bibinfo{author}{M.~{Moe}}, \bibinfo{author}{R.~{Di Stefano}},
\newblock \bibinfo{title}{{Mind Your Ps and Qs: The Interrelation between
  Period (P) and Mass-ratio (Q) Distributions of Binary Stars}},
\newblock \bibinfo{journal}{\apjs} \bibinfo{volume}{230} (\bibinfo{year}{2017})
  \bibinfo{pages}{15}. \DOIprefix\doi{10.3847/1538-4365/aa6fb6}.
\bibitem[{{Raghavan} et~al.(2010){Raghavan}, {McAlister}, {Henry}, {Latham},
  {Marcy}, {Mason}, {Gies}, {White}, and {ten
  Brummelaar}}]{Raghavan2010ApJS..190....1R}
\bibinfo{author}{D.~{Raghavan}}, \bibinfo{author}{H.~A. {McAlister}},
  \bibinfo{author}{T.~J. {Henry}}, et~al.,
\newblock \bibinfo{title}{{A Survey of Stellar Families: Multiplicity of
  Solar-type Stars}},
\newblock \bibinfo{journal}{\apjs} \bibinfo{volume}{190} (\bibinfo{year}{2010})
  \bibinfo{pages}{1--42}. \DOIprefix\doi{10.1088/0067-0049/190/1/1}.
\bibitem[{{Hettinger} et~al.(2015){Hettinger}, {Badenes}, {Strader},
  {Bickerton}, and {Beers}}]{Hettinger2015ApJ...806L...2H}
\bibinfo{author}{T.~{Hettinger}}, \bibinfo{author}{C.~{Badenes}},
  \bibinfo{author}{J.~{Strader}}, et~al.,
\newblock \bibinfo{title}{{Statistical Time-resolved Spectroscopy: A Higher
  Fraction of Short-period Binaries for Metal-rich F-type Dwarfs in SDSS}},
\newblock \bibinfo{journal}{\apjl} \bibinfo{volume}{806} (\bibinfo{year}{2015})
  \bibinfo{pages}{L2}. \DOIprefix\doi{10.1088/2041-8205/806/1/L2}.
\bibitem[{{Liu}(2019)}]{Liu2019MNRAS.490..550L}
\bibinfo{author}{C.~{Liu}},
\newblock \bibinfo{title}{{Smoking gun of the dynamical processing of
  solar-type field binary stars}},
\newblock \bibinfo{journal}{\mnras} \bibinfo{volume}{490}
  (\bibinfo{year}{2019}) \bibinfo{pages}{550--565}.
  \DOIprefix\doi{10.1093/mnras/stz2274}.
\bibitem[{{Eggen}(1967)}]{Eggen1967}
\bibinfo{author}{O.~J. {Eggen}},
\newblock \bibinfo{title}{{Contact binaries, II.}},
\newblock \bibinfo{journal}{\memras}  \bibinfo{volume}{70}
  (\bibinfo{year}{1967}) \bibinfo{pages}{111}.
\bibitem[{{Chen} et~al.(2016){Chen}, {de Grijs}, and {Deng}}]{Chen2016}
\bibinfo{author}{X.~{Chen}}, \bibinfo{author}{R.~{de Grijs}},
  \bibinfo{author}{L.~{Deng}},
\newblock \bibinfo{title}{{Contact Binaries as Viable Distance Indicators: New,
  Competitive (V)JHK $_{s}$ Period-Luminosity Relations}},
\newblock \bibinfo{journal}{\apj} \bibinfo{volume}{832} (\bibinfo{year}{2016})
  \bibinfo{pages}{138}. \DOIprefix\doi{10.3847/0004-637X/832/2/138}.
\bibitem[{{Gettel} et~al.(2006){Gettel}, {Geske}, and {McKay}}]{Gettel2006}
\bibinfo{author}{S.~J. {Gettel}}, \bibinfo{author}{M.~T. {Geske}},
  \bibinfo{author}{T.~A. {McKay}},
\newblock \bibinfo{title}{{A Catalog of 1022 Bright Contact Binary Stars}},
\newblock \bibinfo{journal}{\aj} \bibinfo{volume}{131} (\bibinfo{year}{2006})
  \bibinfo{pages}{621--632}. \DOIprefix\doi{10.1086/498016}.
\bibitem[{{Jayasinghe} et~al.(2020){Jayasinghe}, {Stanek}, {Kochanek},
  {Shappee}, {Pinsonneault}, {Holoien}, {Thompson}, {Prieto}, {Pawlak},
  {Pejcha}, {Pojmanski}, {Otero}, {Hurst}, and {Will}}]{Jayasinghe2020}
\bibinfo{author}{T.~{Jayasinghe}}, \bibinfo{author}{K.~Z. {Stanek}},
  \bibinfo{author}{C.~S. {Kochanek}}, et~al.,
\newblock \bibinfo{title}{{The ASAS-SN catalogue of variable stars - VII.
  Contact binaries are different above and below the Kraft break}},
\newblock \bibinfo{journal}{\mnras} \bibinfo{volume}{493}
  (\bibinfo{year}{2020}) \bibinfo{pages}{4045--4057}.
  \DOIprefix\doi{10.1093/mnras/staa518}.
\bibitem[{{Petrosky} et~al.(2021){Petrosky}, {Hwang}, {Zakamska}, {Chandra},
  and {Hill}}]{Petrosky2021}
\bibinfo{author}{E.~{Petrosky}}, \bibinfo{author}{H.-C. {Hwang}},
  \bibinfo{author}{N.~L. {Zakamska}}, et~al.,
\newblock \bibinfo{title}{{Variability, periodicity, and contact binaries in
  WISE}},
\newblock \bibinfo{journal}{\mnras} \bibinfo{volume}{503}
  (\bibinfo{year}{2021}) \bibinfo{pages}{3975--3991}.
  \DOIprefix\doi{10.1093/mnras/stab592}.
\bibitem[{{Green} et~al.(2022){Green}, {Maoz}, {Mazeh}, {Faigler}, {Shahaf},
  {Gomel}, {El-Badry}, and {Rix}}]{Green2022}
\bibinfo{author}{M.~J. {Green}}, \bibinfo{author}{D.~{Maoz}},
  \bibinfo{author}{T.~{Mazeh}}, et~al.,
\newblock \bibinfo{title}{{15000 Ellipsoidal Binary Candidates in TESS: Orbital
  Periods, Binary Fraction, and Tertiary Companions}},
\newblock \bibinfo{journal}{arXiv e-prints}   (\bibinfo{year}{2022})
  \bibinfo{pages}{arXiv:2211.06194}.
\bibitem[{{Andrews} et~al.(2017){Andrews}, {Chanam{\'e}}, and
  {Ag{\"u}eros}}]{Andrews2017}
\bibinfo{author}{J.~J. {Andrews}}, \bibinfo{author}{J.~{Chanam{\'e}}},
  \bibinfo{author}{M.~A. {Ag{\"u}eros}},
\newblock \bibinfo{title}{{Wide binaries in Tycho-Gaia: search method and the
  distribution of orbital separations}},
\newblock \bibinfo{journal}{\mnras} \bibinfo{volume}{472}
  (\bibinfo{year}{2017}) \bibinfo{pages}{675--699}.
  \DOIprefix\doi{10.1093/mnras/stx2000}.
\bibitem[{{Andrews} et~al.(2018){Andrews}, {Chanam{\'e}}, and
  {Ag{\"u}eros}}]{Andrews2018}
\bibinfo{author}{J.~J. {Andrews}}, \bibinfo{author}{J.~{Chanam{\'e}}},
  \bibinfo{author}{M.~A. {Ag{\"u}eros}},
\newblock \bibinfo{title}{{Wide binaries in Tycho-Gaia II: metallicities,
  abundances and prospects for chemical tagging}},
\newblock \bibinfo{journal}{\mnras} \bibinfo{volume}{473}
  (\bibinfo{year}{2018}) \bibinfo{pages}{5393--5406}.
  \DOIprefix\doi{10.1093/mnras/stx2685}.
\bibitem[{{Han} et~al.(2020){Han}, {Ge}, {Chen}, and {Chen}}]{Han2020RAA}
\bibinfo{author}{Z.-W. {Han}}, \bibinfo{author}{H.-W. {Ge}},
  \bibinfo{author}{X.-F. {Chen}}, \bibinfo{author}{H.-L. {Chen}},
\newblock \bibinfo{title}{{Binary Population Synthesis}},
\newblock \bibinfo{journal}{Research in Astronomy and Astrophysics}
  \bibinfo{volume}{20} (\bibinfo{year}{2020}) \bibinfo{pages}{161}.
  \DOIprefix\doi{10.1088/1674-4527/20/10/161}.
\bibitem[{{Orlov}(1961)}]{Orlov1961SvA.....4..845O}
\bibinfo{author}{A.~A. {Orlov}},
\newblock \bibinfo{title}{{A Generalization of the Roche Model}},
\newblock \bibinfo{journal}{\sovast}  \bibinfo{volume}{4}
  (\bibinfo{year}{1961}) \bibinfo{pages}{845}.
\bibitem[{{Paczynski}(1976)}]{Paczynski1976}
\bibinfo{author}{B.~{Paczynski}},
\newblock \bibinfo{title}{{Common Envelope Binaries}},
\newblock in: \bibinfo{editor}{P.~{Eggleton}}, \bibinfo{editor}{S.~{Mitton}},
  \bibinfo{editor}{J.~{Whelan}} (Eds.), \bibinfo{booktitle}{Structure and
  Evolution of Close Binary Systems}, volume~\bibinfo{volume}{73},
  \bibinfo{year}{1976},  p.~\bibinfo{pages}{75}.
\bibitem[{{Abbott} et~al.(2016){Abbott}, {Abbott}, {Abbott}, {Abernathy},
  {Acernese}, {Ackley}, {Adams}, {Adams}, et~al., {LIGO Scientific
  Collaboration}, and {Virgo Collaboration}}]{Abbott2016PhRvX...6d1015A}
\bibinfo{author}{B.~P. {Abbott}}, \bibinfo{author}{R.~{Abbott}},
  \bibinfo{author}{T.~D. {Abbott}}, et~al.,
\newblock \bibinfo{title}{{Binary Black Hole Mergers in the First Advanced LIGO
  Observing Run}},
\newblock \bibinfo{journal}{Physical Review X} \bibinfo{volume}{6}
  (\bibinfo{year}{2016}) \bibinfo{pages}{041015}.
  \DOIprefix\doi{10.1103/PhysRevX.6.041015}.
\bibitem[{{Ge} et~al.(2010){Ge}, {Webbink}, {Han}, and
  {Chen}}]{Ge2010Ap&SS.329..243G}
\bibinfo{author}{H.~{Ge}}, \bibinfo{author}{R.~F. {Webbink}},
  \bibinfo{author}{Z.~{Han}}, \bibinfo{author}{X.~{Chen}},
\newblock \bibinfo{title}{{Stellar adiabatic mass loss model and
  applications}},
\newblock \bibinfo{journal}{\apss} \bibinfo{volume}{329} (\bibinfo{year}{2010})
  \bibinfo{pages}{243--248}. \DOIprefix\doi{10.1007/s10509-010-0286-1}.
\bibitem[{{Ge} et~al.(2015){Ge}, {Webbink}, {Chen}, and
  {Han}}]{Ge2015ApJ...812...40G}
\bibinfo{author}{H.~{Ge}}, \bibinfo{author}{R.~F. {Webbink}},
  \bibinfo{author}{X.~{Chen}}, \bibinfo{author}{Z.~{Han}},
\newblock \bibinfo{title}{{Adiabatic Mass Loss in Binary Stars. II. From
  Zero-age Main Sequence to the Base of the Giant Branch}},
\newblock \bibinfo{journal}{\apj} \bibinfo{volume}{812} (\bibinfo{year}{2015})
  \bibinfo{pages}{40}. \DOIprefix\doi{10.1088/0004-637X/812/1/40}.
\bibitem[{{Ge} et~al.(2020){Ge}, {Webbink}, and {Han}}]{Ge2020ApJS..249....9G}
\bibinfo{author}{H.~{Ge}}, \bibinfo{author}{R.~F. {Webbink}},
  \bibinfo{author}{Z.~{Han}},
\newblock \bibinfo{title}{{The Thermal Equilibrium Mass-loss Model and Its
  Applications in Binary Evolution}},
\newblock \bibinfo{journal}{\apjs} \bibinfo{volume}{249} (\bibinfo{year}{2020})
  \bibinfo{pages}{9}. \DOIprefix\doi{10.3847/1538-4365/ab98f6}.
\bibitem[{{Ivanova} et~al.(2013){Ivanova}, {Justham}, {Chen}, {De Marco},
  {Fryer}, {Gaburov}, {Ge}, {Glebbeek}, {Han}, {Li}, {Lu}, {Marsh},
  {Podsiadlowski}, {Potter}, {Soker}, {Taam}, {Tauris}, {van den Heuvel}, and
  {Webbink}}]{Ivanova2013A&ARv..21...59I}
\bibinfo{author}{N.~{Ivanova}}, \bibinfo{author}{S.~{Justham}},
  \bibinfo{author}{X.~{Chen}}, et~al.,
\newblock \bibinfo{title}{{Common envelope evolution: where we stand and how we
  can move forward}},
\newblock \bibinfo{journal}{\aapr} \bibinfo{volume}{21} (\bibinfo{year}{2013})
  \bibinfo{pages}{59}. \DOIprefix\doi{10.1007/s00159-013-0059-2}.
\bibitem[{{Roepke} and {De Marco}(2022)}]{Roepke2022}
\bibinfo{author}{F.~K. {Roepke}}, \bibinfo{author}{O.~{De Marco}},
\newblock \bibinfo{title}{{Simulations of common-envelope evolution in binary
  stellar systems: physical models and numerical techniques}},
\newblock \bibinfo{journal}{arXiv e-prints}   (\bibinfo{year}{2022})
  \bibinfo{pages}{arXiv:2212.07308}.
\bibitem[{{Ivanova} et~al.(2013){Ivanova}, {Justham}, {Avendano Nandez}, and
  {Lombardi}}]{Ivanova2013b}
\bibinfo{author}{N.~{Ivanova}}, \bibinfo{author}{S.~{Justham}},
  \bibinfo{author}{J.~L. {Avendano Nandez}}, \bibinfo{author}{J.~C.
  {Lombardi}},
\newblock \bibinfo{title}{{Identification of the Long-Sought Common-Envelope
  Events}},
\newblock \bibinfo{journal}{Science} \bibinfo{volume}{339}
  (\bibinfo{year}{2013}) \bibinfo{pages}{433}.
  \DOIprefix\doi{10.1126/science.1225540}.
\bibitem[{{Blagorodnova} et~al.(2017){Blagorodnova}, {Kotak}, {Polshaw},
  {Kasliwal}, {Cao}, {Cody}, {Doran}, {Elias-Rosa}, {Fraser}, {Fremling},
  {Gonzalez-Fernandez}, {Harmanen}, {Jencson}, {Kankare}, {Kudritzki},
  {Kulkarni}, {Magnier}, {Manulis}, {Masci}, {Mattila}, {Nugent}, {Ochner},
  {Pastorello}, {Reynolds}, {Smith}, {Sollerman}, {Taddia}, {Terreran},
  {Tomasella}, {Turatto}, {Vreeswijk}, {Wozniak}, and
  {Zaggia}}]{Blagorodnova2017}
\bibinfo{author}{N.~{Blagorodnova}}, \bibinfo{author}{R.~{Kotak}},
  \bibinfo{author}{J.~{Polshaw}}, et~al.,
\newblock \bibinfo{title}{{Common Envelope Ejection for a Luminous Red Nova in
  M101}},
\newblock \bibinfo{journal}{\apj} \bibinfo{volume}{834} (\bibinfo{year}{2017})
  \bibinfo{pages}{107}. \DOIprefix\doi{10.3847/1538-4357/834/2/107}.
\bibitem[{{MacLeod} et~al.(2017){MacLeod}, {Macias}, {Ramirez-Ruiz},
  {Grindlay}, {Batta}, and {Montes}}]{MacLeod2017}
\bibinfo{author}{M.~{MacLeod}}, \bibinfo{author}{P.~{Macias}},
  \bibinfo{author}{E.~{Ramirez-Ruiz}}, et~al.,
\newblock \bibinfo{title}{{Lessons from the Onset of a Common Envelope Episode:
  the Remarkable M31 2015 Luminous Red Nova Outburst}},
\newblock \bibinfo{journal}{\apj} \bibinfo{volume}{835} (\bibinfo{year}{2017})
  \bibinfo{pages}{282}. \DOIprefix\doi{10.3847/1538-4357/835/2/282}.
\bibitem[{{Li} et~al.(2022){Li}, {Onken}, {Wolf}, {N{\'e}meth}, {Bessell},
  {Li}, {Zhang}, {Li}, {Wang}, {Li}, {Luo}, {Chen}, {Ji}, {Chen}, and
  {Han}}]{Li2022}
\bibinfo{author}{J.~{Li}}, \bibinfo{author}{C.~A. {Onken}},
  \bibinfo{author}{C.~{Wolf}}, et~al.,
\newblock \bibinfo{title}{{A Roche lobe-filling hot subdwarf and white dwarf
  binary: possible detection of an ejected common envelope}},
\newblock \bibinfo{journal}{\mnras} \bibinfo{volume}{515}
  (\bibinfo{year}{2022}) \bibinfo{pages}{3370--3382}.
  \DOIprefix\doi{10.1093/mnras/stac1768}.
\bibitem[{{Moe} and {Di Stefano}(2017)}]{Moe2017}
\bibinfo{author}{M.~{Moe}}, \bibinfo{author}{R.~{Di Stefano}},
\newblock \bibinfo{title}{{Mind Your Ps and Qs: The Interrelation between
  Period (P) and Mass-ratio (Q) Distributions of Binary Stars}},
\newblock \bibinfo{journal}{\apjs} \bibinfo{volume}{230} (\bibinfo{year}{2017})
  \bibinfo{pages}{15}. \DOIprefix\doi{10.3847/1538-4365/aa6fb6}.
\bibitem[{{El-Badry} and {Rix}(2018)}]{El-Badry2018}
\bibinfo{author}{K.~{El-Badry}}, \bibinfo{author}{H.-W. {Rix}},
\newblock \bibinfo{title}{{Imprints of white dwarf recoil in the separation
  distribution of Gaia wide binaries}},
\newblock \bibinfo{journal}{\mnras} \bibinfo{volume}{480}
  (\bibinfo{year}{2018}) \bibinfo{pages}{4884--4902}.
  \DOIprefix\doi{10.1093/mnras/sty2186}.
\bibitem[{{El-Badry} et~al.(2021){El-Badry}, {Rix}, and
  {Heintz}}]{El-Badry2021}
\bibinfo{author}{K.~{El-Badry}}, \bibinfo{author}{H.-W. {Rix}},
  \bibinfo{author}{T.~M. {Heintz}},
\newblock \bibinfo{title}{{A million binaries from Gaia eDR3: sample selection
  and validation of Gaia parallax uncertainties}},
\newblock \bibinfo{journal}{\mnras} \bibinfo{volume}{506}
  (\bibinfo{year}{2021}) \bibinfo{pages}{2269--2295}.
  \DOIprefix\doi{10.1093/mnras/stab323}.
\bibitem[{{Barrientos} and {Chanam{\'e}}(2021)}]{Barrientos2021}
\bibinfo{author}{M.~{Barrientos}}, \bibinfo{author}{J.~{Chanam{\'e}}},
\newblock \bibinfo{title}{{Improved Constraints on the Initial-to-final Mass
  Relation of White Dwarfs Using Wide Binaries}},
\newblock \bibinfo{journal}{\apj} \bibinfo{volume}{923} (\bibinfo{year}{2021})
  \bibinfo{pages}{181}. \DOIprefix\doi{10.3847/1538-4357/ac2f49}.
\bibitem[{{Andersen}(1991)}]{Andersen1991}
\bibinfo{author}{J.~{Andersen}},
\newblock \bibinfo{title}{{Accurate masses and radii of normal stars}},
\newblock \bibinfo{journal}{\aapr} \bibinfo{volume}{3} (\bibinfo{year}{1991})
  \bibinfo{pages}{91--126}. \DOIprefix\doi{10.1007/BF00873538}.
\bibitem[{{Aubourg} et~al.(1993){Aubourg}, {Bareyre}, {Br{\'e}hin}, {Gros},
  {Lachi{\'e}ze-Rey}, {Laurent}, and et~al.}]{Aubourg1993}
\bibinfo{author}{E.~{Aubourg}}, \bibinfo{author}{P.~{Bareyre}},
  \bibinfo{author}{S.~{Br{\'e}hin}}, et~al.,
\newblock \bibinfo{title}{{Status of EROS (presented by M. MONIEZ)}},
\newblock in: \bibinfo{editor}{J.~{Surdej}},
  \bibinfo{editor}{D.~{Fraipont-Caro}}, \bibinfo{editor}{E.~{Gosset}}, et~al.
  (Eds.), \bibinfo{booktitle}{Liege International Astrophysical Colloquia},
  volume~\bibinfo{volume}{31} of \textit{\bibinfo{series}{Liege International
  Astrophysical Colloquia}}, \bibinfo{year}{1993},  p. \bibinfo{pages}{493}.
\bibitem[{{Alcock} et~al.(1997){Alcock}, {Allsman}, {Alves}, {Axelrod},
  {Becker}, {Bennett}, and et~al.}]{Alcock1997}
\bibinfo{author}{C.~{Alcock}}, \bibinfo{author}{R.~A. {Allsman}},
  \bibinfo{author}{D.~{Alves}}, et~al.,
\newblock \bibinfo{title}{{The MACHO Project Large Magellanic Cloud
  Microlensing Results from the First Two Years and the Nature of the Galactic
  Dark Halo}},
\newblock \bibinfo{journal}{\apj} \bibinfo{volume}{486} (\bibinfo{year}{1997})
  \bibinfo{pages}{697--726}. \DOIprefix\doi{10.1086/304535}.
\bibitem[{{Udalski} et~al.(1992){Udalski}, {Szymanski}, {Kaluzny}, {Kubiak},
  and {Mateo}}]{Udalski1992}
\bibinfo{author}{A.~{Udalski}}, \bibinfo{author}{M.~{Szymanski}},
  \bibinfo{author}{J.~{Kaluzny}}, et~al.,
\newblock \bibinfo{title}{{The Optical Gravitational Lensing Experiment}},
\newblock \bibinfo{journal}{\actaa}  \bibinfo{volume}{42}
  (\bibinfo{year}{1992}) \bibinfo{pages}{253--284}.
\bibitem[{ESA(1997)}]{ESA1997}
\bibinfo{title}{{The HIPPARCOS and TYCHO catalogues. Astrometric and
  photometric star catalogues derived from the ESA HIPPARCOS Space Astrometry
  Mission}}, volume \bibinfo{volume}{1200} of \textit{\bibinfo{series}{ESA
  Special Publication}},  \bibinfo{year}{1997}.
\bibitem[{{Devor} et~al.(2008){Devor}, {Charbonneau}, {O'Donovan}, {Mandushev},
  and {Torres}}]{Devor2008}
\bibinfo{author}{J.~{Devor}}, \bibinfo{author}{D.~{Charbonneau}},
  \bibinfo{author}{F.~T. {O'Donovan}}, et~al.,
\newblock \bibinfo{title}{{Identification, Classifications, and Absolute
  Properties of 773 Eclipsing Binaries Found in the Trans-Atlantic Exoplanet
  Survey}},
\newblock \bibinfo{journal}{\aj} \bibinfo{volume}{135} (\bibinfo{year}{2008})
  \bibinfo{pages}{850--877}. \DOIprefix\doi{10.1088/0004-6256/135/3/850}.
\bibitem[{{Pojmanski}(2002)}]{Pojmanski2002}
\bibinfo{author}{G.~{Pojmanski}},
\newblock \bibinfo{title}{{The All Sky Automated Survey. Catalog of Variable
  Stars. I. 0 h - 6 hQuarter of the Southern Hemisphere}},
\newblock \bibinfo{journal}{\actaa}  \bibinfo{volume}{52}
  (\bibinfo{year}{2002}) \bibinfo{pages}{397--427}.
\bibitem[{{Palaversa} et~al.(2013){Palaversa}, {Ivezi{\'c}}, {Eyer},
  {Ru{\v{z}}djak}, {Sudar}, {Galin}, {Kroflin}, {Mesari{\'c}}, {Munk},
  {Vrbanec}, {Bo{\v{z}}i{\'c}}, {Loebman}, {Sesar}, {Rimoldini}, {Hunt-Walker},
  {VanderPlas}, {Westman}, {Stuart}, {Becker}, {Srdo{\v{c}}}, {Wozniak}, and
  {Oluseyi}}]{Palaversa2013}
\bibinfo{author}{L.~{Palaversa}}, \bibinfo{author}{{\v{Z}}.~{Ivezi{\'c}}},
  \bibinfo{author}{L.~{Eyer}}, et~al.,
\newblock \bibinfo{title}{{Exploring the Variable Sky with LINEAR. III.
  Classification of Periodic Light Curves}},
\newblock \bibinfo{journal}{\aj} \bibinfo{volume}{146} (\bibinfo{year}{2013})
  \bibinfo{pages}{101}. \DOIprefix\doi{10.1088/0004-6256/146/4/101}.
\bibitem[{{Kim} et~al.(2014){Kim}, {Protopapas}, {Bailer-Jones}, {Byun},
  {Chang}, {Marquette}, and {Shin}}]{Kim2014}
\bibinfo{author}{D.-W. {Kim}}, \bibinfo{author}{P.~{Protopapas}},
  \bibinfo{author}{C.~A.~L. {Bailer-Jones}}, et~al.,
\newblock \bibinfo{title}{{The EPOCH Project. I. Periodic variable stars in the
  EROS-2 LMC database}},
\newblock \bibinfo{journal}{\aap} \bibinfo{volume}{566} (\bibinfo{year}{2014})
  \bibinfo{pages}{A43}. \DOIprefix\doi{10.1051/0004-6361/201323252}.
\bibitem[{{Drake} et~al.(2017){Drake}, {Djorgovski}, {Catelan}, {Graham},
  {Mahabal}, {Larson}, {Christensen}, {Torrealba}, {Beshore}, {McNaught},
  {Garradd}, {Belokurov}, and {Koposov}}]{Drake2017}
\bibinfo{author}{A.~J. {Drake}}, \bibinfo{author}{S.~G. {Djorgovski}},
  \bibinfo{author}{M.~{Catelan}}, et~al.,
\newblock \bibinfo{title}{{The Catalina Surveys Southern periodic variable star
  catalogue}},
\newblock \bibinfo{journal}{\mnras} \bibinfo{volume}{469}
  (\bibinfo{year}{2017}) \bibinfo{pages}{3688--3712}.
  \DOIprefix\doi{10.1093/mnras/stx1085}.
\bibitem[{{Heinze} et~al.(2018){Heinze}, {Tonry}, {Denneau}, {Flewelling},
  {Stalder}, {Rest}, {Smith}, {Smartt}, and {Weiland}}]{Heinze2018}
\bibinfo{author}{A.~N. {Heinze}}, \bibinfo{author}{J.~L. {Tonry}},
  \bibinfo{author}{L.~{Denneau}}, et~al.,
\newblock \bibinfo{title}{{A First Catalog of Variable Stars Measured by the
  Asteroid Terrestrial-impact Last Alert System (ATLAS)}},
\newblock \bibinfo{journal}{\aj} \bibinfo{volume}{156} (\bibinfo{year}{2018})
  \bibinfo{pages}{241}. \DOIprefix\doi{10.3847/1538-3881/aae47f}.
\bibitem[{{Chen} et~al.(2020){Chen}, {Wang}, {Deng}, {de Grijs}, {Yang}, and
  {Tian}}]{Chen2020}
\bibinfo{author}{X.~{Chen}}, \bibinfo{author}{S.~{Wang}},
  \bibinfo{author}{L.~{Deng}}, et~al.,
\newblock \bibinfo{title}{{The Zwicky Transient Facility Catalog of Periodic
  Variable Stars}},
\newblock \bibinfo{journal}{\apjs} \bibinfo{volume}{249} (\bibinfo{year}{2020})
  \bibinfo{pages}{18}. \DOIprefix\doi{10.3847/1538-4365/ab9cae}.
\bibitem[{{Pawlak} et~al.(2016){Pawlak}, {Soszy{\'n}ski}, {Udalski},
  {Szyma{\'n}ski}, {Wyrzykowski}, {Ulaczyk}, {Poleski}, {Pietrukowicz},
  {Koz{\l}owski}, {Skowron}, {Skowron}, {Mr{\'o}z}, and
  {Hamanowicz}}]{Pawlak2016}
\bibinfo{author}{M.~{Pawlak}}, \bibinfo{author}{I.~{Soszy{\'n}ski}},
  \bibinfo{author}{A.~{Udalski}}, et~al.,
\newblock \bibinfo{title}{{The OGLE Collection of Variable Stars. Eclipsing
  Binaries in the Magellanic System}},
\newblock \bibinfo{journal}{\actaa}  \bibinfo{volume}{66}
  (\bibinfo{year}{2016}) \bibinfo{pages}{421--432}.
\bibitem[{{Soszy{\'n}ski} et~al.(2016{\natexlab{a}}){Soszy{\'n}ski}, {Udalski},
  {Szyma{\'n}ski}, {Wyrzykowski}, {Ulaczyk}, {Poleski}, {Pietrukowicz},
  {Koz{\l}owski}, {Skowron}, {Skowron}, {Mr{\'o}z}, and
  {Pawlak}}]{Soszynski2016a}
\bibinfo{author}{I.~{Soszy{\'n}ski}}, \bibinfo{author}{A.~{Udalski}},
  \bibinfo{author}{M.~K. {Szyma{\'n}ski}}, et~al.,
\newblock \bibinfo{title}{{The OGLE Collection of Variable Stars. Over 45 000
  RR Lyrae Stars in the Magellanic System}},
\newblock \bibinfo{journal}{\actaa}  \bibinfo{volume}{66}
  (\bibinfo{year}{2016}{\natexlab{a}}) \bibinfo{pages}{131--147}.
\bibitem[{{Soszy{\'n}ski} et~al.(2016{\natexlab{b}}){Soszy{\'n}ski}, {Pawlak},
  {Pietrukowicz}, {Udalski}, {Szyma{\'n}ski}, {Wyrzykowski}, {Ulaczyk},
  {Poleski}, {Koz{\l}owski}, {Skowron}, {Skowron}, {Mr{\'o}z}, and
  {Hamanowicz}}]{Soszynski2016b}
\bibinfo{author}{I.~{Soszy{\'n}ski}}, \bibinfo{author}{M.~{Pawlak}},
  \bibinfo{author}{P.~{Pietrukowicz}}, et~al.,
\newblock \bibinfo{title}{{The OGLE Collection of Variable Stars. Over 450 000
  Eclipsing and Ellipsoidal Binary Systems Toward the Galactic Bulge}},
\newblock \bibinfo{journal}{\actaa}  \bibinfo{volume}{66}
  (\bibinfo{year}{2016}{\natexlab{b}}) \bibinfo{pages}{405--420}.
\bibitem[{{Borucki} et~al.(2010){Borucki}, {Koch}, {Basri}, {Batalha}, {Brown},
  and et~al.}]{Borucki2010}
\bibinfo{author}{W.~J. {Borucki}}, \bibinfo{author}{D.~{Koch}},
  \bibinfo{author}{G.~{Basri}}, et~al.,
\newblock \bibinfo{title}{{Kepler Planet-Detection Mission: Introduction and
  First Results}},
\newblock \bibinfo{journal}{Science} \bibinfo{volume}{327}
  (\bibinfo{year}{2010}) \bibinfo{pages}{977}.
  \DOIprefix\doi{10.1126/science.1185402}.
\bibitem[{{Ricker} et~al.(2014){Ricker}, {Winn}, {Vanderspek}, {Latham},
  {Bakos}, and et~al.}]{Ricker2014}
\bibinfo{author}{G.~R. {Ricker}}, \bibinfo{author}{J.~N. {Winn}},
  \bibinfo{author}{R.~{Vanderspek}}, et~al.,
\newblock \bibinfo{title}{{Transiting Exoplanet Survey Satellite (TESS)}},
\newblock in: \bibinfo{editor}{J.~{Oschmann}, Jacobus~M.},
  \bibinfo{editor}{M.~{Clampin}}, \bibinfo{editor}{G.~G. {Fazio}},
  \bibinfo{editor}{H.~A. {MacEwen}} (Eds.), \bibinfo{booktitle}{Space
  Telescopes and Instrumentation 2014: Optical, Infrared, and Millimeter Wave},
  volume \bibinfo{volume}{9143} of \textit{\bibinfo{series}{Society of
  Photo-Optical Instrumentation Engineers (SPIE) Conference Series}},
  \bibinfo{year}{2014}, p. \bibinfo{pages}{914320}.
  \DOIprefix\doi{10.1117/12.2063489}.
\bibitem[{{Kirk} et~al.(2016){Kirk}, {Conroy}, {Pr{\v{s}}a}, {Abdul-Masih},
  {Kochoska}, and et~al.}]{Kirk2016}
\bibinfo{author}{B.~{Kirk}}, \bibinfo{author}{K.~{Conroy}},
  \bibinfo{author}{A.~{Pr{\v{s}}a}}, et~al.,
\newblock \bibinfo{title}{{Kepler Eclipsing Binary Stars. VII. The Catalog of
  Eclipsing Binaries Found in the Entire Kepler Data Set}},
\newblock \bibinfo{journal}{\aj} \bibinfo{volume}{151} (\bibinfo{year}{2016})
  \bibinfo{pages}{68}. \DOIprefix\doi{10.3847/0004-6256/151/3/68}.
\bibitem[{{Pr{\v{s}}a} et~al.(2022){Pr{\v{s}}a}, {Kochoska}, {Conroy},
  {Eisner}, {Hey}, and et~al.}]{Prsa2022}
\bibinfo{author}{A.~{Pr{\v{s}}a}}, \bibinfo{author}{A.~{Kochoska}},
  \bibinfo{author}{K.~E. {Conroy}}, et~al.,
\newblock \bibinfo{title}{{TESS Eclipsing Binary Stars. I. Short-cadence
  Observations of 4584 Eclipsing Binaries in Sectors 1-26}},
\newblock \bibinfo{journal}{\apjs} \bibinfo{volume}{258} (\bibinfo{year}{2022})
  \bibinfo{pages}{16}. \DOIprefix\doi{10.3847/1538-4365/ac324a}.
\bibitem[{{Gaia Collaboration} et~al.(2016){Gaia Collaboration}, {Prusti}, {de
  Bruijne}, {Brown}, {Vallenari}, {Babusiaux}, {Bailer-Jones}, {Bastian},
  {Biermann}, {Evans}, {Eyer}, and et~al.}]{Gaia2016}
\bibinfo{author}{{Gaia Collaboration}}, \bibinfo{author}{T.~{Prusti}},
  \bibinfo{author}{J.~H.~J. {de Bruijne}}, et~al.,
\newblock \bibinfo{title}{{The Gaia mission}},
\newblock \bibinfo{journal}{\aap} \bibinfo{volume}{595} (\bibinfo{year}{2016})
  \bibinfo{pages}{A1}. \DOIprefix\doi{10.1051/0004-6361/201629272}.
\bibitem[{{Gaia Collaboration} et~al.(2018){Gaia Collaboration}, {Brown},
  {Vallenari}, {Prusti}, {de Bruijne}, {Babusiaux}, {Bailer-Jones}, {Biermann},
  {Evans}, {Eyer}, {Jansen}, and et~al.}]{Gaia2018}
\bibinfo{author}{{Gaia Collaboration}}, \bibinfo{author}{A.~G.~A. {Brown}},
  \bibinfo{author}{A.~{Vallenari}}, et~al.,
\newblock \bibinfo{title}{{Gaia Data Release 2. Summary of the contents and
  survey properties}},
\newblock \bibinfo{journal}{\aap} \bibinfo{volume}{616} (\bibinfo{year}{2018})
  \bibinfo{pages}{A1}. \DOIprefix\doi{10.1051/0004-6361/201833051}.
\bibitem[{{S{\"u}veges} et~al.(2017){S{\"u}veges}, {Barblan},
  {Lecoeur-Ta{\"\i}bi}, {Pr{\v{s}}a}, {Holl}, {Eyer}, {Kochoska}, {Mowlavi},
  and {Rimoldini}}]{Suveges2017}
\bibinfo{author}{M.~{S{\"u}veges}}, \bibinfo{author}{F.~{Barblan}},
  \bibinfo{author}{I.~{Lecoeur-Ta{\"\i}bi}}, et~al.,
\newblock \bibinfo{title}{{Gaia eclipsing binary and multiple systems.
  Supervised classification and self-organizing maps}},
\newblock \bibinfo{journal}{\aap} \bibinfo{volume}{603} (\bibinfo{year}{2017})
  \bibinfo{pages}{A117}. \DOIprefix\doi{10.1051/0004-6361/201629710}.
\bibitem[{{Kochoska} et~al.(2017){Kochoska}, {Mowlavi}, {Pr{\v{s}}a},
  {Lecoeur-Ta{\"\i}bi}, {Holl}, {Rimoldini}, {S{\"u}veges}, and
  {Eyer}}]{Kochoska2017}
\bibinfo{author}{A.~{Kochoska}}, \bibinfo{author}{N.~{Mowlavi}},
  \bibinfo{author}{A.~{Pr{\v{s}}a}}, et~al.,
\newblock \bibinfo{title}{{Gaia eclipsing binary and multiple systems. A study
  of detectability and classification of eclipsing binaries with Gaia}},
\newblock \bibinfo{journal}{\aap} \bibinfo{volume}{602} (\bibinfo{year}{2017})
  \bibinfo{pages}{A110}. \DOIprefix\doi{10.1051/0004-6361/201629957}.
\bibitem[{{Mowlavi} et~al.(2017){Mowlavi}, {Lecoeur-Ta{\"\i}bi}, {Holl},
  {Rimoldini}, {Barblan}, {Pr{\v{s}}a}, {Kochoska}, {S{\"u}veges}, {Eyer},
  {Nienartowicz}, {Jevardat}, {Charnas}, {Guy}, and {Audard}}]{Mowlavi2017}
\bibinfo{author}{N.~{Mowlavi}}, \bibinfo{author}{I.~{Lecoeur-Ta{\"\i}bi}},
  \bibinfo{author}{B.~{Holl}}, et~al.,
\newblock \bibinfo{title}{{Gaia eclipsing binary and multiple systems.
  Two-Gaussian models applied to OGLE-III eclipsing binary light curves in the
  Large Magellanic Cloud}},
\newblock \bibinfo{journal}{\aap} \bibinfo{volume}{606} (\bibinfo{year}{2017})
  \bibinfo{pages}{A92}. \DOIprefix\doi{10.1051/0004-6361/201730613}.
\bibitem[{{Kallrath} et~al.(1998){Kallrath}, {Milone}, {Terrell}, and
  {Young}}]{Kallrath1998ApJ...508..308K}
\bibinfo{author}{J.~{Kallrath}}, \bibinfo{author}{E.~F. {Milone}},
  \bibinfo{author}{D.~{Terrell}}, \bibinfo{author}{A.~T. {Young}},
\newblock \bibinfo{title}{{Recent Improvements to a Version of the
  Wilson-Devinney Program}},
\newblock \bibinfo{journal}{\apj} \bibinfo{volume}{508} (\bibinfo{year}{1998})
  \bibinfo{pages}{308--313}. \DOIprefix\doi{10.1086/306375}.
\bibitem[{{Southworth} et~al.(2004{\natexlab{a}}){Southworth}, {Maxted}, and
  {Smalley}}]{Southworth2004a}
\bibinfo{author}{J.~{Southworth}}, \bibinfo{author}{P.~F.~L. {Maxted}},
  \bibinfo{author}{B.~{Smalley}},
\newblock \bibinfo{title}{{Eclipsing binaries in open clusters - II. V453 Cyg
  in NGC 6871}},
\newblock \bibinfo{journal}{\mnras} \bibinfo{volume}{351}
  (\bibinfo{year}{2004}{\natexlab{a}}) \bibinfo{pages}{1277--1289}.
  \DOIprefix\doi{10.1111/j.1365-2966.2004.07871.x}.
\bibitem[{{Southworth} et~al.(2004{\natexlab{b}}){Southworth}, {Zucker},
  {Maxted}, and {Smalley}}]{Southworth2004b}
\bibinfo{author}{J.~{Southworth}}, \bibinfo{author}{S.~{Zucker}},
  \bibinfo{author}{P.~F.~L. {Maxted}}, \bibinfo{author}{B.~{Smalley}},
\newblock \bibinfo{title}{{Eclipsing binaries in open clusters - III. V621 Per
  in {\ensuremath{\chi}} Persei}},
\newblock \bibinfo{journal}{\mnras} \bibinfo{volume}{355}
  (\bibinfo{year}{2004}{\natexlab{b}}) \bibinfo{pages}{986--994}.
  \DOIprefix\doi{10.1111/j.1365-2966.2004.08389.x}.
\bibitem[{{Conroy} et~al.(2020){Conroy}, {Kochoska}, {Hey}, {Pablo},
  {Hambleton}, {Jones}, {Giammarco}, {Abdul-Masih}, and
  {Pr{\v{s}}a}}]{Conroy2020ApJS..250...34C}
\bibinfo{author}{K.~E. {Conroy}}, \bibinfo{author}{A.~{Kochoska}},
  \bibinfo{author}{D.~{Hey}}, et~al.,
\newblock \bibinfo{title}{{Physics of Eclipsing Binaries. V. General Framework
  for Solving the Inverse Problem}},
\newblock \bibinfo{journal}{\apjs} \bibinfo{volume}{250} (\bibinfo{year}{2020})
  \bibinfo{pages}{34}. \DOIprefix\doi{10.3847/1538-4365/abb4e2}.
\bibitem[{{Foreman-Mackey} et~al.(2019){Foreman-Mackey}, {Farr}, {Sinha},
  {Archibald}, {Hogg}, {Sanders}, {Zuntz}, {Williams}, {Nelson}, {de
  Val-Borro}, {Erhardt}, {Pashchenko}, and {Pla}}]{Foreman-Mackey2019}
\bibinfo{author}{D.~{Foreman-Mackey}}, \bibinfo{author}{W.~{Farr}},
  \bibinfo{author}{M.~{Sinha}}, et~al.,
\newblock \bibinfo{title}{{emcee v3: A Python ensemble sampling toolkit for
  affine-invariant MCMC}},
\newblock \bibinfo{journal}{The Journal of Open Source Software}
  \bibinfo{volume}{4} (\bibinfo{year}{2019}) \bibinfo{pages}{1864}.
  \DOIprefix\doi{10.21105/joss.01864}.
\bibitem[{{Rowan} et~al.(2022){Rowan}, {Jayasinghe}, {Stanek}, {Kochanek},
  {Thompson}, {Shappee}, {Holoien}, {Prieto}, and {Giles}}]{Rowan2022}
\bibinfo{author}{D.~M. {Rowan}}, \bibinfo{author}{T.~{Jayasinghe}},
  \bibinfo{author}{K.~Z. {Stanek}}, et~al.,
\newblock \bibinfo{title}{{The value-added catalogue of ASAS-SN eclipsing
  binaries: parameters of 30 000 detached systems}},
\newblock \bibinfo{journal}{\mnras} \bibinfo{volume}{517}
  (\bibinfo{year}{2022}) \bibinfo{pages}{2190--2213}.
  \DOIprefix\doi{10.1093/mnras/stac2520}.
\bibitem[{{Pr{\v{s}}a} et~al.(2008){Pr{\v{s}}a}, {Guinan}, {Devinney},
  {DeGeorge}, {Bradstreet}, {Giammarco}, {Alcock}, and
  {Engle}}]{Prsa2008ApJ...687..542P}
\bibinfo{author}{A.~{Pr{\v{s}}a}}, \bibinfo{author}{E.~F. {Guinan}},
  \bibinfo{author}{E.~J. {Devinney}}, et~al.,
\newblock \bibinfo{title}{{Artificial Intelligence Approach to the
  Determination of Physical Properties of Eclipsing Binaries. I. The EBAI
  Project}},
\newblock \bibinfo{journal}{\apj} \bibinfo{volume}{687} (\bibinfo{year}{2008})
  \bibinfo{pages}{542--565}. \DOIprefix\doi{10.1086/591783}.
\bibitem[{{Pr{\v{s}}a} et~al.(2011){Pr{\v{s}}a}, {Batalha}, {Slawson}, {Doyle},
  {Welsh}, {Orosz}, {Seager}, {Rucker}, {Mjaseth}, {Engle}, {Conroy},
  {Jenkins}, {Caldwell}, {Koch}, and {Borucki}}]{Prsa2011}
\bibinfo{author}{A.~{Pr{\v{s}}a}}, \bibinfo{author}{N.~{Batalha}},
  \bibinfo{author}{R.~W. {Slawson}}, et~al.,
\newblock \bibinfo{title}{{Kepler Eclipsing Binary Stars. I. Catalog and
  Principal Characterization of 1879 Eclipsing Binaries in the First Data
  Release}},
\newblock \bibinfo{journal}{\aj} \bibinfo{volume}{141} (\bibinfo{year}{2011})
  \bibinfo{pages}{83}. \DOIprefix\doi{10.1088/0004-6256/141/3/83}.
\bibitem[{{Slawson} et~al.(2011){Slawson}, {Pr{\v{s}}a}, {Welsh}, {Orosz}, and
  et~al.}]{Slawson2011}
\bibinfo{author}{R.~W. {Slawson}}, \bibinfo{author}{A.~{Pr{\v{s}}a}},
  \bibinfo{author}{W.~F. {Welsh}}, et~al.,
\newblock \bibinfo{title}{{Kepler Eclipsing Binary Stars. II. 2165 Eclipsing
  Binaries in the Second Data Release}},
\newblock \bibinfo{journal}{\aj} \bibinfo{volume}{142} (\bibinfo{year}{2011})
  \bibinfo{pages}{160}. \DOIprefix\doi{10.1088/0004-6256/142/5/160}.
\bibitem[{{Matijevi{\v{c}}} et~al.(2012){Matijevi{\v{c}}}, {Pr{\v{s}}a},
  {Orosz}, {Welsh}, {Bloemen}, and {Barclay}}]{Matijevic2012}
\bibinfo{author}{G.~{Matijevi{\v{c}}}}, \bibinfo{author}{A.~{Pr{\v{s}}a}},
  \bibinfo{author}{J.~A. {Orosz}}, et~al.,
\newblock \bibinfo{title}{{Kepler Eclipsing Binary Stars. III. Classification
  of Kepler Eclipsing Binary Light Curves with Locally Linear Embedding}},
\newblock \bibinfo{journal}{\aj} \bibinfo{volume}{143} (\bibinfo{year}{2012})
  \bibinfo{pages}{123}. \DOIprefix\doi{10.1088/0004-6256/143/5/123}.
\bibitem[{{Conroy} et~al.(2014){Conroy}, {Pr{\v{s}}a}, {Stassun}, {Orosz},
  {Fabrycky}, and {Welsh}}]{Conroy2014}
\bibinfo{author}{K.~E. {Conroy}}, \bibinfo{author}{A.~{Pr{\v{s}}a}},
  \bibinfo{author}{K.~G. {Stassun}}, et~al.,
\newblock \bibinfo{title}{{Kepler Eclipsing Binary Stars. IV. Precise Eclipse
  Times for Close Binaries and Identification of Candidate Three-body
  Systems}},
\newblock \bibinfo{journal}{\aj} \bibinfo{volume}{147} (\bibinfo{year}{2014})
  \bibinfo{pages}{45}. \DOIprefix\doi{10.1088/0004-6256/147/2/45}.
\bibitem[{{Abdul-Masih} et~al.(2016){Abdul-Masih}, {Pr{\v{s}}a}, {Conroy},
  {Bloemen}, {Boyajian}, {Doyle}, {Johnston}, {Kostov}, {Latham},
  {Matijevi{\v{c}}}, {Shporer}, and {Southworth}}]{Abdul-Masih2016}
\bibinfo{author}{M.~{Abdul-Masih}}, \bibinfo{author}{A.~{Pr{\v{s}}a}},
  \bibinfo{author}{K.~{Conroy}}, et~al.,
\newblock \bibinfo{title}{{Kepler Eclipsing Binary Stars. VIII. Identification
  of False Positive Eclipsing Binaries and Re-extraction of New Light Curves}},
\newblock \bibinfo{journal}{\aj} \bibinfo{volume}{151} (\bibinfo{year}{2016})
  \bibinfo{pages}{101}. \DOIprefix\doi{10.3847/0004-6256/151/4/101}.
\bibitem[{{Holanda} and {da Silva}(2018)}]{Holanda2018}
\bibinfo{author}{N.~{Holanda}}, \bibinfo{author}{J.~R.~P. {da Silva}},
\newblock \bibinfo{title}{{The fidelity of Kepler eclipsing binary parameters
  inferred by the neural network}},
\newblock \bibinfo{journal}{\mnras} \bibinfo{volume}{478}
  (\bibinfo{year}{2018}) \bibinfo{pages}{1272--1280}.
  \DOIprefix\doi{10.1093/mnras/sty956}.
\bibitem[{{Papageorgiou} et~al.(2019){Papageorgiou}, {Catelan},
  {Christopoulou}, {Drake}, and {Djorgovski}}]{Papageorgiou2019}
\bibinfo{author}{A.~{Papageorgiou}}, \bibinfo{author}{M.~{Catelan}},
  \bibinfo{author}{P.-E. {Christopoulou}}, et~al.,
\newblock \bibinfo{title}{{Physical Parameters of Northern Eclipsing Binaries
  in the Catalina Sky Survey}},
\newblock \bibinfo{journal}{\apjs} \bibinfo{volume}{242} (\bibinfo{year}{2019})
  \bibinfo{pages}{6}. \DOIprefix\doi{10.3847/1538-4365/ab13b8}.
\bibitem[{{Ding} et~al.(2022){Ding}, {Ji}, {Li}, {Xiong}, {Cheng}, {Wang}, and
  {Liu}}]{Ding2022}
\bibinfo{author}{X.~{Ding}}, \bibinfo{author}{K.~{Ji}},
  \bibinfo{author}{X.~{Li}}, et~al.,
\newblock \bibinfo{title}{{Fast Derivation of Contact Binary Parameters for
  Large Photometric Surveys}},
\newblock \bibinfo{journal}{\aj} \bibinfo{volume}{164} (\bibinfo{year}{2022})
  \bibinfo{pages}{200}. \DOIprefix\doi{10.3847/1538-3881/ac8e66}.
\bibitem[{{Steinmetz} et~al.(2006){Steinmetz}, {Zwitter}, {Siebert}, {Watson},
  {Freeman}, {Munari}, {Campbell}, and et~al.}]{Steinmetz2006}
\bibinfo{author}{M.~{Steinmetz}}, \bibinfo{author}{T.~{Zwitter}},
  \bibinfo{author}{A.~{Siebert}}, et~al.,
\newblock \bibinfo{title}{{The Radial Velocity Experiment (RAVE): First Data
  Release}},
\newblock \bibinfo{journal}{\aj} \bibinfo{volume}{132} (\bibinfo{year}{2006})
  \bibinfo{pages}{1645--1668}. \DOIprefix\doi{10.1086/506564}.
\bibitem[{{Steinmetz} et~al.(2020){Steinmetz}, {Guiglion}, {McMillan},
  {Matijevi{\v{c}}}, {Enke}, {Kordopatis}, et~al., and {RAVE
  Collaboration}}]{Steinmetz2020}
\bibinfo{author}{M.~{Steinmetz}}, \bibinfo{author}{G.~{Guiglion}},
  \bibinfo{author}{P.~J. {McMillan}}, et~al.,
\newblock \bibinfo{title}{{The Sixth Data Release of the Radial Velocity
  Experiment (RAVE). II. Stellar Atmospheric Parameters, Chemical Abundances,
  and Distances}},
\newblock \bibinfo{journal}{\aj} \bibinfo{volume}{160} (\bibinfo{year}{2020})
  \bibinfo{pages}{83}. \DOIprefix\doi{10.3847/1538-3881/ab9ab8}.
\bibitem[{{York} et~al.(2000){York}, {Adelman}, {Anderson}, {Anderson},
  {Annis}, {Bahcall}, {Bakken}, et~al., and {SDSS Collaboration}}]{York2000}
\bibinfo{author}{D.~G. {York}}, \bibinfo{author}{J.~{Adelman}},
  \bibinfo{author}{J.~{Anderson}, John~E.}, et~al.,
\newblock \bibinfo{title}{{The Sloan Digital Sky Survey: Technical Summary}},
\newblock \bibinfo{journal}{\aj} \bibinfo{volume}{120} (\bibinfo{year}{2000})
  \bibinfo{pages}{1579--1587}. \DOIprefix\doi{10.1086/301513}.
\bibitem[{{Yanny} et~al.(2009){Yanny}, {Rockosi}, {Newberg}, {Knapp},
  {Adelman-McCarthy}, {Alcorn}, {Allam}, {Allende Prieto}, {An}, and
  et~al.}]{Yanny2009}
\bibinfo{author}{B.~{Yanny}}, \bibinfo{author}{C.~{Rockosi}},
  \bibinfo{author}{H.~J. {Newberg}}, et~al.,
\newblock \bibinfo{title}{{SEGUE: A Spectroscopic Survey of 240,000 Stars with
  g = 14-20}},
\newblock \bibinfo{journal}{\aj} \bibinfo{volume}{137} (\bibinfo{year}{2009})
  \bibinfo{pages}{4377--4399}. \DOIprefix\doi{10.1088/0004-6256/137/5/4377}.
\bibitem[{{Cui} et~al.(2012){Cui}, {Zhao}, {Chu}, {Li}, {Li}, {Zhang}, {Su},
  {Yao}, and et~al.}]{Cui2012}
\bibinfo{author}{X.-Q. {Cui}}, \bibinfo{author}{Y.-H. {Zhao}},
  \bibinfo{author}{Y.-Q. {Chu}}, et~al.,
\newblock \bibinfo{title}{{The Large Sky Area Multi-Object Fiber Spectroscopic
  Telescope (LAMOST)}},
\newblock \bibinfo{journal}{Research in Astronomy and Astrophysics}
  \bibinfo{volume}{12} (\bibinfo{year}{2012}) \bibinfo{pages}{1197--1242}.
  \DOIprefix\doi{10.1088/1674-4527/12/9/003}.
\bibitem[{{Deng} et~al.(2012){Deng}, {Newberg}, {Liu}, {Carlin}, {Beers},
  {Chen}, {Chen}, {Christlieb}, {Grillmair}, {Guhathakurta}, {Han}, {Hou},
  {Lee}, {L{\'e}pine}, {Li}, {Liu}, {Pan}, {Sellwood}, {Wang}, {Wang}, {Yang},
  {Yanny}, {Zhang}, {Zhang}, {Zheng}, and {Zhu}}]{Deng2012}
\bibinfo{author}{L.-C. {Deng}}, \bibinfo{author}{H.~J. {Newberg}},
  \bibinfo{author}{C.~{Liu}}, et~al.,
\newblock \bibinfo{title}{{LAMOST Experiment for Galactic Understanding and
  Exploration (LEGUE) {\textemdash} The survey's science plan}},
\newblock \bibinfo{journal}{Research in Astronomy and Astrophysics}
  \bibinfo{volume}{12} (\bibinfo{year}{2012}) \bibinfo{pages}{735--754}.
  \DOIprefix\doi{10.1088/1674-4527/12/7/003}.
\bibitem[{{Zhao} et~al.(2012){Zhao}, {Zhao}, {Chu}, {Jing}, and
  {Deng}}]{Zhao2012}
\bibinfo{author}{G.~{Zhao}}, \bibinfo{author}{Y.-H. {Zhao}},
  \bibinfo{author}{Y.-Q. {Chu}}, et~al.,
\newblock \bibinfo{title}{{LAMOST spectral survey {\textemdash} An overview}},
\newblock \bibinfo{journal}{Research in Astronomy and Astrophysics}
  \bibinfo{volume}{12} (\bibinfo{year}{2012}) \bibinfo{pages}{723--734}.
  \DOIprefix\doi{10.1088/1674-4527/12/7/002}.
\bibitem[{{Luo} et~al.(2015){Luo}, {Zhao}, {Zhao}, {Deng}, {Liu}, {Jing},
  {Wang}, {Zhang}, {Shi}, {Cui}, {Chu}, and et~al.}]{Luo2015RAA}
\bibinfo{author}{A.~L. {Luo}}, \bibinfo{author}{Y.-H. {Zhao}},
  \bibinfo{author}{G.~{Zhao}}, et~al.,
\newblock \bibinfo{title}{{The first data release (DR1) of the LAMOST regular
  survey}},
\newblock \bibinfo{journal}{Research in Astronomy and Astrophysics}
  \bibinfo{volume}{15} (\bibinfo{year}{2015}) \bibinfo{pages}{1095}.
  \DOIprefix\doi{10.1088/1674-4527/15/8/002}.
\bibitem[{{Liu} et~al.(2020){Liu}, {Fu}, {Shi}, {Wu}, {Han}, {Chen}, and
  et~al.}]{Liu2020}
\bibinfo{author}{C.~{Liu}}, \bibinfo{author}{J.~{Fu}},
  \bibinfo{author}{J.~{Shi}}, et~al.,
\newblock \bibinfo{title}{{LAMOST Medium-Resolution Spectroscopic Survey
  (LAMOST-MRS): Scientific goals and survey plan}},
\newblock \bibinfo{journal}{arXiv e-prints}   (\bibinfo{year}{2020})
  \bibinfo{pages}{arXiv:2005.07210}.
\bibitem[{{Katz} et~al.(2004){Katz}, {Munari}, {Cropper}, {Zwitter},
  {Th{\'e}venin}, and et~al.}]{Katz2004}
\bibinfo{author}{D.~{Katz}}, \bibinfo{author}{U.~{Munari}},
  \bibinfo{author}{M.~{Cropper}}, et~al.,
\newblock \bibinfo{title}{{Spectroscopic survey of the Galaxy with Gaia- I.
  Design and performance of the Radial Velocity Spectrometer}},
\newblock \bibinfo{journal}{\mnras} \bibinfo{volume}{354}
  (\bibinfo{year}{2004}) \bibinfo{pages}{1223--1238}.
  \DOIprefix\doi{10.1111/j.1365-2966.2004.08282.x}.
\bibitem[{{Gilmore} et~al.(2012){Gilmore}, {Randich}, {Asplund}, {Binney},
  {Bonifacio}, {Drew}, {Feltzing}, {Ferguson}, {Jeffries}, {Micela},
  {Negueruela}, et~al., and {Gaia-ESO Survey Team}}]{Gilmore2012}
\bibinfo{author}{G.~{Gilmore}}, \bibinfo{author}{S.~{Randich}},
  \bibinfo{author}{M.~{Asplund}}, et~al.,
\newblock \bibinfo{title}{{The Gaia-ESO Public Spectroscopic Survey}},
\newblock \bibinfo{journal}{The Messenger}  \bibinfo{volume}{147}
  (\bibinfo{year}{2012}) \bibinfo{pages}{25--31}.
\bibitem[{{Cropper} et~al.(2018){Cropper}, {Katz}, {Sartoretti}, {Prusti}, {de
  Bruijne}, and et~al.}]{Cropper2018}
\bibinfo{author}{M.~{Cropper}}, \bibinfo{author}{D.~{Katz}},
  \bibinfo{author}{P.~{Sartoretti}}, et~al.,
\newblock \bibinfo{title}{{Gaia Data Release 2. Gaia Radial Velocity
  Spectrometer}},
\newblock \bibinfo{journal}{\aap} \bibinfo{volume}{616} (\bibinfo{year}{2018})
  \bibinfo{pages}{A5}. \DOIprefix\doi{10.1051/0004-6361/201832763}.
\bibitem[{{Merle} et~al.(2017){Merle}, {Van Eck}, {Jorissen}, {Van der
  Swaelmen}, {Masseron}, and et~al.}]{Merle2017}
\bibinfo{author}{T.~{Merle}}, \bibinfo{author}{S.~{Van Eck}},
  \bibinfo{author}{A.~{Jorissen}}, et~al.,
\newblock \bibinfo{title}{{The Gaia-ESO Survey: double-, triple-, and
  quadruple-line spectroscopic binary candidates}},
\newblock \bibinfo{journal}{\aap} \bibinfo{volume}{608} (\bibinfo{year}{2017})
  \bibinfo{pages}{A95}. \DOIprefix\doi{10.1051/0004-6361/201730442}.
\bibitem[{{Merle} et~al.(2020){Merle}, {Van der Swaelmen}, {Van Eck},
  {Jorissen}, {Jackson}, {Traven}, {Zwitter}, {Pourbaix}, and
  et~al.}]{Merle2020}
\bibinfo{author}{T.~{Merle}}, \bibinfo{author}{M.~{Van der Swaelmen}},
  \bibinfo{author}{S.~{Van Eck}}, et~al.,
\newblock \bibinfo{title}{{The Gaia-ESO Survey: detection and characterisation
  of single-line spectroscopic binaries}},
\newblock \bibinfo{journal}{\aap} \bibinfo{volume}{635} (\bibinfo{year}{2020})
  \bibinfo{pages}{A155}. \DOIprefix\doi{10.1051/0004-6361/201935819}.
\bibitem[{{De Silva} et~al.(2015){De Silva}, {Freeman}, {Bland-Hawthorn},
  {Martell}, {de Boer}, {Asplund}, and et~al.}]{De-Silva2015}
\bibinfo{author}{G.~M. {De Silva}}, \bibinfo{author}{K.~C. {Freeman}},
  \bibinfo{author}{J.~{Bland-Hawthorn}}, et~al.,
\newblock \bibinfo{title}{{The GALAH survey: scientific motivation}},
\newblock \bibinfo{journal}{\mnras} \bibinfo{volume}{449}
  (\bibinfo{year}{2015}) \bibinfo{pages}{2604--2617}.
  \DOIprefix\doi{10.1093/mnras/stv327}.
\bibitem[{{Majewski} et~al.(2017){Majewski}, {Schiavon}, {Frinchaboy}, {Allende
  Prieto}, {Barkhouser}, {Bizyaev}, {Blank}, and et~al.}]{Majewski2017}
\bibinfo{author}{S.~R. {Majewski}}, \bibinfo{author}{R.~P. {Schiavon}},
  \bibinfo{author}{P.~M. {Frinchaboy}}, et~al.,
\newblock \bibinfo{title}{{The Apache Point Observatory Galactic Evolution
  Experiment (APOGEE)}},
\newblock \bibinfo{journal}{\aj} \bibinfo{volume}{154} (\bibinfo{year}{2017})
  \bibinfo{pages}{94}. \DOIprefix\doi{10.3847/1538-3881/aa784d}.
\bibitem[{{Zorotovic} et~al.(2010){Zorotovic}, {Schreiber}, {G{\"a}nsicke}, and
  {Nebot G{\'o}mez-Mor{\'a}n}}]{Zorotovic2010A&A...520A..86Z}
\bibinfo{author}{M.~{Zorotovic}}, \bibinfo{author}{M.~R. {Schreiber}},
  \bibinfo{author}{B.~T. {G{\"a}nsicke}}, \bibinfo{author}{A.~{Nebot
  G{\'o}mez-Mor{\'a}n}},
\newblock \bibinfo{title}{{Post-common-envelope binaries from SDSS. IX:
  Constraining the common-envelope efficiency}},
\newblock \bibinfo{journal}{\aap} \bibinfo{volume}{520} (\bibinfo{year}{2010})
  \bibinfo{pages}{A86}. \DOIprefix\doi{10.1051/0004-6361/200913658}.
\bibitem[{{Zorotovic} et~al.(2011){Zorotovic}, {Schreiber}, {G{\"a}nsicke},
  {Rebassa-Mansergas}, {Nebot G{\'o}mez-Mor{\'a}n}, {Southworth}, {Schwope},
  {Pyrzas}, {Rodr{\'\i}guez-Gil}, {Schmidtobreick}, {Schwarz}, {Tappert},
  {Toloza}, and {Vogt}}]{Zorotovic2011A&A...536L...3Z}
\bibinfo{author}{M.~{Zorotovic}}, \bibinfo{author}{M.~R. {Schreiber}},
  \bibinfo{author}{B.~T. {G{\"a}nsicke}}, et~al.,
\newblock \bibinfo{title}{{Post common envelope binaries from SDSS. XIII. Mass
  dependencies of the orbital period distribution}},
\newblock \bibinfo{journal}{\aap} \bibinfo{volume}{536} (\bibinfo{year}{2011})
  \bibinfo{pages}{L3}. \DOIprefix\doi{10.1051/0004-6361/201117803}.
\bibitem[{{Zorotovic} et~al.(2014){Zorotovic}, {Schreiber},
  {Garc{\'\i}a-Berro}, {Camacho}, {Torres}, {Rebassa-Mansergas}, and
  {G{\"a}nsicke}}]{Zorotovic2014A&A...568A..68Z}
\bibinfo{author}{M.~{Zorotovic}}, \bibinfo{author}{M.~R. {Schreiber}},
  \bibinfo{author}{E.~{Garc{\'\i}a-Berro}}, et~al.,
\newblock \bibinfo{title}{{Monte Carlo simulations of post-common-envelope
  white dwarf + main sequence binaries: The effects of including recombination
  energy}},
\newblock \bibinfo{journal}{\aap} \bibinfo{volume}{568} (\bibinfo{year}{2014})
  \bibinfo{pages}{A68}. \DOIprefix\doi{10.1051/0004-6361/201323039}.
\bibitem[{{Rebassa-Mansergas} et~al.(2012{\natexlab{a}}){Rebassa-Mansergas},
  {Nebot G{\'o}mez-Mor{\'a}n}, {Schreiber}, {G{\"a}nsicke}, {Schwope},
  {Gallardo}, and {Koester}}]{Rebassa-Mansergas2012a}
\bibinfo{author}{A.~{Rebassa-Mansergas}}, \bibinfo{author}{A.~{Nebot
  G{\'o}mez-Mor{\'a}n}}, \bibinfo{author}{M.~R. {Schreiber}}, et~al.,
\newblock \bibinfo{title}{{Post-common envelope binaries from SDSS - XIV. The
  DR7 white dwarf-main-sequence binary catalogue}},
\newblock \bibinfo{journal}{\mnras} \bibinfo{volume}{419}
  (\bibinfo{year}{2012}{\natexlab{a}}) \bibinfo{pages}{806--816}.
  \DOIprefix\doi{10.1111/j.1365-2966.2011.19923.x}.
\bibitem[{{Rebassa-Mansergas} et~al.(2012{\natexlab{b}}){Rebassa-Mansergas},
  {Zorotovic}, {Schreiber}, {G{\"a}nsicke}, {Southworth}, {Nebot
  G{\'o}mez-Mor{\'a}n}, {Tappert}, {Koester}, {Pyrzas}, {Papadaki},
  {Schmidtobreick}, {Schwope}, and {Toloza}}]{Rebassa-Mansergas2012b}
\bibinfo{author}{A.~{Rebassa-Mansergas}}, \bibinfo{author}{M.~{Zorotovic}},
  \bibinfo{author}{M.~R. {Schreiber}}, et~al.,
\newblock \bibinfo{title}{{Post-common envelope binaries from SDSS - XVI. Long
  orbital period systems and the energy budget of common envelope evolution}},
\newblock \bibinfo{journal}{\mnras} \bibinfo{volume}{423}
  (\bibinfo{year}{2012}{\natexlab{b}}) \bibinfo{pages}{320--327}.
  \DOIprefix\doi{10.1111/j.1365-2966.2012.20880.x}.
\bibitem[{{Toonen} and {Nelemans}(2013)}]{Toonen2013A&A...557A..87T}
\bibinfo{author}{S.~{Toonen}}, \bibinfo{author}{G.~{Nelemans}},
\newblock \bibinfo{title}{{The effect of common-envelope evolution on the
  visible population of post-common-envelope binaries}},
\newblock \bibinfo{journal}{\aap} \bibinfo{volume}{557} (\bibinfo{year}{2013})
  \bibinfo{pages}{A87}. \DOIprefix\doi{10.1051/0004-6361/201321753}.
\bibitem[{{Camacho} et~al.(2014){Camacho}, {Torres}, {Garc{\'\i}a-Berro},
  {Zorotovic}, {Schreiber}, {Rebassa-Mansergas}, {Nebot G{\'o}mez-Mor{\'a}n},
  and {G{\"a}nsicke}}]{Camacho2014A&A...566A..86C}
\bibinfo{author}{J.~{Camacho}}, \bibinfo{author}{S.~{Torres}},
  \bibinfo{author}{E.~{Garc{\'\i}a-Berro}}, et~al.,
\newblock \bibinfo{title}{{Monte Carlo simulations of post-common-envelope
  white dwarf + main sequence binaries: comparison with the SDSS DR7 observed
  sample}},
\newblock \bibinfo{journal}{\aap} \bibinfo{volume}{566} (\bibinfo{year}{2014})
  \bibinfo{pages}{A86}. \DOIprefix\doi{10.1051/0004-6361/201323052}.
\bibitem[{{Rebassa-Mansergas} et~al.(2010){Rebassa-Mansergas}, {G{\"a}nsicke},
  {Schreiber}, {Koester}, and {Rodr{\'\i}guez-Gil}}]{Rebassa-Mansergas2010}
\bibinfo{author}{A.~{Rebassa-Mansergas}}, \bibinfo{author}{B.~T.
  {G{\"a}nsicke}}, \bibinfo{author}{M.~R. {Schreiber}}, et~al.,
\newblock \bibinfo{title}{{Post-common envelope binaries from SDSS - VII. A
  catalogue of white dwarf-main sequence binaries}},
\newblock \bibinfo{journal}{\mnras} \bibinfo{volume}{402}
  (\bibinfo{year}{2010}) \bibinfo{pages}{620--640}.
  \DOIprefix\doi{10.1111/j.1365-2966.2009.15915.x}.
\bibitem[{{Rebassa-Mansergas} et~al.(2013){Rebassa-Mansergas}, {Agurto-Gangas},
  {Schreiber}, {G{\"a}nsicke}, and {Koester}}]{Rebassa-Mansergas2013}
\bibinfo{author}{A.~{Rebassa-Mansergas}}, \bibinfo{author}{C.~{Agurto-Gangas}},
  \bibinfo{author}{M.~R. {Schreiber}}, et~al.,
\newblock \bibinfo{title}{{White dwarf main-sequence binaries from SDSS DR 8:
  unveiling the cool white dwarf population}},
\newblock \bibinfo{journal}{\mnras} \bibinfo{volume}{433}
  (\bibinfo{year}{2013}) \bibinfo{pages}{3398--3410}.
  \DOIprefix\doi{10.1093/mnras/stt974}.
\bibitem[{{Rebassa-Mansergas} et~al.(2016){Rebassa-Mansergas}, {Ren},
  {Parsons}, {G{\"a}nsicke}, {Schreiber}, {Garc{\'\i}a-Berro}, {Liu}, and
  {Koester}}]{Rebassa-Mansergas2016}
\bibinfo{author}{A.~{Rebassa-Mansergas}}, \bibinfo{author}{J.~J. {Ren}},
  \bibinfo{author}{S.~G. {Parsons}}, et~al.,
\newblock \bibinfo{title}{{The SDSS spectroscopic catalogue of white
  dwarf-main-sequence binaries: new identifications from DR 9-12}},
\newblock \bibinfo{journal}{\mnras} \bibinfo{volume}{458}
  (\bibinfo{year}{2016}) \bibinfo{pages}{3808--3819}.
  \DOIprefix\doi{10.1093/mnras/stw554}.
\bibitem[{{Rebassa-Mansergas} et~al.(2007){Rebassa-Mansergas}, {G{\"a}nsicke},
  {Rodr{\'\i}guez-Gil}, {Schreiber}, and {Koester}}]{Rebassa-Mansergas2007}
\bibinfo{author}{A.~{Rebassa-Mansergas}}, \bibinfo{author}{B.~T.
  {G{\"a}nsicke}}, \bibinfo{author}{P.~{Rodr{\'\i}guez-Gil}}, et~al.,
\newblock \bibinfo{title}{{Post-common-envelope binaries from SDSS - I. 101
  white dwarf main-sequence binaries with multiple Sloan Digital Sky Survey
  spectroscopy}},
\newblock \bibinfo{journal}{\mnras} \bibinfo{volume}{382}
  (\bibinfo{year}{2007}) \bibinfo{pages}{1377--1393}.
  \DOIprefix\doi{10.1111/j.1365-2966.2007.12288.x}.
\bibitem[{{Schreiber} et~al.(2008){Schreiber}, {G{\"a}nsicke}, {Southworth},
  {Schwope}, and {Koester}}]{Schreiber2008}
\bibinfo{author}{M.~R. {Schreiber}}, \bibinfo{author}{B.~T. {G{\"a}nsicke}},
  \bibinfo{author}{J.~{Southworth}}, et~al.,
\newblock \bibinfo{title}{{Post common envelope binaries from SDSS. II:
  Identification of 9 close binaries with VLT/FORS2}},
\newblock \bibinfo{journal}{\aap} \bibinfo{volume}{484} (\bibinfo{year}{2008})
  \bibinfo{pages}{441--450}. \DOIprefix\doi{10.1051/0004-6361:20078765}.
\bibitem[{{Schreiber} et~al.(2010){Schreiber}, {G{\"a}nsicke},
  {Rebassa-Mansergas}, {Nebot Gomez-Moran}, {Southworth}, {Schwope},
  {M{\"u}ller}, {Papadaki}, {Pyrzas}, {Rabitz}, {Rodr{\'\i}guez-Gil},
  {Schmidtobreick}, {Schwarz}, {Tappert}, {Toloza}, {Vogel}, and
  {Zorotovic}}]{Schreiber2010}
\bibinfo{author}{M.~R. {Schreiber}}, \bibinfo{author}{B.~T. {G{\"a}nsicke}},
  \bibinfo{author}{A.~{Rebassa-Mansergas}}, et~al.,
\newblock \bibinfo{title}{{Post common envelope binaries from SDSS. VIII.
  Evidence for disrupted magnetic braking}},
\newblock \bibinfo{journal}{\aap} \bibinfo{volume}{513} (\bibinfo{year}{2010})
  \bibinfo{pages}{L7}. \DOIprefix\doi{10.1051/0004-6361/201013990}.
\bibitem[{{Rebassa-Mansergas} et~al.(2011){Rebassa-Mansergas}, {Nebot
  G{\'o}mez-Mor{\'a}n}, {Schreiber}, {Girven}, and
  {G{\"a}nsicke}}]{Rebassa-Mansergas2011}
\bibinfo{author}{A.~{Rebassa-Mansergas}}, \bibinfo{author}{A.~{Nebot
  G{\'o}mez-Mor{\'a}n}}, \bibinfo{author}{M.~R. {Schreiber}}, et~al.,
\newblock \bibinfo{title}{{Post-common envelope binaries from SDSS-X: the
  origin of low-mass white dwarfs}},
\newblock \bibinfo{journal}{\mnras} \bibinfo{volume}{413}
  (\bibinfo{year}{2011}) \bibinfo{pages}{1121--1131}.
  \DOIprefix\doi{10.1111/j.1365-2966.2011.18200.x}.
\bibitem[{{Rebassa-Mansergas} et~al.(2008){Rebassa-Mansergas}, {G{\"a}nsicke},
  {Schreiber}, {Southworth}, {Schwope}, {Nebot Gomez-Moran}, {Aungwerojwit},
  {Rodr{\'\i}guez-Gil}, {Karamanavis}, {Krumpe}, {Tremou}, {Schwarz}, {Staude},
  and {Vogel}}]{Rebassa-Mansergas2008}
\bibinfo{author}{A.~{Rebassa-Mansergas}}, \bibinfo{author}{B.~T.
  {G{\"a}nsicke}}, \bibinfo{author}{M.~R. {Schreiber}}, et~al.,
\newblock \bibinfo{title}{{Post-common envelope binaries from SDSS - III. Seven
  new orbital periods}},
\newblock \bibinfo{journal}{\mnras} \bibinfo{volume}{390}
  (\bibinfo{year}{2008}) \bibinfo{pages}{1635--1646}.
  \DOIprefix\doi{10.1111/j.1365-2966.2008.13850.x}.
\bibitem[{{Pyrzas} et~al.(2009){Pyrzas}, {G{\"a}nsicke}, {Marsh},
  {Aungwerojwit}, {Rebassa-Mansergas}, {Rodr{\'\i}guez-Gil}, {Southworth},
  {Schreiber}, {Nebot Gomez-Moran}, and {Koester}}]{Pyrzas2009MNRAS.394..978P}
\bibinfo{author}{S.~{Pyrzas}}, \bibinfo{author}{B.~T. {G{\"a}nsicke}},
  \bibinfo{author}{T.~R. {Marsh}}, et~al.,
\newblock \bibinfo{title}{{Post-common-envelope binaries from SDSS - V. Four
  eclipsing white dwarf main-sequence binaries}},
\newblock \bibinfo{journal}{\mnras} \bibinfo{volume}{394}
  (\bibinfo{year}{2009}) \bibinfo{pages}{978--994}.
  \DOIprefix\doi{10.1111/j.1365-2966.2008.14378.x}.
\bibitem[{{Nebot G{\'o}mez-Mor{\'a}n} et~al.(2011){Nebot G{\'o}mez-Mor{\'a}n},
  {G{\"a}nsicke}, {Schreiber}, {Rebassa-Mansergas}, {Schwope}, {Southworth},
  {Aungwerojwit}, {Bothe}, {Davis}, {Kolb}, {M{\"u}ller}, {Papadaki}, {Pyrzas},
  {Rabitz}, {Rodr{\'\i}guez-Gil}, {Schmidtobreick}, {Schwarz}, {Tappert},
  {Toloza}, {Vogel}, and {Zorotovic}}]{Nebot-Gomez-Moran2011A&A...536A..43N}
\bibinfo{author}{A.~{Nebot G{\'o}mez-Mor{\'a}n}}, \bibinfo{author}{B.~T.
  {G{\"a}nsicke}}, \bibinfo{author}{M.~R. {Schreiber}}, et~al.,
\newblock \bibinfo{title}{{Post common envelope binaries from SDSS. XII. The
  orbital period distribution}},
\newblock \bibinfo{journal}{\aap} \bibinfo{volume}{536} (\bibinfo{year}{2011})
  \bibinfo{pages}{A43}. \DOIprefix\doi{10.1051/0004-6361/201117514}.
\bibitem[{{Zorotovic} et~al.(2011){Zorotovic}, {Schreiber}, and
  {G{\"a}nsicke}}]{Zorotovic2011A&A...536A..42Z}
\bibinfo{author}{M.~{Zorotovic}}, \bibinfo{author}{M.~R. {Schreiber}},
  \bibinfo{author}{B.~T. {G{\"a}nsicke}},
\newblock \bibinfo{title}{{Post common envelope binaries from SDSS. XI. The
  white dwarf mass distributions of CVs and pre-CVs}},
\newblock \bibinfo{journal}{\aap} \bibinfo{volume}{536} (\bibinfo{year}{2011})
  \bibinfo{pages}{A42}. \DOIprefix\doi{10.1051/0004-6361/201116626}.
\bibitem[{{Wilson} et~al.(2010){Wilson}, {Hearty}, {Skrutskie}, {Majewski},
  {Schiavon}, and et~al.}]{Wilson2010SPIE.7735E..1CW}
\bibinfo{author}{J.~C. {Wilson}}, \bibinfo{author}{F.~{Hearty}},
  \bibinfo{author}{M.~F. {Skrutskie}}, et~al.,
\newblock \bibinfo{title}{{The Apache Point Observatory Galactic Evolution
  Experiment (APOGEE) high-resolution near-infrared multi-object fiber
  spectrograph}},
\newblock in: \bibinfo{editor}{I.~S. {McLean}}, \bibinfo{editor}{S.~K.
  {Ramsay}}, \bibinfo{editor}{H.~{Takami}} (Eds.),
  \bibinfo{booktitle}{Ground-based and Airborne Instrumentation for Astronomy
  III}, volume \bibinfo{volume}{7735} of \textit{\bibinfo{series}{Society of
  Photo-Optical Instrumentation Engineers (SPIE) Conference Series}},
  \bibinfo{year}{2010}, p. \bibinfo{pages}{77351C}.
  \DOIprefix\doi{10.1117/12.856708}.
\bibitem[{{Wilson} and {Nordhaus}(2019)}]{Wilson2019MNRAS.485.4492W}
\bibinfo{author}{E.~C. {Wilson}}, \bibinfo{author}{J.~{Nordhaus}},
\newblock \bibinfo{title}{{The role of convection in determining the ejection
  efficiency of common envelope interactions}},
\newblock \bibinfo{journal}{\mnras} \bibinfo{volume}{485}
  (\bibinfo{year}{2019}) \bibinfo{pages}{4492--4501}.
  \DOIprefix\doi{10.1093/mnras/stz601}.
\bibitem[{{Garc{\'\i}a P{\'e}rez} et~al.(2016){Garc{\'\i}a P{\'e}rez}, {Allende
  Prieto}, {Holtzman}, {Shetrone}, and et~al}]{Garcia-Perez2016AJ....151..144G}
\bibinfo{author}{A.~E. {Garc{\'\i}a P{\'e}rez}}, \bibinfo{author}{C.~{Allende
  Prieto}}, \bibinfo{author}{J.~A. {Holtzman}}, et~al.,
\newblock \bibinfo{title}{{ASPCAP: The APOGEE Stellar Parameter and Chemical
  Abundances Pipeline}},
\newblock \bibinfo{journal}{\aj} \bibinfo{volume}{151} (\bibinfo{year}{2016})
  \bibinfo{pages}{144}. \DOIprefix\doi{10.3847/0004-6256/151/6/144}.
\bibitem[{{Olney} et~al.(2020){Olney}, {Kounkel}, {Schillinger}, {Scoggins},
  {Yin}, {Howard}, {Covey}, {Hutchinson}, and
  {Stassun}}]{Olney2020AJ....159..182O}
\bibinfo{author}{R.~{Olney}}, \bibinfo{author}{M.~{Kounkel}},
  \bibinfo{author}{C.~{Schillinger}}, et~al.,
\newblock \bibinfo{title}{{APOGEE Net: Improving the Derived Spectral
  Parameters for Young Stars through Deep Learning}},
\newblock \bibinfo{journal}{\aj} \bibinfo{volume}{159} (\bibinfo{year}{2020})
  \bibinfo{pages}{182}. \DOIprefix\doi{10.3847/1538-3881/ab7a97}.
\bibitem[{{El-Badry} et~al.(2018){El-Badry}, {Rix}, {Ting}, {Weisz},
  {Bergemann}, {Cargile}, {Conroy}, and {Eilers}}]{El-Badry2018MNRAS.473.5043E}
\bibinfo{author}{K.~{El-Badry}}, \bibinfo{author}{H.-W. {Rix}},
  \bibinfo{author}{Y.-S. {Ting}}, et~al.,
\newblock \bibinfo{title}{{Signatures of unresolved binaries in stellar
  spectra: implications for spectral fitting}},
\newblock \bibinfo{journal}{\mnras} \bibinfo{volume}{473}
  (\bibinfo{year}{2018}) \bibinfo{pages}{5043--5049}.
  \DOIprefix\doi{10.1093/mnras/stx2758}.
\bibitem[{{El-Badry} and {Rix}(2018)}]{El-Badry2018MNRAS.480.4884E}
\bibinfo{author}{K.~{El-Badry}}, \bibinfo{author}{H.-W. {Rix}},
\newblock \bibinfo{title}{{Imprints of white dwarf recoil in the separation
  distribution of Gaia wide binaries}},
\newblock \bibinfo{journal}{\mnras} \bibinfo{volume}{480}
  (\bibinfo{year}{2018}) \bibinfo{pages}{4884--4902}.
  \DOIprefix\doi{10.1093/mnras/sty2186}.
\bibitem[{{El-Badry} et~al.(2018){El-Badry}, {Ting}, {Rix}, {Quataert},
  {Weisz}, {Cargile}, {Conroy}, {Hogg}, {Bergemann}, and
  {Liu}}]{El-Badry2018MNRAS.476..528E}
\bibinfo{author}{K.~{El-Badry}}, \bibinfo{author}{Y.-S. {Ting}},
  \bibinfo{author}{H.-W. {Rix}}, et~al.,
\newblock \bibinfo{title}{{Discovery and characterization of 3000+
  main-sequence binaries from APOGEE spectra}},
\newblock \bibinfo{journal}{\mnras} \bibinfo{volume}{476}
  (\bibinfo{year}{2018}) \bibinfo{pages}{528--553}.
  \DOIprefix\doi{10.1093/mnras/sty240}.
\bibitem[{{Kovalev} et~al.(2022){Kovalev}, {Chen}, and
  {Han}}]{Kovalev2022MNRAS.517..356K}
\bibinfo{author}{M.~{Kovalev}}, \bibinfo{author}{X.~{Chen}},
  \bibinfo{author}{Z.~{Han}},
\newblock \bibinfo{title}{{Detection of 2460 double-lined spectroscopic binary
  candidates in the LAMOST-MRS using projected rotational velocities and a
  binary spectral model}},
\newblock \bibinfo{journal}{\mnras} \bibinfo{volume}{517}
  (\bibinfo{year}{2022}) \bibinfo{pages}{356--373}.
  \DOIprefix\doi{10.1093/mnras/stac2513}.
\bibitem[{{Kounkel} et~al.(2019){Kounkel}, {Covey}, {Moe}, {Kratter},
  {Su{\'a}rez}, and et~al.}]{Kounkel2019AJ....157..196K}
\bibinfo{author}{M.~{Kounkel}}, \bibinfo{author}{K.~{Covey}},
  \bibinfo{author}{M.~{Moe}}, et~al.,
\newblock \bibinfo{title}{{Close Companions around Young Stars}},
\newblock \bibinfo{journal}{\aj} \bibinfo{volume}{157} (\bibinfo{year}{2019})
  \bibinfo{pages}{196}. \DOIprefix\doi{10.3847/1538-3881/ab13b1}.
\bibitem[{{Kounkel} et~al.(2021){Kounkel}, {Covey}, {Stassun}, {Price-Whelan},
  {Holtzman}, {Chojnowski}, {Longa-Pe{\~n}a}, and
  et~al.}]{Kounke2021AJ....162..184K}
\bibinfo{author}{M.~{Kounkel}}, \bibinfo{author}{K.~R. {Covey}},
  \bibinfo{author}{K.~G. {Stassun}}, et~al.,
\newblock \bibinfo{title}{{Double-lined Spectroscopic Binaries in the APOGEE
  DR16 and DR17 Data}},
\newblock \bibinfo{journal}{\aj} \bibinfo{volume}{162} (\bibinfo{year}{2021})
  \bibinfo{pages}{184}. \DOIprefix\doi{10.3847/1538-3881/ac1798}.
\bibitem[{{Matijevi{\v{c}}} et~al.(2010){Matijevi{\v{c}}}, {Zwitter}, {Munari},
  {Bienaym{\'e}}, {Binney}, and et~al.}]{Matijevic2010AJ....140..184M}
\bibinfo{author}{G.~{Matijevi{\v{c}}}}, \bibinfo{author}{T.~{Zwitter}},
  \bibinfo{author}{U.~{Munari}}, et~al.,
\newblock \bibinfo{title}{{Double-lined Spectroscopic Binary Stars in the
  Radial Velocity Experiment Survey}},
\newblock \bibinfo{journal}{\aj} \bibinfo{volume}{140} (\bibinfo{year}{2010})
  \bibinfo{pages}{184--195}. \DOIprefix\doi{10.1088/0004-6256/140/1/184}.
\bibitem[{{Merle} et~al.(2017){Merle}, {Van Eck}, {Jorissen}, {Van der
  Swaelmen}, and et~al.}]{Merle2017A&A...608A..95M}
\bibinfo{author}{T.~{Merle}}, \bibinfo{author}{S.~{Van Eck}},
  \bibinfo{author}{A.~{Jorissen}}, et~al.,
\newblock \bibinfo{title}{{The Gaia-ESO Survey: double-, triple-, and
  quadruple-line spectroscopic binary candidates}},
\newblock \bibinfo{journal}{\aap} \bibinfo{volume}{608} (\bibinfo{year}{2017})
  \bibinfo{pages}{A95}. \DOIprefix\doi{10.1051/0004-6361/201730442}.
\bibitem[{{Traven} et~al.(2020){Traven}, {Feltzing}, {Merle}, {Van der
  Swaelmen}, {{\v{C}}otar}, {Church}, {Zwitter}, {Ting}, {Sahlholdt},
  {Asplund}, {Bland-Hawthorn}, {De Silva}, {Freeman}, {Martell}, {Sharma},
  {Zucker}, {Buder}, {Casey}, {D'Orazi}, {Kos}, {Lewis}, {Lin}, {Lind},
  {Simpson}, {Stello}, {Munari}, and {Wittenmyer}}]{Traven2020A&A...638A.145T}
\bibinfo{author}{G.~{Traven}}, \bibinfo{author}{S.~{Feltzing}},
  \bibinfo{author}{T.~{Merle}}, et~al.,
\newblock \bibinfo{title}{{The GALAH survey: multiple stars and our Galaxy. I.
  A comprehensive method for deriving properties of FGK binary stars}},
\newblock \bibinfo{journal}{\aap} \bibinfo{volume}{638} (\bibinfo{year}{2020})
  \bibinfo{pages}{A145}. \DOIprefix\doi{10.1051/0004-6361/202037484}.
\bibitem[{{Li} et~al.(2021){Li}, {Shi}, {Yan}, {Fu}, {Li}, and
  {Hou}}]{Li2021ApJS..256...31L}
\bibinfo{author}{C.-q. {Li}}, \bibinfo{author}{J.-r. {Shi}},
  \bibinfo{author}{H.-l. {Yan}}, et~al.,
\newblock \bibinfo{title}{{Double- and Triple-line Spectroscopic Candidates in
  the LAMOST Medium-Resolution Spectroscopic Survey}},
\newblock \bibinfo{journal}{\apjs} \bibinfo{volume}{256} (\bibinfo{year}{2021})
  \bibinfo{pages}{31}. \DOIprefix\doi{10.3847/1538-4365/ac22a8}.
\bibitem[{{Reipurth} and {Mikkola}(2012)}]{Reipurth2012Natur.492..221R}
\bibinfo{author}{B.~{Reipurth}}, \bibinfo{author}{S.~{Mikkola}},
\newblock \bibinfo{title}{{Formation of the widest binary stars from dynamical
  unfolding of triple systems}},
\newblock \bibinfo{journal}{\nat} \bibinfo{volume}{492} (\bibinfo{year}{2012})
  \bibinfo{pages}{221--224}. \DOIprefix\doi{10.1038/nature11662}.
\bibitem[{{Kouwenhoven} et~al.(2010){Kouwenhoven}, {Goodwin}, {Parker},
  {Davies}, {Malmberg}, and {Kroupa}}]{Kouwenhoven2010MNRAS.404.1835K}
\bibinfo{author}{M.~B.~N. {Kouwenhoven}}, \bibinfo{author}{S.~P. {Goodwin}},
  \bibinfo{author}{R.~J. {Parker}}, et~al.,
\newblock \bibinfo{title}{{The formation of very wide binaries during the star
  cluster dissolution phase}},
\newblock \bibinfo{journal}{\mnras} \bibinfo{volume}{404}
  (\bibinfo{year}{2010}) \bibinfo{pages}{1835--1848}.
  \DOIprefix\doi{10.1111/j.1365-2966.2010.16399.x}.
\bibitem[{{Moeckel} and
  {Clarke}(2011{\natexlab{a}})}]{Moeckel2011MNRAS.410.2799M}
\bibinfo{author}{N.~{Moeckel}}, \bibinfo{author}{C.~J. {Clarke}},
\newblock \bibinfo{title}{{Collisional formation of very massive stars in dense
  clusters}},
\newblock \bibinfo{journal}{\mnras} \bibinfo{volume}{410}
  (\bibinfo{year}{2011}{\natexlab{a}}) \bibinfo{pages}{2799--2806}.
  \DOIprefix\doi{10.1111/j.1365-2966.2010.17659.x}.
\bibitem[{{Moeckel} and
  {Clarke}(2011{\natexlab{b}})}]{Moeckel2011MNRAS.415.1179M}
\bibinfo{author}{N.~{Moeckel}}, \bibinfo{author}{C.~J. {Clarke}},
\newblock \bibinfo{title}{{The formation of permanent soft binaries in
  dispersing clusters}},
\newblock \bibinfo{journal}{\mnras} \bibinfo{volume}{415}
  (\bibinfo{year}{2011}{\natexlab{b}}) \bibinfo{pages}{1179--1187}.
  \DOIprefix\doi{10.1111/j.1365-2966.2011.18731.x}.
\bibitem[{{Lee} et~al.(2017){Lee}, {Lee}, {Dunham}, {Tatematsu}, {Choi},
  {Bergin}, and {Evans}}]{Lee2017NatAs...1E.172L}
\bibinfo{author}{J.-E. {Lee}}, \bibinfo{author}{S.~{Lee}},
  \bibinfo{author}{M.~M. {Dunham}}, et~al.,
\newblock \bibinfo{title}{{Formation of wide binaries by turbulent
  fragmentation}},
\newblock \bibinfo{journal}{Nature Astronomy} \bibinfo{volume}{1}
  (\bibinfo{year}{2017}) \bibinfo{pages}{0172}.
  \DOIprefix\doi{10.1038/s41550-017-0172}.
\bibitem[{{Tokovinin}(2017)}]{Tokovinin2017MNRAS.468.3461T}
\bibinfo{author}{A.~{Tokovinin}},
\newblock \bibinfo{title}{{Formation of wide binary stars from adjacent
  cores}},
\newblock \bibinfo{journal}{\mnras} \bibinfo{volume}{468}
  (\bibinfo{year}{2017}) \bibinfo{pages}{3461--3467}.
  \DOIprefix\doi{10.1093/mnras/stx707}.
\bibitem[{{L{\'e}pine} et~al.(2007){L{\'e}pine}, {Rich}, and
  {Shara}}]{Lepine2007ApJ...669.1235L}
\bibinfo{author}{S.~{L{\'e}pine}}, \bibinfo{author}{R.~M. {Rich}},
  \bibinfo{author}{M.~M. {Shara}},
\newblock \bibinfo{title}{{Revised Metallicity Classes for Low-Mass Stars:
  Dwarfs (dM), Subdwarfs (sdM), Extreme Subdwarfs (esdM), and Ultrasubdwarfs
  (usdM)}},
\newblock \bibinfo{journal}{\apj} \bibinfo{volume}{669} (\bibinfo{year}{2007})
  \bibinfo{pages}{1235--1247}. \DOIprefix\doi{10.1086/521614}.
\bibitem[{{Rojas-Ayala} and {Lloyd}(2010)}]{Rojas-Ayala2010ASPC..430..528R}
\bibinfo{author}{B.~{Rojas-Ayala}}, \bibinfo{author}{J.~P. {Lloyd}},
\newblock \bibinfo{title}{{Metallicity of M-dwarfs from NIR Spectra}},
\newblock in: \bibinfo{editor}{V.~{Coud{\'e} du Foresto}},
  \bibinfo{editor}{D.~M. {Gelino}}, \bibinfo{editor}{I.~{Ribas}} (Eds.),
  \bibinfo{booktitle}{Pathways Towards Habitable Planets}, volume
  \bibinfo{volume}{430} of \textit{\bibinfo{series}{Astronomical Society of the
  Pacific Conference Series}}, \bibinfo{year}{2010},  p. \bibinfo{pages}{528}.
\bibitem[{{Montes} et~al.(2018){Montes}, {Gonz{\'a}lez-Peinado}, {Tabernero},
  {Caballero}, {Marfil}, and et~al.}]{Montes2018MNRAS.479.1332M}
\bibinfo{author}{D.~{Montes}}, \bibinfo{author}{R.~{Gonz{\'a}lez-Peinado}},
  \bibinfo{author}{H.~M. {Tabernero}}, et~al.,
\newblock \bibinfo{title}{{Calibrating the metallicity of M dwarfs in wide
  physical binaries with F-, G-, and K-primaries - I: High-resolution
  spectroscopy with HERMES: stellar parameters, abundances, and kinematics}},
\newblock \bibinfo{journal}{\mnras} \bibinfo{volume}{479}
  (\bibinfo{year}{2018}) \bibinfo{pages}{1332--1382}.
  \DOIprefix\doi{10.1093/mnras/sty1295}.
\bibitem[{{Garc{\'e}s} et~al.(2011){Garc{\'e}s}, {Catal{\'a}n}, and
  {Ribas}}]{Garces2011A&A...531A...7G}
\bibinfo{author}{A.~{Garc{\'e}s}}, \bibinfo{author}{S.~{Catal{\'a}n}},
  \bibinfo{author}{I.~{Ribas}},
\newblock \bibinfo{title}{{Time evolution of high-energy emissions of low-mass
  stars. I. Age determination using stellar chronology with white dwarfs in
  wide binaries}},
\newblock \bibinfo{journal}{\aap} \bibinfo{volume}{531} (\bibinfo{year}{2011})
  \bibinfo{pages}{A7}. \DOIprefix\doi{10.1051/0004-6361/201116775}.
\bibitem[{{Chanam{\'e}} and {Ram{\'\i}rez}(2012)}]{Chaname2012ApJ...746..102C}
\bibinfo{author}{J.~{Chanam{\'e}}}, \bibinfo{author}{I.~{Ram{\'\i}rez}},
\newblock \bibinfo{title}{{Toward Precise Ages for Single Stars in the Field.
  Gyrochronology Constraints at Several Gyr Using Wide Binaries. I. Ages for
  Initial Sample}},
\newblock \bibinfo{journal}{\apj} \bibinfo{volume}{746} (\bibinfo{year}{2012})
  \bibinfo{pages}{102}. \DOIprefix\doi{10.1088/0004-637X/746/1/102}.
\bibitem[{{Andrews} et~al.(2015){Andrews}, {Ag{\"u}eros}, {Gianninas}, {Kilic},
  {Dhital}, and {Anderson}}]{Andrews2015ApJ...815...63A}
\bibinfo{author}{J.~J. {Andrews}}, \bibinfo{author}{M.~A. {Ag{\"u}eros}},
  \bibinfo{author}{A.~{Gianninas}}, et~al.,
\newblock \bibinfo{title}{{Constraints on the Initial-Final Mass Relation from
  Wide Double White Dwarfs}},
\newblock \bibinfo{journal}{\apj} \bibinfo{volume}{815} (\bibinfo{year}{2015})
  \bibinfo{pages}{63}. \DOIprefix\doi{10.1088/0004-637X/815/1/63}.
\bibitem[{{Bahcall} et~al.(1985){Bahcall}, {Hut}, and
  {Tremaine}}]{Bahcall1985ApJ...290...15B}
\bibinfo{author}{J.~N. {Bahcall}}, \bibinfo{author}{P.~{Hut}},
  \bibinfo{author}{S.~{Tremaine}},
\newblock \bibinfo{title}{{Maximum mass of objects that constitute unseen disk
  material}},
\newblock \bibinfo{journal}{\apj} \bibinfo{volume}{290} (\bibinfo{year}{1985})
  \bibinfo{pages}{15--20}. \DOIprefix\doi{10.1086/162953}.
\bibitem[{{Yoo} et~al.(2004){Yoo}, {Chanam{\'e}}, and
  {Gould}}]{Yoo2004ApJ...601..311Y}
\bibinfo{author}{J.~{Yoo}}, \bibinfo{author}{J.~{Chanam{\'e}}},
  \bibinfo{author}{A.~{Gould}},
\newblock \bibinfo{title}{{The End of the MACHO Era: Limits on Halo Dark Matter
  from Stellar Halo Wide Binaries}},
\newblock \bibinfo{journal}{\apj} \bibinfo{volume}{601} (\bibinfo{year}{2004})
  \bibinfo{pages}{311--318}. \DOIprefix\doi{10.1086/380562}.
\bibitem[{{Andrews} et~al.(2018){Andrews}, {Chanam{\'e}}, and
  {Ag{\"u}eros}}]{Andrews2018MNRAS.473.5393A}
\bibinfo{author}{J.~J. {Andrews}}, \bibinfo{author}{J.~{Chanam{\'e}}},
  \bibinfo{author}{M.~A. {Ag{\"u}eros}},
\newblock \bibinfo{title}{{Wide binaries in Tycho-Gaia II: metallicities,
  abundances and prospects for chemical tagging}},
\newblock \bibinfo{journal}{\mnras} \bibinfo{volume}{473}
  (\bibinfo{year}{2018}) \bibinfo{pages}{5393--5406}.
  \DOIprefix\doi{10.1093/mnras/stx2685}.
\bibitem[{{Pittordis} and {Sutherland}(2018)}]{Pittordis2018MNRAS.480.1778P}
\bibinfo{author}{C.~{Pittordis}}, \bibinfo{author}{W.~{Sutherland}},
\newblock \bibinfo{title}{{Testing modified-gravity theories via wide binaries
  and GAIA}},
\newblock \bibinfo{journal}{\mnras} \bibinfo{volume}{480}
  (\bibinfo{year}{2018}) \bibinfo{pages}{1778--1795}.
  \DOIprefix\doi{10.1093/mnras/sty1578}.
\bibitem[{{Pittordis} and {Sutherland}(2019)}]{Pittordis2019MNRAS.488.4740P}
\bibinfo{author}{C.~{Pittordis}}, \bibinfo{author}{W.~{Sutherland}},
\newblock \bibinfo{title}{{Testing modified gravity with wide binaries in Gaia
  DR2}},
\newblock \bibinfo{journal}{\mnras} \bibinfo{volume}{488}
  (\bibinfo{year}{2019}) \bibinfo{pages}{4740--4752}.
  \DOIprefix\doi{10.1093/mnras/stz1898}.
\bibitem[{{Pittordis} and {Sutherland}(2022)}]{Pittordis2022arXiv220502846P}
\bibinfo{author}{C.~{Pittordis}}, \bibinfo{author}{W.~{Sutherland}},
\newblock \bibinfo{title}{{Wide Binaries from GAIA EDR3: preference for GR over
  MOND ?}},
\newblock \bibinfo{journal}{arXiv e-prints}   (\bibinfo{year}{2022})
  \bibinfo{pages}{arXiv:2205.02846}.
\bibitem[{{Tian} et~al.(2020){Tian}, {El-Badry}, {Rix}, and
  {Gould}}]{Tian2020ApJS..246....4T}
\bibinfo{author}{H.-J. {Tian}}, \bibinfo{author}{K.~{El-Badry}},
  \bibinfo{author}{H.-W. {Rix}}, \bibinfo{author}{A.~{Gould}},
\newblock \bibinfo{title}{{The Separation Distribution of Ultrawide Binaries
  across Galactic Populations}},
\newblock \bibinfo{journal}{\apjs} \bibinfo{volume}{246} (\bibinfo{year}{2020})
  \bibinfo{pages}{4}. \DOIprefix\doi{10.3847/1538-4365/ab54c4}.
\bibitem[{{L{\'e}pine} and {Shara}(2005)}]{Lepine2005AJ....129.1483L}
\bibinfo{author}{S.~{L{\'e}pine}}, \bibinfo{author}{M.~M. {Shara}},
\newblock \bibinfo{title}{{A Catalog of Northern Stars with Annual Proper
  Motions Larger than 0.15'' (LSPM-NORTH Catalog)}},
\newblock \bibinfo{journal}{\aj} \bibinfo{volume}{129} (\bibinfo{year}{2005})
  \bibinfo{pages}{1483--1522}. \DOIprefix\doi{10.1086/427854}.
\bibitem[{{L{\'e}pine} and {Gaidos}(2011)}]{Lepine2011AJ....142..138L}
\bibinfo{author}{S.~{L{\'e}pine}}, \bibinfo{author}{E.~{Gaidos}},
\newblock \bibinfo{title}{{An All-sky Catalog of Bright M Dwarfs}},
\newblock \bibinfo{journal}{\aj} \bibinfo{volume}{142} (\bibinfo{year}{2011})
  \bibinfo{pages}{138}. \DOIprefix\doi{10.1088/0004-6256/142/4/138}.
\bibitem[{{Hartman} and {L{\'e}pine}(2020)}]{Hartman2020ApJS..247...66H}
\bibinfo{author}{Z.~D. {Hartman}}, \bibinfo{author}{S.~{L{\'e}pine}},
\newblock \bibinfo{title}{{The SUPERWIDE Catalog: A Catalog of 99,203 Wide
  Binaries Found in Gaia and Supplemented by the SUPERBLINK High Proper Motion
  Catalog}},
\newblock \bibinfo{journal}{\apjs} \bibinfo{volume}{247} (\bibinfo{year}{2020})
  \bibinfo{pages}{66}. \DOIprefix\doi{10.3847/1538-4365/ab79a6}.
\bibitem[{{El-Badry} and {Rix}(2019)}]{El-Badry2019MNRAS.482L.139E}
\bibinfo{author}{K.~{El-Badry}}, \bibinfo{author}{H.-W. {Rix}},
\newblock \bibinfo{title}{{The wide binary fraction of solar-type stars:
  emergence of metallicity dependence at a < 200 au}},
\newblock \bibinfo{journal}{\mnras} \bibinfo{volume}{482}
  (\bibinfo{year}{2019}) \bibinfo{pages}{L139--L144}.
  \DOIprefix\doi{10.1093/mnrasl/sly206}.
\bibitem[{{Hwang} et~al.(2021){Hwang}, {Ting}, {Schlaufman}, {Zakamska}, and
  {Wyse}}]{Hwang2021MNRAS.501.4329H}
\bibinfo{author}{H.-C. {Hwang}}, \bibinfo{author}{Y.-S. {Ting}},
  \bibinfo{author}{K.~C. {Schlaufman}}, et~al.,
\newblock \bibinfo{title}{{The non-monotonic, strong metallicity dependence of
  the wide-binary fraction}},
\newblock \bibinfo{journal}{\mnras} \bibinfo{volume}{501}
  (\bibinfo{year}{2021}) \bibinfo{pages}{4329--4343}.
  \DOIprefix\doi{10.1093/mnras/staa3854}.
\bibitem[{{Abt} et~al.(1990){Abt}, {Gomez}, and
  {Levy}}]{Abt1990ApJS...74..551A}
\bibinfo{author}{H.~A. {Abt}}, \bibinfo{author}{A.~E. {Gomez}},
  \bibinfo{author}{S.~G. {Levy}},
\newblock \bibinfo{title}{{The Frequency and Formation Mechanism of B2--B5
  Main-Sequence Binaries}},
\newblock \bibinfo{journal}{\apjs} \bibinfo{volume}{74} (\bibinfo{year}{1990})
  \bibinfo{pages}{551}. \DOIprefix\doi{10.1086/191508}.
\bibitem[{{Tout}(1991)}]{Tout1991MNRAS.250..701T}
\bibinfo{author}{C.~A. {Tout}},
\newblock \bibinfo{title}{{On the relation between the mass-ratio distribution
  in binary stars and the mass function for single stars.}},
\newblock \bibinfo{journal}{\mnras} \bibinfo{volume}{250}
  (\bibinfo{year}{1991}) \bibinfo{pages}{701--706}.
  \DOIprefix\doi{10.1093/mnras/250.4.701}.
\bibitem[{{McDonald} and {Clarke}(1995)}]{McDonald1995MNRAS.275..671M}
\bibinfo{author}{J.~M. {McDonald}}, \bibinfo{author}{C.~J. {Clarke}},
\newblock \bibinfo{title}{{The effect of star-disc interactions on the binary
  mass-ratio distribution}},
\newblock \bibinfo{journal}{\mnras} \bibinfo{volume}{275}
  (\bibinfo{year}{1995}) \bibinfo{pages}{671--684}.
  \DOIprefix\doi{10.1093/mnras/275.3.671}.
\bibitem[{{Tokovinin}(2000)}]{Tokovinin2000A&A...360..997T}
\bibinfo{author}{A.~A. {Tokovinin}},
\newblock \bibinfo{title}{{On the origin of binaries with twin components}},
\newblock \bibinfo{journal}{\aap}  \bibinfo{volume}{360} (\bibinfo{year}{2000})
  \bibinfo{pages}{997--1002}.
\bibitem[{{Sana} et~al.(2012){Sana}, {de Mink}, {de Koter}, {Langer}, {Evans},
  {Gieles}, {Gosset}, {Izzard}, {Le Bouquin}, and
  {Schneider}}]{Sana2012Sci...337..444S}
\bibinfo{author}{H.~{Sana}}, \bibinfo{author}{S.~E. {de Mink}},
  \bibinfo{author}{A.~{de Koter}}, et~al.,
\newblock \bibinfo{title}{{Binary Interaction Dominates the Evolution of
  Massive Stars}},
\newblock \bibinfo{journal}{Science} \bibinfo{volume}{337}
  (\bibinfo{year}{2012}) \bibinfo{pages}{444}.
  \DOIprefix\doi{10.1126/science.1223344}.
\bibitem[{{Bonnell} and {Bastien}(1992)}]{Bonnell1992ApJ...401..654B}
\bibinfo{author}{I.~{Bonnell}}, \bibinfo{author}{P.~{Bastien}},
\newblock \bibinfo{title}{{Fragmentation of Elongated Cylindrical Clouds. V.
  Dependence of Mass Ratios on Initial Conditions}},
\newblock \bibinfo{journal}{\apj} \bibinfo{volume}{401} (\bibinfo{year}{1992})
  \bibinfo{pages}{654}. \DOIprefix\doi{10.1086/172093}.
\bibitem[{{Kroupa}(1995{\natexlab{a}})}]{Kroupa1995MNRAS.277.1507K}
\bibinfo{author}{P.~{Kroupa}},
\newblock \bibinfo{title}{{The dynamical properties of stellar systems in the
  Galactic disc}},
\newblock \bibinfo{journal}{\mnras} \bibinfo{volume}{277}
  (\bibinfo{year}{1995}{\natexlab{a}}) \bibinfo{pages}{1507}.
  \DOIprefix\doi{10.1093/mnras/277.4.1507}.
\bibitem[{{Kroupa}(1995{\natexlab{b}})}]{Kroupa1995MNRAS.277.1491K}
\bibinfo{author}{P.~{Kroupa}},
\newblock \bibinfo{title}{{Inverse dynamical population synthesis and star
  formation}},
\newblock \bibinfo{journal}{\mnras} \bibinfo{volume}{277}
  (\bibinfo{year}{1995}{\natexlab{b}}) \bibinfo{pages}{1491}.
  \DOIprefix\doi{10.1093/mnras/277.4.1491}.
\bibitem[{{Clarke} and {Syer}(1996)}]{Clarke1996MNRAS.278L..23C}
\bibinfo{author}{C.~J. {Clarke}}, \bibinfo{author}{D.~{Syer}},
\newblock \bibinfo{title}{{Low-mass companions to T Tauri stars: a mechanism
  for rapid-rise FU Orionis outbursts}},
\newblock \bibinfo{journal}{\mnras} \bibinfo{volume}{278}
  (\bibinfo{year}{1996}) \bibinfo{pages}{L23--L27}.
  \DOIprefix\doi{10.1093/mnras/278.1.L23}.
\bibitem[{{Bate} and {Bonnell}(1997)}]{Bate1997MNRAS.285...33B}
\bibinfo{author}{M.~R. {Bate}}, \bibinfo{author}{I.~A. {Bonnell}},
\newblock \bibinfo{title}{{Accretion during binary star formation - II. Gaseous
  accretion and disc formation}},
\newblock \bibinfo{journal}{\mnras} \bibinfo{volume}{285}
  (\bibinfo{year}{1997}) \bibinfo{pages}{33--48}.
  \DOIprefix\doi{10.1093/mnras/285.1.33}.
\bibitem[{{Bate}(2012)}]{Bate2012MNRAS.419.3115B}
\bibinfo{author}{M.~R. {Bate}},
\newblock \bibinfo{title}{{Stellar, brown dwarf and multiple star properties
  from a radiation hydrodynamical simulation of star cluster formation}},
\newblock \bibinfo{journal}{\mnras} \bibinfo{volume}{419}
  (\bibinfo{year}{2012}) \bibinfo{pages}{3115--3146}.
  \DOIprefix\doi{10.1111/j.1365-2966.2011.19955.x}.
\bibitem[{{Kratter} and {Matzner}(2006)}]{Kratter2006MNRAS.373.1563K}
\bibinfo{author}{K.~M. {Kratter}}, \bibinfo{author}{C.~D. {Matzner}},
\newblock \bibinfo{title}{{Fragmentation of massive protostellar discs}},
\newblock \bibinfo{journal}{\mnras} \bibinfo{volume}{373}
  (\bibinfo{year}{2006}) \bibinfo{pages}{1563--1576}.
  \DOIprefix\doi{10.1111/j.1365-2966.2006.11103.x}.
\bibitem[{{Kouwenhoven} et~al.(2009){Kouwenhoven}, {Brown}, {Goodwin},
  {Portegies Zwart}, and {Kaper}}]{Kouwenhoven2009A&A...493..979K}
\bibinfo{author}{M.~B.~N. {Kouwenhoven}}, \bibinfo{author}{A.~G.~A. {Brown}},
  \bibinfo{author}{S.~P. {Goodwin}}, et~al.,
\newblock \bibinfo{title}{{Exploring the consequences of pairing algorithms for
  binary stars}},
\newblock \bibinfo{journal}{\aap} \bibinfo{volume}{493} (\bibinfo{year}{2009})
  \bibinfo{pages}{979--1016}. \DOIprefix\doi{10.1051/0004-6361:200810234}.
\bibitem[{{Marks} and {Kroupa}(2011)}]{Marks2011MNRAS.417.1702M}
\bibinfo{author}{M.~{Marks}}, \bibinfo{author}{P.~{Kroupa}},
\newblock \bibinfo{title}{{Dynamical population synthesis: constructing the
  stellar single and binary contents of galactic field populations}},
\newblock \bibinfo{journal}{\mnras} \bibinfo{volume}{417}
  (\bibinfo{year}{2011}) \bibinfo{pages}{1702--1714}.
  \DOIprefix\doi{10.1111/j.1365-2966.2011.19519.x}.
\bibitem[{{de Mink} and {Belczynski}(2015)}]{de-Mink2015ApJ...814...58D}
\bibinfo{author}{S.~E. {de Mink}}, \bibinfo{author}{K.~{Belczynski}},
\newblock \bibinfo{title}{{Merger Rates of Double Neutron Stars and Stellar
  Origin Black Holes: The Impact of Initial Conditions on Binary Evolution
  Predictions}},
\newblock \bibinfo{journal}{\apj} \bibinfo{volume}{814} (\bibinfo{year}{2015})
  \bibinfo{pages}{58}. \DOIprefix\doi{10.1088/0004-637X/814/1/58}.
\bibitem[{{Sana} et~al.(2014){Sana}, {Le Bouquin}, {Lacour}, {Berger},
  {Duvert}, {Gauchet}, {Norris}, {Olofsson}, {Pickel}, {Zins}, {Absil}, {de
  Koter}, {Kratter}, {Schnurr}, and {Zinnecker}}]{Sana2014}
\bibinfo{author}{H.~{Sana}}, \bibinfo{author}{J.~B. {Le Bouquin}},
  \bibinfo{author}{S.~{Lacour}}, et~al.,
\newblock \bibinfo{title}{{Southern Massive Stars at High Angular Resolution:
  Observational Campaign and Companion Detection}},
\newblock \bibinfo{journal}{\apjs} \bibinfo{volume}{215} (\bibinfo{year}{2014})
  \bibinfo{pages}{15}. \DOIprefix\doi{10.1088/0067-0049/215/1/15}.
\bibitem[{{GRAVITY Collaboration} et~al.(2018){GRAVITY Collaboration}, {Karl},
  {Pfuhl}, {Eisenhauer}, {Genzel}, {Grellmann}, {Habibi}, {Abuter}, {Accardo},
  and {Amorim}}]{Gravity2018}
\bibinfo{author}{{GRAVITY Collaboration}}, \bibinfo{author}{M.~{Karl}},
  \bibinfo{author}{O.~{Pfuhl}}, et~al.,
\newblock \bibinfo{title}{{Multiple star systems in the Orion nebula}},
\newblock \bibinfo{journal}{\aap} \bibinfo{volume}{620} (\bibinfo{year}{2018})
  \bibinfo{pages}{A116}. \DOIprefix\doi{10.1051/0004-6361/201833575}.
\bibitem[{{Bordier} et~al.(2022){Bordier}, {Frost}, {Sana}, {Reggiani},
  {M{\'e}rand}, {Rainot}, {Ram{\'\i}rez-Tannus}, and {de Wit}}]{Bordier2022}
\bibinfo{author}{E.~{Bordier}}, \bibinfo{author}{A.~J. {Frost}},
  \bibinfo{author}{H.~{Sana}}, et~al.,
\newblock \bibinfo{title}{{The origin of close massive binaries in the M17
  star-forming region}},
\newblock \bibinfo{journal}{\aap} \bibinfo{volume}{663} (\bibinfo{year}{2022})
  \bibinfo{pages}{A26}. \DOIprefix\doi{10.1051/0004-6361/202141849}.
\bibitem[{{Guo} et~al.(2022){Guo}, {Liu}, {Wang}, {Wang}, {Zhang}, {Ji}, {Han},
  and {Chen}}]{Guo2022A&A...667A..44G}
\bibinfo{author}{Y.~{Guo}}, \bibinfo{author}{C.~{Liu}},
  \bibinfo{author}{L.~{Wang}}, et~al.,
\newblock \bibinfo{title}{{The statistical properties of early-type stars from
  LAMOST DR8}},
\newblock \bibinfo{journal}{\aap} \bibinfo{volume}{667} (\bibinfo{year}{2022})
  \bibinfo{pages}{A44}. \DOIprefix\doi{10.1051/0004-6361/202244300}.
\bibitem[{{Abbott} et~al.(2016{\natexlab{a}}){Abbott}, {Abbott}, {Abbott},
  {Abernathy}, {Acernese}, {Ackley}, {Adams}, {Adams}, {Addesso}, {Adhikari},
  {Adya}, {Affeldt}, et~al., {LIGO Scientific Collaboration}, and {VIRGO
  Collaboration}}]{Abbott2016PhRvL.116x1103A}
\bibinfo{author}{B.~P. {Abbott}}, \bibinfo{author}{R.~{Abbott}},
  \bibinfo{author}{T.~D. {Abbott}}, et~al.,
\newblock \bibinfo{title}{{GW151226: Observation of Gravitational Waves from a
  22-Solar-Mass Binary Black Hole Coalescence}},
\newblock \bibinfo{journal}{\prl} \bibinfo{volume}{116}
  (\bibinfo{year}{2016}{\natexlab{a}}) \bibinfo{pages}{241103}.
  \DOIprefix\doi{10.1103/PhysRevLett.116.241103}.
\bibitem[{{Abbott} et~al.(2016{\natexlab{b}}){Abbott}, {Abbott}, {Abbott},
  {Abernathy}, {Acernese}, {Ackley}, {Adams}, {Adams}, {Addesso}, {Adhikari},
  {Adya}, {Affeldt}, et~al., {LIGO Scientific Collaboration}, and {Virgo
  Collaboration}}]{Abbott2016PhRvL.116f1102A}
\bibinfo{author}{B.~P. {Abbott}}, \bibinfo{author}{R.~{Abbott}},
  \bibinfo{author}{T.~D. {Abbott}}, et~al.,
\newblock \bibinfo{title}{{Observation of Gravitational Waves from a Binary
  Black Hole Merger}},
\newblock \bibinfo{journal}{\prl} \bibinfo{volume}{116}
  (\bibinfo{year}{2016}{\natexlab{b}}) \bibinfo{pages}{061102}.
  \DOIprefix\doi{10.1103/PhysRevLett.116.061102}.
\bibitem[{{Langer} et~al.(2020){Langer}, {Sch{\"u}rmann}, {Stoll}, {Marchant},
  {Lennon}, {Mahy}, {de Mink}, and et~al.}]{Langer2020A&A...638A..39L}
\bibinfo{author}{N.~{Langer}}, \bibinfo{author}{C.~{Sch{\"u}rmann}},
  \bibinfo{author}{K.~{Stoll}}, et~al.,
\newblock \bibinfo{title}{{Properties of OB star-black hole systems derived
  from detailed binary evolution models}},
\newblock \bibinfo{journal}{\aap} \bibinfo{volume}{638} (\bibinfo{year}{2020})
  \bibinfo{pages}{A39}. \DOIprefix\doi{10.1051/0004-6361/201937375}.
\bibitem[{{Kiminki} and {Kobulnicky}(2012)}]{Kiminki2012ApJ...751....4K}
\bibinfo{author}{D.~C. {Kiminki}}, \bibinfo{author}{H.~A. {Kobulnicky}},
\newblock \bibinfo{title}{{An Updated Look at Binary Characteristics of Massive
  Stars in the Cygnus OB2 Association}},
\newblock \bibinfo{journal}{\apj} \bibinfo{volume}{751} (\bibinfo{year}{2012})
  \bibinfo{pages}{4}. \DOIprefix\doi{10.1088/0004-637X/751/1/4}.
\bibitem[{{Kobulnicky} et~al.(2014){Kobulnicky}, {Kiminki}, {Lundquist},
  {Burke}, {Chapman}, {Keller}, {Lester}, {Rolen}, {Topel}, {Bhattacharjee},
  {Smullen}, {Vargas {\'A}lvarez}, {Runnoe}, {Dale}, and
  {Brotherton}}]{Kobulnicky2014ApJS..213...34K}
\bibinfo{author}{H.~A. {Kobulnicky}}, \bibinfo{author}{D.~C. {Kiminki}},
  \bibinfo{author}{M.~J. {Lundquist}}, et~al.,
\newblock \bibinfo{title}{{Toward Complete Statistics of Massive Binary Stars:
  Penultimate Results from the Cygnus OB2 Radial Velocity Survey}},
\newblock \bibinfo{journal}{\apjs} \bibinfo{volume}{213} (\bibinfo{year}{2014})
  \bibinfo{pages}{34}. \DOIprefix\doi{10.1088/0067-0049/213/2/34}.
\bibitem[{{Sota} et~al.(2011){Sota}, {Ma{\'\i}z Apell{\'a}niz}, {Walborn},
  {Alfaro}, {Barb{\'a}}, {Morrell}, {Gamen}, and
  {Arias}}]{Sota2011ApJS..193...24S}
\bibinfo{author}{A.~{Sota}}, \bibinfo{author}{J.~{Ma{\'\i}z Apell{\'a}niz}},
  \bibinfo{author}{N.~R. {Walborn}}, et~al.,
\newblock \bibinfo{title}{{The Galactic O-Star Spectroscopic Survey. I.
  Classification System and Bright Northern Stars in the Blue-violet at R
  \raisebox{-0.5ex}\textasciitilde 2500}},
\newblock \bibinfo{journal}{\apjs} \bibinfo{volume}{193} (\bibinfo{year}{2011})
  \bibinfo{pages}{24}. \DOIprefix\doi{10.1088/0067-0049/193/2/24}.
\bibitem[{{Sota} et~al.(2014){Sota}, {Ma{\'\i}z Apell{\'a}niz}, {Morrell},
  {Barb{\'a}}, {Walborn}, {Gamen}, {Arias}, and
  {Alfaro}}]{Sota2014ApJS..211...10S}
\bibinfo{author}{A.~{Sota}}, \bibinfo{author}{J.~{Ma{\'\i}z Apell{\'a}niz}},
  \bibinfo{author}{N.~I. {Morrell}}, et~al.,
\newblock \bibinfo{title}{{The Galactic O-Star Spectroscopic Survey (GOSSS).
  II. Bright Southern Stars}},
\newblock \bibinfo{journal}{\apjs} \bibinfo{volume}{211} (\bibinfo{year}{2014})
  \bibinfo{pages}{10}. \DOIprefix\doi{10.1088/0067-0049/211/1/10}.
\bibitem[{{Ma{\'\i}z Apell{\'a}niz} et~al.(2016){Ma{\'\i}z Apell{\'a}niz},
  {Sota}, {Arias}, {Barb{\'a}}, {Walborn}, {Sim{\'o}n-D{\'\i}az}, {Negueruela},
  {Marco}, {Le{\~a}o}, {Herrero}, {Gamen}, and
  {Alfaro}}]{Maiz-Apellaniz2016ApJS..224....4M}
\bibinfo{author}{J.~{Ma{\'\i}z Apell{\'a}niz}}, \bibinfo{author}{A.~{Sota}},
  \bibinfo{author}{J.~I. {Arias}}, et~al.,
\newblock \bibinfo{title}{{The Galactic O-Star Spectroscopic Survey (GOSSS).
  III. 142 Additional O-type Systems.}},
\newblock \bibinfo{journal}{\apjs} \bibinfo{volume}{224} (\bibinfo{year}{2016})
  \bibinfo{pages}{4}. \DOIprefix\doi{10.3847/0067-0049/224/1/4}.
\bibitem[{{Chini} et~al.(2012){Chini}, {Hoffmeister}, {Nasseri}, {Stahl}, and
  {Zinnecker}}]{Chini2012MNRAS.424.1925C}
\bibinfo{author}{R.~{Chini}}, \bibinfo{author}{V.~H. {Hoffmeister}},
  \bibinfo{author}{A.~{Nasseri}}, et~al.,
\newblock \bibinfo{title}{{A spectroscopic survey on the multiplicity of
  high-mass stars}},
\newblock \bibinfo{journal}{\mnras} \bibinfo{volume}{424}
  (\bibinfo{year}{2012}) \bibinfo{pages}{1925--1929}.
  \DOIprefix\doi{10.1111/j.1365-2966.2012.21317.x}.
\bibitem[{{Sim{\'o}n-D{\'\i}az} et~al.(2015){Sim{\'o}n-D{\'\i}az},
  {Negueruela}, {Ma{\'\i}z Apell{\'a}niz}, {Castro}, {Herrero}, {Garcia},
  {P{\'e}rez-Prieto}, {Caon}, {Alacid}, {Camacho}, {Dorda}, {Godart},
  {Gonz{\'a}lez-Fern{\'a}ndez}, {Holgado}, and
  {R{\"u}bke}}]{Simon-Diaz2015hsa8.conf..576S}
\bibinfo{author}{S.~{Sim{\'o}n-D{\'\i}az}}, \bibinfo{author}{I.~{Negueruela}},
  \bibinfo{author}{J.~{Ma{\'\i}z Apell{\'a}niz}}, et~al.,
\newblock \bibinfo{title}{{The IACOB spectroscopic database: recent updates and
  first data release}},
\newblock in: \bibinfo{booktitle}{Highlights of Spanish Astrophysics VIII},
  \bibinfo{year}{2015},  pp. \bibinfo{pages}{576--581}.
\bibitem[{{Barb{\'a}} et~al.(2010){Barb{\'a}}, {Gamen}, {Arias}, {Morrell},
  {Ma{\'\i}z Apell{\'a}niz}, {Alfaro}, {Walborn}, and
  {Sota}}]{Barba2010RMxAC..38...30B}
\bibinfo{author}{R.~H. {Barb{\'a}}}, \bibinfo{author}{R.~{Gamen}},
  \bibinfo{author}{J.~I. {Arias}}, et~al.,
\newblock \bibinfo{title}{{Spectroscopic survey of galactic O and WN stars. OWN
  Survey: new binaries and trapezium-like systems}},
\newblock in: \bibinfo{booktitle}{Revista Mexicana de Astronomia y Astrofisica
  Conference Series}, volume~\bibinfo{volume}{38} of
  \textit{\bibinfo{series}{Revista Mexicana de Astronomia y Astrofisica
  Conference Series}}, \bibinfo{year}{2010},  pp. \bibinfo{pages}{30--32}.
\bibitem[{{Evans} et~al.(2011){Evans}, {Taylor}, {H{\'e}nault-Brunet}, {Sana},
  {de Koter}, {Sim{\'o}n-D{\'\i}az}, {Carraro}, {Bagnoli}, {Bastian},
  {Bestenlehner}, {Bonanos}, {Bressert}, {Brott}, {Campbell}, {Cantiello},
  {Clark}, {Costa}, {Crowther}, {de Mink}, {Doran}, {Dufton}, {Dunstall},
  {Friedrich}, {Garcia}, {Gieles}, {Gr{\"a}fener}, {Herrero}, {Howarth},
  {Izzard}, {Langer}, {Lennon}, {Ma{\'\i}z Apell{\'a}niz}, {Markova},
  {Najarro}, {Puls}, {Ramirez}, {Sab{\'\i}n-Sanjuli{\'a}n}, {Smartt}, {Stroud},
  {van Loon}, {Vink}, and {Walborn}}]{Evans2011A&A...530A.108E}
\bibinfo{author}{C.~J. {Evans}}, \bibinfo{author}{W.~D. {Taylor}},
  \bibinfo{author}{V.~{H{\'e}nault-Brunet}}, et~al.,
\newblock \bibinfo{title}{{The VLT-FLAMES Tarantula Survey. I. Introduction and
  observational overview}},
\newblock \bibinfo{journal}{\aap} \bibinfo{volume}{530} (\bibinfo{year}{2011})
  \bibinfo{pages}{A108}. \DOIprefix\doi{10.1051/0004-6361/201116782}.
\bibitem[{{Vink} et~al.(2017){Vink}, {Evans}, {Bestenlehner}, {McEvoy},
  {Ram{\'\i}rez-Agudelo}, {Sana}, {Schneider}, and {VFTS
  Collaboration}}]{Vink2017IAUS..329..279V}
\bibinfo{author}{J.~S. {Vink}}, \bibinfo{author}{C.~J. {Evans}},
  \bibinfo{author}{J.~{Bestenlehner}}, et~al.,
\newblock \bibinfo{title}{{The VLT-FLAMES Tarantula Survey}},
\newblock in: \bibinfo{editor}{J.~J. {Eldridge}}, \bibinfo{editor}{J.~C.
  {Bray}}, \bibinfo{editor}{L.~A.~S. {McClelland}}, \bibinfo{editor}{L.~{Xiao}}
  (Eds.), \bibinfo{booktitle}{The Lives and Death-Throes of Massive Stars},
  volume \bibinfo{volume}{329}, \bibinfo{year}{2017}, pp.
  \bibinfo{pages}{279--286}. \DOIprefix\doi{10.1017/S1743921317002496}.
\bibitem[{{Almeida} et~al.(2017){Almeida}, {Sana}, {Taylor}, {Barb{\'a}},
  {Bonanos}, {Crowther}, and et~al.}]{Almeida2017A&A...598A..84A}
\bibinfo{author}{L.~A. {Almeida}}, \bibinfo{author}{H.~{Sana}},
  \bibinfo{author}{W.~{Taylor}}, et~al.,
\newblock \bibinfo{title}{{The Tarantula Massive Binary Monitoring. I.
  Observational campaign and OB-type spectroscopic binaries}},
\newblock \bibinfo{journal}{\aap} \bibinfo{volume}{598} (\bibinfo{year}{2017})
  \bibinfo{pages}{A84}. \DOIprefix\doi{10.1051/0004-6361/201629844}.
\bibitem[{{Shenar} et~al.(2017){Shenar}, {Richardson}, {Sablowski}, {Hainich},
  {Sana}, and et~al.}]{Shenar2017A&A...598A..85S}
\bibinfo{author}{T.~{Shenar}}, \bibinfo{author}{N.~D. {Richardson}},
  \bibinfo{author}{D.~P. {Sablowski}}, et~al.,
\newblock \bibinfo{title}{{The Tarantula Massive Binary Monitoring. II. First
  SB2 orbital and spectroscopic analysis for the Wolf-Rayet binary R145}},
\newblock \bibinfo{journal}{\aap} \bibinfo{volume}{598} (\bibinfo{year}{2017})
  \bibinfo{pages}{A85}. \DOIprefix\doi{10.1051/0004-6361/201629621}.
\bibitem[{{Shenar} et~al.(2021){Shenar}, {Sana}, {Marchant}, {Pablo},
  {Richardson}, {Moffat}, {Van Reeth}, {Barb{\'a}}, {Bowman}, {Broos},
  {Crowther}, {Clark}, {de Koter}, {de Mink}, {Dsilva}, {Gr{\"a}fener},
  {Howarth}, {Langer}, {Mahy}, {Ma{\'\i}z Apell{\'a}niz}, {Pollock},
  {Schneider}, {Townsley}, and {Vink}}]{Shenar2021A&A...650A.147S}
\bibinfo{author}{T.~{Shenar}}, \bibinfo{author}{H.~{Sana}},
  \bibinfo{author}{P.~{Marchant}}, et~al.,
\newblock \bibinfo{title}{{The Tarantula Massive Binary Monitoring. V. R 144: a
  wind-eclipsing binary with a total mass {\ensuremath{\gtrsim}}140
  M$_{{\ensuremath{\odot}}}$}},
\newblock \bibinfo{journal}{\aap} \bibinfo{volume}{650} (\bibinfo{year}{2021})
  \bibinfo{pages}{A147}. \DOIprefix\doi{10.1051/0004-6361/202140693}.
\bibitem[{{Shenar} et~al.(2022){Shenar}, {Sana}, {Mahy}, {Ma{\'\i}z
  Apell{\'a}niz}, {Crowther}, {Gromadzki}, {Herrero}, {Langer}, {Marchant},
  {Schneider}, {Sen}, {Soszy{\'n}ski}, and
  {Toonen}}]{Shenar2022A&A...665A.148S}
\bibinfo{author}{T.~{Shenar}}, \bibinfo{author}{H.~{Sana}},
  \bibinfo{author}{L.~{Mahy}}, et~al.,
\newblock \bibinfo{title}{{The Tarantula Massive Binary Monitoring. VI.
  Characterisation of hidden companions in 51 single-lined O-type binaries: A
  flat mass-ratio distribution and black-hole binary candidates}},
\newblock \bibinfo{journal}{\aap} \bibinfo{volume}{665} (\bibinfo{year}{2022})
  \bibinfo{pages}{A148}. \DOIprefix\doi{10.1051/0004-6361/202244245}.
\bibitem[{{Mahy} et~al.(2020{\natexlab{a}}){Mahy}, {Sana}, {Abdul-Masih},
  {Almeida}, {Langer}, {Shenar}, {de Koter}, {de Mink}, {de Wit}, {Grin},
  {Evans}, {Moffat}, {Schneider}, {Barb{\'a}}, {Clark}, {Crowther},
  {Gr{\"a}fener}, {Lennon}, {Tramper}, and {Vink}}]{Mahy2020A&A...634A.118M}
\bibinfo{author}{L.~{Mahy}}, \bibinfo{author}{H.~{Sana}},
  \bibinfo{author}{M.~{Abdul-Masih}}, et~al.,
\newblock \bibinfo{title}{{The Tarantula Massive Binary Monitoring. III.
  Atmosphere analysis of double-lined spectroscopic systems}},
\newblock \bibinfo{journal}{\aap} \bibinfo{volume}{634}
  (\bibinfo{year}{2020}{\natexlab{a}}) \bibinfo{pages}{A118}.
  \DOIprefix\doi{10.1051/0004-6361/201936151}.
\bibitem[{{Mahy} et~al.(2020{\natexlab{b}}){Mahy}, {Almeida}, {Sana}, {Clark},
  {de Koter}, {de Mink}, {Evans}, {Grin}, {Langer}, {Moffat}, {Schneider},
  {Shenar}, and {Tramper}}]{Mahy2020A&A...634A.119M}
\bibinfo{author}{L.~{Mahy}}, \bibinfo{author}{L.~A. {Almeida}},
  \bibinfo{author}{H.~{Sana}}, et~al.,
\newblock \bibinfo{title}{{The Tarantula Massive Binary Monitoring. IV.
  Double-lined photometric binaries}},
\newblock \bibinfo{journal}{\aap} \bibinfo{volume}{634}
  (\bibinfo{year}{2020}{\natexlab{b}}) \bibinfo{pages}{A119}.
  \DOIprefix\doi{10.1051/0004-6361/201936152}.
\bibitem[{{Maoz} et~al.(2012){Maoz}, {Badenes}, and
  {Bickerton}}]{Maoz2012ApJ...751..143M}
\bibinfo{author}{D.~{Maoz}}, \bibinfo{author}{C.~{Badenes}},
  \bibinfo{author}{S.~J. {Bickerton}},
\newblock \bibinfo{title}{{Characterizing the Galactic White Dwarf Binary
  Population with Sparsely Sampled Radial Velocity Data}},
\newblock \bibinfo{journal}{\apj} \bibinfo{volume}{751} (\bibinfo{year}{2012})
  \bibinfo{pages}{143}. \DOIprefix\doi{10.1088/0004-637X/751/2/143}.
\bibitem[{{Badenes} and {Maoz}(2012)}]{Badenes2012ApJ...749L..11B}
\bibinfo{author}{C.~{Badenes}}, \bibinfo{author}{D.~{Maoz}},
\newblock \bibinfo{title}{{The Merger Rate of Binary White Dwarfs in the
  Galactic Disk}},
\newblock \bibinfo{journal}{\apjl} \bibinfo{volume}{749} (\bibinfo{year}{2012})
  \bibinfo{pages}{L11}. \DOIprefix\doi{10.1088/2041-8205/749/1/L11}.
\bibitem[{{Guillochon} et~al.(2010){Guillochon}, {Dan}, {Ramirez-Ruiz}, and
  {Rosswog}}]{Guillochon2010ApJ...709L..64G}
\bibinfo{author}{J.~{Guillochon}}, \bibinfo{author}{M.~{Dan}},
  \bibinfo{author}{E.~{Ramirez-Ruiz}}, \bibinfo{author}{S.~{Rosswog}},
\newblock \bibinfo{title}{{Surface Detonations in Double Degenerate Binary
  Systems Triggered by Accretion Stream Instabilities}},
\newblock \bibinfo{journal}{\apjl} \bibinfo{volume}{709} (\bibinfo{year}{2010})
  \bibinfo{pages}{L64--L69}. \DOIprefix\doi{10.1088/2041-8205/709/1/L64}.
\bibitem[{{Pakmor} et~al.(2011){Pakmor}, {Hachinger}, {R{\"o}pke}, and
  {Hillebrandt}}]{Pakmor2011A&A...528A.117P}
\bibinfo{author}{R.~{Pakmor}}, \bibinfo{author}{S.~{Hachinger}},
  \bibinfo{author}{F.~K. {R{\"o}pke}}, \bibinfo{author}{W.~{Hillebrandt}},
\newblock \bibinfo{title}{{Violent mergers of nearly equal-mass white dwarf as
  progenitors of subluminous Type Ia supernovae}},
\newblock \bibinfo{journal}{\aap} \bibinfo{volume}{528} (\bibinfo{year}{2011})
  \bibinfo{pages}{A117}. \DOIprefix\doi{10.1051/0004-6361/201015653}.
\bibitem[{{Badenes} et~al.(2018){Badenes}, {Mazzola}, {Thompson}, {Covey},
  {Freeman}, {Walker}, {Moe}, {Troup}, {Nidever}, {Allende Prieto}, {Andrews},
  {Barb{\'a}}, {Beers}, {Bovy}, {Carlberg}, {De Lee}, {Johnson}, {Lewis},
  {Majewski}, {Pinsonneault}, {Sobeck}, {Stassun}, {Stringfellow}, and
  {Zasowski}}]{Badenes2018ApJ...854..147B}
\bibinfo{author}{C.~{Badenes}}, \bibinfo{author}{C.~{Mazzola}},
  \bibinfo{author}{T.~A. {Thompson}}, et~al.,
\newblock \bibinfo{title}{{Stellar Multiplicity Meets Stellar Evolution and
  Metallicity: The APOGEE View}},
\newblock \bibinfo{journal}{\apj} \bibinfo{volume}{854} (\bibinfo{year}{2018})
  \bibinfo{pages}{147}. \DOIprefix\doi{10.3847/1538-4357/aaa765}.
\bibitem[{{Gao} et~al.(2014){Gao}, {Liu}, {Zhang}, {Justham}, {Deng}, and
  {Yang}}]{Gao2014ApJ...788L..37G}
\bibinfo{author}{S.~{Gao}}, \bibinfo{author}{C.~{Liu}},
  \bibinfo{author}{X.~{Zhang}}, et~al.,
\newblock \bibinfo{title}{{The Binarity of Milky Way F,G,K Stars as a Function
  of Effective Temperature and Metallicity}},
\newblock \bibinfo{journal}{\apjl} \bibinfo{volume}{788} (\bibinfo{year}{2014})
  \bibinfo{pages}{L37}. \DOIprefix\doi{10.1088/2041-8205/788/2/L37}.
\bibitem[{{Chabrier}(2003)}]{Chabrier2003}
\bibinfo{author}{G.~{Chabrier}},
\newblock \bibinfo{title}{{Galactic Stellar and Substellar Initial Mass
  Function}},
\newblock \bibinfo{journal}{\pasp} \bibinfo{volume}{115} (\bibinfo{year}{2003})
  \bibinfo{pages}{763--795}. \DOIprefix\doi{10.1086/376392}.
\bibitem[{{Price-Whelan} et~al.(2020){Price-Whelan}, {Hogg}, {Rix}, {Beaton},
  {Lewis}, and et~al.}]{Price-Whelan2020ApJ...895....2P}
\bibinfo{author}{A.~M. {Price-Whelan}}, \bibinfo{author}{D.~W. {Hogg}},
  \bibinfo{author}{H.-W. {Rix}}, et~al.,
\newblock \bibinfo{title}{{Close Binary Companions to APOGEE DR16 Stars: 20,000
  Binary-star Systems Across the Color-Magnitude Diagram}},
\newblock \bibinfo{journal}{\apj} \bibinfo{volume}{895} (\bibinfo{year}{2020})
  \bibinfo{pages}{2}. \DOIprefix\doi{10.3847/1538-4357/ab8acc}.
\bibitem[{{Li} et~al.(2022){Li}, {Li}, {Liu}, {Li}, {Guo}, {Wang}, {Chen},
  {Xing}, {Hou}, and {Han}}]{Lijiangdan2022}
\bibinfo{author}{J.~{Li}}, \bibinfo{author}{J.~{Li}},
  \bibinfo{author}{C.~{Liu}}, et~al.,
\newblock \bibinfo{title}{{Mass-ratio Distribution of Binaries from the
  LAMOST-MRS Survey}},
\newblock \bibinfo{journal}{\apj} \bibinfo{volume}{933} (\bibinfo{year}{2022})
  \bibinfo{pages}{119}. \DOIprefix\doi{10.3847/1538-4357/ac731d}.
\bibitem[{{Yuan} et~al.(2015){Yuan}, {Liu}, {Xiang}, {Huang}, {Chen}, {Wu},
  {Hou}, and {Zhang}}]{Yuan2015ApJ...799..135Y}
\bibinfo{author}{H.~{Yuan}}, \bibinfo{author}{X.~{Liu}},
  \bibinfo{author}{M.~{Xiang}}, et~al.,
\newblock \bibinfo{title}{{Stellar Loci II. A Model-free Estimate of the Binary
  Fraction for Field FGK Stars}},
\newblock \bibinfo{journal}{\apj} \bibinfo{volume}{799} (\bibinfo{year}{2015})
  \bibinfo{pages}{135}. \DOIprefix\doi{10.1088/0004-637X/799/2/135}.
\bibitem[{{Kuiper}(1941)}]{Kuiper1941ApJ....93..133K}
\bibinfo{author}{G.~P. {Kuiper}},
\newblock \bibinfo{title}{{On the Interpretation of {\ensuremath{\beta}} Lyrae
  and Other Close Binaries.}},
\newblock \bibinfo{journal}{\apj} \bibinfo{volume}{93} (\bibinfo{year}{1941})
  \bibinfo{pages}{133}. \DOIprefix\doi{10.1086/144252}.
\bibitem[{{Kopal}(1954)}]{Kopal1954MNRAS.114..101K}
\bibinfo{author}{Z.~{Kopal}},
\newblock \bibinfo{title}{{Photometric effects of reflection in close binary
  systems}},
\newblock \bibinfo{journal}{\mnras} \bibinfo{volume}{114}
  (\bibinfo{year}{1954}) \bibinfo{pages}{101}.
  \DOIprefix\doi{10.1093/mnras/114.1.101}.
\bibitem[{{Rasio} and {Livio}(1996)}]{Rasio1996ApJ...471..366R}
\bibinfo{author}{F.~A. {Rasio}}, \bibinfo{author}{M.~{Livio}},
\newblock \bibinfo{title}{{On the Formation and Evolution of Common Envelope
  Systems}},
\newblock \bibinfo{journal}{\apj} \bibinfo{volume}{471} (\bibinfo{year}{1996})
  \bibinfo{pages}{366}. \DOIprefix\doi{10.1086/177975}.
\bibitem[{{Taam} and {Ricker}(2006)}]{Taam2006astro.ph.11043T}
\bibinfo{author}{R.~E. {Taam}}, \bibinfo{author}{P.~M. {Ricker}},
\newblock \bibinfo{title}{{Common Envelope Evolution}},
\newblock \bibinfo{journal}{arXiv e-prints}   (\bibinfo{year}{2006})
  \bibinfo{pages}{astro--ph/0611043}.
\bibitem[{{Ricker} and {Taam}(2008)}]{Ricker2008ApJ...672L..41R}
\bibinfo{author}{P.~M. {Ricker}}, \bibinfo{author}{R.~E. {Taam}},
\newblock \bibinfo{title}{{The Interaction of Stellar Objects within a Common
  Envelope}},
\newblock \bibinfo{journal}{\apjl} \bibinfo{volume}{672} (\bibinfo{year}{2008})
  \bibinfo{pages}{L41}. \DOIprefix\doi{10.1086/526343}.
\bibitem[{{Paczynski}(1976)}]{Paczynski1976IAUS...73...75P}
\bibinfo{author}{B.~{Paczynski}},
\newblock \bibinfo{title}{{Common Envelope Binaries}},
\newblock in: \bibinfo{editor}{P.~{Eggleton}}, \bibinfo{editor}{S.~{Mitton}},
  \bibinfo{editor}{J.~{Whelan}} (Eds.), \bibinfo{booktitle}{Structure and
  Evolution of Close Binary Systems}, volume~\bibinfo{volume}{73},
  \bibinfo{year}{1976},  p.~\bibinfo{pages}{75}.
\bibitem[{{Hjellming} and {Webbink}(1987)}]{Hjellming1987ApJ...318..794H}
\bibinfo{author}{M.~S. {Hjellming}}, \bibinfo{author}{R.~F. {Webbink}},
\newblock \bibinfo{title}{{Thresholds for Rapid Mass Transfer in Binary System.
  I. Polytropic Models}},
\newblock \bibinfo{journal}{\apj} \bibinfo{volume}{318} (\bibinfo{year}{1987})
  \bibinfo{pages}{794}. \DOIprefix\doi{10.1086/165412}.
\bibitem[{{Webbink}(1988)}]{Webbink1988covp.conf..403W}
\bibinfo{author}{R.~F. {Webbink}},
\newblock \bibinfo{title}{{Late stages of close binary systems - clues to
  common envelope evolution.}},
\newblock in: \bibinfo{booktitle}{Critical Observations Versus Physical Models
  for Close Binary Systems}, \bibinfo{year}{1988},  pp.
  \bibinfo{pages}{403--446}.
\bibitem[{{Podsiadlowski}(2001)}]{Podsiadlowski2001ASPC..229..239P}
\bibinfo{author}{P.~{Podsiadlowski}},
\newblock \bibinfo{title}{{Common-Envelope Evolution and Stellar Mergers}},
\newblock in: \bibinfo{editor}{P.~{Podsiadlowski}},
  \bibinfo{editor}{S.~{Rappaport}}, \bibinfo{editor}{A.~R. {King}}, et~al.
  (Eds.), \bibinfo{booktitle}{Evolution of Binary and Multiple Star Systems},
  volume \bibinfo{volume}{229} of \textit{\bibinfo{series}{Astronomical Society
  of the Pacific Conference Series}}, \bibinfo{year}{2001},  p.
  \bibinfo{pages}{239}.
\bibitem[{{Tauris} and {Savonije}(1999)}]{Tauris1999A&A...350..928T}
\bibinfo{author}{T.~M. {Tauris}}, \bibinfo{author}{G.~J. {Savonije}},
\newblock \bibinfo{title}{{Formation of millisecond pulsars. I. Evolution of
  low-mass X-ray binaries with P\_orb> 2 days}},
\newblock \bibinfo{journal}{\aap}  \bibinfo{volume}{350} (\bibinfo{year}{1999})
  \bibinfo{pages}{928--944}.
\bibitem[{{Podsiadlowski} et~al.(2002){Podsiadlowski}, {Rappaport}, and
  {Pfahl}}]{Podsiadlowski2002ApJ...565.1107P}
\bibinfo{author}{P.~{Podsiadlowski}}, \bibinfo{author}{S.~{Rappaport}},
  \bibinfo{author}{E.~D. {Pfahl}},
\newblock \bibinfo{title}{{Evolutionary Sequences for Low- and
  Intermediate-Mass X-Ray Binaries}},
\newblock \bibinfo{journal}{\apj} \bibinfo{volume}{565} (\bibinfo{year}{2002})
  \bibinfo{pages}{1107--1133}. \DOIprefix\doi{10.1086/324686}.
\bibitem[{{Podsiadlowski} et~al.(1994){Podsiadlowski}, {Cannon}, and
  {Rees}}]{Podsiadlowski1994AIPC..308..403P}
\bibinfo{author}{P.~{Podsiadlowski}}, \bibinfo{author}{R.~C. {Cannon}},
  \bibinfo{author}{M.~J. {Rees}},
\newblock \bibinfo{title}{{The Fate of Thorne-Zytkow Objects}},
\newblock in: \bibinfo{editor}{S.~{Holt}}, \bibinfo{editor}{C.~S. {Day}}
  (Eds.), \bibinfo{booktitle}{The Evolution of X-ray Binariese}, volume
  \bibinfo{volume}{308} of \textit{\bibinfo{series}{American Institute of
  Physics Conference Series}}, \bibinfo{year}{1994}, p. \bibinfo{pages}{403}.
  \DOIprefix\doi{10.1063/1.45980}.
\bibitem[{{Podsiadlowski} et~al.(1992){Podsiadlowski}, {Joss}, and
  {Hsu}}]{Podsiadlowski1992ApJ...391..246P}
\bibinfo{author}{P.~{Podsiadlowski}}, \bibinfo{author}{P.~C. {Joss}},
  \bibinfo{author}{J.~J.~L. {Hsu}},
\newblock \bibinfo{title}{{Presupernova Evolution in Massive Interacting
  Binaries}},
\newblock \bibinfo{journal}{\apj} \bibinfo{volume}{391} (\bibinfo{year}{1992})
  \bibinfo{pages}{246}. \DOIprefix\doi{10.1086/171341}.
\bibitem[{{Eggleton} et~al.(1989){Eggleton}, {Fitchett}, and
  {Tout}}]{Eggleton1989ApJ...347..998E}
\bibinfo{author}{P.~P. {Eggleton}}, \bibinfo{author}{M.~J. {Fitchett}},
  \bibinfo{author}{C.~A. {Tout}},
\newblock \bibinfo{title}{{The Distribution of Visual Binaries with Two Bright
  Components}},
\newblock \bibinfo{journal}{\apj} \bibinfo{volume}{347} (\bibinfo{year}{1989})
  \bibinfo{pages}{998}. \DOIprefix\doi{10.1086/168190}.
\bibitem[{{Soberman} et~al.(1997){Soberman}, {Phinney}, and {van den
  Heuvel}}]{Soberman1997A&A...327..620S}
\bibinfo{author}{G.~E. {Soberman}}, \bibinfo{author}{E.~S. {Phinney}},
  \bibinfo{author}{E.~P.~J. {van den Heuvel}},
\newblock \bibinfo{title}{{Stability criteria for mass transfer in binary
  stellar evolution.}},
\newblock \bibinfo{journal}{\aap}  \bibinfo{volume}{327} (\bibinfo{year}{1997})
  \bibinfo{pages}{620--635}.
\bibitem[{{Han} et~al.(2001){Han}, {Eggleton}, {Podsiadlowski}, {Tout}, and
  {Webbink}}]{Han2001ASPC..229..205H}
\bibinfo{author}{Z.~{Han}}, \bibinfo{author}{P.~P. {Eggleton}},
  \bibinfo{author}{P.~{Podsiadlowski}}, et~al.,
\newblock \bibinfo{title}{{A Self-Consistent Binary Population Synthesis
  Model}},
\newblock in: \bibinfo{editor}{P.~{Podsiadlowski}},
  \bibinfo{editor}{S.~{Rappaport}}, \bibinfo{editor}{A.~R. {King}}, et~al.
  (Eds.), \bibinfo{booktitle}{Evolution of Binary and Multiple Star Systems},
  volume \bibinfo{volume}{229} of \textit{\bibinfo{series}{Astronomical Society
  of the Pacific Conference Series}}, \bibinfo{year}{2001},  p.
  \bibinfo{pages}{205}.
\bibitem[{{Nelemans} et~al.(2000){Nelemans}, {Verbunt}, {Yungelson}, and
  {Portegies Zwart}}]{Nelemans2000A&A...360.1011N}
\bibinfo{author}{G.~{Nelemans}}, \bibinfo{author}{F.~{Verbunt}},
  \bibinfo{author}{L.~R. {Yungelson}}, \bibinfo{author}{S.~F. {Portegies
  Zwart}},
\newblock \bibinfo{title}{{Reconstructing the evolution of double helium white
  dwarfs: envelope loss without spiral-in}},
\newblock \bibinfo{journal}{\aap}  \bibinfo{volume}{360} (\bibinfo{year}{2000})
  \bibinfo{pages}{1011--1018}.
\bibitem[{{Woods} et~al.(2012){Woods}, {Ivanova}, {van der Sluys}, and
  {Chaichenets}}]{Woods2012ApJ...744...12W}
\bibinfo{author}{T.~E. {Woods}}, \bibinfo{author}{N.~{Ivanova}},
  \bibinfo{author}{M.~V. {van der Sluys}}, \bibinfo{author}{S.~{Chaichenets}},
\newblock \bibinfo{title}{{On the Formation of Double White Dwarfs through
  Stable Mass Transfer and a Common Envelope}},
\newblock \bibinfo{journal}{\apj} \bibinfo{volume}{744} (\bibinfo{year}{2012})
  \bibinfo{pages}{12}. \DOIprefix\doi{10.1088/0004-637X/744/1/12}.
\bibitem[{{Chen} and {Han}(2008)}]{Chen2008MNRAS.387.1416C}
\bibinfo{author}{X.~{Chen}}, \bibinfo{author}{Z.~{Han}},
\newblock \bibinfo{title}{{Mass transfer from a giant star to a main-sequence
  companion and its contribution to long-orbital-period blue stragglers}},
\newblock \bibinfo{journal}{\mnras} \bibinfo{volume}{387}
  (\bibinfo{year}{2008}) \bibinfo{pages}{1416--1430}.
  \DOIprefix\doi{10.1111/j.1365-2966.2008.13334.x}.
\bibitem[{{Hjellming}(1989{\natexlab{a}})}]{Hjellming1989SSRv...50..155H}
\bibinfo{author}{M.~S. {Hjellming}},
\newblock \bibinfo{title}{{ALGOLS as Limits on Binary Evolution Scenarios}},
\newblock \bibinfo{journal}{\ssr} \bibinfo{volume}{50}
  (\bibinfo{year}{1989}{\natexlab{a}}) \bibinfo{pages}{155--164}.
  \DOIprefix\doi{10.1007/BF00215927}.
\bibitem[{{Hjellming}(1989{\natexlab{b}})}]{Hjellming1989PhDT.........7H}
\bibinfo{author}{M.~S. {Hjellming}}, \bibinfo{title}{{Rapid Mass Transfer in
  Binary Systems.}}, Ph.D. thesis, University of Illinois, Urbana-Champaign,
  \bibinfo{year}{1989}{\natexlab{b}}.
\bibitem[{{Han} et~al.(2002){Han}, {Podsiadlowski}, {Maxted}, {Marsh}, and
  {Ivanova}}]{Han2002MNRAS.336..449H}
\bibinfo{author}{Z.~{Han}}, \bibinfo{author}{P.~{Podsiadlowski}},
  \bibinfo{author}{P.~F.~L. {Maxted}}, et~al.,
\newblock \bibinfo{title}{{The origin of subdwarf B stars - I. The formation
  channels}},
\newblock \bibinfo{journal}{\mnras} \bibinfo{volume}{336}
  (\bibinfo{year}{2002}) \bibinfo{pages}{449--466}.
  \DOIprefix\doi{10.1046/j.1365-8711.2002.05752.x}.
\bibitem[{{Rappaport} et~al.(1995){Rappaport}, {Podsiadlowski}, {Joss}, {Di
  Stefano}, and {Han}}]{Rappaport1995MNRAS.273..731R}
\bibinfo{author}{S.~{Rappaport}}, \bibinfo{author}{P.~{Podsiadlowski}},
  \bibinfo{author}{P.~C. {Joss}}, et~al.,
\newblock \bibinfo{title}{{The relation between white dwarf mass and orbital
  period in wide binary radio pulsars}},
\newblock \bibinfo{journal}{\mnras} \bibinfo{volume}{273}
  (\bibinfo{year}{1995}) \bibinfo{pages}{731--741}.
  \DOIprefix\doi{10.1093/mnras/273.3.731}.
\bibitem[{{Chen} and {Han}(2009)}]{Chen2009MNRAS.395.1822C}
\bibinfo{author}{X.~{Chen}}, \bibinfo{author}{Z.~{Han}},
\newblock \bibinfo{title}{{Primordial binary evolution and blue stragglers}},
\newblock \bibinfo{journal}{\mnras} \bibinfo{volume}{395}
  (\bibinfo{year}{2009}) \bibinfo{pages}{1822--1836}.
  \DOIprefix\doi{10.1111/j.1365-2966.2009.14669.x}.
\bibitem[{{Chen} et~al.(2013){Chen}, {Han}, {Deca}, and
  {Podsiadlowski}}]{Chen2013MNRAS.434..186C}
\bibinfo{author}{X.~{Chen}}, \bibinfo{author}{Z.~{Han}},
  \bibinfo{author}{J.~{Deca}}, \bibinfo{author}{P.~{Podsiadlowski}},
\newblock \bibinfo{title}{{The orbital periods of subdwarf B binaries produced
  by the first stable Roche Lobe overflow channel}},
\newblock \bibinfo{journal}{\mnras} \bibinfo{volume}{434}
  (\bibinfo{year}{2013}) \bibinfo{pages}{186--193}.
  \DOIprefix\doi{10.1093/mnras/stt992}.
\bibitem[{{Vos} et~al.(2019){Vos}, {Vu{\v{c}}kovi{\'c}}, {Chen}, {Han},
  {Boudreaux}, {Barlow}, {{\O}stensen}, and
  {N{\'e}meth}}]{Vos2019MNRAS.482.4592V}
\bibinfo{author}{J.~{Vos}}, \bibinfo{author}{M.~{Vu{\v{c}}kovi{\'c}}},
  \bibinfo{author}{X.~{Chen}}, et~al.,
\newblock \bibinfo{title}{{The orbital period-mass ratio relation of wide
  sdB+MS binaries and its application to the stability of RLOF}},
\newblock \bibinfo{journal}{\mnras} \bibinfo{volume}{482}
  (\bibinfo{year}{2019}) \bibinfo{pages}{4592--4605}.
  \DOIprefix\doi{10.1093/mnras/sty3017}.
\bibitem[{{Pavlovskii} and {Ivanova}(2015)}]{Pavlovskii2015MNRAS.449.4415P}
\bibinfo{author}{K.~{Pavlovskii}}, \bibinfo{author}{N.~{Ivanova}},
\newblock \bibinfo{title}{{Mass transfer from giant donors}},
\newblock \bibinfo{journal}{\mnras} \bibinfo{volume}{449}
  (\bibinfo{year}{2015}) \bibinfo{pages}{4415--4427}.
  \DOIprefix\doi{10.1093/mnras/stv619}.
\bibitem[{{Pavlovskii} et~al.(2017){Pavlovskii}, {Ivanova}, {Belczynski}, and
  {Van}}]{Pavlovskii2017MNRAS.465.2092P}
\bibinfo{author}{K.~{Pavlovskii}}, \bibinfo{author}{N.~{Ivanova}},
  \bibinfo{author}{K.~{Belczynski}}, \bibinfo{author}{K.~X. {Van}},
\newblock \bibinfo{title}{{Stability of mass transfer from massive giants:
  double black hole binary formation and ultraluminous X-ray sources}},
\newblock \bibinfo{journal}{\mnras} \bibinfo{volume}{465}
  (\bibinfo{year}{2017}) \bibinfo{pages}{2092--2100}.
  \DOIprefix\doi{10.1093/mnras/stw2786}.
\bibitem[{{Ge} et~al.(2010){Ge}, {Hjellming}, {Webbink}, {Chen}, and
  {Han}}]{Ge2010ApJ...717..724G}
\bibinfo{author}{H.~{Ge}}, \bibinfo{author}{M.~S. {Hjellming}},
  \bibinfo{author}{R.~F. {Webbink}}, et~al.,
\newblock \bibinfo{title}{{Adiabatic Mass Loss in Binary Stars. I.
  Computational Method}},
\newblock \bibinfo{journal}{\apj} \bibinfo{volume}{717} (\bibinfo{year}{2010})
  \bibinfo{pages}{724--738}. \DOIprefix\doi{10.1088/0004-637X/717/2/724}.
\bibitem[{{Ge} et~al.(2020){Ge}, {Webbink}, {Chen}, and
  {Han}}]{Ge2020ApJ...899..132G}
\bibinfo{author}{H.~{Ge}}, \bibinfo{author}{R.~F. {Webbink}},
  \bibinfo{author}{X.~{Chen}}, \bibinfo{author}{Z.~{Han}},
\newblock \bibinfo{title}{{Adiabatic Mass Loss in Binary Stars. III. From the
  Base of the Red Giant Branch to the Tip of the Asymptotic Giant Branch}},
\newblock \bibinfo{journal}{\apj} \bibinfo{volume}{899} (\bibinfo{year}{2020})
  \bibinfo{pages}{132}. \DOIprefix\doi{10.3847/1538-4357/aba7b7}.
\bibitem[{{Ostriker} and {Bodenheimer}(1973)}]{Ostriker1973ApJ...180..171O}
\bibinfo{author}{J.~P. {Ostriker}}, \bibinfo{author}{P.~{Bodenheimer}},
\newblock \bibinfo{title}{{On the Oscillations and Stability of Rapidly
  Rotating Stellar Models. 111. Zero-Viscosity Polytropic Sequences}},
\newblock \bibinfo{journal}{\apj} \bibinfo{volume}{180} (\bibinfo{year}{1973})
  \bibinfo{pages}{171--180}. \DOIprefix\doi{10.1086/151952}.
\bibitem[{{Webbink}(1975)}]{Webbink1975MNRAS.171..555W}
\bibinfo{author}{R.~F. {Webbink}},
\newblock \bibinfo{title}{{Evolution of helium white dwarfs in close
  binaries.}},
\newblock \bibinfo{journal}{\mnras} \bibinfo{volume}{171}
  (\bibinfo{year}{1975}) \bibinfo{pages}{555--568}.
  \DOIprefix\doi{10.1093/mnras/171.3.555}.
\bibitem[{{van den Heuvel}(1976)}]{van-den-Heuvel1976IAUS...73...35V}
\bibinfo{author}{E.~P.~J. {van den Heuvel}},
\newblock \bibinfo{title}{{Late Stages of Close Binary Systems}},
\newblock in: \bibinfo{editor}{P.~{Eggleton}}, \bibinfo{editor}{S.~{Mitton}},
  \bibinfo{editor}{J.~{Whelan}} (Eds.), \bibinfo{booktitle}{Structure and
  Evolution of Close Binary Systems}, volume~\bibinfo{volume}{73},
  \bibinfo{year}{1976},  p.~\bibinfo{pages}{35}.
\bibitem[{{Webbink}(1984)}]{Webbink1984ApJ...277..355W}
\bibinfo{author}{R.~F. {Webbink}},
\newblock \bibinfo{title}{{Double white dwarfs as progenitors of R Coronae
  Borealis stars and type I supernovae.}},
\newblock \bibinfo{journal}{\apj} \bibinfo{volume}{277} (\bibinfo{year}{1984})
  \bibinfo{pages}{355--360}. \DOIprefix\doi{10.1086/161701}.
\bibitem[{{Claeys} et~al.(2014){Claeys}, {Pols}, {Izzard}, {Vink}, and
  {Verbunt}}]{Claeys2014A&A...563A..83C}
\bibinfo{author}{J.~S.~W. {Claeys}}, \bibinfo{author}{O.~R. {Pols}},
  \bibinfo{author}{R.~G. {Izzard}}, et~al.,
\newblock \bibinfo{title}{{Theoretical uncertainties of the Type Ia supernova
  rate}},
\newblock \bibinfo{journal}{\aap} \bibinfo{volume}{563} (\bibinfo{year}{2014})
  \bibinfo{pages}{A83}. \DOIprefix\doi{10.1051/0004-6361/201322714}.
\bibitem[{{Paczy{\'n}ski} and
  {Zi{\'o}{\l}kowski}(1967)}]{Paczynski1967AcA....17....7P}
\bibinfo{author}{B.~{Paczy{\'n}ski}}, \bibinfo{author}{J.~{Zi{\'o}{\l}kowski}},
\newblock \bibinfo{title}{{Evolution of Close Binaries. III.}},
\newblock \bibinfo{journal}{\actaa}  \bibinfo{volume}{17}
  (\bibinfo{year}{1967}) \bibinfo{pages}{7}.
\bibitem[{{Hirai} and {Mandel}(2022)}]{hirai2022}
\bibinfo{author}{R.~{Hirai}}, \bibinfo{author}{I.~{Mandel}},
\newblock \bibinfo{title}{{A Two-stage Formalism for Common-envelope Phases of
  Massive Stars}},
\newblock \bibinfo{journal}{\apjl} \bibinfo{volume}{937} (\bibinfo{year}{2022})
  \bibinfo{pages}{L42}. \DOIprefix\doi{10.3847/2041-8213/ac9519}.
\bibitem[{{Di Stefano} et~al.(2022){Di Stefano}, {Kruckow}, {Gao},
  {Neunteufel}, and {Kobayashi}}]{stefano2022}
\bibinfo{author}{R.~{Di Stefano}}, \bibinfo{author}{M.~U. {Kruckow}},
  \bibinfo{author}{Y.~{Gao}}, et~al.,
\newblock \bibinfo{title}{{SCATTER: A New Common Envelope Formalism}},
\newblock \bibinfo{journal}{arXiv e-prints}   (\bibinfo{year}{2022})
  \bibinfo{pages}{arXiv:2212.06770}.
\bibitem[{{Han} et~al.(1994){Han}, {Podsiadlowski}, and
  {Eggleton}}]{Han1994MNRAS.270..121H}
\bibinfo{author}{Z.~{Han}}, \bibinfo{author}{P.~{Podsiadlowski}},
  \bibinfo{author}{P.~P. {Eggleton}},
\newblock \bibinfo{title}{{A possible criterion for envelope ejection in
  asymptotic giant branch or first giant branch stars.}},
\newblock \bibinfo{journal}{\mnras} \bibinfo{volume}{270}
  (\bibinfo{year}{1994}) \bibinfo{pages}{121--130}.
  \DOIprefix\doi{10.1093/mnras/270.1.121}.
\bibitem[{{Han} et~al.(2003){Han}, {Podsiadlowski}, {Maxted}, and
  {Marsh}}]{Han2003MNRAS.341..669H}
\bibinfo{author}{Z.~{Han}}, \bibinfo{author}{P.~{Podsiadlowski}},
  \bibinfo{author}{P.~F.~L. {Maxted}}, \bibinfo{author}{T.~R. {Marsh}},
\newblock \bibinfo{title}{{The origin of subdwarf B stars - II}},
\newblock \bibinfo{journal}{\mnras} \bibinfo{volume}{341}
  (\bibinfo{year}{2003}) \bibinfo{pages}{669--691}.
  \DOIprefix\doi{10.1046/j.1365-8711.2003.06451.x}.
\bibitem[{{Xu} and {Li}(2010)}]{Xu2010ApJ...716..114X}
\bibinfo{author}{X.-J. {Xu}}, \bibinfo{author}{X.-D. {Li}},
\newblock \bibinfo{title}{{On the Binding Energy Parameter
  {\ensuremath{\lambda}} of Common Envelope Evolution}},
\newblock \bibinfo{journal}{\apj} \bibinfo{volume}{716} (\bibinfo{year}{2010})
  \bibinfo{pages}{114--121}. \DOIprefix\doi{10.1088/0004-637X/716/1/114}.
\bibitem[{{Justham} et~al.(2006){Justham}, {Rappaport}, and
  {Podsiadlowski}}]{Justham2006MNRAS.366.1415J}
\bibinfo{author}{S.~{Justham}}, \bibinfo{author}{S.~{Rappaport}},
  \bibinfo{author}{P.~{Podsiadlowski}},
\newblock \bibinfo{title}{{Magnetic braking of Ap/Bp stars: application to
  compact black-hole X-ray binaries}},
\newblock \bibinfo{journal}{\mnras} \bibinfo{volume}{366}
  (\bibinfo{year}{2006}) \bibinfo{pages}{1415--1423}.
  \DOIprefix\doi{10.1111/j.1365-2966.2005.09907.x}.
\bibitem[{{Ivanova} and {Chaichenets}(2011)}]{Ivanova2011ApJ...731L..36I}
\bibinfo{author}{N.~{Ivanova}}, \bibinfo{author}{S.~{Chaichenets}},
\newblock \bibinfo{title}{{Common Envelope: Enthalpy Consideration}},
\newblock \bibinfo{journal}{\apjl} \bibinfo{volume}{731} (\bibinfo{year}{2011})
  \bibinfo{pages}{L36}. \DOIprefix\doi{10.1088/2041-8205/731/2/L36}.
\bibitem[{{Han} and {Podsiadlowski}(2004)}]{Han2004MNRAS.350.1301H}
\bibinfo{author}{Z.~{Han}}, \bibinfo{author}{P.~{Podsiadlowski}},
\newblock \bibinfo{title}{{The single-degenerate channel for the progenitors of
  Type Ia supernovae}},
\newblock \bibinfo{journal}{\mnras} \bibinfo{volume}{350}
  (\bibinfo{year}{2004}) \bibinfo{pages}{1301--1309}.
  \DOIprefix\doi{10.1111/j.1365-2966.2004.07713.x}.
\bibitem[{{Ivanova}(2011)}]{Ivanova2011ApJ...730...76I}
\bibinfo{author}{N.~{Ivanova}},
\newblock \bibinfo{title}{{Common Envelope: On the Mass and the Fate of the
  Remnant}},
\newblock \bibinfo{journal}{\apj} \bibinfo{volume}{730} (\bibinfo{year}{2011})
  \bibinfo{pages}{76}. \DOIprefix\doi{10.1088/0004-637X/730/2/76}.
\bibitem[{{Vigna-G{\'o}mez} et~al.(2022){Vigna-G{\'o}mez}, {Wassink},
  {Klencki}, {Istrate}, {Nelemans}, and
  {Mandel}}]{Vigna-Gomez2022MNRAS.511.2326V}
\bibinfo{author}{A.~{Vigna-G{\'o}mez}}, \bibinfo{author}{M.~{Wassink}},
  \bibinfo{author}{J.~{Klencki}}, et~al.,
\newblock \bibinfo{title}{{Stellar response after stripping as a model for
  common-envelope outcomes}},
\newblock \bibinfo{journal}{\mnras} \bibinfo{volume}{511}
  (\bibinfo{year}{2022}) \bibinfo{pages}{2326--2338}.
  \DOIprefix\doi{10.1093/mnras/stac237}.
\bibitem[{{Deloye} and {Taam}(2010)}]{Deloye2010ApJ...719L..28D}
\bibinfo{author}{C.~J. {Deloye}}, \bibinfo{author}{R.~E. {Taam}},
\newblock \bibinfo{title}{{Adiabatic Mass Loss and the Outcome of the Common
  Envelope Phase of Binary Evolution}},
\newblock \bibinfo{journal}{\apjl} \bibinfo{volume}{719} (\bibinfo{year}{2010})
  \bibinfo{pages}{L28--L31}. \DOIprefix\doi{10.1088/2041-8205/719/1/L28}.
\bibitem[{{Ge} et~al.(2022){Ge}, {Tout}, {Chen}, {Kruckow}, {Chen}, {Jiang},
  {Li}, {Liu}, and {Han}}]{Ge2022ApJ...933..137G}
\bibinfo{author}{H.~{Ge}}, \bibinfo{author}{C.~A. {Tout}},
  \bibinfo{author}{X.~{Chen}}, et~al.,
\newblock \bibinfo{title}{{The Common Envelope Evolution Outcome-A Case Study
  on Hot Subdwarf B Stars}},
\newblock \bibinfo{journal}{\apj} \bibinfo{volume}{933} (\bibinfo{year}{2022})
  \bibinfo{pages}{137}. \DOIprefix\doi{10.3847/1538-4357/ac75d3}.
\bibitem[{{Raymond} et~al.(2003){Raymond}, {Szkody}, {Hawley}, {Anderson},
  {Brinkmann}, {Covey}, {McGehee}, {Schneider}, {West}, and
  {York}}]{Raymond2003AJ....125.2621R}
\bibinfo{author}{S.~N. {Raymond}}, \bibinfo{author}{P.~{Szkody}},
  \bibinfo{author}{S.~L. {Hawley}}, et~al.,
\newblock \bibinfo{title}{{A First Look at White Dwarf-M Dwarf Pairs in the
  Sloan Digital Sky Survey}},
\newblock \bibinfo{journal}{\aj} \bibinfo{volume}{125} (\bibinfo{year}{2003})
  \bibinfo{pages}{2621--2629}. \DOIprefix\doi{10.1086/374762}.
\bibitem[{{Silvestri} et~al.(2006){Silvestri}, {Hawley}, {West}, {Szkody},
  {Bochanski}, and et~al.}]{Silvestri2006AJ....131.1674S}
\bibinfo{author}{N.~M. {Silvestri}}, \bibinfo{author}{S.~L. {Hawley}},
  \bibinfo{author}{A.~A. {West}}, et~al.,
\newblock \bibinfo{title}{{A Catalog of Spectroscopically Selected Close Binary
  Systems from the Sloan Digital Sky Survey Data Release Four}},
\newblock \bibinfo{journal}{\aj} \bibinfo{volume}{131} (\bibinfo{year}{2006})
  \bibinfo{pages}{1674--1686}. \DOIprefix\doi{10.1086/499494}.
\bibitem[{{Heller} et~al.(2009){Heller}, {Homeier}, {Dreizler}, and
  {{\O}stensen}}]{Heller2009A&A...496..191H}
\bibinfo{author}{R.~{Heller}}, \bibinfo{author}{D.~{Homeier}},
  \bibinfo{author}{S.~{Dreizler}}, \bibinfo{author}{R.~{{\O}stensen}},
\newblock \bibinfo{title}{{Spectral analysis of 636 white dwarf-M star binaries
  from the sloan digital sky survey}},
\newblock \bibinfo{journal}{\aap} \bibinfo{volume}{496} (\bibinfo{year}{2009})
  \bibinfo{pages}{191--205}. \DOIprefix\doi{10.1051/0004-6361:200810632}.
\bibitem[{{Zorotovic} and {Schreiber}(2022)}]{Zorotovic2022MNRAS.513.3587Z}
\bibinfo{author}{M.~{Zorotovic}}, \bibinfo{author}{M.~{Schreiber}},
\newblock \bibinfo{title}{{Close detached white dwarf + brown dwarf binaries:
  further evidence for low values of the common envelope efficiency}},
\newblock \bibinfo{journal}{\mnras} \bibinfo{volume}{513}
  (\bibinfo{year}{2022}) \bibinfo{pages}{3587--3595}.
  \DOIprefix\doi{10.1093/mnras/stac1137}.
\bibitem[{{Maxted} et~al.(2009){Maxted}, {G{\"a}nsicke}, {Burleigh},
  {Southworth}, {Marsh}, {Napiwotzki}, {Nelemans}, and {Wood}}]{maxted09}
\bibinfo{author}{P.~F.~L. {Maxted}}, \bibinfo{author}{B.~T. {G{\"a}nsicke}},
  \bibinfo{author}{M.~R. {Burleigh}}, et~al.,
\newblock \bibinfo{title}{{A survey for post-common-envelope binary stars using
  GALEX and SDSS photometry}},
\newblock \bibinfo{journal}{\mnras} \bibinfo{volume}{400}
  (\bibinfo{year}{2009}) \bibinfo{pages}{2012--2021}.
  \DOIprefix\doi{10.1111/j.1365-2966.2009.15594.x}.
\bibitem[{{Landsman} et~al.(1993){Landsman}, {Simon}, and
  {Bergeron}}]{Landsman93}
\bibinfo{author}{W.~{Landsman}}, \bibinfo{author}{T.~{Simon}},
  \bibinfo{author}{P.~{Bergeron}},
\newblock \bibinfo{title}{{The Hot White Dwarf Companions of HR 1608, HR 8210,
  and HD 15638}},
\newblock \bibinfo{journal}{\pasp} \bibinfo{volume}{105} (\bibinfo{year}{1993})
  \bibinfo{pages}{841}. \DOIprefix\doi{10.1086/133242}.
\bibitem[{{Landsman} et~al.(1996){Landsman}, {Simon}, and
  {Bergeron}}]{Landsman96}
\bibinfo{author}{W.~{Landsman}}, \bibinfo{author}{T.~{Simon}},
  \bibinfo{author}{P.~{Bergeron}},
\newblock \bibinfo{title}{{The White-Dwarf Companions of 56 Persei and HR
  3643}},
\newblock \bibinfo{journal}{\pasp} \bibinfo{volume}{108} (\bibinfo{year}{1996})
  \bibinfo{pages}{250}. \DOIprefix\doi{10.1086/133718}.
\bibitem[{{Barstow} et~al.(1994){Barstow}, {Holberg}, and
  {Koester}}]{Barstow94}
\bibinfo{author}{M.~A. {Barstow}}, \bibinfo{author}{J.~B. {Holberg}},
  \bibinfo{author}{D.~{Koester}},
\newblock \bibinfo{title}{{Extreme-ultraviolet spectrophotometry of HD 15638
  and HR 8210 (IK Peg).}},
\newblock \bibinfo{journal}{\mnras} \bibinfo{volume}{270}
  (\bibinfo{year}{1994}) \bibinfo{pages}{516--522}.
  \DOIprefix\doi{10.1093/mnras/270.3.516}.
\bibitem[{{Vennes} et~al.(1995){Vennes}, {Mathioudakis}, {Doyle},
  {Thorstensen}, and {Byrne}}]{Vennes1995}
\bibinfo{author}{S.~{Vennes}}, \bibinfo{author}{M.~{Mathioudakis}},
  \bibinfo{author}{J.~G. {Doyle}}, et~al.,
\newblock \bibinfo{title}{{Discovery of a white dwarf companion (EUVE
  J0254-053) to the K0 IV star HD18131}},
\newblock \bibinfo{journal}{\aap}  \bibinfo{volume}{299} (\bibinfo{year}{1995})
  \bibinfo{pages}{L29}.
\bibitem[{{Vennes} et~al.(1997){Vennes}, {Christian}, {Mathioudakis}, and
  {Doyle}}]{Vennes1997}
\bibinfo{author}{S.~{Vennes}}, \bibinfo{author}{D.~J. {Christian}},
  \bibinfo{author}{M.~{Mathioudakis}}, \bibinfo{author}{J.~G. {Doyle}},
\newblock \bibinfo{title}{{An active K0 IV-V star and a hot white dwarf (EUVE
  J0702+129) in a wide binary.}},
\newblock \bibinfo{journal}{\aap}  \bibinfo{volume}{318} (\bibinfo{year}{1997})
  \bibinfo{pages}{L9--L12}.
\bibitem[{{Christian} et~al.(1996){Christian}, {Vennes}, {Thorstensen}, and
  {Mathioudakis}}]{Chris1996}
\bibinfo{author}{D.~J. {Christian}}, \bibinfo{author}{S.~{Vennes}},
  \bibinfo{author}{J.~R. {Thorstensen}}, \bibinfo{author}{M.~{Mathioudakis}},
\newblock \bibinfo{title}{{Discovery of a White Dwarf Companion (MS0354.6-3650
  = EUVE J0356-366) to a G2V Star}},
\newblock \bibinfo{journal}{\aj} \bibinfo{volume}{112} (\bibinfo{year}{1996})
  \bibinfo{pages}{258}. \DOIprefix\doi{10.1086/118008}.
\bibitem[{{Burleigh} et~al.(1997){Burleigh}, {Barstow}, and
  {Fleming}}]{Burleigh97}
\bibinfo{author}{M.~R. {Burleigh}}, \bibinfo{author}{M.~A. {Barstow}},
  \bibinfo{author}{T.~A. {Fleming}},
\newblock \bibinfo{title}{{A search for hidden white dwarfs in the ROSAT
  extreme ultraviolet survey}},
\newblock \bibinfo{journal}{\mnras} \bibinfo{volume}{287}
  (\bibinfo{year}{1997}) \bibinfo{pages}{381--401}.
  \DOIprefix\doi{10.1093/mnras/287.2.381}.
\bibitem[{{Burleigh} and {Barstow}(1998)}]{Burleigh98a}
\bibinfo{author}{M.~R. {Burleigh}}, \bibinfo{author}{M.~A. {Barstow}},
\newblock \bibinfo{title}{{HR 2875: spectroscopic discovery of the first B
  star+white dwarf binary}},
\newblock \bibinfo{journal}{\mnras} \bibinfo{volume}{295}
  (\bibinfo{year}{1998}) \bibinfo{pages}{L15--L19}.
  \DOIprefix\doi{10.1046/j.1365-8711.1998.29511471.x}.
\bibitem[{{Burleigh} et~al.(1998){Burleigh}, {Barstow}, and
  {Holberg}}]{Buleigh98b}
\bibinfo{author}{M.~R. {Burleigh}}, \bibinfo{author}{M.~A. {Barstow}},
  \bibinfo{author}{J.~B. {Holberg}},
\newblock \bibinfo{title}{{A search for hidden white dwarfs in theROSATEUV
  survey - II. Discovery of a distant DA+F6/7V binary system in a direction of
  low-density neutral hydrogen}},
\newblock \bibinfo{journal}{\mnras} \bibinfo{volume}{300}
  (\bibinfo{year}{1998}) \bibinfo{pages}{511--527}.
  \DOIprefix\doi{10.1046/j.1365-8711.1998.01914.x}.
\bibitem[{{Burleigh} and {Barstow}(1999)}]{Buleigh99}
\bibinfo{author}{M.~R. {Burleigh}}, \bibinfo{author}{M.~A. {Barstow}},
\newblock \bibinfo{title}{{Theta Hya: spectroscopic identification of a second
  B star+white dwarf binary}},
\newblock \bibinfo{journal}{\aap}  \bibinfo{volume}{341} (\bibinfo{year}{1999})
  \bibinfo{pages}{795--798}.
\bibitem[{{Parsons} et~al.(2015){Parsons}, {Schreiber}, {G{\"a}nsicke},
  {Rebassa-Mansergas}, {Brahm}, {Zorotovic}, {Toloza}, {Pala}, {Tappert},
  {Bayo}, and {Jord{\'a}n}}]{parsons15}
\bibinfo{author}{S.~G. {Parsons}}, \bibinfo{author}{M.~R. {Schreiber}},
  \bibinfo{author}{B.~T. {G{\"a}nsicke}}, et~al.,
\newblock \bibinfo{title}{{The first pre-supersoft X-ray binary}},
\newblock \bibinfo{journal}{\mnras} \bibinfo{volume}{452}
  (\bibinfo{year}{2015}) \bibinfo{pages}{1754--1763}.
  \DOIprefix\doi{10.1093/mnras/stv1395}.
\bibitem[{{Parsons} et~al.(2016){Parsons}, {Rebassa-Mansergas}, {Schreiber},
  {G{\"a}nsicke}, {Zorotovic}, and {Ren}}]{WDFGK-i}
\bibinfo{author}{S.~G. {Parsons}}, \bibinfo{author}{A.~{Rebassa-Mansergas}},
  \bibinfo{author}{M.~R. {Schreiber}}, et~al.,
\newblock \bibinfo{title}{{The white dwarf binary pathways survey - I. A sample
  of FGK stars with white dwarf companions}},
\newblock \bibinfo{journal}{\mnras} \bibinfo{volume}{463}
  (\bibinfo{year}{2016}) \bibinfo{pages}{2125--2136}.
  \DOIprefix\doi{10.1093/mnras/stw2143}.
\bibitem[{{Rebassa-Mansergas} et~al.(2017){Rebassa-Mansergas}, {Ren},
  {Irawati}, {Garc{\'\i}a-Berro}, {Parsons}, {Schreiber}, {G{\"a}nsicke},
  {Rodr{\'\i}guez-Gil}, {Liu}, {Manser}, {Nevado}, {Jim{\'e}nez-Ibarra},
  {Costero}, {Echevarr{\'\i}a}, {Michel}, {Zorotovic}, {Hollands}, {Han},
  {Luo}, {Villaver}, and {Kong}}]{WDFGK-ii}
\bibinfo{author}{A.~{Rebassa-Mansergas}}, \bibinfo{author}{J.~J. {Ren}},
  \bibinfo{author}{P.~{Irawati}}, et~al.,
\newblock \bibinfo{title}{{The white dwarf binary pathways survey - II. Radial
  velocities of 1453 FGK stars with white dwarf companions from LAMOST DR 4}},
\newblock \bibinfo{journal}{\mnras} \bibinfo{volume}{472}
  (\bibinfo{year}{2017}) \bibinfo{pages}{4193--4203}.
  \DOIprefix\doi{10.1093/mnras/stx2259}.
\bibitem[{{Lagos} et~al.(2020){Lagos}, {Schreiber}, {Parsons}, {Zurlo}, {Mesa},
  {G{\"a}nsicke}, {Brahm}, {Caceres}, {Canovas}, {Hernandez}, {Jordan},
  {Koester}, {Schmidtobreick}, {Tappert}, and {Zorotovic}}]{WDFGK-iii}
\bibinfo{author}{F.~{Lagos}}, \bibinfo{author}{M.~R. {Schreiber}},
  \bibinfo{author}{S.~G. {Parsons}}, et~al.,
\newblock \bibinfo{title}{{The White Dwarf Binary Pathways Survey -III.
  Contamination from hierarchical triples containing a white dwarf}},
\newblock \bibinfo{journal}{\mnras} \bibinfo{volume}{494}
  (\bibinfo{year}{2020}) \bibinfo{pages}{915--922}.
  \DOIprefix\doi{10.1093/mnras/staa747}.
\bibitem[{{Hernandez} et~al.(2021){Hernandez}, {Schreiber}, {Parsons},
  {G{\"a}nsicke}, {Lagos}, {Raddi}, {Toloza}, {Tovmassian}, {Zorotovic},
  {Irawati}, {Past{\'e}n}, {Rebassa-Mansergas}, {Ren}, {Rittipruk}, and
  {Tappert}}]{WDFGK-iv}
\bibinfo{author}{M.~S. {Hernandez}}, \bibinfo{author}{M.~R. {Schreiber}},
  \bibinfo{author}{S.~G. {Parsons}}, et~al.,
\newblock \bibinfo{title}{{The White Dwarf Binary Pathways Survey - IV. Three
  close white dwarf binaries with G-type secondary stars}},
\newblock \bibinfo{journal}{\mnras} \bibinfo{volume}{501}
  (\bibinfo{year}{2021}) \bibinfo{pages}{1677--1689}.
  \DOIprefix\doi{10.1093/mnras/staa3815}.
\bibitem[{{Ren} et~al.(2020){Ren}, {Raddi}, {Rebassa-Mansergas}, {Hernandez},
  {Parsons}, {Irawati}, {Rittipruk}, {Schreiber}, {G{\"a}nsicke}, {Torres},
  {Wang}, {Zhang}, {Zhao}, {Zhou}, {Han}, {Wang}, {Liu}, {Liu}, {Wang},
  {Zheng}, {Wang}, {Zhao}, {Cui}, {Shi}, and {Tian}}]{WDFGK-v}
\bibinfo{author}{J.~J. {Ren}}, \bibinfo{author}{R.~{Raddi}},
  \bibinfo{author}{A.~{Rebassa-Mansergas}}, et~al.,
\newblock \bibinfo{title}{{The White Dwarf Binary Pathways Survey. V. The Gaia
  White Dwarf Plus AFGK Binary Sample and the Identification of 23 Close
  Binaries}},
\newblock \bibinfo{journal}{\apj} \bibinfo{volume}{905} (\bibinfo{year}{2020})
  \bibinfo{pages}{38}. \DOIprefix\doi{10.3847/1538-4357/abc017}.
\bibitem[{{Hernandez} et~al.(2022){Hernandez}, {Schreiber}, {Parsons},
  {G{\"a}nsicke}, {Toloza}, {Tovmassian}, {Zorotovic}, {Lagos}, {Raddi},
  {Rebassa-Mansergas}, {Ren}, and {Tappert}}]{WDFGK-vi}
\bibinfo{author}{M.~S. {Hernandez}}, \bibinfo{author}{M.~R. {Schreiber}},
  \bibinfo{author}{S.~G. {Parsons}}, et~al.,
\newblock \bibinfo{title}{{The white dwarf binary pathways survey - VI. Two
  close post-common envelope binaries with TESS light curves}},
\newblock \bibinfo{journal}{\mnras} \bibinfo{volume}{512}
  (\bibinfo{year}{2022}) \bibinfo{pages}{1843--1856}.
  \DOIprefix\doi{10.1093/mnras/stac604}.
\bibitem[{{Lagos} et~al.(2022){Lagos}, {Schreiber}, {Parsons}, {Toloza},
  {G{\"a}nsicke}, {Hernandez}, {Schmidtobreick}, and {Belloni}}]{WDFGK-vii}
\bibinfo{author}{F.~{Lagos}}, \bibinfo{author}{M.~R. {Schreiber}},
  \bibinfo{author}{S.~G. {Parsons}}, et~al.,
\newblock \bibinfo{title}{{The white dwarf binary pathways survey - VII.
  Evidence for a bi-modal distribution of post-mass transfer systems?}},
\newblock \bibinfo{journal}{\mnras} \bibinfo{volume}{512}
  (\bibinfo{year}{2022}) \bibinfo{pages}{2625--2635}.
  \DOIprefix\doi{10.1093/mnras/stac673}.
\bibitem[{{Hernandez} et~al.(2022){Hernandez}, {Schreiber}, {Parsons},
  {G{\"a}nsicke}, {Toloza}, {Zorotovic}, {Raddi}, {Rebassa-Mansergas}, and
  {Ren}}]{WDFGK-viii}
\bibinfo{author}{M.~S. {Hernandez}}, \bibinfo{author}{M.~R. {Schreiber}},
  \bibinfo{author}{S.~G. {Parsons}}, et~al.,
\newblock \bibinfo{title}{{The white dwarf binary pathways survey - VIII. A
  post-common envelope binary with a massive white dwarf and an active G-type
  secondary star}},
\newblock \bibinfo{journal}{\mnras} \bibinfo{volume}{517}
  (\bibinfo{year}{2022}) \bibinfo{pages}{2867--2875}.
  \DOIprefix\doi{10.1093/mnras/stac2837}.
\bibitem[{{Parsons} et~al.(2023){Parsons}, {Hernandez}, {Toloza}, {Zorotovic},
  {Schreiber}, {G{\"a}nsicke}, {Lagos}, {Raddi}, {Rebassa-Mansergas}, {Ren},
  and {Koester}}]{WDFGK-ix}
\bibinfo{author}{S.~G. {Parsons}}, \bibinfo{author}{M.~S. {Hernandez}},
  \bibinfo{author}{O.~{Toloza}}, et~al.,
\newblock \bibinfo{title}{{The white dwarf binary pathways survey - IX. Three
  long period white dwarf plus subgiant binaries}},
\newblock \bibinfo{journal}{\mnras} \bibinfo{volume}{518}
  (\bibinfo{year}{2023}) \bibinfo{pages}{4579--4594}.
  \DOIprefix\doi{10.1093/mnras/stac3368}.
\bibitem[{{O'Brien} et~al.(2001){O'Brien}, {Bond}, and
  {Sion}}]{OBrien2001ApJ...563..971O}
\bibinfo{author}{M.~S. {O'Brien}}, \bibinfo{author}{H.~E. {Bond}},
  \bibinfo{author}{E.~M. {Sion}},
\newblock \bibinfo{title}{{Hubble Space Telescope Spectroscopy of V471 Tauri:
  Oversized K Star, Paradoxical White Dwarf}},
\newblock \bibinfo{journal}{\apj} \bibinfo{volume}{563} (\bibinfo{year}{2001})
  \bibinfo{pages}{971--986}. \DOIprefix\doi{10.1086/324040}.
\bibitem[{{Yungelson} et~al.(1993){Yungelson}, {Tutukov}, and
  {Livio}}]{Yungelson1993ApJ...418..794Y}
\bibinfo{author}{L.~R. {Yungelson}}, \bibinfo{author}{A.~V. {Tutukov}},
  \bibinfo{author}{M.~{Livio}},
\newblock \bibinfo{title}{{The Formation of Binary and Single Nuclei of
  Planetary Nebulae}},
\newblock \bibinfo{journal}{\apj} \bibinfo{volume}{418} (\bibinfo{year}{1993})
  \bibinfo{pages}{794}. \DOIprefix\doi{10.1086/173436}.
\bibitem[{{Han} et~al.(1995){Han}, {Eggleton}, {Podsiadlowski}, and
  {Tout}}]{Han1995MNRAS.277.1443H}
\bibinfo{author}{Z.~{Han}}, \bibinfo{author}{P.~P. {Eggleton}},
  \bibinfo{author}{P.~{Podsiadlowski}}, \bibinfo{author}{C.~A. {Tout}},
\newblock \bibinfo{title}{{The formation of barium and CH stars and related
  objects}},
\newblock \bibinfo{journal}{\mnras} \bibinfo{volume}{277}
  (\bibinfo{year}{1995}) \bibinfo{pages}{1443--1462}.
  \DOIprefix\doi{10.1093/mnras/277.4.1443}.
\bibitem[{{Bond} and {Livio}(1990)}]{Bond1990ApJ...355..568B}
\bibinfo{author}{H.~E. {Bond}}, \bibinfo{author}{M.~{Livio}},
\newblock \bibinfo{title}{{Morphologies of Planetary Nebulae Ejected by
  Close-Binary Nuclei}},
\newblock \bibinfo{journal}{\apj} \bibinfo{volume}{355} (\bibinfo{year}{1990})
  \bibinfo{pages}{568}. \DOIprefix\doi{10.1086/168789}.
\bibitem[{{Bond}(2000)}]{Bond2000ASPC..199..115B}
\bibinfo{author}{H.~E. {Bond}},
\newblock \bibinfo{title}{{Binarity of Central Stars of Planetary Nebulae}},
\newblock in: \bibinfo{editor}{J.~H. {Kastner}}, \bibinfo{editor}{N.~{Soker}},
  \bibinfo{editor}{S.~{Rappaport}} (Eds.), \bibinfo{booktitle}{Asymmetrical
  Planetary Nebulae II: From Origins to Microstructures}, volume
  \bibinfo{volume}{199} of \textit{\bibinfo{series}{Astronomical Society of the
  Pacific Conference Series}}, \bibinfo{year}{2000},  p. \bibinfo{pages}{115}.
\bibitem[{{Kupfer} et~al.(2015){Kupfer}, {Geier}, {Heber}, {{\O}stensen},
  {Barlow}, {Maxted}, {Heuser}, {Schaffenroth}, and
  {G{\"a}nsicke}}]{Kupfer2015A&A...576A..44K}
\bibinfo{author}{T.~{Kupfer}}, \bibinfo{author}{S.~{Geier}},
  \bibinfo{author}{U.~{Heber}}, et~al.,
\newblock \bibinfo{title}{{Hot subdwarf binaries from the MUCHFUSS project.
  Analysis of 12 new systems and a study of the short-period binary
  population}},
\newblock \bibinfo{journal}{\aap} \bibinfo{volume}{576} (\bibinfo{year}{2015})
  \bibinfo{pages}{A44}. \DOIprefix\doi{10.1051/0004-6361/201425213}.
\bibitem[{{De Marco} et~al.(2004){De Marco}, {Bond}, {Harmer}, and
  {Fleming}}]{De-Marco2004ApJ...602L..93D}
\bibinfo{author}{O.~{De Marco}}, \bibinfo{author}{H.~E. {Bond}},
  \bibinfo{author}{D.~{Harmer}}, \bibinfo{author}{A.~J. {Fleming}},
\newblock \bibinfo{title}{{Indications of a Large Fraction of Spectroscopic
  Binaries among Nuclei of Planetary Nebulae}},
\newblock \bibinfo{journal}{\apjl} \bibinfo{volume}{602} (\bibinfo{year}{2004})
  \bibinfo{pages}{L93--L96}. \DOIprefix\doi{10.1086/382156}.
\bibitem[{{De Marco} et~al.(2008){De Marco}, {Hillwig}, and
  {Smith}}]{De-Marco2008AJ....136..323D}
\bibinfo{author}{O.~{De Marco}}, \bibinfo{author}{T.~C. {Hillwig}},
  \bibinfo{author}{A.~J. {Smith}},
\newblock \bibinfo{title}{{Binary Central Stars of Planetary Nebulae Discovered
  Through Photometric Variability. I. What we Know and what we would like to
  Find Out}},
\newblock \bibinfo{journal}{\aj} \bibinfo{volume}{136} (\bibinfo{year}{2008})
  \bibinfo{pages}{323--336}. \DOIprefix\doi{10.1088/0004-6256/136/1/323}.
\bibitem[{{Han} et~al.(1995){Han}, {Podsiadlowski}, and
  {Eggleton}}]{Han1995MNRAS.272..800H}
\bibinfo{author}{Z.~{Han}}, \bibinfo{author}{P.~{Podsiadlowski}},
  \bibinfo{author}{P.~P. {Eggleton}},
\newblock \bibinfo{title}{{The formation of bipolar planetary nebulae and close
  white dwarf binaries}},
\newblock \bibinfo{journal}{\mnras} \bibinfo{volume}{272}
  (\bibinfo{year}{1995}) \bibinfo{pages}{800--820}.
  \DOIprefix\doi{10.1093/mnras/272.4.800}.
\bibitem[{{Miszalski} et~al.(2009){Miszalski}, {Acker}, {Moffat}, {Parker}, and
  {Udalski}}]{Miszalski2009A&A...496..813M}
\bibinfo{author}{B.~{Miszalski}}, \bibinfo{author}{A.~{Acker}},
  \bibinfo{author}{A.~F.~J. {Moffat}}, et~al.,
\newblock \bibinfo{title}{{Binary planetary nebulae nuclei towards the Galactic
  bulge. I. Sample discovery, period distribution, and binary fraction}},
\newblock \bibinfo{journal}{\aap} \bibinfo{volume}{496} (\bibinfo{year}{2009})
  \bibinfo{pages}{813--825}. \DOIprefix\doi{10.1051/0004-6361/200811380}.
\bibitem[{{de Kool}(1992)}]{de-Kool1992A&A...261..188D}
\bibinfo{author}{M.~{de Kool}},
\newblock \bibinfo{title}{{Statistics of cataclysmic variable formation}},
\newblock \bibinfo{journal}{\aap}  \bibinfo{volume}{261} (\bibinfo{year}{1992})
  \bibinfo{pages}{188--202}.
\bibitem[{{Nelemans} et~al.(2001){Nelemans}, {Yungelson}, {Portegies Zwart},
  and {Verbunt}}]{Nelemans2001A&A...365..491N}
\bibinfo{author}{G.~{Nelemans}}, \bibinfo{author}{L.~R. {Yungelson}},
  \bibinfo{author}{S.~F. {Portegies Zwart}}, \bibinfo{author}{F.~{Verbunt}},
\newblock \bibinfo{title}{{Population synthesis for double white dwarfs . I.
  Close detached systems}},
\newblock \bibinfo{journal}{\aap} \bibinfo{volume}{365} (\bibinfo{year}{2001})
  \bibinfo{pages}{491--507}. \DOIprefix\doi{10.1051/0004-6361:20000147}.
\bibitem[{{Portegies Zwart}(2013)}]{Portegies-Zwart2013MNRAS.429L..45P}
\bibinfo{author}{S.~{Portegies Zwart}},
\newblock \bibinfo{title}{{Planet-mediated precision reconstruction of the
  evolution of the cataclysmic variable HU Aquarii.}},
\newblock \bibinfo{journal}{\mnras} \bibinfo{volume}{429}
  (\bibinfo{year}{2013}) \bibinfo{pages}{L45--L49}.
  \DOIprefix\doi{10.1093/mnrasl/sls022}.
\bibitem[{{Wilson} and {Nordhaus}(2022)}]{Wilson2022MNRAS.516.2189W}
\bibinfo{author}{E.~C. {Wilson}}, \bibinfo{author}{J.~{Nordhaus}},
\newblock \bibinfo{title}{{Convection reconciles the difference in efficiencies
  between low-mass and high-mass common envelopes}},
\newblock \bibinfo{journal}{\mnras} \bibinfo{volume}{516}
  (\bibinfo{year}{2022}) \bibinfo{pages}{2189--2195}.
  \DOIprefix\doi{10.1093/mnras/stac2300}.
\bibitem[{{Ivanova} and {Nandez}(2016)}]{Ivanova16}
\bibinfo{author}{N.~{Ivanova}}, \bibinfo{author}{J.~L.~A. {Nandez}},
\newblock \bibinfo{title}{{Common envelope events with low-mass giants:
  understanding the transition to the slow spiral-in}},
\newblock \bibinfo{journal}{\mnras} \bibinfo{volume}{462}
  (\bibinfo{year}{2016}) \bibinfo{pages}{362--381}.
  \DOIprefix\doi{10.1093/mnras/stw1676}.
\bibitem[{{Taam} et~al.(1978){Taam}, {Bodenheimer}, and
  {Ostriker}}]{Taam1978ApJ...222..269T}
\bibinfo{author}{R.~E. {Taam}}, \bibinfo{author}{P.~{Bodenheimer}},
  \bibinfo{author}{J.~P. {Ostriker}},
\newblock \bibinfo{title}{{Double core evolution. I. A 16 M sun star with a 1 M
  sun neutron-star companion.}},
\newblock \bibinfo{journal}{\apj} \bibinfo{volume}{222} (\bibinfo{year}{1978})
  \bibinfo{pages}{269--280}. \DOIprefix\doi{10.1086/156142}.
\bibitem[{{Meyer} and {Meyer-Hofmeister}(1979)}]{Meyer1979A&A....78..167M}
\bibinfo{author}{F.~{Meyer}}, \bibinfo{author}{E.~{Meyer-Hofmeister}},
\newblock \bibinfo{title}{{Formation of cataclysmic binaries through common
  envelope evolution.}},
\newblock \bibinfo{journal}{\aap}  \bibinfo{volume}{78} (\bibinfo{year}{1979})
  \bibinfo{pages}{167--176}.
\bibitem[{{Taam} and {Bodenheimer}(1989)}]{Taam1989ApJ...337..849T}
\bibinfo{author}{R.~E. {Taam}}, \bibinfo{author}{P.~{Bodenheimer}},
\newblock \bibinfo{title}{{Double-Core Evolution. III. The Evolution of a 5 M
  Red Giant with a 1 M Companion}},
\newblock \bibinfo{journal}{\apj} \bibinfo{volume}{337} (\bibinfo{year}{1989})
  \bibinfo{pages}{849}. \DOIprefix\doi{10.1086/167155}.
\bibitem[{{Taam} and {Bodenheimer}(1991)}]{Taam1991ApJ...373..246T}
\bibinfo{author}{R.~E. {Taam}}, \bibinfo{author}{P.~{Bodenheimer}},
\newblock \bibinfo{title}{{Double Core Evolution. IV. The Late Stages of
  Evolution of a 2 M$_{sun}$ Red Giant with a 1 M$_{sun}$ Companion}},
\newblock \bibinfo{journal}{\apj} \bibinfo{volume}{373} (\bibinfo{year}{1991})
  \bibinfo{pages}{246}. \DOIprefix\doi{10.1086/170043}.
\bibitem[{{de Kool}(1987)}]{de-Kool1987PhDT.......112D}
\bibinfo{author}{M.~{de Kool}}, \bibinfo{title}{{Models of interacting binary
  stars}}, Ph.D. thesis, -,  \bibinfo{year}{1987}.
\bibitem[{{Terman} et~al.(1994){Terman}, {Taam}, and
  {Hernquist}}]{Terman1994ApJ...422..729T}
\bibinfo{author}{J.~L. {Terman}}, \bibinfo{author}{R.~E. {Taam}},
  \bibinfo{author}{L.~{Hernquist}},
\newblock \bibinfo{title}{{Double-Core Evolution. V. Three-dimensional Effects
  in the Merger of a Red Giant with a Dwarf Companion}},
\newblock \bibinfo{journal}{\apj} \bibinfo{volume}{422} (\bibinfo{year}{1994})
  \bibinfo{pages}{729}. \DOIprefix\doi{10.1086/173765}.
\bibitem[{{Terman} et~al.(1995){Terman}, {Taam}, and
  {Hernquist}}]{Terman1995ApJ...445..367T}
\bibinfo{author}{J.~L. {Terman}}, \bibinfo{author}{R.~E. {Taam}},
  \bibinfo{author}{L.~{Hernquist}},
\newblock \bibinfo{title}{{Double Core Evolution. VII. The Infall of a Neutron
  Star through the Envelope of Its Massive Star Companion}},
\newblock \bibinfo{journal}{\apj} \bibinfo{volume}{445} (\bibinfo{year}{1995})
  \bibinfo{pages}{367}. \DOIprefix\doi{10.1086/175702}.
\bibitem[{{Terman} and {Taam}(1996)}]{Terman1996ApJ...458..692T}
\bibinfo{author}{J.~L. {Terman}}, \bibinfo{author}{R.~E. {Taam}},
\newblock \bibinfo{title}{{Double-Core Evolution. IX. The Infall of a
  Main-Sequence Star through the Envelope of Its Intermediate-Mass Red Giant
  Companion}},
\newblock \bibinfo{journal}{\apj} \bibinfo{volume}{458} (\bibinfo{year}{1996})
  \bibinfo{pages}{692}. \DOIprefix\doi{10.1086/176850}.
\bibitem[{{Sandquist} et~al.(1998){Sandquist}, {Taam}, {Chen}, {Bodenheimer},
  and {Burkert}}]{Sandquist1998ApJ...500..909S}
\bibinfo{author}{E.~L. {Sandquist}}, \bibinfo{author}{R.~E. {Taam}},
  \bibinfo{author}{X.~{Chen}}, et~al.,
\newblock \bibinfo{title}{{Double Core Evolution. X. Through the Envelope
  Ejection Phase}},
\newblock \bibinfo{journal}{\apj} \bibinfo{volume}{500} (\bibinfo{year}{1998})
  \bibinfo{pages}{909--922}. \DOIprefix\doi{10.1086/305778}.
\bibitem[{{Sandquist} et~al.(2000){Sandquist}, {Taam}, and
  {Burkert}}]{Sandquist2000ApJ...533..984S}
\bibinfo{author}{E.~L. {Sandquist}}, \bibinfo{author}{R.~E. {Taam}},
  \bibinfo{author}{A.~{Burkert}},
\newblock \bibinfo{title}{{On the Formation of Helium Double Degenerate Stars
  and Pre-Cataclysmic Variables}},
\newblock \bibinfo{journal}{\apj} \bibinfo{volume}{533} (\bibinfo{year}{2000})
  \bibinfo{pages}{984--997}. \DOIprefix\doi{10.1086/308687}.
\bibitem[{{Hut}(1980)}]{Hut1980A&A....92..167H}
\bibinfo{author}{P.~{Hut}},
\newblock \bibinfo{title}{{Stability of tidal equilibrium}},
\newblock \bibinfo{journal}{\aap}  \bibinfo{volume}{92} (\bibinfo{year}{1980})
  \bibinfo{pages}{167--170}.
\bibitem[{{Lai} et~al.(1993){Lai}, {Rasio}, and
  {Shapiro}}]{Lai1993ApJ...406L..63L}
\bibinfo{author}{D.~{Lai}}, \bibinfo{author}{F.~A. {Rasio}},
  \bibinfo{author}{S.~L. {Shapiro}},
\newblock \bibinfo{title}{{Hydrodynamic Instability and Coalescence of Close
  Binary Systems}},
\newblock \bibinfo{journal}{\apjl} \bibinfo{volume}{406} (\bibinfo{year}{1993})
  \bibinfo{pages}{L63}. \DOIprefix\doi{10.1086/186787}.
\bibitem[{{Eggleton} and
  {Kiseleva-Eggleton}(2001)}]{Eggleton2001ApJ...562.1012E}
\bibinfo{author}{P.~P. {Eggleton}}, \bibinfo{author}{L.~{Kiseleva-Eggleton}},
\newblock \bibinfo{title}{{Orbital Evolution in Binary and Triple Stars, with
  an Application to SS Lacertae}},
\newblock \bibinfo{journal}{\apj} \bibinfo{volume}{562} (\bibinfo{year}{2001})
  \bibinfo{pages}{1012--1030}. \DOIprefix\doi{10.1086/323843}.
\bibitem[{{Passy} et~al.(2012){Passy}, {De Marco}, {Fryer}, {Herwig}, {Diehl},
  {Oishi}, {Mac Low}, {Bryan}, and {Rockefeller}}]{Passy2012ApJ...744...52P}
\bibinfo{author}{J.-C. {Passy}}, \bibinfo{author}{O.~{De Marco}},
  \bibinfo{author}{C.~L. {Fryer}}, et~al.,
\newblock \bibinfo{title}{{Simulating the Common Envelope Phase of a Red Giant
  Using Smoothed-particle Hydrodynamics and Uniform-grid Codes}},
\newblock \bibinfo{journal}{\apj} \bibinfo{volume}{744} (\bibinfo{year}{2012})
  \bibinfo{pages}{52}. \DOIprefix\doi{10.1088/0004-637X/744/1/52}.
\bibitem[{{Ivanova} et~al.(2013){Ivanova}, {Justham}, {Chen}, {De Marco},
  {Fryer}, {Gaburov}, {Ge}, {Glebbeek}, {Han}, {Li}, {Lu}, {Marsh},
  {Podsiadlowski}, {Potter}, {Soker}, {Taam}, {Tauris}, {van den Heuvel}, and
  {Webbink}}]{Ivanova2013a}
\bibinfo{author}{N.~{Ivanova}}, \bibinfo{author}{S.~{Justham}},
  \bibinfo{author}{X.~{Chen}}, et~al.,
\newblock \bibinfo{title}{{Common envelope evolution: where we stand and how we
  can move forward}},
\newblock \bibinfo{journal}{\aapr} \bibinfo{volume}{21} (\bibinfo{year}{2013})
  \bibinfo{pages}{59}. \DOIprefix\doi{10.1007/s00159-013-0059-2}.
\bibitem[{{Nandez} and {Ivanova}(2016)}]{Nandez}
\bibinfo{author}{J.~L.~A. {Nandez}}, \bibinfo{author}{N.~{Ivanova}},
\newblock \bibinfo{title}{{Common envelope events with low-mass giants:
  understanding the energy budget}},
\newblock \bibinfo{journal}{\mnras} \bibinfo{volume}{460}
  (\bibinfo{year}{2016}) \bibinfo{pages}{3992--4002}.
  \DOIprefix\doi{10.1093/mnras/stw1266}.
\bibitem[{{Ivanova} et~al.(2015){Ivanova}, {Justham}, and
  {Podsiadlowski}}]{Ivanova2015MNRAS.447.2181I}
\bibinfo{author}{N.~{Ivanova}}, \bibinfo{author}{S.~{Justham}},
  \bibinfo{author}{P.~{Podsiadlowski}},
\newblock \bibinfo{title}{{On the role of recombination in common-envelope
  ejections}},
\newblock \bibinfo{journal}{\mnras} \bibinfo{volume}{447}
  (\bibinfo{year}{2015}) \bibinfo{pages}{2181--2197}.
  \DOIprefix\doi{10.1093/mnras/stu2582}.
\bibitem[{{Ricker} and {Taam}(2012)}]{Ricker2012ApJ...746...74R}
\bibinfo{author}{P.~M. {Ricker}}, \bibinfo{author}{R.~E. {Taam}},
\newblock \bibinfo{title}{{An AMR Study of the Common-envelope Phase of Binary
  Evolution}},
\newblock \bibinfo{journal}{\apj} \bibinfo{volume}{746} (\bibinfo{year}{2012})
  \bibinfo{pages}{74}. \DOIprefix\doi{10.1088/0004-637X/746/1/74}.
\bibitem[{{Nandez} and {Ivanova}(2016)}]{Nandez2016MNRAS.460.3992N}
\bibinfo{author}{J.~L.~A. {Nandez}}, \bibinfo{author}{N.~{Ivanova}},
\newblock \bibinfo{title}{{Common envelope events with low-mass giants:
  understanding the energy budget}},
\newblock \bibinfo{journal}{\mnras} \bibinfo{volume}{460}
  (\bibinfo{year}{2016}) \bibinfo{pages}{3992--4002}.
  \DOIprefix\doi{10.1093/mnras/stw1266}.
\bibitem[{{Lau} et~al.(2022){Lau}, {Hirai}, {Gonz{\'a}lez-Bol{\'\i}var},
  {Price}, {De Marco}, and {Mandel}}]{lau2022}
\bibinfo{author}{M.~Y.~M. {Lau}}, \bibinfo{author}{R.~{Hirai}},
  \bibinfo{author}{M.~{Gonz{\'a}lez-Bol{\'\i}var}}, et~al.,
\newblock \bibinfo{title}{{Common envelopes in massive stars: towards the role
  of radiation pressure and recombination energy in ejecting red supergiant
  envelopes}},
\newblock \bibinfo{journal}{\mnras} \bibinfo{volume}{512}
  (\bibinfo{year}{2022}) \bibinfo{pages}{5462--5480}.
  \DOIprefix\doi{10.1093/mnras/stac049}.
\bibitem[{{Moreno} et~al.(2022){Moreno}, {Schneider}, {R{\"o}pke}, {Ohlmann},
  {Pakmor}, {Podsiadlowski}, and {Sand}}]{moreno2022}
\bibinfo{author}{M.~M. {Moreno}}, \bibinfo{author}{F.~R.~N. {Schneider}},
  \bibinfo{author}{F.~K. {R{\"o}pke}}, et~al.,
\newblock \bibinfo{title}{{From 3D hydrodynamic simulations of common-envelope
  interaction to gravitational-wave mergers}},
\newblock \bibinfo{journal}{\aap} \bibinfo{volume}{667} (\bibinfo{year}{2022})
  \bibinfo{pages}{A72}. \DOIprefix\doi{10.1051/0004-6361/202142731}.
\bibitem[{{Ivanova} et~al.(2002){Ivanova}, {Podsiadlowski}, and
  {Spruit}}]{Ivanova2002MNRAS.334..819I}
\bibinfo{author}{N.~{Ivanova}}, \bibinfo{author}{P.~{Podsiadlowski}},
  \bibinfo{author}{H.~{Spruit}},
\newblock \bibinfo{title}{{Hydrodynamical simulations of the stream-core
  interaction in the slow merger of massive stars}},
\newblock \bibinfo{journal}{\mnras} \bibinfo{volume}{334}
  (\bibinfo{year}{2002}) \bibinfo{pages}{819--832}.
  \DOIprefix\doi{10.1046/j.1365-8711.2002.05543.x}.
\bibitem[{{Bondi} and {Hoyle}(1944)}]{Bondi1944MNRAS.104..273B}
\bibinfo{author}{H.~{Bondi}}, \bibinfo{author}{F.~{Hoyle}},
\newblock \bibinfo{title}{{On the mechanism of accretion by stars}},
\newblock \bibinfo{journal}{\mnras} \bibinfo{volume}{104}
  (\bibinfo{year}{1944}) \bibinfo{pages}{273}.
  \DOIprefix\doi{10.1093/mnras/104.5.273}.
\bibitem[{{Shiber} et~al.(2019){Shiber}, {Iaconi}, {De Marco}, and
  {Soker}}]{Shiber19}
\bibinfo{author}{S.~{Shiber}}, \bibinfo{author}{R.~{Iaconi}},
  \bibinfo{author}{O.~{De Marco}}, \bibinfo{author}{N.~{Soker}},
\newblock \bibinfo{title}{{Companion-launched jets and their effect on the
  dynamics of common envelope interaction simulations}},
\newblock \bibinfo{journal}{\mnras} \bibinfo{volume}{488}
  (\bibinfo{year}{2019}) \bibinfo{pages}{5615--5632}.
  \DOIprefix\doi{10.1093/mnras/stz2013}.
\bibitem[{{L{\'o}pez-C{\'a}mara} et~al.(2022){L{\'o}pez-C{\'a}mara}, {De
  Colle}, {Moreno M{\'e}ndez}, {Shiber}, and {Iaconi}}]{jet2022}
\bibinfo{author}{D.~{L{\'o}pez-C{\'a}mara}}, \bibinfo{author}{F.~{De Colle}},
  \bibinfo{author}{E.~{Moreno M{\'e}ndez}}, et~al.,
\newblock \bibinfo{title}{{Jets in common envelopes: a low-mass main-sequence
  star in a red giant}},
\newblock \bibinfo{journal}{\mnras} \bibinfo{volume}{513}
  (\bibinfo{year}{2022}) \bibinfo{pages}{3634--3645}.
  \DOIprefix\doi{10.1093/mnras/stac932}.
\bibitem[{{Clayton} et~al.(2017){Clayton}, {Podsiadlowski}, {Ivanova}, and
  {Justham}}]{clay17}
\bibinfo{author}{M.~{Clayton}}, \bibinfo{author}{P.~{Podsiadlowski}},
  \bibinfo{author}{N.~{Ivanova}}, \bibinfo{author}{S.~{Justham}},
\newblock \bibinfo{title}{{Episodic mass ejections from common-envelope
  objects}},
\newblock \bibinfo{journal}{\mnras} \bibinfo{volume}{470}
  (\bibinfo{year}{2017}) \bibinfo{pages}{1788--1808}.
  \DOIprefix\doi{10.1093/mnras/stx1290}.
\bibitem[{{Glanz} and {Perets}(2018)}]{glanz18}
\bibinfo{author}{H.~{Glanz}}, \bibinfo{author}{H.~B. {Perets}},
\newblock \bibinfo{title}{{Efficient common-envelope ejection through
  dust-driven winds}},
\newblock \bibinfo{journal}{\mnras} \bibinfo{volume}{478}
  (\bibinfo{year}{2018}) \bibinfo{pages}{L12--L17}.
  \DOIprefix\doi{10.1093/mnrasl/sly065}.
\bibitem[{{Prust} and {Chang}(2019)}]{prust19}
\bibinfo{author}{L.~J. {Prust}}, \bibinfo{author}{P.~{Chang}},
\newblock \bibinfo{title}{{Common envelope evolution on a moving mesh}},
\newblock \bibinfo{journal}{\mnras} \bibinfo{volume}{486}
  (\bibinfo{year}{2019}) \bibinfo{pages}{5809--5818}.
  \DOIprefix\doi{10.1093/mnras/stz1219}.
\bibitem[{{Reg{\H{o}}s} and {Tout}(1995)}]{Regos1995MNRAS.273..146R}
\bibinfo{author}{E.~{Reg{\H{o}}s}}, \bibinfo{author}{C.~A. {Tout}},
\newblock \bibinfo{title}{{The effect of magnetic fields in common-envelope
  evolution on the formation of cataclysmic variables}},
\newblock \bibinfo{journal}{\mnras} \bibinfo{volume}{273}
  (\bibinfo{year}{1995}) \bibinfo{pages}{146--156}.
  \DOIprefix\doi{10.1093/mnras/273.1.146}.
\bibitem[{{Nordhaus} et~al.(2011){Nordhaus}, {Wellons}, {Spiegel}, {Metzger},
  and {Blackman}}]{Nordhaus2011PNAS..108.3135N}
\bibinfo{author}{J.~{Nordhaus}}, \bibinfo{author}{S.~{Wellons}},
  \bibinfo{author}{D.~S. {Spiegel}}, et~al.,
\newblock \bibinfo{title}{{Formation of high-field magnetic white dwarfs from
  common envelopes}},
\newblock \bibinfo{journal}{Proceedings of the National Academy of Science}
  \bibinfo{volume}{108} (\bibinfo{year}{2011}) \bibinfo{pages}{3135--3140}.
  \DOIprefix\doi{10.1073/pnas.1015005108}.
\bibitem[{{Ohlmann} et~al.(2016){Ohlmann}, {R{\"o}pke}, {Pakmor}, {Springel},
  and {M{\"u}ller}}]{ohl2016b}
\bibinfo{author}{S.~T. {Ohlmann}}, \bibinfo{author}{F.~K. {R{\"o}pke}},
  \bibinfo{author}{R.~{Pakmor}}, et~al.,
\newblock \bibinfo{title}{{Magnetic field amplification during the common
  envelope phase}},
\newblock \bibinfo{journal}{\mnras} \bibinfo{volume}{462}
  (\bibinfo{year}{2016}) \bibinfo{pages}{L121--L125}.
  \DOIprefix\doi{10.1093/mnrasl/slw144}.
\bibitem[{{Potter} and {Tout}(2010)}]{Potter2010MNRAS.402.1072P}
\bibinfo{author}{A.~T. {Potter}}, \bibinfo{author}{C.~A. {Tout}},
\newblock \bibinfo{title}{{Magnetic field evolution of white dwarfs in strongly
  interacting binary star systems}},
\newblock \bibinfo{journal}{\mnras} \bibinfo{volume}{402}
  (\bibinfo{year}{2010}) \bibinfo{pages}{1072--1080}.
  \DOIprefix\doi{10.1111/j.1365-2966.2009.15935.x}.
\bibitem[{{Ondratschek} et~al.(2022){Ondratschek}, {R{\"o}pke}, {Schneider},
  {Fendt}, {Sand}, {Ohlmann}, {Pakmor}, and {Springel}}]{ond2022}
\bibinfo{author}{P.~A. {Ondratschek}}, \bibinfo{author}{F.~K. {R{\"o}pke}},
  \bibinfo{author}{F.~R.~N. {Schneider}}, et~al.,
\newblock \bibinfo{title}{{Bipolar planetary nebulae from common-envelope
  evolution of binary stars}},
\newblock \bibinfo{journal}{\aap} \bibinfo{volume}{660} (\bibinfo{year}{2022})
  \bibinfo{pages}{L8}. \DOIprefix\doi{10.1051/0004-6361/202142478}.
\bibitem[{{Ivanova} et~al.(2013){Ivanova}, {Justham}, {Avendano Nandez}, and
  {Lombardi}}]{Ivanova2013Sci...339..433I}
\bibinfo{author}{N.~{Ivanova}}, \bibinfo{author}{S.~{Justham}},
  \bibinfo{author}{J.~L. {Avendano Nandez}}, \bibinfo{author}{J.~C.
  {Lombardi}},
\newblock \bibinfo{title}{{Identification of the Long-Sought Common-Envelope
  Events}},
\newblock \bibinfo{journal}{Science} \bibinfo{volume}{339}
  (\bibinfo{year}{2013}) \bibinfo{pages}{433}.
  \DOIprefix\doi{10.1126/science.1225540}.
\bibitem[{{Popov}(1993)}]{Popov1993ApJ...414..712P}
\bibinfo{author}{D.~V. {Popov}},
\newblock \bibinfo{title}{{An Analytical Model for the Plateau Stage of Type II
  Supernovae}},
\newblock \bibinfo{journal}{\apj} \bibinfo{volume}{414} (\bibinfo{year}{1993})
  \bibinfo{pages}{712}. \DOIprefix\doi{10.1086/173117}.
\bibitem[{{Pastorello} et~al.(2019){Pastorello}, {Mason}, {Taubenberger},
  {Fraser}, {Cortini}, {Tomasella}, {Botticella}, {Elias-Rosa}, {Kotak},
  {Smartt}, {Benetti}, {Cappellaro}, {Turatto}, {Tartaglia}, {Djorgovski},
  {Drake}, {Berton}, {Briganti}, {Brimacombe}, {Bufano}, {Cai}, {Chen},
  {Christensen}, {Ciabattari}, {Congiu}, {Dimai}, {Inserra}, {Kankare},
  {Magill}, {Maguire}, {Martinelli}, {Morales-Garoffolo}, {Ochner}, {Pignata},
  {Reguitti}, {Sollerman}, {Spiro}, {Terreran}, and {Wright}}]{Pastorello2019}
\bibinfo{author}{A.~{Pastorello}}, \bibinfo{author}{E.~{Mason}},
  \bibinfo{author}{S.~{Taubenberger}}, et~al.,
\newblock \bibinfo{title}{{Luminous red novae: Stellar mergers or giant
  eruptions?}},
\newblock \bibinfo{journal}{\aap} \bibinfo{volume}{630} (\bibinfo{year}{2019})
  \bibinfo{pages}{A75}. \DOIprefix\doi{10.1051/0004-6361/201935999}.
\bibitem[{{Cai} et~al.(2019){Cai}, {Pastorello}, {Fraser}, {Prentice},
  {Reynolds}, {Cappellaro}, {Benetti}, {Morales-Garoffolo}, {Reguitti},
  {Elias-Rosa}, {Brennan}, {Callis}, {Cannizzaro}, {Fiore}, {Gromadzki},
  {Galindo-Guil}, {Gall}, {Heikkil{\"a}}, {Mason}, {Moran}, {Onori},
  {Sagu{\'e}s Carracedo}, and {Valerin}}]{Cai2019}
\bibinfo{author}{Y.~Z. {Cai}}, \bibinfo{author}{A.~{Pastorello}},
  \bibinfo{author}{M.~{Fraser}}, et~al.,
\newblock \bibinfo{title}{{The transitional gap transient AT 2018hso: new
  insights into the luminous red nova phenomenon}},
\newblock \bibinfo{journal}{\aap} \bibinfo{volume}{632} (\bibinfo{year}{2019})
  \bibinfo{pages}{L6}. \DOIprefix\doi{10.1051/0004-6361/201936749}.
\bibitem[{{Matsumoto} and {Metzger}(2022{\natexlab{a}})}]{Matsumoto2022}
\bibinfo{author}{T.~{Matsumoto}}, \bibinfo{author}{B.~D. {Metzger}},
\newblock \bibinfo{title}{{Light-curve Model for Luminous Red Novae and
  Inferences about the Ejecta of Stellar Mergers}},
\newblock \bibinfo{journal}{\apj} \bibinfo{volume}{938}
  (\bibinfo{year}{2022}{\natexlab{a}}) \bibinfo{pages}{5}.
  \DOIprefix\doi{10.3847/1538-4357/ac6269}.
\bibitem[{{Matsumoto} and
  {Metzger}(2022{\natexlab{b}})}]{Matsumoto2022ApJ...938....5M}
\bibinfo{author}{T.~{Matsumoto}}, \bibinfo{author}{B.~D. {Metzger}},
\newblock \bibinfo{title}{{Light-curve Model for Luminous Red Novae and
  Inferences about the Ejecta of Stellar Mergers}},
\newblock \bibinfo{journal}{\apj} \bibinfo{volume}{938}
  (\bibinfo{year}{2022}{\natexlab{b}}) \bibinfo{pages}{5}.
  \DOIprefix\doi{10.3847/1538-4357/ac6269}.
\bibitem[{{Cai} et~al.(2022){Cai}, {Pastorello}, {Fraser}, {Wang},
  {Filippenko}, {Reguitti}, {Patra}, {Goranskij}, {Barsukova}, {Brink},
  {Elias-Rosa}, and et~al.}]{Cai2022A&A...667A...4C}
\bibinfo{author}{Y.~Z. {Cai}}, \bibinfo{author}{A.~{Pastorello}},
  \bibinfo{author}{M.~{Fraser}}, et~al.,
\newblock \bibinfo{title}{{Forbidden hugs in pandemic times. III. Observations
  of the luminous red nova AT 2021biy in the nearby galaxy NGC 4631}},
\newblock \bibinfo{journal}{\aap} \bibinfo{volume}{667} (\bibinfo{year}{2022})
  \bibinfo{pages}{A4}. \DOIprefix\doi{10.1051/0004-6361/202244393}.
\bibitem[{{Hoadley} et~al.(2020){Hoadley}, {Martin}, {Metzger}, {Seibert},
  {McWilliam}, {Shen}, {Neill}, {Stefansson}, {Monson}, and
  {Schaefer}}]{Hoadley2020Natur.587..387H}
\bibinfo{author}{K.~{Hoadley}}, \bibinfo{author}{D.~C. {Martin}},
  \bibinfo{author}{B.~D. {Metzger}}, et~al.,
\newblock \bibinfo{title}{{A blue ring nebula from a stellar merger several
  thousand years ago}},
\newblock \bibinfo{journal}{\nat} \bibinfo{volume}{587} (\bibinfo{year}{2020})
  \bibinfo{pages}{387--391}. \DOIprefix\doi{10.1038/s41586-020-2893-5}.
\bibitem[{{Refsdal} et~al.(1974){Refsdal}, {Roth}, and
  {Weigert}}]{Refsdal1974A&A....36..113R}
\bibinfo{author}{S.~{Refsdal}}, \bibinfo{author}{M.~L. {Roth}},
  \bibinfo{author}{A.~{Weigert}},
\newblock \bibinfo{title}{{On the binary system AS Eri.}},
\newblock \bibinfo{journal}{\aap}  \bibinfo{volume}{36} (\bibinfo{year}{1974})
  \bibinfo{pages}{113--122}.
\bibitem[{{Warner}(1978)}]{Warner1978AcA....28..303W}
\bibinfo{author}{B.~{Warner}},
\newblock \bibinfo{title}{{Apsidal motion and evolution of cataclysmic
  variables.}},
\newblock \bibinfo{journal}{\actaa}  \bibinfo{volume}{28}
  (\bibinfo{year}{1978}) \bibinfo{pages}{303--326}.
\bibitem[{{Eggleton}(2000)}]{Eggleton2000NewAR..44..111E}
\bibinfo{author}{P.~P. {Eggleton}},
\newblock \bibinfo{title}{{New labour on Algols: conservative or liberal?}},
\newblock \bibinfo{journal}{\nar} \bibinfo{volume}{44} (\bibinfo{year}{2000})
  \bibinfo{pages}{111--117}. \DOIprefix\doi{10.1016/S1387-6473(00)00023-3}.
\bibitem[{{Sytov} et~al.(2007){Sytov}, {Kaigorodov}, {Bisikalo}, {Kuznetsov},
  and {Boyarchuk}}]{Sytov2007ARep...51..836S}
\bibinfo{author}{A.~Y. {Sytov}}, \bibinfo{author}{P.~V. {Kaigorodov}},
  \bibinfo{author}{D.~V. {Bisikalo}}, et~al.,
\newblock \bibinfo{title}{{The mechanism of circumbinary envelope formation in
  close binaries}},
\newblock \bibinfo{journal}{Astronomy Reports} \bibinfo{volume}{51}
  (\bibinfo{year}{2007}) \bibinfo{pages}{836--846}.
  \DOIprefix\doi{10.1134/S1063772907100083}.
\bibitem[{{Nanouris} et~al.(2015){Nanouris}, {Kalimeris}, {Antonopoulou}, and
  {Rovithis-Livaniou}}]{Nanouris2015A&A...575A..64N}
\bibinfo{author}{N.~{Nanouris}}, \bibinfo{author}{A.~{Kalimeris}},
  \bibinfo{author}{E.~{Antonopoulou}},
  \bibinfo{author}{H.~{Rovithis-Livaniou}},
\newblock \bibinfo{title}{{Efficiency of ETV diagrams as diagnostic tools for
  long-term period variations. II. Non-conservative mass transfer, and
  gravitational radiation}},
\newblock \bibinfo{journal}{\aap} \bibinfo{volume}{575} (\bibinfo{year}{2015})
  \bibinfo{pages}{A64}. \DOIprefix\doi{10.1051/0004-6361/201323136}.
\bibitem[{{De Marco} and {Izzard}(2017)}]{De-Marco2017PASA...34....1D}
\bibinfo{author}{O.~{De Marco}}, \bibinfo{author}{R.~G. {Izzard}},
\newblock \bibinfo{title}{{Dawes Review 6: The Impact of Companions on Stellar
  Evolution}},
\newblock \bibinfo{journal}{\pasa} \bibinfo{volume}{34} (\bibinfo{year}{2017})
  \bibinfo{pages}{e001}. \DOIprefix\doi{10.1017/pasa.2016.52}.
\bibitem[{{Lu} et~al.(2023){Lu}, {Fuller}, {Quataert}, and
  {Bonnerot}}]{Lu2023MNRAS.519.1409L}
\bibinfo{author}{W.~{Lu}}, \bibinfo{author}{J.~{Fuller}},
  \bibinfo{author}{E.~{Quataert}}, \bibinfo{author}{C.~{Bonnerot}},
\newblock \bibinfo{title}{{On rapid binary mass transfer - I. Physical model}},
\newblock \bibinfo{journal}{\mnras} \bibinfo{volume}{519}
  (\bibinfo{year}{2023}) \bibinfo{pages}{1409--1424}.
  \DOIprefix\doi{10.1093/mnras/stac3621}.
\bibitem[{{van Rensbergen} et~al.(2006){van Rensbergen}, {De Loore}, and
  {Jansen}}]{van-Rensbergen2006A&A...446.1071V}
\bibinfo{author}{W.~{van Rensbergen}}, \bibinfo{author}{C.~{De Loore}},
  \bibinfo{author}{K.~{Jansen}},
\newblock \bibinfo{title}{{Evolution of interacting binaries with a B type
  primary at birth}},
\newblock \bibinfo{journal}{\aap} \bibinfo{volume}{446} (\bibinfo{year}{2006})
  \bibinfo{pages}{1071--1079}. \DOIprefix\doi{10.1051/0004-6361:20053543}.
\bibitem[{{van Rensbergen} et~al.(2008){van Rensbergen}, {De Greve}, {De
  Loore}, and {Mennekens}}]{van-Rensbergen2008A&A...487.1129V}
\bibinfo{author}{W.~{van Rensbergen}}, \bibinfo{author}{J.~P. {De Greve}},
  \bibinfo{author}{C.~{De Loore}}, \bibinfo{author}{N.~{Mennekens}},
\newblock \bibinfo{title}{{Spin-up and hot spots can drive mass out of a
  binary}},
\newblock \bibinfo{journal}{\aap} \bibinfo{volume}{487} (\bibinfo{year}{2008})
  \bibinfo{pages}{1129--1138}. \DOIprefix\doi{10.1051/0004-6361:200809943}.
\bibitem[{{van Rensbergen} et~al.(2011){van Rensbergen}, {de Greve},
  {Mennekens}, {Jansen}, and {de Loore}}]{van-Rensbergen2011A&A...528A..16V}
\bibinfo{author}{W.~{van Rensbergen}}, \bibinfo{author}{J.~P. {de Greve}},
  \bibinfo{author}{N.~{Mennekens}}, et~al.,
\newblock \bibinfo{title}{{Mass loss out of close binaries. The formation of
  Algol-type systems, completed with case B RLOF}},
\newblock \bibinfo{journal}{\aap} \bibinfo{volume}{528} (\bibinfo{year}{2011})
  \bibinfo{pages}{A16}. \DOIprefix\doi{10.1051/0004-6361/201015596}.
\bibitem[{{Erdem} and {{\"O}zt{\"u}rk}(2014)}]{Erdem2014MNRAS.441.1166E}
\bibinfo{author}{A.~{Erdem}}, \bibinfo{author}{O.~{{\"O}zt{\"u}rk}},
\newblock \bibinfo{title}{{Non-conservative mass transfers in Algols}},
\newblock \bibinfo{journal}{\mnras} \bibinfo{volume}{441}
  (\bibinfo{year}{2014}) \bibinfo{pages}{1166--1176}.
  \DOIprefix\doi{10.1093/mnras/stu630}.
\bibitem[{{Marino} et~al.(2019){Marino}, {Di Salvo}, {Burderi}, {Sanna},
  {Riggio}, {Papitto}, {Del Santo}, {Gambino}, {Iaria}, and
  {Mazzola}}]{Marino2019A&A...627A.125M}
\bibinfo{author}{A.~{Marino}}, \bibinfo{author}{T.~{Di Salvo}},
  \bibinfo{author}{L.~{Burderi}}, et~al.,
\newblock \bibinfo{title}{{Indications of non-conservative mass transfer in
  AMXPs}},
\newblock \bibinfo{journal}{\aap} \bibinfo{volume}{627} (\bibinfo{year}{2019})
  \bibinfo{pages}{A125}. \DOIprefix\doi{10.1051/0004-6361/201834460}.
\bibitem[{{Iaria} et~al.(2021){Iaria}, {Sanna}, {Di Salvo}, {Gambino},
  {Mazzola}, {Riggio}, {Marino}, and {Burderi}}]{Iaria2021A&A...646A.120I}
\bibinfo{author}{R.~{Iaria}}, \bibinfo{author}{A.~{Sanna}},
  \bibinfo{author}{T.~{Di Salvo}}, et~al.,
\newblock \bibinfo{title}{{Evidence of a non-conservative mass transfer in the
  ultra-compact X-ray source XB 1916-053}},
\newblock \bibinfo{journal}{\aap} \bibinfo{volume}{646} (\bibinfo{year}{2021})
  \bibinfo{pages}{A120}. \DOIprefix\doi{10.1051/0004-6361/202039225}.
\bibitem[{{Sarna}(1993)}]{Sarna1993MNRAS.262..534S}
\bibinfo{author}{M.~J. {Sarna}},
\newblock \bibinfo{title}{{The evolutionary status of beta Per.}},
\newblock \bibinfo{journal}{\mnras} \bibinfo{volume}{262}
  (\bibinfo{year}{1993}) \bibinfo{pages}{534--542}.
  \DOIprefix\doi{10.1093/mnras/262.2.534}.
\bibitem[{{Abbott} et~al.(2017){Abbott}, {Abbott}, {Abbott}, {Acernese},
  {Ackley}, {Adams}, {Adams}, {Addesso}, {Adhikari}, and
  et~al.}]{Abbott2017PhRvL.119p1101A}
\bibinfo{author}{B.~P. {Abbott}}, \bibinfo{author}{R.~{Abbott}},
  \bibinfo{author}{T.~D. {Abbott}}, et~al.,
\newblock \bibinfo{title}{{GW170817: Observation of Gravitational Waves from a
  Binary Neutron Star Inspiral}},
\newblock \bibinfo{journal}{\prl} \bibinfo{volume}{119} (\bibinfo{year}{2017})
  \bibinfo{pages}{161101}.
  \DOIprefix\doi{10.1103/PhysRevLett.119.16110110.48550/arXiv.1710.05832}.
\bibitem[{{Landau} and {Lifshitz}(1975)}]{Landau1975ctf..book.....L}
\bibinfo{author}{L.~D. {Landau}}, \bibinfo{author}{E.~M. {Lifshitz}},
  \bibinfo{title}{{The classical theory of fields}},  \bibinfo{year}{1975}.
\bibitem[{{Skumanich}(1972)}]{Skumanich1972ApJ...171..565S}
\bibinfo{author}{A.~{Skumanich}},
\newblock \bibinfo{title}{{Time Scales for Ca II Emission Decay, Rotational
  Braking, and Lithium Depletion}},
\newblock \bibinfo{journal}{\apj} \bibinfo{volume}{171} (\bibinfo{year}{1972})
  \bibinfo{pages}{565}. \DOIprefix\doi{10.1086/151310}.
\bibitem[{{Verbunt} and {Zwaan}(1981)}]{Verbunt1981A&A...100L...7V}
\bibinfo{author}{F.~{Verbunt}}, \bibinfo{author}{C.~{Zwaan}},
\newblock \bibinfo{title}{{Magnetic braking in low-mass X-ray binaries.}},
\newblock \bibinfo{journal}{\aap}  \bibinfo{volume}{100} (\bibinfo{year}{1981})
  \bibinfo{pages}{L7--L9}.
\bibitem[{{Rappaport} et~al.(1983){Rappaport}, {Verbunt}, and
  {Joss}}]{Rappaport1983ApJ...275..713R}
\bibinfo{author}{S.~{Rappaport}}, \bibinfo{author}{F.~{Verbunt}},
  \bibinfo{author}{P.~C. {Joss}},
\newblock \bibinfo{title}{{A new technique for calculations of binary stellar
  evolution application to magnetic braking.}},
\newblock \bibinfo{journal}{\apj} \bibinfo{volume}{275} (\bibinfo{year}{1983})
  \bibinfo{pages}{713--731}. \DOIprefix\doi{10.1086/161569}.
\bibitem[{{Van} and {Ivanova}(2019)}]{Van2019ApJ...886L..31V}
\bibinfo{author}{K.~X. {Van}}, \bibinfo{author}{N.~{Ivanova}},
\newblock \bibinfo{title}{{Evolving LMXBs: CARB Magnetic Braking}},
\newblock \bibinfo{journal}{\apjl} \bibinfo{volume}{886} (\bibinfo{year}{2019})
  \bibinfo{pages}{L31}. \DOIprefix\doi{10.3847/2041-8213/ab571c}.
\bibitem[{{Carroll} and {Ostlie}(2006)}]{Carroll2006ima..book.....C}
\bibinfo{author}{B.~W. {Carroll}}, \bibinfo{author}{D.~A. {Ostlie}},
  \bibinfo{title}{{An introduction to modern astrophysics and cosmology}},
  \bibinfo{year}{2006}.
\bibitem[{{Van} et~al.(2019){Van}, {Ivanova}, and
  {Heinke}}]{Van2019MNRAS.483.5595V}
\bibinfo{author}{K.~X. {Van}}, \bibinfo{author}{N.~{Ivanova}},
  \bibinfo{author}{C.~O. {Heinke}},
\newblock \bibinfo{title}{{Low-mass X-ray binaries: the effects of the magnetic
  braking prescription}},
\newblock \bibinfo{journal}{\mnras} \bibinfo{volume}{483}
  (\bibinfo{year}{2019}) \bibinfo{pages}{5595--5613}.
  \DOIprefix\doi{10.1093/mnras/sty3489}.
\bibitem[{{Matt} and {Pudritz}(2008)}]{Matt2008ApJ...678.1109M}
\bibinfo{author}{S.~{Matt}}, \bibinfo{author}{R.~E. {Pudritz}},
\newblock \bibinfo{title}{{Accretion-powered Stellar Winds. II. Numerical
  Solutions for Stellar Wind Torques}},
\newblock \bibinfo{journal}{\apj} \bibinfo{volume}{678} (\bibinfo{year}{2008})
  \bibinfo{pages}{1109--1118}. \DOIprefix\doi{10.1086/533428}.
\bibitem[{{Matt} et~al.(2012){Matt}, {MacGregor}, {Pinsonneault}, and
  {Greene}}]{Matt2012ApJ...754L..26M}
\bibinfo{author}{S.~P. {Matt}}, \bibinfo{author}{K.~B. {MacGregor}},
  \bibinfo{author}{M.~H. {Pinsonneault}}, \bibinfo{author}{T.~P. {Greene}},
\newblock \bibinfo{title}{{Magnetic Braking Formulation for Sun-like Stars:
  Dependence on Dipole Field Strength and Rotation Rate}},
\newblock \bibinfo{journal}{\apjl} \bibinfo{volume}{754} (\bibinfo{year}{2012})
  \bibinfo{pages}{L26}. \DOIprefix\doi{10.1088/2041-8205/754/2/L26}.
\bibitem[{{Reiners} and {Mohanty}(2012)}]{Reiners2012ApJ...746...43R}
\bibinfo{author}{A.~{Reiners}}, \bibinfo{author}{S.~{Mohanty}},
\newblock \bibinfo{title}{{Radius-dependent Angular Momentum Evolution in
  Low-mass Stars. I}},
\newblock \bibinfo{journal}{\apj} \bibinfo{volume}{746} (\bibinfo{year}{2012})
  \bibinfo{pages}{43}. \DOIprefix\doi{10.1088/0004-637X/746/1/43}.
\bibitem[{{Deng} et~al.(2021){Deng}, {Li}, {Gao}, and
  {Shao}}]{Deng2021ApJ...909..174D}
\bibinfo{author}{Z.-L. {Deng}}, \bibinfo{author}{X.-D. {Li}},
  \bibinfo{author}{Z.-F. {Gao}}, \bibinfo{author}{Y.~{Shao}},
\newblock \bibinfo{title}{{Evolution of LMXBs under Different Magnetic Braking
  Prescriptions}},
\newblock \bibinfo{journal}{\apj} \bibinfo{volume}{909} (\bibinfo{year}{2021})
  \bibinfo{pages}{174}. \DOIprefix\doi{10.3847/1538-4357/abe0b2}.
\bibitem[{{Chen} et~al.(2021){Chen}, {Tauris}, {Han}, and
  {Chen}}]{Chen2021MNRAS.503.3540C}
\bibinfo{author}{H.-L. {Chen}}, \bibinfo{author}{T.~M. {Tauris}},
  \bibinfo{author}{Z.~{Han}}, \bibinfo{author}{X.~{Chen}},
\newblock \bibinfo{title}{{Formation of millisecond pulsars with helium white
  dwarfs, ultra-compact X-ray binaries, and gravitational wave sources}},
\newblock \bibinfo{journal}{\mnras} \bibinfo{volume}{503}
  (\bibinfo{year}{2021}) \bibinfo{pages}{3540--3551}.
  \DOIprefix\doi{10.1093/mnras/stab670}.
\bibitem[{{van der Sluys} et~al.(2005{\natexlab{a}}){van der Sluys}, {Verbunt},
  and {Pols}}]{van-der-Sluys2005A&A...431..647V}
\bibinfo{author}{M.~V. {van der Sluys}}, \bibinfo{author}{F.~{Verbunt}},
  \bibinfo{author}{O.~R. {Pols}},
\newblock \bibinfo{title}{{Creating ultra-compact binaries in globular clusters
  through stable mass transfer}},
\newblock \bibinfo{journal}{\aap} \bibinfo{volume}{431}
  (\bibinfo{year}{2005}{\natexlab{a}}) \bibinfo{pages}{647--658}.
  \DOIprefix\doi{10.1051/0004-6361:20041777}.
\bibitem[{{van der Sluys} et~al.(2005{\natexlab{b}}){van der Sluys}, {Verbunt},
  and {Pols}}]{van-der-Sluys2005A&A...440..973V}
\bibinfo{author}{M.~V. {van der Sluys}}, \bibinfo{author}{F.~{Verbunt}},
  \bibinfo{author}{O.~R. {Pols}},
\newblock \bibinfo{title}{{Reduced magnetic braking and the magnetic capture
  model for the formation of ultra-compact binaries}},
\newblock \bibinfo{journal}{\aap} \bibinfo{volume}{440}
  (\bibinfo{year}{2005}{\natexlab{b}}) \bibinfo{pages}{973--979}.
  \DOIprefix\doi{10.1051/0004-6361:20052696}.
\bibitem[{{Istrate} et~al.(2014){Istrate}, {Tauris}, and
  {Langer}}]{Istrate2014A&A...571A..45I}
\bibinfo{author}{A.~G. {Istrate}}, \bibinfo{author}{T.~M. {Tauris}},
  \bibinfo{author}{N.~{Langer}},
\newblock \bibinfo{title}{{The formation of low-mass helium white dwarfs
  orbiting pulsars . Evolution of low-mass X-ray binaries below the bifurcation
  period}},
\newblock \bibinfo{journal}{\aap} \bibinfo{volume}{571} (\bibinfo{year}{2014})
  \bibinfo{pages}{A45}. \DOIprefix\doi{10.1051/0004-6361/201424680}.
\bibitem[{{Soethe} and {Kepler}(2021)}]{Soethe2021MNRAS.506.3266S}
\bibinfo{author}{L.~T.~T. {Soethe}}, \bibinfo{author}{S.~O. {Kepler}},
\newblock \bibinfo{title}{{Convection and rotation boosted prescription of
  magnetic braking: application to the formation of extremely low-mass white
  dwarfs}},
\newblock \bibinfo{journal}{\mnras} \bibinfo{volume}{506}
  (\bibinfo{year}{2021}) \bibinfo{pages}{3266--3281}.
  \DOIprefix\doi{10.1093/mnras/stab1916}.
\bibitem[{{Li} et~al.(2021){Li}, {Chen}, {Chen}, and
  {Han}}]{Li2021ApJ...922..158L}
\bibinfo{author}{Z.~{Li}}, \bibinfo{author}{X.~{Chen}}, \bibinfo{author}{H.-L.
  {Chen}}, \bibinfo{author}{Z.~{Han}},
\newblock \bibinfo{title}{{The Maximum Accreted Mass of Recycled Pulsars}},
\newblock \bibinfo{journal}{\apj} \bibinfo{volume}{922} (\bibinfo{year}{2021})
  \bibinfo{pages}{158}. \DOIprefix\doi{10.3847/1538-4357/ac1b2e}.
\bibitem[{{Mandel} and {de Mink}(2016)}]{Mandel2016}
\bibinfo{author}{I.~{Mandel}}, \bibinfo{author}{S.~E. {de Mink}},
\newblock \bibinfo{title}{{Merging binary black holes formed through chemically
  homogeneous evolution in short-period stellar binaries}},
\newblock \bibinfo{journal}{\mnras} \bibinfo{volume}{458}
  (\bibinfo{year}{2016}) \bibinfo{pages}{2634--2647}.
  \DOIprefix\doi{10.1093/mnras/stw379}.
\bibitem[{{de Mink} et~al.(2008){de Mink}, {Cantiello}, {Langer}, {Yoon},
  {Brott}, {Glebbeek}, {Verkoulen}, and {Pols}}]{deMink2008}
\bibinfo{author}{S.~E. {de Mink}}, \bibinfo{author}{M.~{Cantiello}},
  \bibinfo{author}{N.~{Langer}}, et~al.,
\newblock \bibinfo{title}{{Rotational mixing in close binaries}},
\newblock in: \bibinfo{editor}{L.~{Deng}}, \bibinfo{editor}{K.~L. {Chan}}
  (Eds.), \bibinfo{booktitle}{The Art of Modeling Stars in the 21st Century},
  volume \bibinfo{volume}{252}, \bibinfo{year}{2008}, pp.
  \bibinfo{pages}{365--370}. \DOIprefix\doi{10.1017/S1743921308023223}.
\bibitem[{{Abbott} et~al.(2016){Abbott}, {Abbott}, {Abbott}, {Abernathy},
  {Acernese}, {Ackley}, {Adams}, {Adams}, {Addesso}, and
  {Adhikari}}]{Abbott2016ApJ...832L..21A}
\bibinfo{author}{B.~P. {Abbott}}, \bibinfo{author}{R.~{Abbott}},
  \bibinfo{author}{T.~D. {Abbott}}, et~al.,
\newblock \bibinfo{title}{{Upper Limits on the Rates of Binary Neutron Star and
  Neutron Star-Black Hole Mergers from Advanced LIGO{\textquoteright}s First
  Observing Run}},
\newblock \bibinfo{journal}{\apjl} \bibinfo{volume}{832} (\bibinfo{year}{2016})
  \bibinfo{pages}{L21}.
  \DOIprefix\doi{10.3847/2041-8205/832/2/L2110.48550/arXiv.1607.07456}.
\bibitem[{{Belczynski} et~al.(2010){Belczynski}, {Dominik}, {Bulik},
  {O'Shaughnessy}, {Fryer}, and {Holz}}]{Belczynski2010ApJ...715L.138B}
\bibinfo{author}{K.~{Belczynski}}, \bibinfo{author}{M.~{Dominik}},
  \bibinfo{author}{T.~{Bulik}}, et~al.,
\newblock \bibinfo{title}{{The Effect of Metallicity on the Detection Prospects
  for Gravitational Waves}},
\newblock \bibinfo{journal}{\apjl} \bibinfo{volume}{715} (\bibinfo{year}{2010})
  \bibinfo{pages}{L138--L141}. \DOIprefix\doi{10.1088/2041-8205/715/2/L138}.
\bibitem[{{Spera} et~al.(2015){Spera}, {Mapelli}, and
  {Bressan}}]{Spera2015MNRAS.451.4086S}
\bibinfo{author}{M.~{Spera}}, \bibinfo{author}{M.~{Mapelli}},
  \bibinfo{author}{A.~{Bressan}},
\newblock \bibinfo{title}{{The mass spectrum of compact remnants from the
  PARSEC stellar evolution tracks}},
\newblock \bibinfo{journal}{\mnras} \bibinfo{volume}{451}
  (\bibinfo{year}{2015}) \bibinfo{pages}{4086--4103}.
  \DOIprefix\doi{10.1093/mnras/stv1161}.
\bibitem[{{Rodriguez} et~al.(2016){Rodriguez}, {Haster}, {Chatterjee},
  {Kalogera}, and {Rasio}}]{Rodriguez2016ApJ...824L...8R}
\bibinfo{author}{C.~L. {Rodriguez}}, \bibinfo{author}{C.-J. {Haster}},
  \bibinfo{author}{S.~{Chatterjee}}, et~al.,
\newblock \bibinfo{title}{{Dynamical Formation of the GW150914 Binary Black
  Hole}},
\newblock \bibinfo{journal}{\apjl} \bibinfo{volume}{824} (\bibinfo{year}{2016})
  \bibinfo{pages}{L8}.
  \DOIprefix\doi{10.3847/2041-8205/824/1/L810.48550/arXiv.1604.04254}.
\bibitem[{{O'Leary} et~al.(2016){O'Leary}, {Meiron}, and
  {Kocsis}}]{OLeary2016ApJ...824L..12O}
\bibinfo{author}{R.~M. {O'Leary}}, \bibinfo{author}{Y.~{Meiron}},
  \bibinfo{author}{B.~{Kocsis}},
\newblock \bibinfo{title}{{Dynamical Formation Signatures of Black Hole
  Binaries in the First Detected Mergers by LIGO}},
\newblock \bibinfo{journal}{\apjl} \bibinfo{volume}{824} (\bibinfo{year}{2016})
  \bibinfo{pages}{L12}.
  \DOIprefix\doi{10.3847/2041-8205/824/1/L1210.48550/arXiv.1602.02809}.
\bibitem[{{Askar} et~al.(2017){Askar}, {Szkudlarek}, {Gondek-Rosi{\'n}ska},
  {Giersz}, and {Bulik}}]{Askar2017MNRAS.464L..36A}
\bibinfo{author}{A.~{Askar}}, \bibinfo{author}{M.~{Szkudlarek}},
  \bibinfo{author}{D.~{Gondek-Rosi{\'n}ska}}, et~al.,
\newblock \bibinfo{title}{{MOCCA-SURVEY Database - I. Coalescing binary black
  holes originating from globular clusters}},
\newblock \bibinfo{journal}{\mnras} \bibinfo{volume}{464}
  (\bibinfo{year}{2017}) \bibinfo{pages}{L36--L40}.
  \DOIprefix\doi{10.1093/mnrasl/slw17710.48550/arXiv.1608.02520}.
\bibitem[{{Belczynski} et~al.(2016){Belczynski}, {Holz}, {Bulik}, and
  {O'Shaughnessy}}]{Belczynski2016Natur.534..512B}
\bibinfo{author}{K.~{Belczynski}}, \bibinfo{author}{D.~E. {Holz}},
  \bibinfo{author}{T.~{Bulik}}, \bibinfo{author}{R.~{O'Shaughnessy}},
\newblock \bibinfo{title}{{The first gravitational-wave source from the
  isolated evolution of two stars in the 40-100 solar mass range}},
\newblock \bibinfo{journal}{\nat} \bibinfo{volume}{534} (\bibinfo{year}{2016})
  \bibinfo{pages}{512--515}.
  \DOIprefix\doi{10.1038/nature1832210.48550/arXiv.1602.04531}.
\bibitem[{{Eldridge} and {Stanway}(2016)}]{Eldridge2016MNRAS.462.3302E}
\bibinfo{author}{J.~J. {Eldridge}}, \bibinfo{author}{E.~R. {Stanway}},
\newblock \bibinfo{title}{{BPASS predictions for binary black hole mergers}},
\newblock \bibinfo{journal}{\mnras} \bibinfo{volume}{462}
  (\bibinfo{year}{2016}) \bibinfo{pages}{3302--3313}.
  \DOIprefix\doi{10.1093/mnras/stw177210.48550/arXiv.1602.03790}.
\bibitem[{{Marchant} et~al.(2016){Marchant}, {Langer}, {Podsiadlowski},
  {Tauris}, and {Moriya}}]{Marchant2016A&A...588A..50M}
\bibinfo{author}{P.~{Marchant}}, \bibinfo{author}{N.~{Langer}},
  \bibinfo{author}{P.~{Podsiadlowski}}, et~al.,
\newblock \bibinfo{title}{{A new route towards merging massive black holes}},
\newblock \bibinfo{journal}{\aap} \bibinfo{volume}{588} (\bibinfo{year}{2016})
  \bibinfo{pages}{A50}.
  \DOIprefix\doi{10.1051/0004-6361/20162813310.48550/arXiv.1601.03718}.
\bibitem[{{Mandel} and {de Mink}(2016)}]{Mandel2016MNRAS.458.2634M}
\bibinfo{author}{I.~{Mandel}}, \bibinfo{author}{S.~E. {de Mink}},
\newblock \bibinfo{title}{{Merging binary black holes formed through chemically
  homogeneous evolution in short-period stellar binaries}},
\newblock \bibinfo{journal}{\mnras} \bibinfo{volume}{458}
  (\bibinfo{year}{2016}) \bibinfo{pages}{2634--2647}.
  \DOIprefix\doi{10.1093/mnras/stw37910.48550/arXiv.1601.00007}.
\bibitem[{{de Mink} and {Mandel}(2016)}]{de-Mink2016MNRAS.460.3545D}
\bibinfo{author}{S.~E. {de Mink}}, \bibinfo{author}{I.~{Mandel}},
\newblock \bibinfo{title}{{The chemically homogeneous evolutionary channel for
  binary black hole mergers: rates and properties of gravitational-wave events
  detectable by advanced LIGO}},
\newblock \bibinfo{journal}{\mnras} \bibinfo{volume}{460}
  (\bibinfo{year}{2016}) \bibinfo{pages}{3545--3553}.
  \DOIprefix\doi{10.1093/mnras/stw121910.48550/arXiv.1603.02291}.
\bibitem[{{Sasaki} et~al.(2016){Sasaki}, {Suyama}, {Tanaka}, and
  {Yokoyama}}]{Sasaki2016PhRvL.117f1101S}
\bibinfo{author}{M.~{Sasaki}}, \bibinfo{author}{T.~{Suyama}},
  \bibinfo{author}{T.~{Tanaka}}, \bibinfo{author}{S.~{Yokoyama}},
\newblock \bibinfo{title}{{Primordial Black Hole Scenario for the
  Gravitational-Wave Event GW150914}},
\newblock \bibinfo{journal}{\prl} \bibinfo{volume}{117} (\bibinfo{year}{2016})
  \bibinfo{pages}{061101}.
  \DOIprefix\doi{10.1103/PhysRevLett.117.06110110.48550/arXiv.1603.08338}.
\bibitem[{{Maeder}(1987)}]{Maeder1987A&A...178..159M}
\bibinfo{author}{A.~{Maeder}},
\newblock \bibinfo{title}{{Evidences for a bifurcation in massive star
  evolution. The ON-blue stragglers.}},
\newblock \bibinfo{journal}{\aap}  \bibinfo{volume}{178} (\bibinfo{year}{1987})
  \bibinfo{pages}{159--169}.
\bibitem[{{Maeder} and {Meynet}(2000)}]{Maeder2000ARA&A..38..143M}
\bibinfo{author}{A.~{Maeder}}, \bibinfo{author}{G.~{Meynet}},
\newblock \bibinfo{title}{{The Evolution of Rotating Stars}},
\newblock \bibinfo{journal}{\araa} \bibinfo{volume}{38} (\bibinfo{year}{2000})
  \bibinfo{pages}{143--190}.
  \DOIprefix\doi{10.1146/annurev.astro.38.1.14310.48550/arXiv.astro-ph/0004204}.
\bibitem[{{Langer}(1992)}]{Langer1992A&A...265L..17L}
\bibinfo{author}{N.~{Langer}},
\newblock \bibinfo{title}{{Helium enrichment in massive early type stars.}},
\newblock \bibinfo{journal}{\aap}  \bibinfo{volume}{265} (\bibinfo{year}{1992})
  \bibinfo{pages}{L17--L20}.
\bibitem[{{Heger} and {Langer}(2000)}]{Heger2000ApJ...544.1016H}
\bibinfo{author}{A.~{Heger}}, \bibinfo{author}{N.~{Langer}},
\newblock \bibinfo{title}{{Presupernova Evolution of Rotating Massive Stars.
  II. Evolution of the Surface Properties}},
\newblock \bibinfo{journal}{\apj} \bibinfo{volume}{544} (\bibinfo{year}{2000})
  \bibinfo{pages}{1016--1035}. \DOIprefix\doi{10.1086/317239}.
\bibitem[{{Yoon} and {Langer}(2005)}]{Yoon2005A&A...443..643Y}
\bibinfo{author}{S.~C. {Yoon}}, \bibinfo{author}{N.~{Langer}},
\newblock \bibinfo{title}{{Evolution of rapidly rotating metal-poor massive
  stars towards gamma-ray bursts}},
\newblock \bibinfo{journal}{\aap} \bibinfo{volume}{443} (\bibinfo{year}{2005})
  \bibinfo{pages}{643--648}. \DOIprefix\doi{10.1051/0004-6361:20054030}.
\bibitem[{{Woosley} and {Kasen}(2011)}]{Woosley2011ApJ...734...38W}
\bibinfo{author}{S.~E. {Woosley}}, \bibinfo{author}{D.~{Kasen}},
\newblock \bibinfo{title}{{Sub-Chandrasekhar Mass Models for Supernovae}},
\newblock \bibinfo{journal}{\apj} \bibinfo{volume}{734} (\bibinfo{year}{2011})
  \bibinfo{pages}{38}. \DOIprefix\doi{10.1088/0004-637X/734/1/38}.
\bibitem[{{Brott} et~al.(2011){Brott}, {de Mink}, {Cantiello}, {Langer}, {de
  Koter}, {Evans}, {Hunter}, {Trundle}, and {Vink}}]{Brott2011A&A...530A.115B}
\bibinfo{author}{I.~{Brott}}, \bibinfo{author}{S.~E. {de Mink}},
  \bibinfo{author}{M.~{Cantiello}}, et~al.,
\newblock \bibinfo{title}{{Rotating massive main-sequence stars. I. Grids of
  evolutionary models and isochrones}},
\newblock \bibinfo{journal}{\aap} \bibinfo{volume}{530} (\bibinfo{year}{2011})
  \bibinfo{pages}{A115}.
  \DOIprefix\doi{10.1051/0004-6361/20101611310.48550/arXiv.1102.0530}.
\bibitem[{{K{\"o}hler} et~al.(2015){K{\"o}hler}, {Langer}, {de Koter}, {de
  Mink}, {Crowther}, {Evans}, {Gr{\"a}fener}, {Sana}, {Sanyal}, {Schneider},
  and {Vink}}]{Koehler2015A&A...573A..71K}
\bibinfo{author}{K.~{K{\"o}hler}}, \bibinfo{author}{N.~{Langer}},
  \bibinfo{author}{A.~{de Koter}}, et~al.,
\newblock \bibinfo{title}{{The evolution of rotating very massive stars with
  LMC composition}},
\newblock \bibinfo{journal}{\aap} \bibinfo{volume}{573} (\bibinfo{year}{2015})
  \bibinfo{pages}{A71}. \DOIprefix\doi{10.1051/0004-6361/201424356}.
\bibitem[{{Sz{\'e}csi} et~al.(2015){Sz{\'e}csi}, {Langer}, {Yoon}, {Sanyal},
  {de Mink}, {Evans}, and {Dermine}}]{Szecsi2015A&A...581A..15S}
\bibinfo{author}{D.~{Sz{\'e}csi}}, \bibinfo{author}{N.~{Langer}},
  \bibinfo{author}{S.-C. {Yoon}}, et~al.,
\newblock \bibinfo{title}{{Low-metallicity massive single stars with rotation.
  Evolutionary models applicable to I Zwicky 18}},
\newblock \bibinfo{journal}{\aap} \bibinfo{volume}{581} (\bibinfo{year}{2015})
  \bibinfo{pages}{A15}. \DOIprefix\doi{10.1051/0004-6361/201526617}.
\bibitem[{{Woosley} and {Heger}(2006)}]{Woosley2006ApJ...637..914W}
\bibinfo{author}{S.~E. {Woosley}}, \bibinfo{author}{A.~{Heger}},
\newblock \bibinfo{title}{{The Progenitor Stars of Gamma-Ray Bursts}},
\newblock \bibinfo{journal}{\apj} \bibinfo{volume}{637} (\bibinfo{year}{2006})
  \bibinfo{pages}{914--921}. \DOIprefix\doi{10.1086/498500}.
\bibitem[{{de Mink} et~al.(2009){de Mink}, {Cantiello}, {Langer}, {Pols},
  {Brott}, and {Yoon}}]{de-Mink2009A&A...497..243D}
\bibinfo{author}{S.~E. {de Mink}}, \bibinfo{author}{M.~{Cantiello}},
  \bibinfo{author}{N.~{Langer}}, et~al.,
\newblock \bibinfo{title}{{Rotational mixing in massive binaries. Detached
  short-period systems}},
\newblock \bibinfo{journal}{\aap} \bibinfo{volume}{497} (\bibinfo{year}{2009})
  \bibinfo{pages}{243--253}.
  \DOIprefix\doi{10.1051/0004-6361/20081143910.48550/arXiv.0902.1751}.
\bibitem[{{Song} et~al.(2016){Song}, {Meynet}, {Maeder}, {Ekstr{\"o}m}, and
  {Eggenberger}}]{Song2016A&A...585A.120S}
\bibinfo{author}{H.~F. {Song}}, \bibinfo{author}{G.~{Meynet}},
  \bibinfo{author}{A.~{Maeder}}, et~al.,
\newblock \bibinfo{title}{{Massive star evolution in close binaries. Conditions
  for homogeneous chemical evolution}},
\newblock \bibinfo{journal}{\aap} \bibinfo{volume}{585} (\bibinfo{year}{2016})
  \bibinfo{pages}{A120}.
  \DOIprefix\doi{10.1051/0004-6361/20152607410.48550/arXiv.1508.06094}.
\bibitem[{{Stevenson} et~al.(2017){Stevenson}, {Vigna-G{\'o}mez}, {Mandel},
  {Barrett}, {Neijssel}, {Perkins}, and {de
  Mink}}]{Stevenson2017NatCo...814906S}
\bibinfo{author}{S.~{Stevenson}}, \bibinfo{author}{A.~{Vigna-G{\'o}mez}},
  \bibinfo{author}{I.~{Mandel}}, et~al.,
\newblock \bibinfo{title}{{Formation of the first three gravitational-wave
  observations through isolated binary evolution}},
\newblock \bibinfo{journal}{Nature Communications} \bibinfo{volume}{8}
  (\bibinfo{year}{2017}) \bibinfo{pages}{14906}.
  \DOIprefix\doi{10.1038/ncomms14906}.
\bibitem[{{Belczynski} et~al.(2020){Belczynski}, {Klencki}, {Fields}, {Olejak},
  {Berti}, and et~al.}]{Belczynski2020A&A...636A.104B}
\bibinfo{author}{K.~{Belczynski}}, \bibinfo{author}{J.~{Klencki}},
  \bibinfo{author}{C.~E. {Fields}}, et~al.,
\newblock \bibinfo{title}{{Evolutionary roads leading to low effective spins,
  high black hole masses, and O1/O2 rates for LIGO/Virgo binary black holes}},
\newblock \bibinfo{journal}{\aap} \bibinfo{volume}{636} (\bibinfo{year}{2020})
  \bibinfo{pages}{A104}. \DOIprefix\doi{10.1051/0004-6361/201936528}.
\bibitem[{{Koenigsberger} et~al.(2014){Koenigsberger}, {Morrell}, {Hillier},
  {Gamen}, {Schneider}, {Gonz{\'a}lez-Jim{\'e}nez}, {Langer}, and
  {Barb{\'a}}}]{Koenigsberger2014AJ....148...62K}
\bibinfo{author}{G.~{Koenigsberger}}, \bibinfo{author}{N.~{Morrell}},
  \bibinfo{author}{D.~J. {Hillier}}, et~al.,
\newblock \bibinfo{title}{{The HD 5980 Multiple System: Masses and Evolutionary
  Status}},
\newblock \bibinfo{journal}{\aj} \bibinfo{volume}{148} (\bibinfo{year}{2014})
  \bibinfo{pages}{62}. \DOIprefix\doi{10.1088/0004-6256/148/4/62}.
\bibitem[{{Almeida} et~al.(2015){Almeida}, {Sana}, {de Mink}, {Tramper},
  {Soszy{\'n}ski}, {Langer}, {Barb{\'a}}, {Cantiello}, {Damineli}, {de Koter},
  {Garcia}, {Gr{\"a}fener}, {Herrero}, {Howarth}, {Ma{\'\i}z Apell{\'a}niz},
  {Norman}, {Ram{\'\i}rez-Agudelo}, and {Vink}}]{Almeida2015ApJ...812..102A}
\bibinfo{author}{L.~A. {Almeida}}, \bibinfo{author}{H.~{Sana}},
  \bibinfo{author}{S.~E. {de Mink}}, et~al.,
\newblock \bibinfo{title}{{Discovery of the Massive Overcontact Binary VFTS352:
  Evidence for Enhanced Internal Mixing}},
\newblock \bibinfo{journal}{\apj} \bibinfo{volume}{812} (\bibinfo{year}{2015})
  \bibinfo{pages}{102}. \DOIprefix\doi{10.1088/0004-637X/812/2/102}.
\bibitem[{{Mohamed} and {Podsiadlowski}(2007)}]{Mohamed2007ASPC..372..397M}
\bibinfo{author}{S.~{Mohamed}}, \bibinfo{author}{P.~{Podsiadlowski}},
\newblock \bibinfo{title}{{Wind Roche-Lobe Overflow: a New Mass-Transfer Mode
  for Wide Binaries}},
\newblock in: \bibinfo{editor}{R.~{Napiwotzki}}, \bibinfo{editor}{M.~R.
  {Burleigh}} (Eds.), \bibinfo{booktitle}{15th European Workshop on White
  Dwarfs}, volume \bibinfo{volume}{372} of
  \textit{\bibinfo{series}{Astronomical Society of the Pacific Conference
  Series}}, \bibinfo{year}{2007},  p. \bibinfo{pages}{397}.
\bibitem[{{Vassiliadis} and {Wood}(1993)}]{Vassiliadis1993ApJ...413..641V}
\bibinfo{author}{E.~{Vassiliadis}}, \bibinfo{author}{P.~R. {Wood}},
\newblock \bibinfo{title}{{Evolution of Low- and Intermediate-Mass Stars to the
  End of the Asymptotic Giant Branch with Mass Loss}},
\newblock \bibinfo{journal}{\apj} \bibinfo{volume}{413} (\bibinfo{year}{1993})
  \bibinfo{pages}{641}. \DOIprefix\doi{10.1086/173033}.
\bibitem[{{Knapp} et~al.(1998){Knapp}, {Young}, {Lee}, and
  {Jorissen}}]{Knapp1998ApJS..117..209K}
\bibinfo{author}{G.~R. {Knapp}}, \bibinfo{author}{K.~{Young}},
  \bibinfo{author}{E.~{Lee}}, \bibinfo{author}{A.~{Jorissen}},
\newblock \bibinfo{title}{{Multiple Molecular Winds in Evolved Stars. I. A
  Survey of CO(2-1) and CO(3-2) Emission from 45 Nearby Asymptotic Giant Branch
  Stars}},
\newblock \bibinfo{journal}{\apjs} \bibinfo{volume}{117} (\bibinfo{year}{1998})
  \bibinfo{pages}{209--231}. \DOIprefix\doi{10.1086/313111}.
\bibitem[{{H{\"o}fner} and {Olofsson}(2018)}]{Hoefner2018A&ARv..26....1H}
\bibinfo{author}{S.~{H{\"o}fner}}, \bibinfo{author}{H.~{Olofsson}},
\newblock \bibinfo{title}{{Mass loss of stars on the asymptotic giant branch.
  Mechanisms, models and measurements}},
\newblock \bibinfo{journal}{\aapr} \bibinfo{volume}{26} (\bibinfo{year}{2018})
  \bibinfo{pages}{1}. \DOIprefix\doi{10.1007/s00159-017-0106-5}.
\bibitem[{{Theuns} and {Jorissen}(1993)}]{Theuns1993MNRAS.265..946T}
\bibinfo{author}{T.~{Theuns}}, \bibinfo{author}{A.~{Jorissen}},
\newblock \bibinfo{title}{{Wind accretion in binary stars - I. Intricacies of
  the flow structure.}},
\newblock \bibinfo{journal}{\mnras} \bibinfo{volume}{265}
  (\bibinfo{year}{1993}) \bibinfo{pages}{946--967}.
  \DOIprefix\doi{10.1093/mnras/265.4.946}.
\bibitem[{{Theuns} et~al.(1996){Theuns}, {Boffin}, and
  {Jorissen}}]{Theuns1996MNRAS.280.1264T}
\bibinfo{author}{T.~{Theuns}}, \bibinfo{author}{H.~M.~J. {Boffin}},
  \bibinfo{author}{A.~{Jorissen}},
\newblock \bibinfo{title}{{Wind accretion in binary stars - II. Accretion
  rates}},
\newblock \bibinfo{journal}{\mnras} \bibinfo{volume}{280}
  (\bibinfo{year}{1996}) \bibinfo{pages}{1264--1276}.
  \DOIprefix\doi{10.1093/mnras/280.4.1264}.
\bibitem[{{de Val-Borro} et~al.(2009){de Val-Borro}, {Karovska}, and
  {Sasselov}}]{de-Val-Borro2009ApJ...700.1148D}
\bibinfo{author}{M.~{de Val-Borro}}, \bibinfo{author}{M.~{Karovska}},
  \bibinfo{author}{D.~{Sasselov}},
\newblock \bibinfo{title}{{Numerical Simulations of Wind Accretion in Symbiotic
  Binaries}},
\newblock \bibinfo{journal}{\apj} \bibinfo{volume}{700} (\bibinfo{year}{2009})
  \bibinfo{pages}{1148--1160}. \DOIprefix\doi{10.1088/0004-637X/700/2/1148}.
\bibitem[{{Shagatova} et~al.(2016){Shagatova}, {Skopal}, and
  {Carikov{\'a}}}]{Shagatova2016A&A...588A..83S}
\bibinfo{author}{N.~{Shagatova}}, \bibinfo{author}{A.~{Skopal}},
  \bibinfo{author}{Z.~{Carikov{\'a}}},
\newblock \bibinfo{title}{{Wind mass transfer in S-type symbiotic binaries. II.
  Indication of wind focusing}},
\newblock \bibinfo{journal}{\aap} \bibinfo{volume}{588} (\bibinfo{year}{2016})
  \bibinfo{pages}{A83}. \DOIprefix\doi{10.1051/0004-6361/201525645}.
\bibitem[{{Liu} et~al.(2017){Liu}, {Stancliffe}, {Abate}, and
  {Matrozis}}]{Liu2017ApJ...846..117L}
\bibinfo{author}{Z.-W. {Liu}}, \bibinfo{author}{R.~J. {Stancliffe}},
  \bibinfo{author}{C.~{Abate}}, \bibinfo{author}{E.~{Matrozis}},
\newblock \bibinfo{title}{{Three-dimensional Hydrodynamical Simulations of Mass
  Transfer in Binary Systems by a Free Wind}},
\newblock \bibinfo{journal}{\apj} \bibinfo{volume}{846} (\bibinfo{year}{2017})
  \bibinfo{pages}{117}. \DOIprefix\doi{10.3847/1538-4357/aa8622}.
\bibitem[{{Chen} et~al.(2017){Chen}, {Frank}, {Blackman}, {Nordhaus}, and
  {Carroll-Nellenback}}]{Chen2017MNRAS.468.4465C}
\bibinfo{author}{Z.~{Chen}}, \bibinfo{author}{A.~{Frank}},
  \bibinfo{author}{E.~G. {Blackman}}, et~al.,
\newblock \bibinfo{title}{{Mass transfer and disc formation in AGB binary
  systems}},
\newblock \bibinfo{journal}{\mnras} \bibinfo{volume}{468}
  (\bibinfo{year}{2017}) \bibinfo{pages}{4465--4477}.
  \DOIprefix\doi{10.1093/mnras/stx680}.
\bibitem[{{Saladino} et~al.(2018){Saladino}, {Pols}, {van der Helm},
  {Pelupessy}, and {Portegies Zwart}}]{Saladino2018A&A...618A..50S}
\bibinfo{author}{M.~I. {Saladino}}, \bibinfo{author}{O.~R. {Pols}},
  \bibinfo{author}{E.~{van der Helm}}, et~al.,
\newblock \bibinfo{title}{{Gone with the wind: the impact of wind mass transfer
  on the orbital evolution of AGB binary systems}},
\newblock \bibinfo{journal}{\aap} \bibinfo{volume}{618} (\bibinfo{year}{2018})
  \bibinfo{pages}{A50}. \DOIprefix\doi{10.1051/0004-6361/201832967}.
\bibitem[{{Siess} et~al.(2022){Siess}, {Homan}, {Toupin}, and
  {Price}}]{Siess2022A&A...667A..75S}
\bibinfo{author}{L.~{Siess}}, \bibinfo{author}{W.~{Homan}},
  \bibinfo{author}{S.~{Toupin}}, \bibinfo{author}{D.~J. {Price}},
\newblock \bibinfo{title}{{3D simulations of AGB stellar winds. I. Steady winds
  and dust formation}},
\newblock \bibinfo{journal}{\aap} \bibinfo{volume}{667} (\bibinfo{year}{2022})
  \bibinfo{pages}{A75}. \DOIprefix\doi{10.1051/0004-6361/202243540}.
\bibitem[{{Mohamed} and {Podsiadlowski}(2012)}]{Mohamed2012BaltA..21...88M}
\bibinfo{author}{S.~{Mohamed}}, \bibinfo{author}{P.~{Podsiadlowski}},
\newblock \bibinfo{title}{{Mass Transfer in Mira-type Binaries}},
\newblock \bibinfo{journal}{Baltic Astronomy} \bibinfo{volume}{21}
  (\bibinfo{year}{2012}) \bibinfo{pages}{88--96}.
  \DOIprefix\doi{10.1515/astro-2017-0362}.
\bibitem[{{Abate} et~al.(2013){Abate}, {Pols}, {Izzard}, {Mohamed}, and {de
  Mink}}]{Abate2013A&A...552A..26A}
\bibinfo{author}{C.~{Abate}}, \bibinfo{author}{O.~R. {Pols}},
  \bibinfo{author}{R.~G. {Izzard}}, et~al.,
\newblock \bibinfo{title}{{Wind Roche-lobe overflow: Application to
  carbon-enhanced metal-poor stars}},
\newblock \bibinfo{journal}{\aap} \bibinfo{volume}{552} (\bibinfo{year}{2013})
  \bibinfo{pages}{A26}. \DOIprefix\doi{10.1051/0004-6361/201220007}.
\bibitem[{{Abate}(2017)}]{Abate2017MmSAI..88..308A}
\bibinfo{author}{C.~{Abate}},
\newblock \bibinfo{title}{{What is the role of wind mass transfer in the
  progenitor evolution of Type Ia supernovae?}},
\newblock \bibinfo{journal}{\memsai}  \bibinfo{volume}{88}
  (\bibinfo{year}{2017}) \bibinfo{pages}{308}.
\bibitem[{{I{\l}kiewicz} et~al.(2019){I{\l}kiewicz}, {Miko{\l}ajewska},
  {Belczy{\'n}ski}, {Wiktorowicz}, and
  {Karczmarek}}]{Ikiewicz2019MNRAS.485.5468I}
\bibinfo{author}{K.~{I{\l}kiewicz}}, \bibinfo{author}{J.~{Miko{\l}ajewska}},
  \bibinfo{author}{K.~{Belczy{\'n}ski}}, et~al.,
\newblock \bibinfo{title}{{Wind Roche lobe overflow as a way to make Type Ia
  supernovae from the widest symbiotic systems}},
\newblock \bibinfo{journal}{\mnras} \bibinfo{volume}{485}
  (\bibinfo{year}{2019}) \bibinfo{pages}{5468--5473}.
  \DOIprefix\doi{10.1093/mnras/stz760}.
\bibitem[{{El Mellah} et~al.(2019){El Mellah}, {Sundqvist}, and
  {Keppens}}]{El-Mellah2019A&A...622L...3E}
\bibinfo{author}{I.~{El Mellah}}, \bibinfo{author}{J.~O. {Sundqvist}},
  \bibinfo{author}{R.~{Keppens}},
\newblock \bibinfo{title}{{Wind Roche lobe overflow in high-mass X-ray
  binaries. A possible mass-transfer mechanism for ultraluminous X-ray
  sources}},
\newblock \bibinfo{journal}{\aap} \bibinfo{volume}{622} (\bibinfo{year}{2019})
  \bibinfo{pages}{L3}. \DOIprefix\doi{10.1051/0004-6361/201834543}.
\bibitem[{{Zuo} et~al.(2021){Zuo}, {Song}, and {Xue}}]{Zuo2021A&A...649L...2Z}
\bibinfo{author}{Z.-Y. {Zuo}}, \bibinfo{author}{H.-T. {Song}},
  \bibinfo{author}{H.-C. {Xue}},
\newblock \bibinfo{title}{{Population synthesis on ultra-luminous X-ray sources
  with an accreting neutron star: Wind Roche-lobe overflow cases}},
\newblock \bibinfo{journal}{\aap} \bibinfo{volume}{649} (\bibinfo{year}{2021})
  \bibinfo{pages}{L2}. \DOIprefix\doi{10.1051/0004-6361/202140792}.
\bibitem[{{Kruckow} et~al.(2018){Kruckow}, {Tauris}, {Langer}, {Kramer}, and
  {Izzard}}]{Kruckow2018MNRAS.481.1908K}
\bibinfo{author}{M.~U. {Kruckow}}, \bibinfo{author}{T.~M. {Tauris}},
  \bibinfo{author}{N.~{Langer}}, et~al.,
\newblock \bibinfo{title}{{Progenitors of gravitational wave mergers: binary
  evolution with the stellar grid-based code COMBINE}},
\newblock \bibinfo{journal}{\mnras} \bibinfo{volume}{481}
  (\bibinfo{year}{2018}) \bibinfo{pages}{1908--1949}.
  \DOIprefix\doi{10.1093/mnras/sty2190}.
\bibitem[{{Breivik} et~al.(2020){Breivik}, {Coughlin}, {Zevin}, {Rodriguez},
  {Kremer}, {Ye}, {Andrews}, {Kurkowski}, {Digman}, {Larson}, and
  {Rasio}}]{Breivik2020ApJ...898...71B}
\bibinfo{author}{K.~{Breivik}}, \bibinfo{author}{S.~{Coughlin}},
  \bibinfo{author}{M.~{Zevin}}, et~al.,
\newblock \bibinfo{title}{{COSMIC Variance in Binary Population Synthesis}},
\newblock \bibinfo{journal}{\apj} \bibinfo{volume}{898} (\bibinfo{year}{2020})
  \bibinfo{pages}{71}. \DOIprefix\doi{10.3847/1538-4357/ab9d85}.
\bibitem[{{Belczynski} et~al.(2002){Belczynski}, {Kalogera}, and
  {Bulik}}]{Belczynski2002ApJ...572..407B}
\bibinfo{author}{K.~{Belczynski}}, \bibinfo{author}{V.~{Kalogera}},
  \bibinfo{author}{T.~{Bulik}},
\newblock \bibinfo{title}{{A Comprehensive Study of Binary Compact Objects as
  Gravitational Wave Sources: Evolutionary Channels, Rates, and Physical
  Properties}},
\newblock \bibinfo{journal}{\apj} \bibinfo{volume}{572} (\bibinfo{year}{2002})
  \bibinfo{pages}{407--431}. \DOIprefix\doi{10.1086/340304}.
\bibitem[{{Os{\l}owski} et~al.(2011){Os{\l}owski}, {Bulik},
  {Gondek-Rosi{\'n}ska}, and {Belczy{\'n}ski}}]{Oslowski2011MNRAS.413..461O}
\bibinfo{author}{S.~{Os{\l}owski}}, \bibinfo{author}{T.~{Bulik}},
  \bibinfo{author}{D.~{Gondek-Rosi{\'n}ska}},
  \bibinfo{author}{K.~{Belczy{\'n}ski}},
\newblock \bibinfo{title}{{Population synthesis of double neutron stars}},
\newblock \bibinfo{journal}{\mnras} \bibinfo{volume}{413}
  (\bibinfo{year}{2011}) \bibinfo{pages}{461--479}.
  \DOIprefix\doi{10.1111/j.1365-2966.2010.18147.x}.
\bibitem[{{Kruckow}(2020)}]{Kruckow2020A&A...639A.123K}
\bibinfo{author}{M.~U. {Kruckow}},
\newblock \bibinfo{title}{{Masses of double neutron star mergers}},
\newblock \bibinfo{journal}{\aap} \bibinfo{volume}{639} (\bibinfo{year}{2020})
  \bibinfo{pages}{A123}. \DOIprefix\doi{10.1051/0004-6361/202037519}.
\bibitem[{{Han}(1998)}]{Han1998MNRAS.296.1019H}
\bibinfo{author}{Z.~{Han}},
\newblock \bibinfo{title}{{The formation of double degenerates and related
  objects}},
\newblock \bibinfo{journal}{\mnras} \bibinfo{volume}{296}
  (\bibinfo{year}{1998}) \bibinfo{pages}{1019--1040}.
  \DOIprefix\doi{10.1046/j.1365-8711.1998.01475.x}.
\bibitem[{{Zhenwei} et~al.(2023){Zhenwei}, {Xuefei}, {Hongwei}, {Hai-Liang},
  and {Zhanwen}}]{Li2023A&A...669A..82Z}
\bibinfo{author}{L.~{Zhenwei}}, \bibinfo{author}{C.~{Xuefei}},
  \bibinfo{author}{G.~{Hongwei}}, et~al.,
\newblock \bibinfo{title}{{Influence of a mass transfer stability criterion on
  double white dwarf populations}},
\newblock \bibinfo{journal}{\aap} \bibinfo{volume}{669} (\bibinfo{year}{2023})
  \bibinfo{pages}{A82}. \DOIprefix\doi{10.1051/0004-6361/202243893}.
\bibitem[{{Yungelson} et~al.(1994){Yungelson}, {Livio}, {Tutukov}, and
  {Saffer}}]{Yungelson1994ApJ...420..336Y}
\bibinfo{author}{L.~R. {Yungelson}}, \bibinfo{author}{M.~{Livio}},
  \bibinfo{author}{A.~V. {Tutukov}}, \bibinfo{author}{R.~A. {Saffer}},
\newblock \bibinfo{title}{{Are the Observed Frequencies of Double Degenerates
  and SN IA Contradictory?}},
\newblock \bibinfo{journal}{\apj} \bibinfo{volume}{420} (\bibinfo{year}{1994})
  \bibinfo{pages}{336}. \DOIprefix\doi{10.1086/173563}.
\bibitem[{{Ruiter} et~al.(2009){Ruiter}, {Belczynski}, and
  {Fryer}}]{Ruiter2009ApJ...699.2026R}
\bibinfo{author}{A.~J. {Ruiter}}, \bibinfo{author}{K.~{Belczynski}},
  \bibinfo{author}{C.~{Fryer}},
\newblock \bibinfo{title}{{Rates and Delay Times of Type Ia Supernovae}},
\newblock \bibinfo{journal}{\apj} \bibinfo{volume}{699} (\bibinfo{year}{2009})
  \bibinfo{pages}{2026--2036}. \DOIprefix\doi{10.1088/0004-637X/699/2/2026}.
\bibitem[{{Wang} et~al.(2009){Wang}, {Meng}, {Chen}, and
  {Han}}]{Wang2009MNRAS.395..847W}
\bibinfo{author}{B.~{Wang}}, \bibinfo{author}{X.~{Meng}},
  \bibinfo{author}{X.~{Chen}}, \bibinfo{author}{Z.~{Han}},
\newblock \bibinfo{title}{{The helium star donor channel for the progenitors of
  Type Ia supernovae}},
\newblock \bibinfo{journal}{\mnras} \bibinfo{volume}{395}
  (\bibinfo{year}{2009}) \bibinfo{pages}{847--854}.
  \DOIprefix\doi{10.1111/j.1365-2966.2009.14545.x}.
\bibitem[{{Meng} et~al.(2009){Meng}, {Chen}, and
  {Han}}]{Meng2009MNRAS.395.2103M}
\bibinfo{author}{X.~{Meng}}, \bibinfo{author}{X.~{Chen}},
  \bibinfo{author}{Z.~{Han}},
\newblock \bibinfo{title}{{A single-degenerate channel for the progenitors of
  Type Ia supernovae with different metallicities}},
\newblock \bibinfo{journal}{\mnras} \bibinfo{volume}{395}
  (\bibinfo{year}{2009}) \bibinfo{pages}{2103--2116}.
  \DOIprefix\doi{10.1111/j.1365-2966.2009.14636.x}.
\bibitem[{{Toonen} et~al.(2012){Toonen}, {Nelemans}, and {Portegies
  Zwart}}]{Toonen2012A&A...546A..70T}
\bibinfo{author}{S.~{Toonen}}, \bibinfo{author}{G.~{Nelemans}},
  \bibinfo{author}{S.~{Portegies Zwart}},
\newblock \bibinfo{title}{{Supernova Type Ia progenitors from merging double
  white dwarfs. Using a new population synthesis model}},
\newblock \bibinfo{journal}{\aap} \bibinfo{volume}{546} (\bibinfo{year}{2012})
  \bibinfo{pages}{A70}. \DOIprefix\doi{10.1051/0004-6361/201218966}.
\bibitem[{{Liu} and {Stancliffe}(2018)}]{Liu2018MNRAS.475.5257L}
\bibinfo{author}{Z.-W. {Liu}}, \bibinfo{author}{R.~J. {Stancliffe}},
\newblock \bibinfo{title}{{Rates and delay times of Type Ia supernovae in the
  helium-enriched main-sequence donor scenario}},
\newblock \bibinfo{journal}{\mnras} \bibinfo{volume}{475}
  (\bibinfo{year}{2018}) \bibinfo{pages}{5257--5267}.
  \DOIprefix\doi{10.1093/mnras/sty172}.
\bibitem[{{Chrimes} et~al.(2020){Chrimes}, {Stanway}, and
  {Eldridge}}]{Chrimes2020MNRAS.491.3479C}
\bibinfo{author}{A.~A. {Chrimes}}, \bibinfo{author}{E.~R. {Stanway}},
  \bibinfo{author}{J.~J. {Eldridge}},
\newblock \bibinfo{title}{{Binary population synthesis models for core-collapse
  gamma-ray burst progenitors}},
\newblock \bibinfo{journal}{\mnras} \bibinfo{volume}{491}
  (\bibinfo{year}{2020}) \bibinfo{pages}{3479--3495}.
  \DOIprefix\doi{10.1093/mnras/stz3246}.
\bibitem[{{Belczynski} et~al.(2006){Belczynski}, {Perna}, {Bulik}, {Kalogera},
  {Ivanova}, and {Lamb}}]{Belczynski2006ApJ...648.1110B}
\bibinfo{author}{K.~{Belczynski}}, \bibinfo{author}{R.~{Perna}},
  \bibinfo{author}{T.~{Bulik}}, et~al.,
\newblock \bibinfo{title}{{A Study of Compact Object Mergers as Short Gamma-Ray
  Burst Progenitors}},
\newblock \bibinfo{journal}{\apj} \bibinfo{volume}{648} (\bibinfo{year}{2006})
  \bibinfo{pages}{1110--1116}. \DOIprefix\doi{10.1086/505169}.
\bibitem[{{Willems} and {Kolb}(2002)}]{Willems2002MNRAS.337.1004W}
\bibinfo{author}{B.~{Willems}}, \bibinfo{author}{U.~{Kolb}},
\newblock \bibinfo{title}{{Population synthesis of wide binary millisecond
  pulsars}},
\newblock \bibinfo{journal}{\mnras} \bibinfo{volume}{337}
  (\bibinfo{year}{2002}) \bibinfo{pages}{1004--1016}.
  \DOIprefix\doi{10.1046/j.1365-8711.2002.05985.x}.
\bibitem[{{Zhu} et~al.(2015){Zhu}, {L{\"u}}, and
  {Wang}}]{Zhu2015MNRAS.454.1725Z}
\bibinfo{author}{C.~{Zhu}}, \bibinfo{author}{G.~{L{\"u}}},
  \bibinfo{author}{Z.~{Wang}},
\newblock \bibinfo{title}{{Population synthesis of millisecond X-ray pulsars}},
\newblock \bibinfo{journal}{\mnras} \bibinfo{volume}{454}
  (\bibinfo{year}{2015}) \bibinfo{pages}{1725--1735}.
  \DOIprefix\doi{10.1093/mnras/stv2100}.
\bibitem[{{Chen} et~al.(2014){Chen}, {Woods}, {Yungelson}, {Gilfanov}, and
  {Han}}]{Chen2014MNRAS.445.1912C}
\bibinfo{author}{H.-L. {Chen}}, \bibinfo{author}{T.~E. {Woods}},
  \bibinfo{author}{L.~R. {Yungelson}}, et~al.,
\newblock \bibinfo{title}{{Next generation population synthesis of accreting
  white dwarfs - I. Hybrid calculations using BSE+MESA}},
\newblock \bibinfo{journal}{\mnras} \bibinfo{volume}{445}
  (\bibinfo{year}{2014}) \bibinfo{pages}{1912--1923}.
  \DOIprefix\doi{10.1093/mnras/stu1884}.
\bibitem[{{Chen} et~al.(2015){Chen}, {Woods}, {Yungelson}, {Gilfanov}, and
  {Han}}]{Chen2015MNRAS.453.3024C}
\bibinfo{author}{H.-L. {Chen}}, \bibinfo{author}{T.~E. {Woods}},
  \bibinfo{author}{L.~R. {Yungelson}}, et~al.,
\newblock \bibinfo{title}{{Population synthesis of accreting white dwarfs - II.
  X-ray and UV emission}},
\newblock \bibinfo{journal}{\mnras} \bibinfo{volume}{453}
  (\bibinfo{year}{2015}) \bibinfo{pages}{3024--3034}.
  \DOIprefix\doi{10.1093/mnras/stv1865}.
\bibitem[{{Chen} et~al.(2016){Chen}, {Woods}, {Yungelson}, {Gilfanov}, and
  {Han}}]{Chen2016MNRAS.458.2916C}
\bibinfo{author}{H.-L. {Chen}}, \bibinfo{author}{T.~E. {Woods}},
  \bibinfo{author}{L.~R. {Yungelson}}, et~al.,
\newblock \bibinfo{title}{{Modelling nova populations in galaxies}},
\newblock \bibinfo{journal}{\mnras} \bibinfo{volume}{458}
  (\bibinfo{year}{2016}) \bibinfo{pages}{2916--2927}.
  \DOIprefix\doi{10.1093/mnras/stw458}.
\bibitem[{{Nelemans} et~al.(2001){Nelemans}, {Portegies Zwart}, {Verbunt}, and
  {Yungelson}}]{Nelemans2001A&A...368..939N}
\bibinfo{author}{G.~{Nelemans}}, \bibinfo{author}{S.~F. {Portegies Zwart}},
  \bibinfo{author}{F.~{Verbunt}}, \bibinfo{author}{L.~R. {Yungelson}},
\newblock \bibinfo{title}{{Population synthesis for double white dwarfs. II.
  Semi-detached systems: AM CVn stars}},
\newblock \bibinfo{journal}{\aap} \bibinfo{volume}{368} (\bibinfo{year}{2001})
  \bibinfo{pages}{939--949}. \DOIprefix\doi{10.1051/0004-6361:20010049}.
\bibitem[{{Goliasch} and {Nelson}(2015)}]{Goliasch2015ApJ...809...80G}
\bibinfo{author}{J.~{Goliasch}}, \bibinfo{author}{L.~{Nelson}},
\newblock \bibinfo{title}{{Population Synthesis of Cataclysmic Variables. I.
  Inclusion of Detailed Nuclear Evolution}},
\newblock \bibinfo{journal}{\apj} \bibinfo{volume}{809} (\bibinfo{year}{2015})
  \bibinfo{pages}{80}. \DOIprefix\doi{10.1088/0004-637X/809/1/80}.
\bibitem[{{Wu} et~al.(2018){Wu}, {Chen}, {Li}, and
  {Han}}]{Wu2018A&A...618A..14W}
\bibinfo{author}{Y.~{Wu}}, \bibinfo{author}{X.~{Chen}},
  \bibinfo{author}{Z.~{Li}}, \bibinfo{author}{Z.~{Han}},
\newblock \bibinfo{title}{{Formation of hot subdwarf B stars with neutron star
  components}},
\newblock \bibinfo{journal}{\aap} \bibinfo{volume}{618} (\bibinfo{year}{2018})
  \bibinfo{pages}{A14}. \DOIprefix\doi{10.1051/0004-6361/201832686}.
\bibitem[{{Izzard} et~al.(2009){Izzard}, {Glebbeek}, {Stancliffe}, and
  {Pols}}]{Izzard2009A&A...508.1359I}
\bibinfo{author}{R.~G. {Izzard}}, \bibinfo{author}{E.~{Glebbeek}},
  \bibinfo{author}{R.~J. {Stancliffe}}, \bibinfo{author}{O.~R. {Pols}},
\newblock \bibinfo{title}{{Population synthesis of binary carbon-enhanced
  metal-poor stars}},
\newblock \bibinfo{journal}{\aap} \bibinfo{volume}{508} (\bibinfo{year}{2009})
  \bibinfo{pages}{1359--1374}. \DOIprefix\doi{10.1051/0004-6361/200912827}.
\bibitem[{{Zhang} et~al.(2004){Zhang}, {Han}, {Li}, and
  {Hurley}}]{Zhang2004A&A...415..117Z}
\bibinfo{author}{F.~{Zhang}}, \bibinfo{author}{Z.~{Han}},
  \bibinfo{author}{L.~{Li}}, \bibinfo{author}{J.~R. {Hurley}},
\newblock \bibinfo{title}{{Evolutionary population synthesis for binary stellar
  populations}},
\newblock \bibinfo{journal}{\aap} \bibinfo{volume}{415} (\bibinfo{year}{2004})
  \bibinfo{pages}{117--122}. \DOIprefix\doi{10.1051/0004-6361:20031268}.
\bibitem[{{Han} et~al.(2007){Han}, {Podsiadlowski}, and
  {Lynas-Gray}}]{Han2007MNRAS.380.1098H}
\bibinfo{author}{Z.~{Han}}, \bibinfo{author}{P.~{Podsiadlowski}},
  \bibinfo{author}{A.~E. {Lynas-Gray}},
\newblock \bibinfo{title}{{A binary model for the UV-upturn of elliptical
  galaxies}},
\newblock \bibinfo{journal}{\mnras} \bibinfo{volume}{380}
  (\bibinfo{year}{2007}) \bibinfo{pages}{1098--1118}.
  \DOIprefix\doi{10.1111/j.1365-2966.2007.12151.x}.
\bibitem[{{Lipunov} et~al.(1996){Lipunov}, {Postnov}, and
  {Prokhorov}}]{Lipunov1996A&A...310..489L}
\bibinfo{author}{V.~M. {Lipunov}}, \bibinfo{author}{K.~A. {Postnov}},
  \bibinfo{author}{M.~E. {Prokhorov}},
\newblock \bibinfo{title}{{The Scenario Machine: restrictions on key parameters
  of binary evolution.}},
\newblock \bibinfo{journal}{\aap}  \bibinfo{volume}{310} (\bibinfo{year}{1996})
  \bibinfo{pages}{489--507}.
\bibitem[{{Lipunov} et~al.(2009){Lipunov}, {Postnov}, {Prokhorov}, and
  {Bogomazov}}]{Lipunov2009ARep...53..915L}
\bibinfo{author}{V.~M. {Lipunov}}, \bibinfo{author}{K.~A. {Postnov}},
  \bibinfo{author}{M.~E. {Prokhorov}}, \bibinfo{author}{A.~I. {Bogomazov}},
\newblock \bibinfo{title}{{Description of the ``Scenario Machine''}},
\newblock \bibinfo{journal}{Astronomy Reports} \bibinfo{volume}{53}
  (\bibinfo{year}{2009}) \bibinfo{pages}{915--940}.
  \DOIprefix\doi{10.1134/S1063772909100047}.
\bibitem[{{Portegies Zwart} and
  {Verbunt}(1996)}]{Portegies-Zwart1996A&A...309..179P}
\bibinfo{author}{S.~F. {Portegies Zwart}}, \bibinfo{author}{F.~{Verbunt}},
\newblock \bibinfo{title}{{Population synthesis of high-mass binaries.}},
\newblock \bibinfo{journal}{\aap}  \bibinfo{volume}{309} (\bibinfo{year}{1996})
  \bibinfo{pages}{179--196}.
\bibitem[{{Zhang} et~al.(2002){Zhang}, {Han}, {Li}, and
  {Hurley}}]{Zhang2002MNRAS.334..883Z}
\bibinfo{author}{F.~{Zhang}}, \bibinfo{author}{Z.~{Han}},
  \bibinfo{author}{L.~{Li}}, \bibinfo{author}{J.~R. {Hurley}},
\newblock \bibinfo{title}{{Colour indices of single stellar populations}},
\newblock \bibinfo{journal}{\mnras} \bibinfo{volume}{334}
  (\bibinfo{year}{2002}) \bibinfo{pages}{883--904}.
  \DOIprefix\doi{10.1046/j.1365-8711.2002.05568.x}.
\bibitem[{{Zhang} et~al.(2005){Zhang}, {Han}, {Li}, and
  {Hurley}}]{Zhang2005MNRAS.357.1088Z}
\bibinfo{author}{F.~{Zhang}}, \bibinfo{author}{Z.~{Han}},
  \bibinfo{author}{L.~{Li}}, \bibinfo{author}{J.~R. {Hurley}},
\newblock \bibinfo{title}{{Inclusion of binaries in evolutionary population
  synthesis}},
\newblock \bibinfo{journal}{\mnras} \bibinfo{volume}{357}
  (\bibinfo{year}{2005}) \bibinfo{pages}{1088--1103}.
  \DOIprefix\doi{10.1111/j.1365-2966.2005.08739.x}.
\bibitem[{{Hurley} et~al.(2002){Hurley}, {Tout}, and
  {Pols}}]{Hurley2002MNRAS.329..897H}
\bibinfo{author}{J.~R. {Hurley}}, \bibinfo{author}{C.~A. {Tout}},
  \bibinfo{author}{O.~R. {Pols}},
\newblock \bibinfo{title}{{Evolution of binary stars and the effect of tides on
  binary populations}},
\newblock \bibinfo{journal}{\mnras} \bibinfo{volume}{329}
  (\bibinfo{year}{2002}) \bibinfo{pages}{897--928}.
  \DOIprefix\doi{10.1046/j.1365-8711.2002.05038.x}.
\bibitem[{{Eldridge} et~al.(2008){Eldridge}, {Izzard}, and
  {Tout}}]{Eldridge2008MNRAS.384.1109E}
\bibinfo{author}{J.~J. {Eldridge}}, \bibinfo{author}{R.~G. {Izzard}},
  \bibinfo{author}{C.~A. {Tout}},
\newblock \bibinfo{title}{{The effect of massive binaries on stellar
  populations and supernova progenitors}},
\newblock \bibinfo{journal}{\mnras} \bibinfo{volume}{384}
  (\bibinfo{year}{2008}) \bibinfo{pages}{1109--1118}.
  \DOIprefix\doi{10.1111/j.1365-2966.2007.12738.x}.
\bibitem[{{Georgy} et~al.(2014){Georgy}, {Granada}, {Ekstr{\"o}m}, {Meynet},
  {Anderson}, {Wyttenbach}, {Eggenberger}, and
  {Maeder}}]{Georgy2014A&A...566A..21G}
\bibinfo{author}{C.~{Georgy}}, \bibinfo{author}{A.~{Granada}},
  \bibinfo{author}{S.~{Ekstr{\"o}m}}, et~al.,
\newblock \bibinfo{title}{{Populations of rotating stars. III. SYCLIST, the new
  Geneva population synthesis code}},
\newblock \bibinfo{journal}{\aap} \bibinfo{volume}{566} (\bibinfo{year}{2014})
  \bibinfo{pages}{A21}. \DOIprefix\doi{10.1051/0004-6361/201423881}.
\bibitem[{{Giacobbo} et~al.(2018){Giacobbo}, {Mapelli}, and
  {Spera}}]{Giacobbo2018MNRAS.474.2959G}
\bibinfo{author}{N.~{Giacobbo}}, \bibinfo{author}{M.~{Mapelli}},
  \bibinfo{author}{M.~{Spera}},
\newblock \bibinfo{title}{{Merging black hole binaries: the effects of
  progenitor's metallicity, mass-loss rate and Eddington factor}},
\newblock \bibinfo{journal}{\mnras} \bibinfo{volume}{474}
  (\bibinfo{year}{2018}) \bibinfo{pages}{2959--2974}.
  \DOIprefix\doi{10.1093/mnras/stx2933}.
\bibitem[{{Andrews} et~al.(2018){Andrews}, {Zezas}, and
  {Fragos}}]{Andrews2018ApJS..237....1A}
\bibinfo{author}{J.~J. {Andrews}}, \bibinfo{author}{A.~{Zezas}},
  \bibinfo{author}{T.~{Fragos}},
\newblock \bibinfo{title}{{dart\_board: Binary Population Synthesis with Markov
  Chain Monte Carlo}},
\newblock \bibinfo{journal}{\apjs} \bibinfo{volume}{237} (\bibinfo{year}{2018})
  \bibinfo{pages}{1}. \DOIprefix\doi{10.3847/1538-4365/aaca30}.
\bibitem[{{Salpeter}(1955)}]{Salpeter1955ApJ...121..161S}
\bibinfo{author}{E.~E. {Salpeter}},
\newblock \bibinfo{title}{{The Luminosity Function and Stellar Evolution.}},
\newblock \bibinfo{journal}{\apj} \bibinfo{volume}{121} (\bibinfo{year}{1955})
  \bibinfo{pages}{161}. \DOIprefix\doi{10.1086/145971}.
\bibitem[{{Miller} and {Scalo}(1979)}]{Miller1979ApJS...41..513M}
\bibinfo{author}{G.~E. {Miller}}, \bibinfo{author}{J.~M. {Scalo}},
\newblock \bibinfo{title}{{The Initial Mass Function and Stellar Birthrate in
  the Solar Neighborhood}},
\newblock \bibinfo{journal}{\apjs} \bibinfo{volume}{41} (\bibinfo{year}{1979})
  \bibinfo{pages}{513}. \DOIprefix\doi{10.1086/190629}.
\bibitem[{{Kroupa} et~al.(1993){Kroupa}, {Tout}, and
  {Gilmore}}]{Kroupa1993MNRAS.262..545K}
\bibinfo{author}{P.~{Kroupa}}, \bibinfo{author}{C.~A. {Tout}},
  \bibinfo{author}{G.~{Gilmore}},
\newblock \bibinfo{title}{{The Distribution of Low-Mass Stars in the Galactic
  Disc}},
\newblock \bibinfo{journal}{\mnras} \bibinfo{volume}{262}
  (\bibinfo{year}{1993}) \bibinfo{pages}{545--587}.
  \DOIprefix\doi{10.1093/mnras/262.3.545}.
\bibitem[{{Kroupa}(2001)}]{Kroupa2001MNRAS.322..231K}
\bibinfo{author}{P.~{Kroupa}},
\newblock \bibinfo{title}{{On the variation of the initial mass function}},
\newblock \bibinfo{journal}{\mnras} \bibinfo{volume}{322}
  (\bibinfo{year}{2001}) \bibinfo{pages}{231--246}.
  \DOIprefix\doi{10.1046/j.1365-8711.2001.04022.x}.
\bibitem[{{Mazeh} et~al.(1992){Mazeh}, {Goldberg}, {Duquennoy}, and
  {Mayor}}]{Mazeh1992ApJ...401..265M}
\bibinfo{author}{T.~{Mazeh}}, \bibinfo{author}{D.~{Goldberg}},
  \bibinfo{author}{A.~{Duquennoy}}, \bibinfo{author}{M.~{Mayor}},
\newblock \bibinfo{title}{{On the Mass-Ratio Distribution of Spectroscopic
  Binaries with Solar-Type Primaries}},
\newblock \bibinfo{journal}{\apj} \bibinfo{volume}{401} (\bibinfo{year}{1992})
  \bibinfo{pages}{265}. \DOIprefix\doi{10.1086/172058}.
\bibitem[{{Duquennoy} and {Mayor}(1991)}]{Duquennoy1991A&A...248..485D}
\bibinfo{author}{A.~{Duquennoy}}, \bibinfo{author}{M.~{Mayor}},
\newblock \bibinfo{title}{{Multiplicity among Solar Type Stars in the Solar
  Neighbourhood - Part Two - Distribution of the Orbital Elements in an
  Unbiased Sample}},
\newblock \bibinfo{journal}{\aap}  \bibinfo{volume}{248} (\bibinfo{year}{1991})
  \bibinfo{pages}{485}.
\bibitem[{{Cheng} et~al.(2020){Cheng}, {Cummings}, {M{\'e}nard}, and
  {Toonen}}]{Cheng2020ApJ...891..160C}
\bibinfo{author}{S.~{Cheng}}, \bibinfo{author}{J.~D. {Cummings}},
  \bibinfo{author}{B.~{M{\'e}nard}}, \bibinfo{author}{S.~{Toonen}},
\newblock \bibinfo{title}{{Double White Dwarf Merger Products among High-mass
  White Dwarfs}},
\newblock \bibinfo{journal}{\apj} \bibinfo{volume}{891} (\bibinfo{year}{2020})
  \bibinfo{pages}{160}. \DOIprefix\doi{10.3847/1538-4357/ab733c}.
\bibitem[{{Li} et~al.(2011){Li}, {Chornock}, {Leaman}, {Filippenko},
  {Poznanski}, {Wang}, {Ganeshalingam}, and {Mannucci}}]{Li2011MNRAS.412.1473L}
\bibinfo{author}{W.~{Li}}, \bibinfo{author}{R.~{Chornock}},
  \bibinfo{author}{J.~{Leaman}}, et~al.,
\newblock \bibinfo{title}{{Nearby supernova rates from the Lick Observatory
  Supernova Search - III. The rate-size relation, and the rates as a function
  of galaxy Hubble type and colour}},
\newblock \bibinfo{journal}{\mnras} \bibinfo{volume}{412}
  (\bibinfo{year}{2011}) \bibinfo{pages}{1473--1507}.
  \DOIprefix\doi{10.1111/j.1365-2966.2011.18162.x}.
\bibitem[{{Brown} et~al.(2020){Brown}, {Kilic}, {Kosakowski}, {Andrews},
  {Heinke}, {Ag{\"u}eros}, {Camilo}, {Gianninas}, {Hermes}, and
  {Kenyon}}]{Brown2020ApJ...889...49B}
\bibinfo{author}{W.~R. {Brown}}, \bibinfo{author}{M.~{Kilic}},
  \bibinfo{author}{A.~{Kosakowski}}, et~al.,
\newblock \bibinfo{title}{{The ELM Survey. VIII. Ninety-eight Double White
  Dwarf Binaries}},
\newblock \bibinfo{journal}{\apj} \bibinfo{volume}{889} (\bibinfo{year}{2020})
  \bibinfo{pages}{49}. \DOIprefix\doi{10.3847/1538-4357/ab63cd}.
\bibitem[{{Maoz} et~al.(2018){Maoz}, {Hallakoun}, and
  {Badenes}}]{Maoz2018MNRAS.476.2584M}
\bibinfo{author}{D.~{Maoz}}, \bibinfo{author}{N.~{Hallakoun}},
  \bibinfo{author}{C.~{Badenes}},
\newblock \bibinfo{title}{{The separation distribution and merger rate of
  double white dwarfs: improved constraints}},
\newblock \bibinfo{journal}{\mnras} \bibinfo{volume}{476}
  (\bibinfo{year}{2018}) \bibinfo{pages}{2584--2590}.
  \DOIprefix\doi{10.1093/mnras/sty339}.
\bibitem[{{Karakas} and {Lattanzio}(2014)}]{Karakas2014}
\bibinfo{author}{A.~I. {Karakas}}, \bibinfo{author}{J.~C. {Lattanzio}},
\newblock \bibinfo{title}{{The Dawes Review 2: Nucleosynthesis and Stellar
  Yields of Low- and Intermediate-Mass Single Stars}},
\newblock \bibinfo{journal}{\pasa} \bibinfo{volume}{31} (\bibinfo{year}{2014})
  \bibinfo{pages}{e030}. \DOIprefix\doi{10.1017/pasa.2014.21}.
\bibitem[{{Gehrz} et~al.(1998){Gehrz}, {Truran}, {Williams}, and
  {Starrfield}}]{Gehrz1998}
\bibinfo{author}{R.~D. {Gehrz}}, \bibinfo{author}{J.~W. {Truran}},
  \bibinfo{author}{R.~E. {Williams}}, \bibinfo{author}{S.~{Starrfield}},
\newblock \bibinfo{title}{{Nucleosynthesis in Classical Novae and Its
  Contribution to the Interstellar Medium}},
\newblock \bibinfo{journal}{\pasp} \bibinfo{volume}{110} (\bibinfo{year}{1998})
  \bibinfo{pages}{3--26}. \DOIprefix\doi{10.1086/316107}.
\bibitem[{{Yaron} et~al.(2005){Yaron}, {Prialnik}, {Shara}, and
  {Kovetz}}]{Yaron2005}
\bibinfo{author}{O.~{Yaron}}, \bibinfo{author}{D.~{Prialnik}},
  \bibinfo{author}{M.~M. {Shara}}, \bibinfo{author}{A.~{Kovetz}},
\newblock \bibinfo{title}{{An Extended Grid of Nova Models. II. The Parameter
  Space of Nova Outbursts}},
\newblock \bibinfo{journal}{\apj} \bibinfo{volume}{623} (\bibinfo{year}{2005})
  \bibinfo{pages}{398--410}. \DOIprefix\doi{10.1086/428435}.
\bibitem[{{Denissenkov} et~al.(2013){Denissenkov}, {Herwig}, {Bildsten}, and
  {Paxton}}]{Denissenkov2013}
\bibinfo{author}{P.~A. {Denissenkov}}, \bibinfo{author}{F.~{Herwig}},
  \bibinfo{author}{L.~{Bildsten}}, \bibinfo{author}{B.~{Paxton}},
\newblock \bibinfo{title}{{MESA Models of Classical Nova Outbursts: The
  Multicycle Evolution and Effects of Convective Boundary Mixing}},
\newblock \bibinfo{journal}{\apj} \bibinfo{volume}{762} (\bibinfo{year}{2013})
  \bibinfo{pages}{8}. \DOIprefix\doi{10.1088/0004-637X/762/1/8}.
\bibitem[{{Jos{\'e}} and {Hernanz}(1998)}]{Jose1998}
\bibinfo{author}{J.~{Jos{\'e}}}, \bibinfo{author}{M.~{Hernanz}},
\newblock \bibinfo{title}{{Nucleosynthesis in Classical Novae: CO versus ONe
  White Dwarfs}},
\newblock \bibinfo{journal}{\apj} \bibinfo{volume}{494} (\bibinfo{year}{1998})
  \bibinfo{pages}{680--690}. \DOIprefix\doi{10.1086/305244}.
\bibitem[{{Guo} et~al.(2022){Guo}, {Wu}, and {Wang}}]{Guo2022aa}
\bibinfo{author}{Y.~{Guo}}, \bibinfo{author}{C.~{Wu}},
  \bibinfo{author}{B.~{Wang}},
\newblock \bibinfo{title}{{Mixing fraction in classical novae}},
\newblock \bibinfo{journal}{\aap} \bibinfo{volume}{660} (\bibinfo{year}{2022})
  \bibinfo{pages}{A53}. \DOIprefix\doi{10.1051/0004-6361/202142163}.
\bibitem[{{Jos{\'e}} et~al.(2006){Jos{\'e}}, {Hernanz}, and
  {Iliadis}}]{Jose2006}
\bibinfo{author}{J.~{Jos{\'e}}}, \bibinfo{author}{M.~{Hernanz}},
  \bibinfo{author}{C.~{Iliadis}},
\newblock \bibinfo{title}{{Nucleosynthesis in classical novae}},
\newblock \bibinfo{journal}{\nphysa} \bibinfo{volume}{777}
  (\bibinfo{year}{2006}) \bibinfo{pages}{550--578}.
  \DOIprefix\doi{10.1016/j.nuclphysa.2005.02.121}.
\bibitem[{{Jos{\'e}}(2017)}]{Jose2017}
\bibinfo{author}{J.~{Jos{\'e}}},
\newblock \bibinfo{title}{{Nucleosynthesis in Novae}},
\newblock in: \bibinfo{editor}{S.~{Kubono}}, \bibinfo{editor}{T.~{Kajino}},
  \bibinfo{editor}{S.~{Nishimura}}, et~al. (Eds.), \bibinfo{booktitle}{14th
  International Symposium on Nuclei in the Cosmos (NIC2016)},
  \bibinfo{year}{2017}, p. \bibinfo{pages}{010501}.
  \DOIprefix\doi{10.7566/JPSCP.14.010501}.
\bibitem[{{Lattimer} and {Schramm}(1974)}]{Lattimer1974ApJ...192L.145L}
\bibinfo{author}{J.~M. {Lattimer}}, \bibinfo{author}{D.~N. {Schramm}},
\newblock \bibinfo{title}{{Black-Hole-Neutron-Star Collisions}},
\newblock \bibinfo{journal}{\apjl} \bibinfo{volume}{192} (\bibinfo{year}{1974})
  \bibinfo{pages}{L145}. \DOIprefix\doi{10.1086/181612}.
\bibitem[{{Eichler} et~al.(1989){Eichler}, {Livio}, {Piran}, and
  {Schramm}}]{Eichler1989Natur.340..126E}
\bibinfo{author}{D.~{Eichler}}, \bibinfo{author}{M.~{Livio}},
  \bibinfo{author}{T.~{Piran}}, \bibinfo{author}{D.~N. {Schramm}},
\newblock \bibinfo{title}{{Nucleosynthesis, neutrino bursts and
  {\ensuremath{\gamma}}-rays from coalescing neutron stars}},
\newblock \bibinfo{journal}{\nat} \bibinfo{volume}{340} (\bibinfo{year}{1989})
  \bibinfo{pages}{126--128}. \DOIprefix\doi{10.1038/340126a0}.
\bibitem[{{Freiburghaus} et~al.(1999){Freiburghaus}, {Rosswog}, and
  {Thielemann}}]{Freiburghaus1999ApJ...525L.121F}
\bibinfo{author}{C.~{Freiburghaus}}, \bibinfo{author}{S.~{Rosswog}},
  \bibinfo{author}{F.~K. {Thielemann}},
\newblock \bibinfo{title}{{R-Process in Neutron Star Mergers}},
\newblock \bibinfo{journal}{\apjl} \bibinfo{volume}{525} (\bibinfo{year}{1999})
  \bibinfo{pages}{L121--L124}. \DOIprefix\doi{10.1086/312343}.
\bibitem[{{Rosswog} et~al.(1999){Rosswog}, {Liebend{\"o}rfer}, {Thielemann},
  {Davies}, {Benz}, and {Piran}}]{Rosswog1999A&A...341..499R}
\bibinfo{author}{S.~{Rosswog}}, \bibinfo{author}{M.~{Liebend{\"o}rfer}},
  \bibinfo{author}{F.~K. {Thielemann}}, et~al.,
\newblock \bibinfo{title}{{Mass ejection in neutron star mergers}},
\newblock \bibinfo{journal}{\aap} \bibinfo{volume}{341} (\bibinfo{year}{1999})
  \bibinfo{pages}{499--526}. \DOIprefix\doi{10.48550/arXiv.astro-ph/9811367}.
\bibitem[{{Goriely} et~al.(2011){Goriely}, {Bauswein}, and
  {Janka}}]{Goriely2011ApJ...738L..32G}
\bibinfo{author}{S.~{Goriely}}, \bibinfo{author}{A.~{Bauswein}},
  \bibinfo{author}{H.-T. {Janka}},
\newblock \bibinfo{title}{{r-process Nucleosynthesis in Dynamically Ejected
  Matter of Neutron Star Mergers}},
\newblock \bibinfo{journal}{\apjl} \bibinfo{volume}{738} (\bibinfo{year}{2011})
  \bibinfo{pages}{L32}.
  \DOIprefix\doi{10.1088/2041-8205/738/2/L3210.48550/arXiv.1107.0899}.
\bibitem[{{Perego} et~al.(2014){Perego}, {Rosswog}, {Cabez{\'o}n}, {Korobkin},
  {K{\"a}ppeli}, {Arcones}, and {Liebend{\"o}rfer}}]{Perego2014MNRAS.443.3134P}
\bibinfo{author}{A.~{Perego}}, \bibinfo{author}{S.~{Rosswog}},
  \bibinfo{author}{R.~M. {Cabez{\'o}n}}, et~al.,
\newblock \bibinfo{title}{{Neutrino-driven winds from neutron star merger
  remnants}},
\newblock \bibinfo{journal}{\mnras} \bibinfo{volume}{443}
  (\bibinfo{year}{2014}) \bibinfo{pages}{3134--3156}.
  \DOIprefix\doi{10.1093/mnras/stu135210.48550/arXiv.1405.6730}.
\bibitem[{{Just} et~al.(2015){Just}, {Bauswein}, {Ardevol Pulpillo}, {Goriely},
  and {Janka}}]{Just2015MNRAS.448..541J}
\bibinfo{author}{O.~{Just}}, \bibinfo{author}{A.~{Bauswein}},
  \bibinfo{author}{R.~{Ardevol Pulpillo}}, et~al.,
\newblock \bibinfo{title}{{Comprehensive nucleosynthesis analysis for ejecta of
  compact binary mergers}},
\newblock \bibinfo{journal}{\mnras} \bibinfo{volume}{448}
  (\bibinfo{year}{2015}) \bibinfo{pages}{541--567}.
  \DOIprefix\doi{10.1093/mnras/stv00910.48550/arXiv.1406.2687}.
\bibitem[{{Sekiguchi} et~al.(2016){Sekiguchi}, {Kiuchi}, {Kyutoku}, {Shibata},
  and {Taniguchi}}]{Sekiguchi2016PhRvD..93l4046S}
\bibinfo{author}{Y.~{Sekiguchi}}, \bibinfo{author}{K.~{Kiuchi}},
  \bibinfo{author}{K.~{Kyutoku}}, et~al.,
\newblock \bibinfo{title}{{Dynamical mass ejection from the merger of
  asymmetric binary neutron stars: Radiation-hydrodynamics study in general
  relativity}},
\newblock \bibinfo{journal}{\prd} \bibinfo{volume}{93} (\bibinfo{year}{2016})
  \bibinfo{pages}{124046}.
  \DOIprefix\doi{10.1103/PhysRevD.93.12404610.48550/arXiv.1603.01918}.
\bibitem[{{Kasen} et~al.(2017){Kasen}, {Metzger}, {Barnes}, {Quataert}, and
  {Ramirez-Ruiz}}]{Kasen2017Natur.551...80K}
\bibinfo{author}{D.~{Kasen}}, \bibinfo{author}{B.~{Metzger}},
  \bibinfo{author}{J.~{Barnes}}, et~al.,
\newblock \bibinfo{title}{{Origin of the heavy elements in binary neutron-star
  mergers from a gravitational-wave event}},
\newblock \bibinfo{journal}{\nat} \bibinfo{volume}{551} (\bibinfo{year}{2017})
  \bibinfo{pages}{80--84}.
  \DOIprefix\doi{10.1038/nature2445310.48550/arXiv.1710.05463}.
\bibitem[{{Pian} et~al.(2017){Pian}, {D'Avanzo}, {Benetti}, {Branchesi},
  {Brocato}, {Campana}, {Cappellaro}, {Covino}, {D'Elia}, {Fynbo}, {Getman},
  {Ghirlanda}, and et~al.}]{Pian2017Natur.551...67P}
\bibinfo{author}{E.~{Pian}}, \bibinfo{author}{P.~{D'Avanzo}},
  \bibinfo{author}{S.~{Benetti}}, et~al.,
\newblock \bibinfo{title}{{Spectroscopic identification of r-process
  nucleosynthesis in a double neutron-star merger}},
\newblock \bibinfo{journal}{\nat} \bibinfo{volume}{551} (\bibinfo{year}{2017})
  \bibinfo{pages}{67--70}.
  \DOIprefix\doi{10.1038/nature2429810.48550/arXiv.1710.05858}.
\bibitem[{{Cowan} et~al.(2021){Cowan}, {Sneden}, {Lawler}, {Aprahamian},
  {Wiescher}, {Langanke}, {Mart{\'\i}nez-Pinedo}, and
  {Thielemann}}]{Cowan2021RvMP...93a5002C}
\bibinfo{author}{J.~J. {Cowan}}, \bibinfo{author}{C.~{Sneden}},
  \bibinfo{author}{J.~E. {Lawler}}, et~al.,
\newblock \bibinfo{title}{{Origin of the heaviest elements: The rapid
  neutron-capture process}},
\newblock \bibinfo{journal}{Reviews of Modern Physics} \bibinfo{volume}{93}
  (\bibinfo{year}{2021}) \bibinfo{pages}{015002}.
  \DOIprefix\doi{10.1103/RevModPhys.93.01500210.48550/arXiv.1901.01410}.
\bibitem[{{Li} and {Paczy{\'n}ski}(1998)}]{Li1998ApJ...507L..59L}
\bibinfo{author}{L.-X. {Li}}, \bibinfo{author}{B.~{Paczy{\'n}ski}},
\newblock \bibinfo{title}{{Transient Events from Neutron Star Mergers}},
\newblock \bibinfo{journal}{\apjl} \bibinfo{volume}{507} (\bibinfo{year}{1998})
  \bibinfo{pages}{L59--L62}.
  \DOIprefix\doi{10.1086/31168010.48550/arXiv.astro-ph/9807272}.
\bibitem[{{Metzger} et~al.(2010){Metzger}, {Mart{\'\i}nez-Pinedo}, {Darbha},
  {Quataert}, {Arcones}, {Kasen}, {Thomas}, {Nugent}, {Panov}, and
  {Zinner}}]{Metzger2010MNRAS.406.2650M}
\bibinfo{author}{B.~D. {Metzger}}, \bibinfo{author}{G.~{Mart{\'\i}nez-Pinedo}},
  \bibinfo{author}{S.~{Darbha}}, et~al.,
\newblock \bibinfo{title}{{Electromagnetic counterparts of compact object
  mergers powered by the radioactive decay of r-process nuclei}},
\newblock \bibinfo{journal}{\mnras} \bibinfo{volume}{406}
  (\bibinfo{year}{2010}) \bibinfo{pages}{2650--2662}.
  \DOIprefix\doi{10.1111/j.1365-2966.2010.16864.x10.48550/arXiv.1001.5029}.
\bibitem[{{Roberts} et~al.(2011){Roberts}, {Kasen}, {Lee}, and
  {Ramirez-Ruiz}}]{Roberts2011ApJ...736L..21R}
\bibinfo{author}{L.~F. {Roberts}}, \bibinfo{author}{D.~{Kasen}},
  \bibinfo{author}{W.~H. {Lee}}, \bibinfo{author}{E.~{Ramirez-Ruiz}},
\newblock \bibinfo{title}{{Electromagnetic Transients Powered by Nuclear Decay
  in the Tidal Tails of Coalescing Compact Binaries}},
\newblock \bibinfo{journal}{\apjl} \bibinfo{volume}{736} (\bibinfo{year}{2011})
  \bibinfo{pages}{L21}.
  \DOIprefix\doi{10.1088/2041-8205/736/1/L2110.48550/arXiv.1104.5504}.
\bibitem[{{Kasen} et~al.(2013){Kasen}, {Badnell}, and
  {Barnes}}]{Kasen2013ApJ...774...25K}
\bibinfo{author}{D.~{Kasen}}, \bibinfo{author}{N.~R. {Badnell}},
  \bibinfo{author}{J.~{Barnes}},
\newblock \bibinfo{title}{{Opacities and Spectra of the r-process Ejecta from
  Neutron Star Mergers}},
\newblock \bibinfo{journal}{\apj} \bibinfo{volume}{774} (\bibinfo{year}{2013})
  \bibinfo{pages}{25}.
  \DOIprefix\doi{10.1088/0004-637X/774/1/2510.48550/arXiv.1303.5788}.
\bibitem[{{Kasen} and {Barnes}(2019)}]{Kasen2019ApJ...876..128K}
\bibinfo{author}{D.~{Kasen}}, \bibinfo{author}{J.~{Barnes}},
\newblock \bibinfo{title}{{Radioactive Heating and Late Time Kilonova Light
  Curves}},
\newblock \bibinfo{journal}{\apj} \bibinfo{volume}{876} (\bibinfo{year}{2019})
  \bibinfo{pages}{128}.
  \DOIprefix\doi{10.3847/1538-4357/ab06c210.48550/arXiv.1807.03319}.
\bibitem[{{Abbott} et~al.(2017{\natexlab{a}}){Abbott}, {Abbott}, {Abbott},
  {Acernese}, {Ackley}, {Adams}, {Adams}, {Addesso}, {Adhikari}, and
  et~al.}]{Abbott2017ApJ...848L..12A}
\bibinfo{author}{B.~P. {Abbott}}, \bibinfo{author}{R.~{Abbott}},
  \bibinfo{author}{T.~D. {Abbott}}, et~al.,
\newblock \bibinfo{title}{{Multi-messenger Observations of a Binary Neutron
  Star Merger}},
\newblock \bibinfo{journal}{\apjl} \bibinfo{volume}{848}
  (\bibinfo{year}{2017}{\natexlab{a}}) \bibinfo{pages}{L12}.
  \DOIprefix\doi{10.3847/2041-8213/aa91c910.48550/arXiv.1710.05833}.
\bibitem[{{Abbott} et~al.(2017{\natexlab{b}}){Abbott}, {Abbott}, {Abbott},
  {Acernese}, {Ackley}, {Adams}, {Adams}, {Addesso}, {Adhikari}, and
  et~al.}]{Abbott2017ApJ...848L..13A}
\bibinfo{author}{B.~P. {Abbott}}, \bibinfo{author}{R.~{Abbott}},
  \bibinfo{author}{T.~D. {Abbott}}, et~al.,
\newblock \bibinfo{title}{{Gravitational Waves and Gamma-Rays from a Binary
  Neutron Star Merger: GW170817 and GRB 170817A}},
\newblock \bibinfo{journal}{\apjl} \bibinfo{volume}{848}
  (\bibinfo{year}{2017}{\natexlab{b}}) \bibinfo{pages}{L13}.
  \DOIprefix\doi{10.3847/2041-8213/aa920c10.48550/arXiv.1710.05834}.
\bibitem[{{Margalit} and {Metzger}(2017)}]{Margalit2017ApJ...850L..19M}
\bibinfo{author}{B.~{Margalit}}, \bibinfo{author}{B.~D. {Metzger}},
\newblock \bibinfo{title}{{Constraining the Maximum Mass of Neutron Stars from
  Multi-messenger Observations of GW170817}},
\newblock \bibinfo{journal}{\apjl} \bibinfo{volume}{850} (\bibinfo{year}{2017})
  \bibinfo{pages}{L19}.
  \DOIprefix\doi{10.3847/2041-8213/aa991c10.48550/arXiv.1710.05938}.
\bibitem[{{Radice} et~al.(2018){Radice}, {Perego}, {Zappa}, and
  {Bernuzzi}}]{Radice2018ApJ...852L..29R}
\bibinfo{author}{D.~{Radice}}, \bibinfo{author}{A.~{Perego}},
  \bibinfo{author}{F.~{Zappa}}, \bibinfo{author}{S.~{Bernuzzi}},
\newblock \bibinfo{title}{{GW170817: Joint Constraint on the Neutron Star
  Equation of State from Multimessenger Observations}},
\newblock \bibinfo{journal}{\apjl} \bibinfo{volume}{852} (\bibinfo{year}{2018})
  \bibinfo{pages}{L29}.
  \DOIprefix\doi{10.3847/2041-8213/aaa40210.48550/arXiv.1711.03647}.
\bibitem[{{Coulter} et~al.(2017){Coulter}, {Foley}, {Kilpatrick}, {Drout},
  {Piro}, {Shappee}, {Siebert}, {Simon}, {Ulloa}, {Kasen}, {Madore},
  {Murguia-Berthier}, {Pan}, {Prochaska}, {Ramirez-Ruiz}, {Rest}, and
  {Rojas-Bravo}}]{Coulter2017Sci...358.1556C}
\bibinfo{author}{D.~A. {Coulter}}, \bibinfo{author}{R.~J. {Foley}},
  \bibinfo{author}{C.~D. {Kilpatrick}}, et~al.,
\newblock \bibinfo{title}{{Swope Supernova Survey 2017a (SSS17a), the optical
  counterpart to a gravitational wave source}},
\newblock \bibinfo{journal}{Science} \bibinfo{volume}{358}
  (\bibinfo{year}{2017}) \bibinfo{pages}{1556--1558}.
  \DOIprefix\doi{10.1126/science.aap981110.48550/arXiv.1710.05452}.
\bibitem[{{Soares-Santos} et~al.(2017){Soares-Santos}, {Holz}, {Annis},
  {Chornock}, {Herner}, {Berger}, {Brout}, {Chen}, {Kessler}, et~al., {Dark
  Energy Survey}, and {Dark Energy Camera GW-EM
  Collaboration}}]{Soares-Santos2017ApJ...848L..16S}
\bibinfo{author}{M.~{Soares-Santos}}, \bibinfo{author}{D.~E. {Holz}},
  \bibinfo{author}{J.~{Annis}}, et~al.,
\newblock \bibinfo{title}{{The Electromagnetic Counterpart of the Binary
  Neutron Star Merger LIGO/Virgo GW170817. I. Discovery of the Optical
  Counterpart Using the Dark Energy Camera}},
\newblock \bibinfo{journal}{\apjl} \bibinfo{volume}{848} (\bibinfo{year}{2017})
  \bibinfo{pages}{L16}.
  \DOIprefix\doi{10.3847/2041-8213/aa905910.48550/arXiv.1710.05459}.
\bibitem[{{Arcavi} et~al.(2017){Arcavi}, {Hosseinzadeh}, {Howell}, {McCully},
  {Poznanski}, {Kasen}, {Barnes}, {Zaltzman}, {Vasylyev}, {Maoz}, and
  {Valenti}}]{Arcavi2017Natur.551...64A}
\bibinfo{author}{I.~{Arcavi}}, \bibinfo{author}{G.~{Hosseinzadeh}},
  \bibinfo{author}{D.~A. {Howell}}, et~al.,
\newblock \bibinfo{title}{{Optical emission from a kilonova following a
  gravitational-wave-detected neutron-star merger}},
\newblock \bibinfo{journal}{\nat} \bibinfo{volume}{551} (\bibinfo{year}{2017})
  \bibinfo{pages}{64--66}.
  \DOIprefix\doi{10.1038/nature2429110.48550/arXiv.1710.05843}.
\bibitem[{{Cowperthwaite} et~al.(2017){Cowperthwaite}, {Berger}, {Villar},
  {Metzger}, {Nicholl}, {Chornock}, {Blanchard}, {Fong}, and
  et~al.}]{Cowperthwaite2017ApJ...848L..17C}
\bibinfo{author}{P.~S. {Cowperthwaite}}, \bibinfo{author}{E.~{Berger}},
  \bibinfo{author}{V.~A. {Villar}}, et~al.,
\newblock \bibinfo{title}{{The Electromagnetic Counterpart of the Binary
  Neutron Star Merger LIGO/Virgo GW170817. II. UV, Optical, and Near-infrared
  Light Curves and Comparison to Kilonova Models}},
\newblock \bibinfo{journal}{\apjl} \bibinfo{volume}{848} (\bibinfo{year}{2017})
  \bibinfo{pages}{L17}.
  \DOIprefix\doi{10.3847/2041-8213/aa8fc710.48550/arXiv.1710.05840}.
\bibitem[{{Drout} et~al.(2017){Drout}, {Piro}, {Shappee}, {Kilpatrick},
  {Simon}, {Contreras}, {Coulter}, {Foley}, {Siebert}, and
  et~al.}]{Drout2017Sci...358.1570D}
\bibinfo{author}{M.~R. {Drout}}, \bibinfo{author}{A.~L. {Piro}},
  \bibinfo{author}{B.~J. {Shappee}}, et~al.,
\newblock \bibinfo{title}{{Light curves of the neutron star merger
  GW170817/SSS17a: Implications for r-process nucleosynthesis}},
\newblock \bibinfo{journal}{Science} \bibinfo{volume}{358}
  (\bibinfo{year}{2017}) \bibinfo{pages}{1570--1574}.
  \DOIprefix\doi{10.1126/science.aaq004910.48550/arXiv.1710.05443}.
\bibitem[{{Curtis} et~al.(2023){Curtis}, {M{\"o}sta}, {Wu}, {Radice},
  {Roberts}, {Ricigliano}, and {Perego}}]{Curtis2023MNRAS.518.5313C}
\bibinfo{author}{S.~{Curtis}}, \bibinfo{author}{P.~{M{\"o}sta}},
  \bibinfo{author}{Z.~{Wu}}, et~al.,
\newblock \bibinfo{title}{{r-process nucleosynthesis and kilonovae from
  hypermassive neutron star post-merger remnants}},
\newblock \bibinfo{journal}{\mnras} \bibinfo{volume}{518}
  (\bibinfo{year}{2023}) \bibinfo{pages}{5313--5322}.
  \DOIprefix\doi{10.1093/mnras/stac312810.48550/arXiv.2112.00772}.
\bibitem[{{Qian}(2000)}]{Qian2000ApJ...534L..67Q}
\bibinfo{author}{Y.~Z. {Qian}},
\newblock \bibinfo{title}{{Supernovae versus Neutron Star Mergers as the Major
  R-Process Sources}},
\newblock \bibinfo{journal}{\apjl} \bibinfo{volume}{534} (\bibinfo{year}{2000})
  \bibinfo{pages}{L67--L70}.
  \DOIprefix\doi{10.1086/31265910.48550/arXiv.astro-ph/0003242}.
\bibitem[{{Kalogera} et~al.(2004){Kalogera}, {Kim}, {Lorimer}, {Burgay},
  {D'Amico}, {Possenti}, {Manchester}, {Lyne}, {Joshi}, {McLaughlin}, {Kramer},
  {Sarkissian}, and {Camilo}}]{Kalogera2004ApJ...601L.179K}
\bibinfo{author}{V.~{Kalogera}}, \bibinfo{author}{C.~{Kim}},
  \bibinfo{author}{D.~R. {Lorimer}}, et~al.,
\newblock \bibinfo{title}{{The Cosmic Coalescence Rates for Double Neutron Star
  Binaries}},
\newblock \bibinfo{journal}{\apjl} \bibinfo{volume}{601} (\bibinfo{year}{2004})
  \bibinfo{pages}{L179--L182}. \DOIprefix\doi{10.1086/382155}.
\bibitem[{{Abadie} et~al.(2010){Abadie}, {Abbott}, {Abbott}, {Abernathy},
  {Accadia}, {Acernese}, {Adams}, {Adhikari}, {Ajith}, {Allen}, {Allen},
  {Amador Ceron}, and {Amin}}]{Abadie2010CQGra..27q3001A}
\bibinfo{author}{J.~{Abadie}}, \bibinfo{author}{B.~P. {Abbott}},
  \bibinfo{author}{R.~{Abbott}}, et~al.,
\newblock \bibinfo{title}{{TOPICAL REVIEW: Predictions for the rates of compact
  binary coalescences observable by ground-based gravitational-wave
  detectors}},
\newblock \bibinfo{journal}{Classical and Quantum Gravity} \bibinfo{volume}{27}
  (\bibinfo{year}{2010}) \bibinfo{pages}{173001}.
  \DOIprefix\doi{10.1088/0264-9381/27/17/17300110.48550/arXiv.1003.2480}.
\bibitem[{{Kim} et~al.(2015){Kim}, {Perera}, and
  {McLaughlin}}]{Kim2015MNRAS.448..928K}
\bibinfo{author}{C.~{Kim}}, \bibinfo{author}{B.~B.~P. {Perera}},
  \bibinfo{author}{M.~A. {McLaughlin}},
\newblock \bibinfo{title}{{Implications of PSR J0737-3039B for the Galactic
  NS-NS binary merger rate}},
\newblock \bibinfo{journal}{\mnras} \bibinfo{volume}{448}
  (\bibinfo{year}{2015}) \bibinfo{pages}{928--938}.
  \DOIprefix\doi{10.1093/mnras/stu272910.48550/arXiv.1308.4676}.
\bibitem[{{Riess} et~al.(1998){Riess}, {Filippenko}, {Challis}, {Clocchiatti},
  {Diercks}, and et~al.}]{Riess1998AJ....116.1009R}
\bibinfo{author}{A.~G. {Riess}}, \bibinfo{author}{A.~V. {Filippenko}},
  \bibinfo{author}{P.~{Challis}}, et~al.,
\newblock \bibinfo{title}{{Observational Evidence from Supernovae for an
  Accelerating Universe and a Cosmological Constant}},
\newblock \bibinfo{journal}{\aj} \bibinfo{volume}{116} (\bibinfo{year}{1998})
  \bibinfo{pages}{1009--1038}. \DOIprefix\doi{10.1086/300499}.
\bibitem[{{Schmidt} et~al.(1998){Schmidt}, {Suntzeff}, {Phillips}, {Schommer},
  and et~al.}]{Schmidt1998ApJ...507...46S}
\bibinfo{author}{B.~P. {Schmidt}}, \bibinfo{author}{N.~B. {Suntzeff}},
  \bibinfo{author}{M.~M. {Phillips}}, et~al.,
\newblock \bibinfo{title}{{The High-Z Supernova Search: Measuring Cosmic
  Deceleration and Global Curvature of the Universe Using Type IA Supernovae}},
\newblock \bibinfo{journal}{\apj} \bibinfo{volume}{507} (\bibinfo{year}{1998})
  \bibinfo{pages}{46--63}. \DOIprefix\doi{10.1086/306308}.
\bibitem[{{Perlmutter} et~al.(1999){Perlmutter}, {Aldering}, {Goldhaber},
  {Knop}, {Nugent}, {Castro}, and et~al}]{Perlmutter1999ApJ...517..565P}
\bibinfo{author}{S.~{Perlmutter}}, \bibinfo{author}{G.~{Aldering}},
  \bibinfo{author}{G.~{Goldhaber}}, et~al.,
\newblock \bibinfo{title}{{Measurements of {\ensuremath{\Omega}} and
  {\ensuremath{\Lambda}} from 42 High-Redshift Supernovae}},
\newblock \bibinfo{journal}{\apj} \bibinfo{volume}{517} (\bibinfo{year}{1999})
  \bibinfo{pages}{565--586}. \DOIprefix\doi{10.1086/307221}.
\bibitem[{{Arnett}(1969)}]{Arnett1969Ap&SS...5..180A}
\bibinfo{author}{W.~D. {Arnett}},
\newblock \bibinfo{title}{{A Possible Model of Supernovae: Detonation of
  $^{12}$C}},
\newblock \bibinfo{journal}{\apss} \bibinfo{volume}{5} (\bibinfo{year}{1969})
  \bibinfo{pages}{180--212}. \DOIprefix\doi{10.1007/BF00650291}.
\bibitem[{{Nomoto} et~al.(1984){Nomoto}, {Thielemann}, and
  {Yokoi}}]{Nomoto1984ApJ...286..644N}
\bibinfo{author}{K.~{Nomoto}}, \bibinfo{author}{F.~K. {Thielemann}},
  \bibinfo{author}{K.~{Yokoi}},
\newblock \bibinfo{title}{{Accreting white dwarf models for type I supern. III.
  Carbon deflagration supernovae.}},
\newblock \bibinfo{journal}{\apj} \bibinfo{volume}{286} (\bibinfo{year}{1984})
  \bibinfo{pages}{644--658}. \DOIprefix\doi{10.1086/162639}.
\bibitem[{{Thielemann} et~al.(1986){Thielemann}, {Nomoto}, and
  {Yokoi}}]{Thielemann1986A&A...158...17T}
\bibinfo{author}{F.~K. {Thielemann}}, \bibinfo{author}{K.~{Nomoto}},
  \bibinfo{author}{K.~{Yokoi}},
\newblock \bibinfo{title}{{Explosive nucleosynthesis in carbon deflagration
  models of Type I supernovae}},
\newblock \bibinfo{journal}{\aap}  \bibinfo{volume}{158} (\bibinfo{year}{1986})
  \bibinfo{pages}{17--33}.
\bibitem[{{Seitenzahl} and {Townsley}(2017)}]{Seitenzahl2017hsn..book.1955S}
\bibinfo{author}{I.~R. {Seitenzahl}}, \bibinfo{author}{D.~M. {Townsley}},
\newblock \bibinfo{title}{{Nucleosynthesis in Thermonuclear Supernovae}},
\newblock in: \bibinfo{editor}{A.~W. {Alsabti}}, \bibinfo{editor}{P.~{Murdin}}
  (Eds.), \bibinfo{booktitle}{Handbook of Supernovae}, \bibinfo{year}{2017}, p.
  \bibinfo{pages}{1955}. \DOIprefix\doi{10.1007/978-3-319-21846-5_87}.
\bibitem[{{Hillebrandt} et~al.(2013){Hillebrandt}, {Kromer}, {R{\"o}pke}, and
  {Ruiter}}]{Hillebrandt2013FrPhy...8..116H}
\bibinfo{author}{W.~{Hillebrandt}}, \bibinfo{author}{M.~{Kromer}},
  \bibinfo{author}{F.~K. {R{\"o}pke}}, \bibinfo{author}{A.~J. {Ruiter}},
\newblock \bibinfo{title}{{Towards an understanding of Type Ia supernovae from
  a synthesis of theory and observations}},
\newblock \bibinfo{journal}{Frontiers of Physics} \bibinfo{volume}{8}
  (\bibinfo{year}{2013}) \bibinfo{pages}{116--143}.
  \DOIprefix\doi{10.1007/s11467-013-0303-2}.
\bibitem[{{Lach} et~al.(2020){Lach}, {R{\"o}pke}, {Seitenzahl}, {Cot{\'e}},
  {Gronow}, and {Ruiter}}]{Lach2020A&A...644A.118L}
\bibinfo{author}{F.~{Lach}}, \bibinfo{author}{F.~K. {R{\"o}pke}},
  \bibinfo{author}{I.~R. {Seitenzahl}}, et~al.,
\newblock \bibinfo{title}{{Nucleosynthesis imprints from different Type Ia
  supernova explosion scenarios and implications for galactic chemical
  evolution}},
\newblock \bibinfo{journal}{\aap} \bibinfo{volume}{644} (\bibinfo{year}{2020})
  \bibinfo{pages}{A118}. \DOIprefix\doi{10.1051/0004-6361/202038721}.
\bibitem[{{Liu} et~al.(2023){Liu}, {R{\"o}pke}, and {Han}}]{Liu2023RAA}
\bibinfo{author}{Z.-W. {Liu}}, \bibinfo{author}{F.~K. {R{\"o}pke}},
  \bibinfo{author}{Z.~{Han}},
\newblock \bibinfo{title}{{Type Ia Supernova Explosions in Binary Systems: A
  Review}},
\newblock \bibinfo{journal}{Research in Astronomy and Astrophysics}
  \bibinfo{volume}{23} (\bibinfo{year}{2023}) \bibinfo{pages}{082001}.
  \DOIprefix\doi{10.1088/1674-4527/acd89e}.
\bibitem[{{Kromer} et~al.(2013){Kromer}, {Fink}, {Stanishev}, {Taubenberger},
  {Ciaraldi-Schoolman}, {Pakmor}, {R{\"o}pke}, {Ruiter}, {Seitenzahl}, {Sim},
  {Blanc}, {Elias-Rosa}, and {Hillebrandt}}]{Kromer2013MNRAS.429.2287K}
\bibinfo{author}{M.~{Kromer}}, \bibinfo{author}{M.~{Fink}},
  \bibinfo{author}{V.~{Stanishev}}, et~al.,
\newblock \bibinfo{title}{{3D deflagration simulations leaving bound remnants:
  a model for 2002cx-like Type Ia supernovae}},
\newblock \bibinfo{journal}{\mnras} \bibinfo{volume}{429}
  (\bibinfo{year}{2013}) \bibinfo{pages}{2287--2297}.
  \DOIprefix\doi{10.1093/mnras/sts498}.
\bibitem[{{Kromer} et~al.(2015){Kromer}, {Ohlmann}, {Pakmor}, {Ruiter},
  {Hillebrandt}, {Marquardt}, {R{\"o}pke}, {Seitenzahl}, {Sim}, and
  {Taubenberger}}]{Kromer2015MNRAS.450.3045K}
\bibinfo{author}{M.~{Kromer}}, \bibinfo{author}{S.~T. {Ohlmann}},
  \bibinfo{author}{R.~{Pakmor}}, et~al.,
\newblock \bibinfo{title}{{Deflagrations in hybrid CONe white dwarfs: a route
  to explain the faint Type Iax supernova 2008ha}},
\newblock \bibinfo{journal}{\mnras} \bibinfo{volume}{450}
  (\bibinfo{year}{2015}) \bibinfo{pages}{3045--3053}.
  \DOIprefix\doi{10.1093/mnras/stv886}.
\bibitem[{{Khokhlov}(1991{\natexlab{a}})}]{Khokhlov1991A&A...245L..25K}
\bibinfo{author}{A.~M. {Khokhlov}},
\newblock \bibinfo{title}{{Nucleosynthesis in delayed detonation models of type
  IA supernovae.}},
\newblock \bibinfo{journal}{\aap}  \bibinfo{volume}{245}
  (\bibinfo{year}{1991}{\natexlab{a}}) \bibinfo{pages}{L25}.
\bibitem[{{Khokhlov}(1991{\natexlab{b}})}]{Khokhlov1991A&A...245..114K}
\bibinfo{author}{A.~M. {Khokhlov}},
\newblock \bibinfo{title}{{Delayed detonation model for type IA supernovae}},
\newblock \bibinfo{journal}{\aap}  \bibinfo{volume}{245}
  (\bibinfo{year}{1991}{\natexlab{b}}) \bibinfo{pages}{114--128}.
\bibitem[{{Hillebrandt} and {Niemeyer}(2000)}]{Hillebrandt2000ARA&A..38..191H}
\bibinfo{author}{W.~{Hillebrandt}}, \bibinfo{author}{J.~C. {Niemeyer}},
\newblock \bibinfo{title}{{Type IA Supernova Explosion Models}},
\newblock \bibinfo{journal}{\araa} \bibinfo{volume}{38} (\bibinfo{year}{2000})
  \bibinfo{pages}{191--230}. \DOIprefix\doi{10.1146/annurev.astro.38.1.191}.
\bibitem[{{Taam}(1980)}]{Taam1980}
\bibinfo{author}{R.~E. {Taam}},
\newblock \bibinfo{title}{{The long-term evolution of accreting carbon white
  dwarfs}},
\newblock \bibinfo{journal}{\apj} \bibinfo{volume}{242} (\bibinfo{year}{1980})
  \bibinfo{pages}{749--755}. \DOIprefix\doi{10.1086/158509}.
\bibitem[{{Livne} and {Glasner}(1990)}]{Livne1990}
\bibinfo{author}{E.~{Livne}}, \bibinfo{author}{A.~S. {Glasner}},
\newblock \bibinfo{title}{{Geometrical effects in off-center detonation of
  helium shells}},
\newblock \bibinfo{journal}{\apj} \bibinfo{volume}{361} (\bibinfo{year}{1990})
  \bibinfo{pages}{244--250}. \DOIprefix\doi{10.1086/169189}.
\bibitem[{{Woosley} and {Weaver}(1994)}]{Woosley1994}
\bibinfo{author}{S.~E. {Woosley}}, \bibinfo{author}{T.~A. {Weaver}},
\newblock \bibinfo{title}{{Sub-Chandrasekhar mass models for Type IA
  supernovae}},
\newblock \bibinfo{journal}{\apj} \bibinfo{volume}{423} (\bibinfo{year}{1994})
  \bibinfo{pages}{371--379}. \DOIprefix\doi{10.1086/173813}.
\bibitem[{{Bildsten} et~al.(2007){Bildsten}, {Shen}, {Weinberg}, and
  {Nelemans}}]{Bildsten2007}
\bibinfo{author}{L.~{Bildsten}}, \bibinfo{author}{K.~J. {Shen}},
  \bibinfo{author}{N.~N. {Weinberg}}, \bibinfo{author}{G.~{Nelemans}},
\newblock \bibinfo{title}{{Faint Thermonuclear Supernovae from AM Canum
  Venaticorum Binaries}},
\newblock \bibinfo{journal}{\apjl} \bibinfo{volume}{662} (\bibinfo{year}{2007})
  \bibinfo{pages}{L95--L98}. \DOIprefix\doi{10.1086/519489}.
\bibitem[{{Fink} et~al.(2010){Fink}, {R{\"o}pke}, {Hillebrandt}, {Seitenzahl},
  {Sim}, and {Kromer}}]{Fink2010}
\bibinfo{author}{M.~{Fink}}, \bibinfo{author}{F.~K. {R{\"o}pke}},
  \bibinfo{author}{W.~{Hillebrandt}}, et~al.,
\newblock \bibinfo{title}{{Double-detonation sub-Chandrasekhar supernovae: can
  minimum helium shell masses detonate the core?}},
\newblock \bibinfo{journal}{\aap} \bibinfo{volume}{514} (\bibinfo{year}{2010})
  \bibinfo{pages}{A53}. \DOIprefix\doi{10.1051/0004-6361/200913892}.
\bibitem[{{Gronow} et~al.(2021){Gronow}, {C{\^o}t{\'e}}, {Lach}, {Seitenzahl},
  {Collins}, {Sim}, and {R{\"o}pke}}]{Gronow2021A&A...656A..94G}
\bibinfo{author}{S.~{Gronow}}, \bibinfo{author}{B.~{C{\^o}t{\'e}}},
  \bibinfo{author}{F.~{Lach}}, et~al.,
\newblock \bibinfo{title}{{Metallicity-dependent nucleosynthetic yields of Type
  Ia supernovae originating from double detonations of sub-M$_{Ch}$ white
  dwarfs}},
\newblock \bibinfo{journal}{\aap} \bibinfo{volume}{656} (\bibinfo{year}{2021})
  \bibinfo{pages}{A94}. \DOIprefix\doi{10.1051/0004-6361/202140881}.
\bibitem[{{Shen} et~al.(2021){Shen}, {Boos}, {Townsley}, and
  {Kasen}}]{Shen2021ApJ...922...68S}
\bibinfo{author}{K.~J. {Shen}}, \bibinfo{author}{S.~J. {Boos}},
  \bibinfo{author}{D.~M. {Townsley}}, \bibinfo{author}{D.~{Kasen}},
\newblock \bibinfo{title}{{Multidimensional Radiative Transfer Calculations of
  Double Detonations of Sub-Chandrasekhar-mass White Dwarfs}},
\newblock \bibinfo{journal}{\apj} \bibinfo{volume}{922} (\bibinfo{year}{2021})
  \bibinfo{pages}{68}. \DOIprefix\doi{10.3847/1538-4357/ac2304}.
\bibitem[{{Kromer} et~al.(2010){Kromer}, {Sim}, {Fink}, {R{\"o}pke},
  {Seitenzahl}, and {Hillebrandt}}]{Kromer2010ApJ...719.1067K}
\bibinfo{author}{M.~{Kromer}}, \bibinfo{author}{S.~A. {Sim}},
  \bibinfo{author}{M.~{Fink}}, et~al.,
\newblock \bibinfo{title}{{Double-detonation Sub-Chandrasekhar Supernovae:
  Synthetic Observables for Minimum Helium Shell Mass Models}},
\newblock \bibinfo{journal}{\apj} \bibinfo{volume}{719} (\bibinfo{year}{2010})
  \bibinfo{pages}{1067--1082}. \DOIprefix\doi{10.1088/0004-637X/719/2/1067}.
\bibitem[{{Sim} et~al.(2012){Sim}, {Fink}, {Kromer}, {R{\"o}pke}, {Ruiter}, and
  {Hillebrandt}}]{Sim2012MNRAS.420.3003S}
\bibinfo{author}{S.~A. {Sim}}, \bibinfo{author}{M.~{Fink}},
  \bibinfo{author}{M.~{Kromer}}, et~al.,
\newblock \bibinfo{title}{{2D simulations of the double-detonation model for
  thermonuclear transients from low-mass carbon-oxygen white dwarfs}},
\newblock \bibinfo{journal}{\mnras} \bibinfo{volume}{420}
  (\bibinfo{year}{2012}) \bibinfo{pages}{3003--3016}.
  \DOIprefix\doi{10.1111/j.1365-2966.2011.20162.x}.
\bibitem[{{Pakmor} et~al.(2013){Pakmor}, {Kromer}, {Taubenberger}, and
  {Springel}}]{Pakmor2013ApJ...770L...8P}
\bibinfo{author}{R.~{Pakmor}}, \bibinfo{author}{M.~{Kromer}},
  \bibinfo{author}{S.~{Taubenberger}}, \bibinfo{author}{V.~{Springel}},
\newblock \bibinfo{title}{{Helium-ignited Violent Mergers as a Unified Model
  for Normal and Rapidly Declining Type Ia Supernovae}},
\newblock \bibinfo{journal}{\apjl} \bibinfo{volume}{770} (\bibinfo{year}{2013})
  \bibinfo{pages}{L8}. \DOIprefix\doi{10.1088/2041-8205/770/1/L8}.
\bibitem[{{Townsley} et~al.(2019){Townsley}, {Miles}, {Shen}, and
  {Kasen}}]{Townsley2019}
\bibinfo{author}{D.~M. {Townsley}}, \bibinfo{author}{B.~J. {Miles}},
  \bibinfo{author}{K.~J. {Shen}}, \bibinfo{author}{D.~{Kasen}},
\newblock \bibinfo{title}{{Double Detonations with Thin, Modestly Enriched
  Helium Layers can Make Normal Type Ia Supernovae}},
\newblock \bibinfo{journal}{\apjl} \bibinfo{volume}{878} (\bibinfo{year}{2019})
  \bibinfo{pages}{L38}. \DOIprefix\doi{10.3847/2041-8213/ab27cd}.
\bibitem[{{Roy} et~al.(2022){Roy}, {Tiwari}, {Bobrick}, {Kosakowski}, {Fisher},
  {Perets}, {Kashyap}, {Lor{\'e}n-Aguilar}, and
  {Garc{\'\i}a-Berro}}]{Roy2022ApJ...932L..24R}
\bibinfo{author}{N.~C. {Roy}}, \bibinfo{author}{V.~{Tiwari}},
  \bibinfo{author}{A.~{Bobrick}}, et~al.,
\newblock \bibinfo{title}{{3D Hydrodynamical Simulations of Helium-ignited
  Double-degenerate White Dwarf Mergers}},
\newblock \bibinfo{journal}{\apjl} \bibinfo{volume}{932} (\bibinfo{year}{2022})
  \bibinfo{pages}{L24}. \DOIprefix\doi{10.3847/2041-8213/ac75e7}.
\bibitem[{{Pakmor} et~al.(2010){Pakmor}, {Kromer}, {R{\"o}pke}, {Sim},
  {Ruiter}, and {Hillebrandt}}]{Pakmor2010Natur.463...61P}
\bibinfo{author}{R.~{Pakmor}}, \bibinfo{author}{M.~{Kromer}},
  \bibinfo{author}{F.~K. {R{\"o}pke}}, et~al.,
\newblock \bibinfo{title}{{Sub-luminous type Ia supernovae from the mergers of
  equal-mass white dwarfs with mass
  \raisebox{-0.5ex}\textasciitilde0.9M$_{solar}$}},
\newblock \bibinfo{journal}{\nat} \bibinfo{volume}{463} (\bibinfo{year}{2010})
  \bibinfo{pages}{61--64}. \DOIprefix\doi{10.1038/nature08642}.
\bibitem[{{Pakmor} et~al.(2012){Pakmor}, {Kromer}, {Taubenberger}, {Sim},
  {R{\"o}pke}, and {Hillebrandt}}]{Pakmor2012ApJ...747L..10P}
\bibinfo{author}{R.~{Pakmor}}, \bibinfo{author}{M.~{Kromer}},
  \bibinfo{author}{S.~{Taubenberger}}, et~al.,
\newblock \bibinfo{title}{{Normal Type Ia Supernovae from Violent Mergers of
  White Dwarf Binaries}},
\newblock \bibinfo{journal}{\apjl} \bibinfo{volume}{747} (\bibinfo{year}{2012})
  \bibinfo{pages}{L10}. \DOIprefix\doi{10.1088/2041-8205/747/1/L10}.
\bibitem[{{Aoki} et~al.(2007){Aoki}, {Beers}, {Christlieb}, {Norris}, {Ryan},
  and {Tsangarides}}]{Aoki2007ApJ...655..492A}
\bibinfo{author}{W.~{Aoki}}, \bibinfo{author}{T.~C. {Beers}},
  \bibinfo{author}{N.~{Christlieb}}, et~al.,
\newblock \bibinfo{title}{{Carbon-enhanced Metal-poor Stars. I. Chemical
  Compositions of 26 Stars}},
\newblock \bibinfo{journal}{\apj} \bibinfo{volume}{655} (\bibinfo{year}{2007})
  \bibinfo{pages}{492--521}. \DOIprefix\doi{10.1086/509817}.
\bibitem[{{Lucatello} et~al.(2006){Lucatello}, {Beers}, {Christlieb},
  {Barklem}, {Rossi}, {Marsteller}, {Sivarani}, and
  {Lee}}]{Lucatello2006ApJ...652L..37L}
\bibinfo{author}{S.~{Lucatello}}, \bibinfo{author}{T.~C. {Beers}},
  \bibinfo{author}{N.~{Christlieb}}, et~al.,
\newblock \bibinfo{title}{{The Frequency of Carbon-enhanced Metal-poor Stars in
  the Galaxy from the HERES Sample}},
\newblock \bibinfo{journal}{\apjl} \bibinfo{volume}{652} (\bibinfo{year}{2006})
  \bibinfo{pages}{L37--L40}. \DOIprefix\doi{10.1086/509780}.
\bibitem[{{Allen} et~al.(2012){Allen}, {Ryan}, {Rossi}, {Beers}, and
  {Tsangarides}}]{Allen2012A&A...548A..34A}
\bibinfo{author}{D.~M. {Allen}}, \bibinfo{author}{S.~G. {Ryan}},
  \bibinfo{author}{S.~{Rossi}}, et~al.,
\newblock \bibinfo{title}{{Elemental abundances and classification of
  carbon-enhanced metal-poor stars}},
\newblock \bibinfo{journal}{\aap} \bibinfo{volume}{548} (\bibinfo{year}{2012})
  \bibinfo{pages}{A34}. \DOIprefix\doi{10.1051/0004-6361/201015615}.
\bibitem[{{Lee} et~al.(2013){Lee}, {Beers}, {Masseron}, {Plez}, {Rockosi},
  {Sobeck}, {Yanny}, {Lucatello}, {Sivarani}, {Placco}, and
  {Carollo}}]{Lee2013AJ....146..132L}
\bibinfo{author}{Y.~S. {Lee}}, \bibinfo{author}{T.~C. {Beers}},
  \bibinfo{author}{T.~{Masseron}}, et~al.,
\newblock \bibinfo{title}{{Carbon-enhanced Metal-poor Stars in SDSS/SEGUE. I.
  Carbon Abundance Estimation and Frequency of CEMP Stars}},
\newblock \bibinfo{journal}{\aj} \bibinfo{volume}{146} (\bibinfo{year}{2013})
  \bibinfo{pages}{132}. \DOIprefix\doi{10.1088/0004-6256/146/5/132}.
\bibitem[{{Placco} et~al.(2014){Placco}, {Frebel}, {Beers}, and
  {Stancliffe}}]{Placco2014ApJ...797...21P}
\bibinfo{author}{V.~M. {Placco}}, \bibinfo{author}{A.~{Frebel}},
  \bibinfo{author}{T.~C. {Beers}}, \bibinfo{author}{R.~J. {Stancliffe}},
\newblock \bibinfo{title}{{Carbon-enhanced Metal-poor Star Frequencies in the
  Galaxy: Corrections for the Effect of Evolutionary Status on Carbon
  Abundances}},
\newblock \bibinfo{journal}{\apj} \bibinfo{volume}{797} (\bibinfo{year}{2014})
  \bibinfo{pages}{21}. \DOIprefix\doi{10.1088/0004-637X/797/1/21}.
\bibitem[{{Beers} and {Christlieb}(2005)}]{Beers2005ARA&A..43..531B}
\bibinfo{author}{T.~C. {Beers}}, \bibinfo{author}{N.~{Christlieb}},
\newblock \bibinfo{title}{{The Discovery and Analysis of Very Metal-Poor Stars
  in the Galaxy}},
\newblock \bibinfo{journal}{\araa} \bibinfo{volume}{43} (\bibinfo{year}{2005})
  \bibinfo{pages}{531--580}.
  \DOIprefix\doi{10.1146/annurev.astro.42.053102.134057}.
\bibitem[{{Arentsen} et~al.(2022){Arentsen}, {Placco}, {Lee}, {Aguado},
  {Martin}, {Starkenburg}, and {Yoon}}]{Arentsen2022MNRAS.515.4082A}
\bibinfo{author}{A.~{Arentsen}}, \bibinfo{author}{V.~M. {Placco}},
  \bibinfo{author}{Y.~S. {Lee}}, et~al.,
\newblock \bibinfo{title}{{On the inconsistency of [C/Fe] abundances and the
  fractions of carbon-enhanced metal-poor stars among various stellar
  surveys}},
\newblock \bibinfo{journal}{\mnras} \bibinfo{volume}{515}
  (\bibinfo{year}{2022}) \bibinfo{pages}{4082--4098}.
  \DOIprefix\doi{10.1093/mnras/stac2062}.
\bibitem[{{Lugaro} et~al.(2012){Lugaro}, {Karakas}, {Stancliffe}, and
  {Rijs}}]{Lugaro2012ApJ...747....2L}
\bibinfo{author}{M.~{Lugaro}}, \bibinfo{author}{A.~I. {Karakas}},
  \bibinfo{author}{R.~J. {Stancliffe}}, \bibinfo{author}{C.~{Rijs}},
\newblock \bibinfo{title}{{The s-process in Asymptotic Giant Branch Stars of
  Low Metallicity and the Composition of Carbon-enhanced Metal-poor Stars}},
\newblock \bibinfo{journal}{\apj} \bibinfo{volume}{747} (\bibinfo{year}{2012})
  \bibinfo{pages}{2}. \DOIprefix\doi{10.1088/0004-637X/747/1/2}.
\bibitem[{{Jonsell} et~al.(2006){Jonsell}, {Barklem}, {Gustafsson},
  {Christlieb}, {Hill}, {Beers}, and {Holmberg}}]{Jonsell2006A&A...451..651J}
\bibinfo{author}{K.~{Jonsell}}, \bibinfo{author}{P.~S. {Barklem}},
  \bibinfo{author}{B.~{Gustafsson}}, et~al.,
\newblock \bibinfo{title}{{The Hamburg/ESO R-process enhanced star survey
  (HERES). III. HE 0338-3945 and the formation of the r + s stars}},
\newblock \bibinfo{journal}{\aap} \bibinfo{volume}{451} (\bibinfo{year}{2006})
  \bibinfo{pages}{651--670}. \DOIprefix\doi{10.1051/0004-6361:20054470}.
\bibitem[{{Lugaro} et~al.(2009){Lugaro}, {Campbell}, and {de
  Mink}}]{Lugaro2009PASA...26..322L}
\bibinfo{author}{M.~{Lugaro}}, \bibinfo{author}{S.~W. {Campbell}},
  \bibinfo{author}{S.~E. {de Mink}},
\newblock \bibinfo{title}{{The Mystery of CEMPs+r Stars and the Dual Core-Flash
  Neutron Superburst}},
\newblock \bibinfo{journal}{\pasa} \bibinfo{volume}{26} (\bibinfo{year}{2009})
  \bibinfo{pages}{322--326}. \DOIprefix\doi{10.1071/AS08068}.
\bibitem[{{Cooke} and {Madau}(2014)}]{Cooke2014ApJ...791..116C}
\bibinfo{author}{R.~J. {Cooke}}, \bibinfo{author}{P.~{Madau}},
\newblock \bibinfo{title}{{Carbon-enhanced Metal-poor Stars: Relics from the
  Dark Ages}},
\newblock \bibinfo{journal}{\apj} \bibinfo{volume}{791} (\bibinfo{year}{2014})
  \bibinfo{pages}{116}. \DOIprefix\doi{10.1088/0004-637X/791/2/116}.
\bibitem[{{Frebel} and {Norris}(2015)}]{Frebel2015ARA&A..53..631F}
\bibinfo{author}{A.~{Frebel}}, \bibinfo{author}{J.~E. {Norris}},
\newblock \bibinfo{title}{{Near-Field Cosmology with Extremely Metal-Poor
  Stars}},
\newblock \bibinfo{journal}{\araa} \bibinfo{volume}{53} (\bibinfo{year}{2015})
  \bibinfo{pages}{631--688}.
  \DOIprefix\doi{10.1146/annurev-astro-082214-122423}.
\bibitem[{{Hansen} et~al.(2015){Hansen}, {Andersen}, {Nordstr{\"o}m}, {Beers},
  {Yoon}, and {Buchhave}}]{Hansen2015A&A...583A..49H}
\bibinfo{author}{T.~T. {Hansen}}, \bibinfo{author}{J.~{Andersen}},
  \bibinfo{author}{B.~{Nordstr{\"o}m}}, et~al.,
\newblock \bibinfo{title}{{The role of binaries in the enrichment of the early
  Galactic halo. I. r-process-enhanced metal-poor stars}},
\newblock \bibinfo{journal}{\aap} \bibinfo{volume}{583} (\bibinfo{year}{2015})
  \bibinfo{pages}{A49}. \DOIprefix\doi{10.1051/0004-6361/201526812}.
\bibitem[{{Komiya} et~al.(2007){Komiya}, {Suda}, {Minaguchi}, {Shigeyama},
  {Aoki}, and {Fujimoto}}]{Komiya2007ApJ...658..367K}
\bibinfo{author}{Y.~{Komiya}}, \bibinfo{author}{T.~{Suda}},
  \bibinfo{author}{H.~{Minaguchi}}, et~al.,
\newblock \bibinfo{title}{{The Origin of Carbon Enhancement and the Initial
  Mass Function of Extremely Metal-poor Stars in the Galactic Halo}},
\newblock \bibinfo{journal}{\apj} \bibinfo{volume}{658} (\bibinfo{year}{2007})
  \bibinfo{pages}{367--390}. \DOIprefix\doi{10.1086/510826}.
\bibitem[{{Masseron} et~al.(2010){Masseron}, {Johnson}, {Plez}, {van Eck},
  {Primas}, {Goriely}, and {Jorissen}}]{Masseron2010A&A...509A..93M}
\bibinfo{author}{T.~{Masseron}}, \bibinfo{author}{J.~A. {Johnson}},
  \bibinfo{author}{B.~{Plez}}, et~al.,
\newblock \bibinfo{title}{{A holistic approach to carbon-enhanced metal-poor
  stars}},
\newblock \bibinfo{journal}{\aap} \bibinfo{volume}{509} (\bibinfo{year}{2010})
  \bibinfo{pages}{A93}. \DOIprefix\doi{10.1051/0004-6361/200911744}.
\bibitem[{{Starkenburg} et~al.(2014){Starkenburg}, {Shetrone}, {McConnachie},
  and {Venn}}]{Starkenburg2014MNRAS.441.1217S}
\bibinfo{author}{E.~{Starkenburg}}, \bibinfo{author}{M.~D. {Shetrone}},
  \bibinfo{author}{A.~W. {McConnachie}}, \bibinfo{author}{K.~A. {Venn}},
\newblock \bibinfo{title}{{Binarity in carbon-enhanced metal-poor stars}},
\newblock \bibinfo{journal}{\mnras} \bibinfo{volume}{441}
  (\bibinfo{year}{2014}) \bibinfo{pages}{1217--1229}.
  \DOIprefix\doi{10.1093/mnras/stu623}.
\bibitem[{{Hansen} et~al.(2016){Hansen}, {Andersen}, {Nordstr{\"o}m}, {Beers},
  {Placco}, {Yoon}, and {Buchhave}}]{Hansen2016A&A...588A...3H}
\bibinfo{author}{T.~T. {Hansen}}, \bibinfo{author}{J.~{Andersen}},
  \bibinfo{author}{B.~{Nordstr{\"o}m}}, et~al.,
\newblock \bibinfo{title}{{The role of binaries in the enrichment of the early
  Galactic halo. III. Carbon-enhanced metal-poor stars - CEMP-s stars}},
\newblock \bibinfo{journal}{\aap} \bibinfo{volume}{588} (\bibinfo{year}{2016})
  \bibinfo{pages}{A3}. \DOIprefix\doi{10.1051/0004-6361/201527409}.
\bibitem[{{Abate} et~al.(2015){Abate}, {Pols}, {Stancliffe}, {Izzard},
  {Karakas}, {Beers}, and {Lee}}]{Abate2015A&A...581A..62A}
\bibinfo{author}{C.~{Abate}}, \bibinfo{author}{O.~R. {Pols}},
  \bibinfo{author}{R.~J. {Stancliffe}}, et~al.,
\newblock \bibinfo{title}{{Modelling the observed properties of carbon-enhanced
  metal-poor stars using binary population synthesis}},
\newblock \bibinfo{journal}{\aap} \bibinfo{volume}{581} (\bibinfo{year}{2015})
  \bibinfo{pages}{A62}. \DOIprefix\doi{10.1051/0004-6361/201526200}.
\bibitem[{{Abate} et~al.(2016){Abate}, {Stancliffe}, and
  {Liu}}]{Abate2016A&A...587A..50A}
\bibinfo{author}{C.~{Abate}}, \bibinfo{author}{R.~J. {Stancliffe}},
  \bibinfo{author}{Z.-W. {Liu}},
\newblock \bibinfo{title}{{How plausible are the proposed formation scenarios
  of CEMP-r/s stars?}},
\newblock \bibinfo{journal}{\aap} \bibinfo{volume}{587} (\bibinfo{year}{2016})
  \bibinfo{pages}{A50}. \DOIprefix\doi{10.1051/0004-6361/201527864}.
\bibitem[{{Hampel} et~al.(2016){Hampel}, {Stancliffe}, {Lugaro}, and
  {Meyer}}]{Hampel2016ApJ...831..171H}
\bibinfo{author}{M.~{Hampel}}, \bibinfo{author}{R.~J. {Stancliffe}},
  \bibinfo{author}{M.~{Lugaro}}, \bibinfo{author}{B.~S. {Meyer}},
\newblock \bibinfo{title}{{The Intermediate Neutron-capture Process and
  Carbon-enhanced Metal-poor Stars}},
\newblock \bibinfo{journal}{\apj} \bibinfo{volume}{831} (\bibinfo{year}{2016})
  \bibinfo{pages}{171}. \DOIprefix\doi{10.3847/0004-637X/831/2/171}.
\bibitem[{{Lucatello} et~al.(2005){Lucatello}, {Tsangarides}, {Beers},
  {Carretta}, {Gratton}, and {Ryan}}]{Lucatello2005ApJ...625..825L}
\bibinfo{author}{S.~{Lucatello}}, \bibinfo{author}{S.~{Tsangarides}},
  \bibinfo{author}{T.~C. {Beers}}, et~al.,
\newblock \bibinfo{title}{{The Binary Frequency Among Carbon-enhanced,
  s-Process-rich, Metal-poor Stars}},
\newblock \bibinfo{journal}{\apj} \bibinfo{volume}{625} (\bibinfo{year}{2005})
  \bibinfo{pages}{825--832}. \DOIprefix\doi{10.1086/428104}.
\bibitem[{{Bisterzo} et~al.(2011){Bisterzo}, {Gallino}, {Straniero},
  {Cristallo}, and {K{\"a}ppeler}}]{Bisterzo2011MNRAS.418..284B}
\bibinfo{author}{S.~{Bisterzo}}, \bibinfo{author}{R.~{Gallino}},
  \bibinfo{author}{O.~{Straniero}}, et~al.,
\newblock \bibinfo{title}{{The s-process in low-metallicity stars - II.
  Interpretation of high-resolution spectroscopic observations with asymptotic
  giant branch models}},
\newblock \bibinfo{journal}{\mnras} \bibinfo{volume}{418}
  (\bibinfo{year}{2011}) \bibinfo{pages}{284--319}.
  \DOIprefix\doi{10.1111/j.1365-2966.2011.19484.x}.
\bibitem[{{Dardelet} et~al.(2014){Dardelet}, {Ritter}, {Prado}, {Heringer},
  {Higgs}, {Sandalski}, {Jones}, {Denisenkov}, {Venn}, {Bertolli}, {Pignatari},
  {Woodward}, and {Herwig}}]{Dardelet2014nic..confE.145D}
\bibinfo{author}{L.~{Dardelet}}, \bibinfo{author}{C.~{Ritter}},
  \bibinfo{author}{P.~{Prado}}, et~al.,
\newblock \bibinfo{title}{{i process and CEMP-s+r stars}},
\newblock in: \bibinfo{booktitle}{XIII Nuclei in the Cosmos (NIC XIII)},
  \bibinfo{year}{2014}, p. \bibinfo{pages}{145}.
  \DOIprefix\doi{10.22323/1.204.0145}.
\bibitem[{{Karinkuzhi} et~al.(2021){Karinkuzhi}, {Van Eck}, {Goriely}, {Siess},
  {Jorissen}, {Merle}, {Escorza}, and
  {Masseron}}]{Karinkuzhi2021A&A...645A..61K}
\bibinfo{author}{D.~{Karinkuzhi}}, \bibinfo{author}{S.~{Van Eck}},
  \bibinfo{author}{S.~{Goriely}}, et~al.,
\newblock \bibinfo{title}{{Low-mass low-metallicity AGB stars as an efficient
  i-process site explaining CEMP-rs stars}},
\newblock \bibinfo{journal}{\aap} \bibinfo{volume}{645} (\bibinfo{year}{2021})
  \bibinfo{pages}{A61}. \DOIprefix\doi{10.1051/0004-6361/202038891}.
\bibitem[{{Frebel} et~al.(2006){Frebel}, {Christlieb}, {Norris}, {Beers},
  {Bessell}, {Rhee}, {Fechner}, {Marsteller}, {Rossi}, {Thom}, {Wisotzki}, and
  {Reimers}}]{Frebel2006ApJ...652.1585F}
\bibinfo{author}{A.~{Frebel}}, \bibinfo{author}{N.~{Christlieb}},
  \bibinfo{author}{J.~E. {Norris}}, et~al.,
\newblock \bibinfo{title}{{Bright Metal-poor Stars from the Hamburg/ESO Survey.
  I. Selection and Follow-up Observations from 329 Fields}},
\newblock \bibinfo{journal}{\apj} \bibinfo{volume}{652} (\bibinfo{year}{2006})
  \bibinfo{pages}{1585--1603}. \DOIprefix\doi{10.1086/508506}.
\bibitem[{{Marsteller} et~al.(2005){Marsteller}, {Beers}, {Rossi},
  {Christlieb}, {Bessell}, and {Rhee}}]{Marsteller2005NuPhA.758..312M}
\bibinfo{author}{B.~{Marsteller}}, \bibinfo{author}{T.~C. {Beers}},
  \bibinfo{author}{S.~{Rossi}}, et~al.,
\newblock \bibinfo{title}{{Carbon-Enhanced Metal-Poor Stars in the Early
  Galaxy}},
\newblock \bibinfo{journal}{\nphysa} \bibinfo{volume}{758}
  (\bibinfo{year}{2005}) \bibinfo{pages}{312--315}.
  \DOIprefix\doi{10.1016/j.nuclphysa.2005.05.056}.
\bibitem[{{Abate} et~al.(2015){Abate}, {Pols}, {Izzard}, and
  {Karakas}}]{Abate2015A&A...581A..22A}
\bibinfo{author}{C.~{Abate}}, \bibinfo{author}{O.~R. {Pols}},
  \bibinfo{author}{R.~G. {Izzard}}, \bibinfo{author}{A.~I. {Karakas}},
\newblock \bibinfo{title}{{Carbon-enhanced metal-poor stars: a window on AGB
  nucleosynthesis and binary evolution. II. Statistical analysis of a sample of
  67 CEMP-s stars}},
\newblock \bibinfo{journal}{\aap} \bibinfo{volume}{581} (\bibinfo{year}{2015})
  \bibinfo{pages}{A22}. \DOIprefix\doi{10.1051/0004-6361/201525876}.
\bibitem[{{Karakas} and {Lattanzio}(2014)}]{Karakas2014PASA...31...30K}
\bibinfo{author}{A.~I. {Karakas}}, \bibinfo{author}{J.~C. {Lattanzio}},
\newblock \bibinfo{title}{{The Dawes Review 2: Nucleosynthesis and Stellar
  Yields of Low- and Intermediate-Mass Single Stars}},
\newblock \bibinfo{journal}{\pasa} \bibinfo{volume}{31} (\bibinfo{year}{2014})
  \bibinfo{pages}{e030}. \DOIprefix\doi{10.1017/pasa.2014.21}.
\bibitem[{{Arcones} and {Thielemann}(2023)}]{Arcones2023A&ARv..31....1A}
\bibinfo{author}{A.~{Arcones}}, \bibinfo{author}{F.-K. {Thielemann}},
\newblock \bibinfo{title}{{Origin of the elements}},
\newblock \bibinfo{journal}{\aapr} \bibinfo{volume}{31} (\bibinfo{year}{2023})
  \bibinfo{pages}{1}. \DOIprefix\doi{10.1007/s00159-022-00146-x}.
\bibitem[{{Truran}(1981)}]{Truran1981A&A....97..391T}
\bibinfo{author}{J.~W. {Truran}},
\newblock \bibinfo{title}{{A new interpretation of the heavy element abundances
  in metal-deficient stars.}},
\newblock \bibinfo{journal}{\aap}  \bibinfo{volume}{97} (\bibinfo{year}{1981})
  \bibinfo{pages}{391--393}.
\bibitem[{{Gallino} et~al.(1998){Gallino}, {Arlandini}, {Busso}, {Lugaro},
  {Travaglio}, {Straniero}, {Chieffi}, and
  {Limongi}}]{Gallino1998ApJ...497..388G}
\bibinfo{author}{R.~{Gallino}}, \bibinfo{author}{C.~{Arlandini}},
  \bibinfo{author}{M.~{Busso}}, et~al.,
\newblock \bibinfo{title}{{Evolution and Nucleosynthesis in Low-Mass Asymptotic
  Giant Branch Stars. II. Neutron Capture and the S-Process}},
\newblock \bibinfo{journal}{\apj} \bibinfo{volume}{497} (\bibinfo{year}{1998})
  \bibinfo{pages}{388--403}. \DOIprefix\doi{10.1086/305437}.
\bibitem[{{Busso} et~al.(1999){Busso}, {Gallino}, and
  {Wasserburg}}]{Busso1999ARA&A..37..239B}
\bibinfo{author}{M.~{Busso}}, \bibinfo{author}{R.~{Gallino}},
  \bibinfo{author}{G.~J. {Wasserburg}},
\newblock \bibinfo{title}{{Nucleosynthesis in Asymptotic Giant Branch Stars:
  Relevance for Galactic Enrichment and Solar System Formation}},
\newblock \bibinfo{journal}{\araa} \bibinfo{volume}{37} (\bibinfo{year}{1999})
  \bibinfo{pages}{239--309}. \DOIprefix\doi{10.1146/annurev.astro.37.1.239}.
\bibitem[{{Frischknecht} et~al.(2012){Frischknecht}, {Hirschi}, and
  {Thielemann}}]{Frischknecht2012A&A...538L...2F}
\bibinfo{author}{U.~{Frischknecht}}, \bibinfo{author}{R.~{Hirschi}},
  \bibinfo{author}{F.~K. {Thielemann}},
\newblock \bibinfo{title}{{Non-standard s-process in low metallicity massive
  rotating stars}},
\newblock \bibinfo{journal}{\aap} \bibinfo{volume}{538} (\bibinfo{year}{2012})
  \bibinfo{pages}{L2}. \DOIprefix\doi{10.1051/0004-6361/201117794}.
\bibitem[{{Thielemann} et~al.(2017){Thielemann}, {Eichler}, {Panov}, and
  {Wehmeyer}}]{Thielemann2017ARNPS..67..253T}
\bibinfo{author}{F.~K. {Thielemann}}, \bibinfo{author}{M.~{Eichler}},
  \bibinfo{author}{I.~V. {Panov}}, \bibinfo{author}{B.~{Wehmeyer}},
\newblock \bibinfo{title}{{Neutron Star Mergers and Nucleosynthesis of Heavy
  Elements}},
\newblock \bibinfo{journal}{Annual Review of Nuclear and Particle Science}
  \bibinfo{volume}{67} (\bibinfo{year}{2017}) \bibinfo{pages}{253--274}.
  \DOIprefix\doi{10.1146/annurev-nucl-101916-123246}.
\bibitem[{{Cowan} and {Rose}(1977)}]{Cowan1977ApJ...212..149C}
\bibinfo{author}{J.~J. {Cowan}}, \bibinfo{author}{W.~K. {Rose}},
\newblock \bibinfo{title}{{Production of $^{14}$C and neutrons in red
  giants.}},
\newblock \bibinfo{journal}{\apj} \bibinfo{volume}{212} (\bibinfo{year}{1977})
  \bibinfo{pages}{149--158}. \DOIprefix\doi{10.1086/155030}.
\bibitem[{{Choplin} et~al.(2021){Choplin}, {Siess}, and
  {Goriely}}]{Choplin2021A&A...648A.119C}
\bibinfo{author}{A.~{Choplin}}, \bibinfo{author}{L.~{Siess}},
  \bibinfo{author}{S.~{Goriely}},
\newblock \bibinfo{title}{{The intermediate neutron capture process. I.
  Development of the i-process in low-metallicity low-mass AGB stars}},
\newblock \bibinfo{journal}{\aap} \bibinfo{volume}{648} (\bibinfo{year}{2021})
  \bibinfo{pages}{A119}. \DOIprefix\doi{10.1051/0004-6361/202040170}.
\bibitem[{{Choplin} et~al.(2022){Choplin}, {Siess}, and
  {Goriely}}]{Choplin2022A&A...667A.155C}
\bibinfo{author}{A.~{Choplin}}, \bibinfo{author}{L.~{Siess}},
  \bibinfo{author}{S.~{Goriely}},
\newblock \bibinfo{title}{{The intermediate neutron capture process. III. The
  i-process in AGB stars of different masses and metallicities without
  overshoot}},
\newblock \bibinfo{journal}{\aap} \bibinfo{volume}{667} (\bibinfo{year}{2022})
  \bibinfo{pages}{A155}. \DOIprefix\doi{10.1051/0004-6361/202244360}.
\bibitem[{{Hampel} et~al.(2019){Hampel}, {Karakas}, {Stancliffe}, {Meyer}, and
  {Lugaro}}]{Hampel2019ApJ...887...11H}
\bibinfo{author}{M.~{Hampel}}, \bibinfo{author}{A.~I. {Karakas}},
  \bibinfo{author}{R.~J. {Stancliffe}}, et~al.,
\newblock \bibinfo{title}{{Learning about the Intermediate Neutron-capture
  Process from Lead Abundances}},
\newblock \bibinfo{journal}{\apj} \bibinfo{volume}{887} (\bibinfo{year}{2019})
  \bibinfo{pages}{11}. \DOIprefix\doi{10.3847/1538-4357/ab4fe8}.
\bibitem[{{Hansen} et~al.(2016){Hansen}, {Andersen}, {Nordstr{\"o}m}, {Beers},
  {Placco}, {Yoon}, and {Buchhave}}]{Hansen2016A&A...586A.160H}
\bibinfo{author}{T.~T. {Hansen}}, \bibinfo{author}{J.~{Andersen}},
  \bibinfo{author}{B.~{Nordstr{\"o}m}}, et~al.,
\newblock \bibinfo{title}{{The role of binaries in the enrichment of the early
  Galactic halo. II. Carbon-enhanced metal-poor stars: CEMP-no stars}},
\newblock \bibinfo{journal}{\aap} \bibinfo{volume}{586} (\bibinfo{year}{2016})
  \bibinfo{pages}{A160}. \DOIprefix\doi{10.1051/0004-6361/201527235}.
\bibitem[{{Gardner} et~al.(2006){Gardner}, {Mather}, {Clampin}, {Doyon}, and
  et~al.}]{Gardner2006SSRv..123..485G}
\bibinfo{author}{J.~P. {Gardner}}, \bibinfo{author}{J.~C. {Mather}},
  \bibinfo{author}{M.~{Clampin}}, et~al.,
\newblock \bibinfo{title}{{The James Webb Space Telescope}},
\newblock \bibinfo{journal}{\ssr} \bibinfo{volume}{123} (\bibinfo{year}{2006})
  \bibinfo{pages}{485--606}. \DOIprefix\doi{10.1007/s11214-006-8315-7}.
\bibitem[{{Ivezi{\'c}} et~al.(2019){Ivezi{\'c}}, {Kahn}, {Tyson}, {Abel},
  {Acosta}, {Allsman}, {Alonso}, {AlSayyad}, and
  et~al.}]{Ivezic2019ApJ...873..111I}
\bibinfo{author}{{\v{Z}}.~{Ivezi{\'c}}}, \bibinfo{author}{S.~M. {Kahn}},
  \bibinfo{author}{J.~A. {Tyson}}, et~al.,
\newblock \bibinfo{title}{{LSST: From Science Drivers to Reference Design and
  Anticipated Data Products}},
\newblock \bibinfo{journal}{\apj} \bibinfo{volume}{873} (\bibinfo{year}{2019})
  \bibinfo{pages}{111}. \DOIprefix\doi{10.3847/1538-4357/ab042c}.
\bibitem[{{Zhan}(2011)}]{Zhan2011SSPMA..41.1441Z}
\bibinfo{author}{H.~{Zhan}},
\newblock \bibinfo{title}{{Consideration for a large-scale multi-color imaging
  and slitless spectroscopy survey on the Chinese space station and its
  application in dark energy research}},
\newblock \bibinfo{journal}{Scientia Sinica Physica, Mechanica \& Astronomica}
  \bibinfo{volume}{41} (\bibinfo{year}{2011}) \bibinfo{pages}{1441}.
  \DOIprefix\doi{10.1360/132011-961}.
\bibitem[{{Laureijs} et~al.(2011){Laureijs}, {Amiaux}, {Arduini},
  {Augu{\`e}res}, {Brinchmann}, {Cole}, {Cropper}, {Dabin}, {Duvet}, and
  et~al.}]{Laureijs2011arXiv1110.3193L}
\bibinfo{author}{R.~{Laureijs}}, \bibinfo{author}{J.~{Amiaux}},
  \bibinfo{author}{S.~{Arduini}}, et~al.,
\newblock \bibinfo{title}{{Euclid Definition Study Report}},
\newblock \bibinfo{journal}{arXiv e-prints}  (\bibinfo{year}{2011})
  \bibinfo{pages}{arXiv:1110.3193}. \DOIprefix\doi{10.48550/arXiv.1110.3193}.

\end{thebibliography}





\newpage


\end{document}